\documentclass{amsbook}
\usepackage{amsfonts}
\usepackage{amsmath}
\usepackage[french]{babel}
\usepackage{amscd}
\usepackage{amsthm}
\theoremstyle{plain}
\newtheorem{proposition}{Proposition}[section]

\begin{document}

\title{Recherches autour \\
de \\
la th\'{e}orie de Markoff }
\author{Serge Perrine}
\date{6 avril 2003 (version 6)}
\begin{abstract}Le texte concerne des g\'{e}n\'{e}ralisations de
l'\'{e}quation de Markoff en th\'{e}orie des nombres, d\'{e}duites
des fractions continues. Il d\'{e}crit la m\'{e}thode pour une
r\'{e}solution compl\`{e}te de ces nouvelles \'{e}quations, ainsi
que leur interpr\'{e}tation en alg\'{e}bre et en g\'{e}om\'{e}trie
alg\'{e}brique. Cette approche alg\'{e}brique est
compl\'{e}t\'{e}e par un d\'{e}veloppement analytique concernant
les groupes fuchsiens. Le lien avec la th\'{e}orie de
Teichm\"{u}ller des tores perc\'{e}s est compl\`{e}tement
d\'{e}crit, les classifiant au moyen d'une th\'{e}orie de la
r\'{e}duction. Des consid\'{e}rations plus g\'{e}n\'{e}rales au
sujet des surfaces de Riemann, les g\'{e}od\'{e}siques et leur
\'{e}tude hamiltonienne sont cit\'{e}es, de m\^{e}me que des
applications \`{a} la physique, au bruit en $1/f$ et \`{a} la
fonction z\'{e}ta. Des id\'{e}es relatives \`{a} d'importantes
conjectures sont pr\'{e}sent\'{e}es. On donne aussi des raisons
pour lesquelles la th\'{e}orie de Markoff appara\^{i}t dans
diff\'{e}rents contextes g\'{e}om\'{e}triques, gr\^{a}ce \`{a} des
r\'{e}sultats de d\'{e}composition valables dans le groupe
$GL(2,\mathbb{Z})$.
\\
\\
\\
\\
\textsc{Abstract.} The text deals with generalizations of the
Markoff equation in number theory, arising from continued
fractions. It gives the method for the complete resolution of such
new equations, and their interpretation in algebra and algebraic
geometry. This algebraic approach is completed with an analytical
development concerning fuchsian groups. The link with the
Teichm\"{u}ller theory for punctured toruses is completely
described, giving their classification with a reduction theory.
More general considerations about Riemann surfaces, geodesics and
their hamiltonian study are quoted, together with applications in
physics, $1/f$-noise and zeta function. Ideas about important
conjectures are presented. Reasons why the Markoff theory appears
in different geometrical contexts are given, thanks to
decomposition results in the group $GL(2,\mathbb{Z})$.
\end{abstract}
\maketitle

\[
\]
\[
\]
\[
\]
\[
\]
\[
\]
\[
\]
\[
\]
\[
\]
\[
\]
\[
\]
\[
\]
\[
\]
\[
\]
\[
\]
\[
\]
\[
\]
\[
\]
\[
\]
\[
\]
\[
\]
\[
\]
\[
\]
\[
\]
\[
\]
\[
\]
\[
\]
\[
\]
\[
\]

\[
\text{''Tout voir, tout entendre, ne perdre aucune id\'{e}e'' \ }
\]
\[
\text{\bfseries{Evariste Galois}}
\]
\[
\]
\[
\]
\[
\text{''Saisir les propri\'{e}t\'{e}s des choses, } \\
\text{d'apr\`{e}s leur mode d'existence dans l'infiniment petit''}
\]
\[
\text{\bfseries{Discours de F\'{e}lix Klein sur Bernhard Riemann
et son influence}}
\]
\[
\]
\[
\]
\[
\text{''Sans l'esp\'{e}rance, on ne trouvera pas l'inesp\'{e}r\'{e}, } \\
\text{qui est introuvable et inaccessible''}
\]
\[
\text{\bfseries{H\'{e}raclite}}
\]
\[
\
\begin{array}{cc}
&  \\
&  \\
&  \\
&  \\
&  \\
&  \\
&  \\
&  \\
&  \\
&  \\
&  \\
&  \\
&
\end{array}
\]
\section{Remerciements}

Mes remerciements s'adressent \`{a} diff\'{e}rentes personnes sans
lesquelles ce texte n'aurait jamais vu le jour, et \`{a} tous ceux
qui m'ont aid\'{e} pour sa mise en forme. Je pense en particulier
aux personnes suivantes :

- Georges Rhin qui tout au long de ces derni\`{e}res ann\'{e}es a
pr\^{e}t\'{e} attention aux diff\'{e}rents documents que je lui adressais
p\'{e}riodiquement.

- Michel Planat avec qui une coop\'{e}ration r\'{e}guli\`{e}re et
des discussions passionnantes autour d'observations physiques
qu'il avait faites ont beaucoup soutenu ma curiosit\'{e} pour la
th\'{e}orie de Markoff. Mon int\'{e}r\^{e}t pour ce sujet venait
de consid\'{e}rations sur le codage de l'information.\ Mais voir
appara\^{i}tre le spectre de Markoff dans les caract\'{e}ristiques
physiques d'un oscillateur \`{a} v\'{e}rouillage de phase a
consid\'{e}rablement relanc\'{e} mes travaux. En observant le
comportement d'oscillateurs construits sur mesure, pourrions-nous
comprendre certaines parties de cette th\'{e}orie restant encore
\'{e}nigmatiques, pourrions-nous inversement construire certains
mod\`{e}les de bruit utiles \`{a} la physique? Ces questions ont
orient\'{e} mes travaux.

- Michel Mend\`{e}s France et Michel Waldschmidt qui se sont \`{a}
diff\'{e}rentes reprises int\'{e}ress\'{e}s \`{a} mes travaux, et m'ont
fourni l'occasion de les perfectionner et de les exposer. Je les remercie
tr\`{e}s chaleureusement de leurs encouragements et de leurs commentaires
sans concession que j'ai toujours consid\'{e}r\'{e}s comme une source de
progr\`{e}s.

Je voudrais aussi remercier C\'{e}cile et les enfants pour leur grande
patience \`{a} supporter le temps consid\'{e}rable que j'ai pass\'{e} sur ce
travail.
\[
\begin{array}{cc}
&  \\
&  \\
&
\end{array}
\]
\[
\begin{array}{cc}
&  \\
&  \\
&  \\
&  \\
&  \\
&  \\
&  \\
&  \\
&  \\
&  \\
&  \\
&  \\
&  \\
&  \\
&  \\
&  \\
&  \\
&  \\
&  \\
&  \\
&
\end{array}
\]

\section{Pr\'{e}sentation g\'{e}n\'{e}rale}

Le but du pr\'{e}sent travail est d'exposer une d\'{e}marche de recherche
conduite autour de la th\'{e}orie de Markoff, ainsi que les r\'{e}sultats
qu'elle a fournis. Cette th\'{e}orie est une branche de ce que Hermann
Minkowski a appel\'{e} la ''g\'{e}om\'{e}trie des nombres'' \cite{Minkowski}%
\cite{Cassels2}.\ Elle fournit une r\'{e}ponse partielle au probl\`{e}me
suivant :

Une forme quadratique r\'{e}elle \'{e}tant donn\'{e}e $f(x,y)=ax^2+bxy+cy^2\in \mathbb{R}[x,y]$%
, quelle est la valeur minimale du nombre $\mid f(x,y)\mid $ lorsque $x$ et $%
y$ sont des entiers non tous deux simultan\'{e}ment nuls ?

Pour une forme d\'{e}finie $f(x,y)$, c'est-\`{a}-dire telle que $%
\Delta (f)=b^2-4ac<0$, ce probl\`{e}me a \'{e}t\'{e} r\'{e}solu par Joseph
Louis Lagrange. Sa solution se d\'{e}duit aussi d'un r\'{e}sultat plus
g\'{e}n\'{e}ral de Charles Hermite \cite{Hermite} donnant :
\[
C(f)=\frac{\inf_{(x,y)\in \mathbb{Z}^2-\{(0,0)\}}\mid f(x,y)\mid
}{\sqrt{\mid \Delta (f)\mid }}\leq \frac
1{\sqrt{3}}=C(x^2+xy+y^2).
\]
Il a aussi \'{e}t\'{e} d\'{e}montr\'{e} (\cite{Cassels2} p.33) que
pour tout nombre $\rho \in ]0,(1/\sqrt{3})]$, on peut trouver une
forme quadratique d\'{e}finie $f(x,y)\in \mathbb{R}[x,y]$ telle
que :
\[
\rho =C(f).
\]
Si la forme $f(x,y)$ est ind\'{e}finie, c'est-\`{a}-dire telle que $%
\Delta (f)=b^2-4ac>0$, on sait depuis \cite{Korkine} que l'on a :
\[
C(f)\leq \frac 1{\sqrt{5}}=C(x^2-xy-y^2).
\]
Pour les autres valeurs, on a \cite{Korkine} :
\[
C(f)\leq \frac 1{\sqrt{8}}=C(x^2-2y^2).
\]
C'est pour mieux comprendre le cas ind\'{e}fini qu'Andrei A. Markoff a
d\'{e}velopp\'{e} sa th\'{e}orie \cite{Markoff}.\ Celle-ci identifie
l'infinit\'{e} des valeurs $C(f)$ comprises entre $(1/\sqrt{5})$ et $(1/3)$
et les trous sans constante qui les s\'{e}parent. Ces valeurs sont
isol\'{e}es et convergent vers $(1/3)$. Pour les valeurs inf\'{e}rieures
\`{a} $(1/3)$, il n'existait jusqu'\`{a} une date r\'{e}cente aucune
approche comparable \`{a} la th\'{e}orie de Markoff. Des r\'{e}sultats
lacunaires existent sur des trous sans constante, mais la situation reste
globalement m\'{e}connue aujourd'hui encore.\ Une synth\`{e}se de ce qui
\'{e}tait connu en 1988 a \'{e}t\'{e} r\'{e}alis\'{e}e par Thomas W. Cusick
et Mary E.\ Flahive \cite{Cusick}, au moment o\`{u} l'auteur soutenait sa
th\`{e}se sur le m\^{e}me sujet. La recherche men\'{e}e depuis cette
p\'{e}riode s'est appuy\'{e}e sur les deux derni\`{e}res contributions
cit\'{e}es.\ Il s'agissait d'aller au del\`{a} des r\'{e}sultats connus sur
le sujet. On a trouv\'{e} quelques r\'{e}sultats relatifs \`{a} de nouveaux
trous du spectre, mais assez rapidement l'id\'{e}e a germ\'{e} de chercher
\`{a} disposer d'une g\'{e}n\'{e}ralisation de la th\'{e}orie de Markoff
pour essayer d'en d\'{e}duire des r\'{e}sultats analogues \`{a} ceux
disponibles au dessus de $(1/3)$.

Parall\`{e}lement la mise en \'{e}vidence en physique, autour d'oscillateurs
sp\'{e}ciaux, de valeurs physiques \'{e}gales aux constantes $C(f)$
donn\'{e}es par la th\'{e}orie de Markoff a \'{e}t\'{e} particuli\`{e}rement
motivante. Cet accomplissement du \`{a} Michel Planat \cite{Planat1} a
conduit \`{a} envisager la construction d'oscillateurs particuliers
permettant de ''voir'' la structure du spectre de Markoff en des endroits
o\`{u} sa structure est suffisamment chaotique pour rester \`{a} ce jour
m\'{e}connue. L'exploration de ce sujet, et son lien possible avec une
mod\'{e}lisation du bruit en $(1/f)$ qui reste \`{a} ce jour assez
\'{e}nigmatique, est devenu progressivement un projet important. Construire
dans ce contexte de nouvelles th\'{e}ories analogues \`{a} celle de Markoff
est apparu utile

On a donc mis au point des notations destin\'{e}es \`{a} permettre
l'appr\'{e}hension de nouvelles th\'{e}ories plus g\'{e}n\'{e}rales que la
th\'{e}orie originale de Markoff.\ Cet objectif, entrevu a l'issue du
travail de th\`{e}se de l'auteur, n'avait pas d\'{e}bouch\'{e} \`{a} ce
moment sur des exemples significatifs et complets. La d\'{e}marche a
consist\'{e} \`{a} comprendre comment construire de fa\c {c}on directe sur
des suites de nombres entiers positifs un processus de cr\'{e}ation
arborescente qui fournisse toujours des suites attach\'{e}es \`{a} une
m\^{e}me \'{e}quation diophantienne du type de celle de Markoff. A cet
\'{e}gard, l'article \cite{Perrine} s'est av\'{e}r\'{e} d\'{e}terminant. Il
a permis de disposer de ce mode de construction pour certaines suites assez
g\'{e}n\'{e}rales, en faisant en sorte qu'elles restent attach\'{e}es \`{a}
l'\'{e}quation
\[
x^2+y^2+z^2=4xyz-x.
\]
On a ainsi pu disposer d'une th\'{e}orie compl\`{e}te permettant d'obtenir des constantes d'approximations convergentes vers la valeur $%
(1/4) $ ainsi que quelques trous du spectre.

Il est ensuite apparu que le mode de construction d\'{e}couvert laissait
invariantes des \'{e}quations de forme plus g\'{e}n\'{e}rale. A cette
occasion le lien naturel qui existe avec les sommes de Dedekind \cite
{Rademacher} a \'{e}t\'{e} mis en \'{e}vidence. Ceci a permis d'identifier
d'autres \'{e}quations permettant de construire des constantes
d'approximations qui convergent vers $(1/3)$ comme dans la th\'{e}orie de
Markoff classique, mais cette fois par valeurs inf\'{e}rieures. On a ainsi
pu obtenir des informations sur une partie totalement m\'{e}connue du
spectre. Un exemple complet a \'{e}t\'{e} d\'{e}taill\'{e} \cite{Perrine4}
concernant l'\'{e}quation
\[
x^2+y^2+z^2=3xyz+2x.
\]
Pour cette derni\`{e}re, on a fourni toutes les solutions enti\`{e}res dans $%
\mathbb{N}$ ou $\mathbb{Z}$. Il est remarquable qu'\`{a} la
diff\'{e}rence de la th\'{e}orie de Markoff classique, les
solutions enti\`{e}res positives se r\'{e}partissent en deux
classes, et non pas en une seule.\ On a montr\'{e} cependant
comment ces deux classes donnent naissance \`{a} un arbre unique
de triplets de Cohn, pour lesquels la construction sur les suites
d'entiers s'applique compl\`{e}tement. Les triplets de Cohn sont
d\'{e}finis de fa\c {c}on g\'{e}n\'{e}rale par la condition
$x>y>z$. Les constantes donn\'{e}es par l'\'{e}quation
pr\'{e}c\'{e}dentes sont diff\'{e}rentes de celles mises en
\'{e}vidence dans la m\^{e}me zone du spectre de Markoff par David
J.\ Crisp et William Moran \cite{Crisp}. C'est ainsi que le
mod\`{e}le g\'{e}om\'{e}trique construit par Harvey Cohn \`{a}
partir du demi-plan de Poincar\'{e} $\mathcal{H}$, prolong\'{e}
par l'\'{e}tude des g\'{e}od\'{e}siques ferm\'{e}es du tore
perc\'{e} se coupant elles-m\^{e}mes \cite{Series}, est devenu
insuffisant pour d\'{e}crire la complexit\'{e} du spectre de
Markoff au voisinage de $(1/3)$. Le projet a donc \'{e}t\'{e} fait
de revisiter cette interpr\'{e}tation g\'{e}om\'{e}trique. Ceci a
\'{e}t\'{e} men\'{e} \`{a} bien et a permis de comprendre la
nature des \'{e}quations que l'on identifiait progressivement.

Avant cela, dans \cite{Perrine7} on a \'{e}tendu le mode de construction
arborescent de suites de nombres entiers positifs pour mettre en
\'{e}vidence d'autres \'{e}quations de forme l\'{e}g\`{e}rement plus
g\'{e}n\'{e}rale donnant des constantes d'approximation dans le voisinage de
$(1/3)$ :
\[
x^2+y^2+z^2=3xyz+sx,\;\;s>0.
\]
Sur de telles \'{e}quations, o\`{u} $s>0$, on a pu montrer dans \cite
{Perrine} l'existence d'un nombre fini de classes de solutions. Le m\^{e}me
r\'{e}sultat est valable aussi pour $s\leq 0$. Mais alors que dans un cas $%
(s>0)$ il convient d'introduire une notion de solution fondamentale pour
obtenir ce r\'{e}sultat, dans l'autre cas $(s\leq 0)$ c'est une notion
diff\'{e}rente de solution minimale qui permet de conclure. Au demeurant,
ces derni\`{e}res \'{e}quations apparaissent li\'{e}es entre elles compte
tenu de l'expression des minima arithm\'{e}tiques des formes quadratiques
binaires associ\'{e}es.

L'approche pr\'{e}c\'{e}dente qui donne des valeurs $C(f)$
inf\'{e}rieures s'accumulant sur $(1/3)$, c'est-\`{a}-dire \`{a}
nouveau dans la partie haute et m\'{e}connue du spectre, a aussi
\'{e}t\'{e} \'{e}tendue \`{a} d'autres situations. Ainsi un nouvel
exemple de th\'{e}orie de Markoff g\'{e}n\'{e}ralis\'{e}e a
\'{e}t\'{e} trait\'{e} avec l'\'{e}quation
\[
x^2+y^2+z^2=3xyz+yz-2x.
\]
Il a permis de donner une nouvelle interpr\'{e}tation \`{a} d'anciens
travaux de Collin J.\ Hightower \cite{Hightower}. Le point d'accumulation
correspondant est \'{e}gal \`{a} $1/(1+\sqrt{5})$. On a aussi compris
comment la connaissance d'une partie du spectre permettait d'obtenir des
informations sur une partie plus basse du spectre.\ Dans le dernier cas
cit\'{e}, c'est la valeur maximale du spectre $(1/\sqrt{5})$ qui est
d\'{e}terminante.

Au final on a consid\'{e}r\'{e} que la bonne g\'{e}n\'{e}ralisation de la
th\'{e}orie de Markoff \'{e}tait relative \`{a} des \'{e}quations
diophantiennes not\'{e}es $M^{s_1s_2}(a,\partial K,u_\theta )$, o\`{u} $s_1$
et $s_2$ signes respectifs de $\varepsilon _1$ et $\varepsilon _2\in
\{-1,+1\}$, $a\in \mathbb{N}\backslash \{0\}$, $\partial K\in \mathbb{Z}$, $%
u_\theta \in \mathbb{Z}$ :
\[
x^2+\varepsilon _2y^2+\varepsilon _1z^2=(a+1)xyz+(\varepsilon
_2\partial K)yz-u_\theta x,\;\;x,y,z\in \mathbb{N}\backslash
\{0\}.
\]
Une telle forme d'\'{e}quation recouvre celles \'{e}voqu\'{e}es ci-dessus.
Il a donc sembl\'{e} que ce type d'\'{e}quation \'{e}tait la bonne. Et en
r\'{e}alit\'{e} on a pu montrer comment elles apparaissaient naturellement
par un calcul relatif aux fractions continues. On a montr\'{e} \'{e}galement
qu'elles correspondent \`{a} une formule de trace ainsi qu'\`{a} une
propri\'{e}t\'{e} remarquable de la fonction $\eta $ de Dedekind \cite
{Perrine9}. Pour ces \'{e}quations on a pu mettre au point une m\'{e}thode
g\'{e}n\'{e}rale de r\'{e}solution qui s'apparente \`{a} la descente infinie
ch\`{e}re aux arithm\'{e}ticiens.\ Elle fait jouer un r\^{o}le essentiel au
groupe du triangle $\mathbf{T}_3$ qui classe les solutions. On a aussi
montr\'{e} comment le recours \`{a} des triplets de Cohn permettait dans
l'essentiel des cas de conclure \`{a} l'existence d'une classe contenant une
infinit\'{e} de solutions, ainsi qu'un nombre fini de telles classes. Ce
nombre de classes a d'ailleurs un lien avec le nombre de classes des corps
quadratiques, mais le travail reste \`{a} faire pour mettre cette
observation en \'{e}tat pr\'{e}sentable.

On a pu \'{e}tudier de fa\c {c}on directe les surfaces ayant pour
\'{e}quation la forme que l'on vient de donner.\ Ces surfaces cubiques sont
rationnelles, on en a donn\'{e} une repr\'{e}sentation rationnelle.\
Coup\'{e}es par un plan, elles donnent des courbes elliptiques dans de
nombreux cas.\ Toutes les courbes elliptiques \`{a} coefficients rationnels
sont obtenues ainsi. Ceci permet d'avoir une id\'{e}e quant \`{a} des
ph\'{e}nom\`{e}nes pouvant affecter des courbes elliptiques diff\'{e}rentes
port\'{e}es par une m\^{e}me surface cubique. Un sujet arithm\'{e}tique
prometteur qui s'est ainsi d\'{e}gag\'{e} concerne le lien entre les
th\'{e}ories de Markoff g\'{e}n\'{e}ralis\'{e}es et la structure des points
entiers sur les courbes elliptiques \cite{Perrine8}. Les r\'{e}flexions dans
ce dernier domaine ne sont pas achev\'{e}es.On a \'{e}galement pu montrer
que tout r\'{e}seau complet d'un corps quadratique permet de construire une
\'{e}quation cubique du type pr\'{e}c\'{e}dent.\ Ce r\'{e}sultat important
donne un sens alg\'{e}brique aux \'{e}quations que l'on \'{e}tudie. Il
permet facilement de comprendre ce que l'on vient d'indiquer sur le nombre
de classes de solutions.

Toutes les constructions qui pr\'{e}c\`{e}dent ont aussi un
support analytique commun analogue \`{a} celui d\'{e}couvert par
Harvey Cohn pour la th\'{e}orie de Markoff classique \cite{Cohn}.
Pour mieux comprendre cette interpr\'{e}tation
g\'{e}om\'{e}trique, on a \'{e}tudi\'{e} de fa\c {c}on directe les
tores perc\'{e}s. Ceci a introduit une distinction entre les tores
perc\'{e}s conformes paraboliques et hyperboliques.\ Le cas
parabolique donne une g\'{e}n\'{e}ralisalisation tr\`{e}s
satisfaisante de la th\'{e}orie de Markoff, mettant en
\'{e}vidence des groupes fuchsiens dont on a \'{e}tabli qu'ils
sont libres \`{a} deux g\'{e}n\'{e}rateurs.\ Ce sont les groupes
de Fricke qui sont ainsi tous obtenus, mais ils correspondent
seulement \`{a} l'\'{e}quation de la th\'{e}orie de Markoff
classique qui les caract\'{e}rise tous. Pour le cas hyperbolique,
on a pu construire un exemple original illustrant le fait
d\'{e}couvert que les groupes fuchsiens correspondants ne sont pas
libres.\ Comme les surfaces intervenant dans ce contexte, des
tores perc\'{e}s, sont des quotients du demi plan de Poincar\'{e}
par un groupe fuchsien agissant sur lui, la th\'{e}orie de
Teichm\"{u}ller \cite{Schneps} constitue un cadre bien adapt\'{e}
pour appr\'{e}hender le sujet. On l'a donc approfondie jusqu'\`{a}
en donner une pr\'{e}sentation qui montre clairement comment elle
g\'{e}n\'{e}ralise la th\'{e}orie de Markoff.\ La th\'{e}orie de
Teichm\"{u}ller d\'{e}crit les propri\'{e}t\'{e}s des
diff\'{e}rentes structures conformes d\'{e}finissant une surface
de Riemann donn\'{e}e sur un m\^{e}me support topologique. Elle
d\'{e}termine par r\'{e}duction une structure cristalline pour
laquelle on a donn\'{e} quelques \'{e}l\'{e}ments d'information
dans l'ouvrage \cite {Perrine9}. On a pu comprendre pourquoi il
n'y a pas \`{a} consid\'{e}rer de tores perc\'{e}s elliptiques,
ainsi que la nature du lien entre nos \'{e}quations de Markoff
g\'{e}n\'{e}ralis\'{e}es et la th\'{e}orie de Teichm\"{u}ller.

On a aussi vu que tous les tores perc\'{e}s conformes paraboliques
d\'{e}finis sur un m\^{e}me tore topologique perc\'{e} peuvent \^{e}tre
distingu\'{e}s par deux nombres r\'{e}els positifs.\ Ce type de r\'{e}sultat
est connu depuis les travaux de R.\ Fricke \cite{Cohn}.\ Mais les
m\'{e}thodes issues de la th\'{e}orie de Markoff conduisent \`{a} se
restreindre \`{a} un premier nombre, un module compris entre 1 et 2. Le
module 1 correspond au tore perc\'{e} d'un groupe dit de Klein.\ Le module 2
correspond au tore perc\'{e} du groupe de Hecke \cite{Hecke}. On voit ainsi
apparaitre de fa\c {c}on naturelle les deux tores \'{e}tudi\'{e}s dans \cite
{Cohn}. Tous les modules interm\'{e}diaires correspondent \`{a} d'autres
tores perc\'{e}s conformes paraboliques isomorphes en tant qu'espaces
topologiques mais non en tant que surfaces de Riemann. Le fait que ces tores
ne soient pas conform\'{e}ment \'{e}quivalents a des cons\'{e}quences
g\'{e}om\'{e}triques int\'{e}ressantes pour les classement des groupes
fuchsiens associ\'{e}s. Ce r\'{e}sultat a \'{e}t\'{e} compl\'{e}t\'{e} en
montrant que tous les tores perc\'{e}s paraboliques sont class\'{e}s au
moyen de deux param\`{e}tres r\'{e}els, tous deux d\'{e}finis \`{a} partir
de la seule \'{e}quation de Markoff classique :
\[
x^2+y^2+z^2=xyz.
\]
Le module d\'{e}finit le domaine fondamental et un second param\`{e}tre
r\'{e}el dit accessoire d\'{e}crit la fa\c {c}on dont ses bords sont
identifi\'{e}s.

Les th\'{e}ories de Markoff des \'{e}quations
$M^{s_1s_2}(a,\partial K,u_\theta )$ conduisent tr\`{e}s
naturellement \`{a} d\'{e}finir des g\'{e}n\'{e}rateurs $A$ et $B$
de groupes fuchsiens \`{a} deux g\'{e}n\'{e}rateurs. Elles donnent
dans le cas parabolique les groupes de Fricke bien connus
\cite{Rosenberger} \cite{Matelski}. On l'a d\'{e}montr\'{e} de
fa\c {c}on rigoureuse. Les cas qui correspondent \`{a} des
matrices $A$ et $B$ \`{a} coefficients entiers ont \'{e}t\'{e}
compl\`{e}tement d\'{e}crits. Il en r\'{e}sulte la possibilit\'{e}
de caract\'{e}riser les tores perc\'{e}s paraboliques
correspondants. La th\'{e}orie de la r\'{e}duction valable pour
les nombres alg\'{e}briques de degr\'{e} 2 s'\'{e}tend alors aux
syst\`{e}mes g\'{e}n\'{e}rateurs de ces groupes de Fricke. Un
r\'{e}sultat qui en d\'{e}coule \cite{Horowitz} concerne la
d\'{e}termination des repr\'{e}sentations du groupe \`{a} deux
g\'{e}n\'{e}rateurs $\mathbf{F}_2$ dans les groupes
$GL(2,\mathbb{Z})$. En approfondissant cette question, on a mis en
\'{e}vidence le lien avec le th\'{e}or\`{e}me de Dyer et Formanek
\cite{Formanek}. Sa d\'{e}monstration classique repose sur des
propri\'{e}t\'{e}s des repr\'{e}sentations $\rho
:Aut(\mathbf{F}_2)\longrightarrow GL(m,\mathbb{Z})$. Les
th\'{e}ories de Markoff correspondantes donnent de telles
repr\'{e}sentations issues du
groupe \`{a} deux g\'{e}n\'{e}rateurs $\mathbf{F}_2$ dans le groupe $GL(2,%
\mathbb{Z})$. Caract\'{e}riser ces repr\'{e}sentations est
essentiel et on a pu comprendre comment ceci revenait \`{a}
consid\'{e}rer dans l'essentiel des cas des structures conformes
sur des tores perc\'{e}s. Le lien esquiss\'{e} \`{a} cette
occasion avec la th\'{e}orie de noeuds m\'{e}riterait d'\^{e}tre
creus\'{e} plus avant \cite{Brumfiel}, comme si au del\`{a} des
noeuds toriques on pouvait introduire une nouvelle
sous-cat\'{e}gorie de noeuds li\'{e}s aux tores perc\'{e}s. A
partir de ces r\'{e}flexions, on a surtout obtenu une meilleure
connaissance du groupe $GL(2,\mathbb{Z})$.\ Deux
d\'{e}compositions ternaires qui semblent nouvelles ont
\'{e}t\'{e} donn\'{e}es dans \cite{Perrine1b} pour toute matrice
de $GL(2,\mathbb{Z}).$ Ceci permet notamment de relier la
th\'{e}orie de Markoff classique \`{a} la structure du groupe du
triangle $\mathbf{T}_3$ et de repr\'{e}senter ce dernier dans
$GL(2,\mathbb{Z})$ \`{a} l'aide d'un groupe di\'{e}dral. Il est
probable que tous les groupes finis donnent des r\'{e}sultats
analogues et permettent de construire des structures
arborescentes, et on conjecture que tous peuvent \^{e}tre
repr\'{e}sent\'{e}s dans $GL(2,\mathbb{Z})$.\ L'auteur pense que
l'on obtient par un tel proc\'{e}d\'{e} toutes ses
g\'{e}n\'{e}ralisations de l'\'{e}quation de Markoff. Quelques
r\'{e}sultats ont \'{e}t\'{e} obtenus en ce sens mais il ne sont
pas encore pr\'{e}sentables.\ Une cons\'{e}quence importante qui
pourrait en d\'{e}couler est la conjecture que tout groupe fini
est obtenu comme groupe des classes d'un corps quadratique
r\'{e}el.

Mais y a-t-il un lien entre ces derni\`{e}res th\'{e}ories de Markoff et les
g\'{e}od\'{e}siques des tores perc\'{e}s conformes associ\'{e}s?\ En y
r\'{e}fl\'{e}chissant l'auteur a envisag\'{e} \`{a} partir de cette question
un domaine d'application pour ses g\'{e}n\'{e}ralisations de la th\'{e}orie
de Markoff au codage des g\'{e}od\'{e}siques des surfaces de Riemann \cite
{Schmutz2}. Il a approfondi la dualit\'{e} naturelle qui existe entre points
et g\'{e}od\'{e}siques sur une telle surface. Malheureusement cette
\'{e}tude apparemment nouvelle n'a pas suffisamment d\'{e}bouch\'{e} pour
donner lieu \`{a} publication. On a cependant donn\'{e} quelques
\'{e}l\'{e}ments au chapitre 7 de l'ouvrage \cite{Perrine9}. La question
particuli\`{e}re de la caract\'{e}risation des g\'{e}od\'{e}siques
ferm\'{e}es par des suites finies d'entiers qui les codent, puis
construisent des propri\'{e}t\'{e}s alg\'{e}briques diverses, est tr\`{e}s
int\'{e}ressante.\ Elle est aussi importante pour comprendre l'approche
ergodique \cite{Series} \cite{Series2} \cite{Schmutz2}. Les
g\'{e}od\'{e}siques d\'{e}pendent de la structure conforme adopt\'{e}e sur
le tore topologique perc\'{e} qui la porte. Les transformations conformes
qui changent une g\'{e}od\'{e}sique ferm\'{e}e en une autre d\'{e}finissent
des op\'{e}rations de transcodage sur les suites d'entiers associ\'{e}es.\
Il y a l\`{a} une perspective d'application dans le codage de l'information,
en particulier le codage en flot (stream cyphering) et les
g\'{e}n\'{e}rateurs pseudo-al\'{e}atoires.

Tout changement de g\'{e}od\'{e}sique se traduit par une d\'{e}formation de
la structure alg\'{e}brique de ces suites. Les r\'{e}flexions sur ce sujet
ont \'{e}t\'{e} nombreuses, mais restent assez lacunaires. On a donn\'{e} au
chapitre 7 de \cite{Perrine9} des pistes pour approfondir le probl\`{e}me.
On a en particulier rappel\'{e} comment se d\'{e}veloppe dans un tel
contexte l'approche hamiltonienne de la m\'{e}canique, en mettant l'accent
sur son caract\`{e}re quasi fonctoriel. Quelques cons\'{e}quences en
r\'{e}sultent pour la compr\'{e}hension m\^{e}me de ce que constituent le
calcul math\'{e}matique \cite{Feynman} et certains objets physiques. Un
point qui tourne librement sur une g\'{e}od\'{e}sique ferm\'{e}e peut
repr\'{e}senter un syst\`{e}me physique stable, donc observable. Les
changements de solutions dans nos \'{e}quations diophantiennes correspondent
alors \`{a} des sauts quantiques dans l'\'{e}volution d'un tel syst\`{e}me
selon des g\'{e}od\'{e}siques diff\'{e}rentes sur un tore perc\'{e}. Cette
id\'{e}e donne une structuration quantique au syst\`{e}me consid\'{e}r\'{e},
structure que l'on peut esp\'{e}rer retrouver dans des syst\`{e}mes
r\'{e}els. On a un ph\'{e}nom\`{e}ne comparable sur les courbes elliptiques
d'une m\^{e}me surface donn\'{e}e par nos \'{e}quations. De l\`{a} \`{a}
\'{e}tendre la probl\'{e}matique pour se poser des probl\`{e}mes de
m\'{e}canique statistique et de th\'{e}orie ergodique, il n'y a qu'un pas
que les travaux de dynamique symbolique de Caroline Series \cite{Series}
\cite{Series2} ont depuis longtemps franchi. Le lien est aussi \'{e}vident
avec le probl\`{e}me des ''petits diviseurs'', les r\'{e}sonances proches de
fr\'{e}quences dans un mouvement quasi p\'{e}riodique, et certains
mod\`{e}les de bruit en $1/f$ (voir \cite{Arnold} \cite{Yoccoz} \cite{Herman}
\cite{Dodson} \cite{Planat3}).

Qui dit g\'{e}od\'{e}sique \'{e}voque le calcul des vatiations
d'Euler-Lagrange, la propagation des ondes, mais aussi le th\'{e}or\`{e}me
KAM et les tores invariants. C'est ce dernier point qui est aussi \`{a} la
base de l'int\'{e}r\^{e}t de diff\'{e}rents physiciens pour la th\'{e}orie
de Markoff \cite{Gutzwiller}. Si un syst\`{e}me physique \'{e}volue
librement selon des trajectoires g\'{e}od\'{e}siques qui peuvent \^{e}tre
repr\'{e}sent\'{e}es sur un tore, par identification de deux mouvements
p\'{e}riodiques fondamentaux, et si un point de ce tore ne peut jamais
\^{e}tre atteint, une th\'{e}orie de Markoff g\'{e}n\'{e}ralis\'{e}e
apparait naturellement.

Les trois derniers th\`{e}mes que l'on vient d'\'{e}voquer ne sont pas
compl\`{e}tement \'{e}puis\'{e}s par les recherches r\'{e}sum\'{e}es. Par
contre elles ont aussi conduit \`{a} approfondir de fa\c {c}on tr\`{e}s
syst\'{e}matique le sujet de l'interpr\'{e}tation de Harvey Cohn de la
th\'{e}orie de Markoff classique. C'est ainsi qu'il a \'{e}t\'{e} \'{e}tabli
qu'on rencontre cette th\'{e}orie d\`{e}s qu'intervient le groupe $GL(2,\mathbb{%
Z})$ des matrices $2\times 2$ de d\'{e}terminant $\pm 1$.\ La
raison essentielle mise en \'{e}vidence est l'existence dans
$GL(2,\mathbb{Z})$ d'un sous-groupe di\'{e}dral $\mathbf{D}_6$
\`{a} $12$ \'{e}l\'{e}ments non
normal d\'{e}finissant intrins\`{e}quement un quotient \`{a} droite $GL(2,%
\mathbb{Z})/\Re _{\mathbf{D}_6}$ qui s'identifie \`{a} l'arbre
complet de la
th\'{e}orie de Markoff (respectivement un quotient \`{a} gauche $GL(2,\mathbb{Z}%
)/_{\mathbf{D}_6}\Re $ \'{e}quipotent). Ce r\'{e}sultat assure
l'ubiquit\'{e} du groupe du triangle $\mathbf{T}_3=\mathbf{C}_2*\mathbf{C}_2*%
\mathbf{C}_2$ produit libre de trois groupes cycliques \`{a} deux
\'{e}l\'{e}ments $\mathbf{C}_2$ dans des situations aussi diverses que les
fibr\'{e}s vectoriels, les ordres des anneaux de quaternions, le topographe
de Conway... \cite{Rudakov} \cite{Hirzebruch} \cite{Vigneras} \cite{Conway}.
L'article \cite{Perrine1b} d\'{e}veloppe cet aspect et a \'{e}t\'{e} repris
en tant que chapitre 6 dans l'ouvrage \cite{Perrine9}.

Tout au long des travaux men\'{e}s on a conserv\'{e} le souci
d'une coh\'{e}rence globale. Il s'agissait de sortir du cadre trop
contraignant de la seule \'{e}quation de Markoff classique pour
construire d'autres exemples mais en cherchant simultan\'{e}ment
\`{a} comprendre comment appr\'{e}hender le ''chaos'' du spectre
des constantes d'approximation des nombres alg\'{e}briques de
degr\'{e} 2.\ On voulait \'{e}galement permettre de maitriser les
applications \`{a} la physique. Ces deux pr\'{e}occupations ont
constitu\'{e} les fils conducteurs de la d\'{e}marche
d\'{e}velopp\'{e}e tout au long de ces derni\`{e}res ann\'{e}es.
C'est ainsi que l'on a recherch\'{e} et finalement trouv\'{e} un
op\'{e}rateur diff\'{e}rentiel intrins\`{e}quement li\'{e} \`{a}
la th\'{e}orie de Markoff classique, la question restant ouverte
de calculer son spectre et de le comparer au spectre de Markoff.
La m\'{e}thode utilis\'{e}e pour le construire est transposable
aux \'{e}quations $M^{s_1s_2}(a,\partial K,u_\theta )$. Elle a
conduit \`{a} s'int\'{e}resser aux \'{e}quations
hyperg\'{e}om\'{e}triques, aux \'{e}quations de Lam\'{e} qui
interviennent sur les param\`{e}tres accessoires des tores
perc\'{e}s \cite{Keen5}, et qui ne sont que des \'{e}quations de
Schr\"{o}dinger particuli\`{e}res dont le groupe de monodromie
associ\'{e} peut \^{e}tre \'{e}tudi\'{e} \cite{Waall}.

Une pr\'{e}sentation d\'{e}velopp\'{e}e des travaux que l'on vient
d'\'{e}voquer a \'{e}t\'{e} donn\'{e}e dans l'ouvrage
\cite{Perrine9}.\ Celui-ci peut \^{e}tre r\'{e}sum\'{e} comme
suit.\ On a mis au point un formalisme g\'{e}n\'{e}ral et
d\'{e}crit ses liens avec les sommes de Dedekind. On a
d\'{e}gag\'{e} les \'{e}quations qui g\'{e}n\'{e}ralisent
l'\'{e}quation de Markoff classique, et on les a
interpr\'{e}t\'{e}es avec une formule de trace et les sommes
li\'{e}es \`{a} la fonction $\eta $ de Dedekind. Partant de ces
\'{e}quations, on en a \'{e}tudi\'{e} de fa\c {c}on directe les
solutions. Ceci a fait appara\^{i}tre des structures
g\'{e}n\'{e}ralisant celle d\'{e}couverte par A. A. Markoff.\ Dans
quelques exemples particuliers, on a d\'{e}crit les classes de
solutions pour l'action du groupe $\mathbf{T}_3$. On a
d\'{e}taill\'{e} l'application \`{a} l'\'{e}tude du spectre de
Markoff.\ On a fait le lien avec des sujets classiques
d'arithm\'{e}tique quadratique, notamment la recherche des points
entiers sur les courbes elliptiques. On a \'{e}tudi\'{e} les
groupes fuchsiens agissant sur le demi-plan de Poincar\'{e}
$\mathcal{H}$ et consid\'{e}r\'{e} le cas des groupes libres \`{a}
deux g\'{e}n\'{e}rateurs, ainsi que les cons\'{e}quences pour la
structure du groupe $GL(2,\mathbb{Z})$. Ceci a montr\'{e}
l'importance alg\'{e}brique de la th\'{e}orie de Markoff classique
et son lien avec la $K$-th\'{e}orie et le th\'{e}or\`{e}me de Dyer
Formanek relatif au groupe des automorphismes d'un groupe libre.
Etudiant de fa\c {c}on g\'{e}n\'{e}rale les surfaces de Riemann et
la th\'{e}orie de Teichm\"{u}ller relative aux m\'{e}triques sur
une m\^{e}me surface, on a fourni de nombreuses perspectives dans
le chapitre 7 de l'ouvrage \cite {Perrine9} en cherchant \`{a}
pr\'{e}ciser le contexte qui leur donne naissance. L'un des points
qui para\^{i}t le plus important \`{a} l'auteur concerne les
d\'{e}veloppements relatifs \`{a} la fonction $\eta $ de Dedekind,
\`{a} son lien avec le laplacien d'objets \`{a} g\'{e}om\'{e}trie
hyperbolique, et \`{a} ses g\'{e}n\'{e}ralisations en physique
nucl\'{e}aire.\ On a aussi donn\'{e} quelques pistes pour
r\'{e}fl\'{e}chir \`{a} d'importantes conjectures.

Le texte qui suit condense l'ouvrage que l'on vient de r\'{e}sumer, en
identifiant les r\'{e}sultats nouveaux obtenus.\ Dans chaque chapitre on
pr\'{e}cise dans le premier paragraphe la probl\'{e}matique envisag\'{e}e
dans le texte qui suit, et on r\'{e}sume dans le dernier paragraphe les
perspectives de recherches futures \`{a} mener. Le lecteur d\'{e}sireux
d'aller \`{a} l'essentiel peut donc, au del\`{a} de la pr\'{e}sente
introduction passer tous les d\'{e}tails techniques qui sont
pr\'{e}sent\'{e}s dans chaque chapitre en ne lisant que les introductions et
les conclusions. Dans les paragraphes d\'{e}taill\'{e}s, on a \'{e}t\'{e}
\`{a} l'essentiel en n'insistant ni sur les d\'{e}finitions donn\'{e}es ni
sur les calculs men\'{e}s.\ On a renvoy\'{e} pour l'essentiel \`{a}
l'ouvrage \cite{Perrine9}, sachant que les d\'{e}finitions qu'il adopte sont
les plus g\'{e}n\'{e}ralement admises. Tout ce qui est relatif aux
d\'{e}finitions classiques et aux r\'{e}sultats bien connus a \'{e}t\'{e}
extrait dans la mesure du possible.\ Le chapitre 5 est consacr\'{e} \`{a} la
g\'{e}n\'{e}ralisation de la th\'{e}orie de Markoff aux surfaces de Riemann
hyperboliques. On a voulu bien identifier des th\`{e}mes qui ont un sens par
rapport \`{a} une probl\'{e}matique de codage et de quantification de
l'information port\'{e}e par une telle surface, et plus g\'{e}n\'{e}ralement
par rapport aux limitations du calcul qui mod\'{e}lise la physique.\ Le
chapitre comprend peu de r\'{e}sultats nouveaux hors l'\'{e}quation
diff\'{e}rentielle intrins\`{e}quement li\'{e}e \`{a} la th\'{e}orie de
Markoff.\ Il fournit le point de vue \'{e}labor\'{e} par l'auteur pour
comprendre la signification de grandes conjectures encore d'actualit\'{e}.\
Il d\'{e}veloppe aussi une signification profonde de la fonction \'{e}ta de
Dedekind expliquant sa d\'{e}composition en produit infini, et les produits
infinis qui en r\'{e}sultent pour d'autres fonctions classiques, telles que
les fonctions th\^{e}ta ou les fonctions elliptiques. On a \'{e}galement
voulu jeter quelques bases pour faire le lien avec les solitons et les
travaux d'actualit\'{e} en g\'{e}om\'{e}trie non commutative (\cite{Connes}
\`{a} \cite{Connes6}) et en th\'{e}orie du chaos quantique.

Dans le texte on utilise le m\^{e}me syst\`{e}me d'indexation des
propositions que dans l'ouvrage \cite{Perrine9}.\ Elles sont
rep\'{e}r\'{e}es dans chaque chapitre avec deux nombres, mais
cit\'{e}es en faisant pr\'{e}c\'{e}der ces derniers d'un nombre
indiquant le chapitre o\`{u} elles se trouvent. On a aussi
ajout\'{e} quelques \'{e}l\'{e}ments nouveaux d\'{e}couverts
depuis la publication de l'ouvrage \cite{Perrine9}, ainsi que
quelques r\'{e}f\'{e}rences compl\'{e}mentaires qui paraissent
importantes. La bibliographie est l\'{e}g\`{e}rement plus large
que ce qui est strictement utilis\'{e} dans le texte, pour
facilter des travaux ult\'{e}rieurs en cours. \
\[
\]

\chapter{G\'{e}n\'{e}ralisation de la th\'{e}orie de Markoff}

\section{Introduction}

Historiquement, la th\'{e}orie de Markoff a \'{e}t\'{e} construite vers 1880
gr\^{a}ce aux fractions continues \cite{Markoff}.\ Puis elle a \'{e}t\'{e}
progressivement reconsid\'{e}r\'{e}e en mettant en avant les formes
quadratiques correspondantes \cite{Cassels}. Aujourd'hui, elle est
usuellement pr\'{e}sent\'{e}e \`{a} l'envers en partant de la r\'{e}solution
de l'\'{e}quation diophantienne qui concluait les deux articles fondateurs
\cite{Cusick} :
\[
x^2+y^2+z^2=3xyz,\;\;x,y,z\in \mathbb{N}\backslash \{0\}.
\]
On a cherch\'{e} au d\'{e}but du 20$^{\grave{e}me}$ si\`{e}cle, et de fa\c
{c}on infructueuse, les \'{e}quations \`{a} \'{e}tudier pour construire une
g\'{e}n\'{e}ralisation de cette th\'{e}orie \cite{Frobenius}.\ Reprenant ce
probl\`{e}me, l'auteur a consid\'{e}r\'{e} que le retour aux fractions
continues \'{e}tait la m\'{e}thode la plus r\'{e}aliste pour atteindre un
tel objectif.\ Il a ainsi pu construire un formalisme g\'{e}n\'{e}ralis\'{e}
et les \'{e}quations diophantiennes qui en r\'{e}sultent \cite{Perrine9} en
partant des suites d'entiers strictement positifs les plus g\'{e}n\'{e}rales
\[
S=(a_0,a_1,...,a_n).
\]

\section{Pr\'{e}sentation de la th\'{e}orie}

\subsection{Notations}

La matrice de la suite $S$ et son d\'{e}terminant sont donn\'{e}s par
\[
M_S=M_{(a_0,a_1,...,a_n)}=\left[
\begin{array}{cc}
a_0 & 1 \\
1 & 0
\end{array}
\right] \left[
\begin{array}{cc}
a_1 & 1 \\
1 & 0
\end{array}
\right] ...\left[
\begin{array}{cc}
a_n & 1 \\
1 & 0
\end{array}
\right] =\left[
\begin{array}{cc}
m & K_1 \\
m-K_2 & K_1-l
\end{array}
\right] ,
\]
\[
\varepsilon _S=\det (M_S)=(-1)^{n+1}.
\]
La suite miroir de $S$ est $S^{*}=(a_n,a_{n-1},...,a_0)$, et on associe
\`{a} $S$ deux suites \'{e}tendues sur la gauche et sur la droite avec $%
S\rhd =(\lhd S^{*})^{*}$ et :
\[
\lhd S=\left\{
\begin{array}{cc}
(1,a_0-1,a_1,...,a_n) & \text{si }a_0\neq 1 \\
(a_1+1,...,a_n) & \text{si }a_0=1
\end{array}
\right\} .
\]
Les matrices $M_S$ engendrent le groupe $GL(2,\mathbb{Z})$ des
matrices de
d\'{e}terminant $\pm 1$. Elles agissent sur la droite projective r\'{e}elle $%
P^1(\mathbb{R})=\mathbb{R}\cup \{\infty \}$ ou la droite complexe $P^1(\mathbb{C})=%
\mathbb{C}\cup \{\infty \}$ par
\[
\left[
\begin{array}{cc}
\alpha & \beta \\
\gamma & \delta
\end{array}
\right] (z)=\frac{\alpha z+\beta }{\gamma z+\delta },
\]
avec des notations classiques pour les fractions continues :
\[
M_S(\infty )=[S]=[a_0,a_1,...,a_n]=a_0+\frac 1{a_1+\dfrac 1{...+\dfrac
1{a_n}}}.
\]
Les nombres alg\'{e}briques de degr\'{e} 2, dits nombres de Markoff, dont le
d\'{e}veloppement en fraction continue est p\'{e}riodique et peut \^{e}tre
\'{e}crit avec une p\'{e}riode $(S^{*},a)$ sont not\'{e}s $\theta _a(S)=[0,%
\underline{S^{*},a}].$ On peut en donner une expression
alg\'{e}brique. La th\'{e}orie de Markoff g\'{e}n\'{e}ralis\'{e}e
s'appuie sur une d\'{e}composition de forme :
\[
S^{*}=(a_n,a_{n-1},...,a_0)=(X_1,b,X_2),
\]
o\`{u} les suites $X_1$ et $X_2$ d\'{e}finissent des matrices de
suites dans $GL(2,\mathbb{Z})$ :
\[
M_{X_1}=\left[
\begin{array}{cc}
m_1 & m_1-k_{12} \\
k_1 & k_1-l_1
\end{array}
\right] \text{ avec }\det (M_{X_1})=\varepsilon _1\in \{-1,+1\},
\]
\[
M_{X_2}=\left[
\begin{array}{cc}
m_2 & m_2-k_2 \\
k_{21} & k_{21}-l_2
\end{array}
\right] \text{ avec }\det (M_{X_2})=\varepsilon _2{}{}{}\in \{-1,+1\},\text{
}
\]
On obtient ainsi les expressions suivantes :
\[
m=(b+1)m_1m_2+m_1k_{21}-m_2k_{12},\;\;\varepsilon _S=-\varepsilon
_1\varepsilon _2.
\]
On d\'{e}finit deux param\`{e}tres auxiliaires $t_1$, $t_2$, et deux nombres
$u$ et $\partial K$ importants :
\[
t_1=k_1+k_{12}-m_1,\;\;t_2=k_2+k_{21}-m_2,
\]
\[
u=m_2t_1-m_1t_2,\;\;\partial K=\varepsilon _2(K_1-K_2).
\]
Ils permettent d'\'{e}valuer :
\[
m_1k_2-m_2k_1=(b+1)m_1m_2-m-u,
\]
\[
\varepsilon _1m_2=K_1m_1-k_1m,\;\;\;\varepsilon _2m_1=k_2m-K_2m_2.
\]
\ La r\'{e}solution des deux derni\`{e}res \'{e}quations de Bezout calcule $%
K_1,K_2,k_1,k_2,$ \`{a} partir du seul triplet $(m,m_1,m_2)$ et de $%
(\varepsilon _1,\varepsilon _2)$.\ On en d\'{e}duit les autres
param\`{e}tres. Ceci permet de reconstruire la suite $S^{*}$ et sa
d\'{e}composition avec $X_1$ et $X_2$. Cette m\'{e}thode a \'{e}t\'{e}
utilis\'{e}e pour construire les premiers exemples de th\'{e}ories de
Markoff g\'{e}n\'{e}ralis\'{e}es \cite{Perrine3}. Le point d\'{e}couvert a
\'{e}t\'{e} que pour $(\varepsilon _1,\varepsilon _2)=(\pm 1,\pm 1)$
donn\'{e}, et \`{a} la r\'{e}solution d'\'{e}quations de Bezout pr\`{e}s, le
triplet $(m,m_1,m_2)$ contient toute l'information n\'{e}cessaire pour
reconstruire les suites $X_1$ et $X_2$, ainsi que $b$ et la suite $S^{*}$,
puis la d\'{e}composition matricielle associ\'{e}e pour $M_{S^{*}}$. On a pu
s'assurer qu'il existe une suite $T$ \'{e}ventuellement vide, telle que l'on
ait $X_1=(\lhd X_2^{*},c,T)$. Ceci impose une propri\'{e}t\'{e} de miroir
partielle \`{a} la suite $S$ :
\[
\lhd S^{*}=(X_2^{*},c,T,b,X_2).
\]
Comme les cas $T=\emptyset $ et $X_2=\emptyset $ sont
envisageables, on a obtenu ainsi un r\'{e}sultat essentiel pour la
construction de la g\'{e}n\'{e}ralisation de la th\'{e}orie de
Markoff que l'on recherche :

\begin{proposition} Hors le cas des suites $(1)$ et $(b,1)$, toute
suite $S$ admet une d\'{e}composition
\[
S^{*}=(\lhd X_2^{*},c,T,b,X_2),
\]
avec $X_2$ et $T$ suites d'entiers strictement positifs, \'{e}ventuellement
vides, ainsi que $b$ et $c$ entiers strictement positifs.
\end{proposition}

\subsection{Forme de Markoff}

Dans le cas le plus g\'{e}n\'{e}ral, on dispose d'une matrice $M_{(S^{*},a)}$
correspondant \`{a} la p\'{e}riode du nombre
\[
\theta _a(S)=[0,\underline{S^{*},a}]=[0,\underline{\lhd X_2^{*},c,T,b,X_2,a}%
].
\]
Cette matrice d\'{e}finit une forme quadratique issue de la recherche des
points fixes de la transformation de M\"{o}bius d\'{e}finie par la matrice $%
M_{(S^{*},a)}$ \cite{Cohn2} \cite{Series}. Cette forme quadratique
binaire enti\`{e}re ind\'{e}finie dite forme de Markoff
s'\'{e}crit :
\begin{eqnarray*}
mF_\theta (x,y) &=&mx^2+(((a+1)m-K_2)-K_1)xy-((a+1)K_1-l)y^2 \\
&=&m(x-\theta _a(S)y)(x-\overline{\theta _a(S)}y).
\end{eqnarray*}
Un calcul direct donne \cite{Perrine}\cite{Perrine1} :

\begin{proposition}
On a :
\[
F_\theta (K_1,m)=F_\theta (K_2-(a+1)m,m)=\varepsilon _1\varepsilon
_2=-\varepsilon _S,
\]
\[
F_\theta (K_1x+((a+1)K_1-l)y,mx+((a+1)m-K_2)y)=-\varepsilon _SF_\theta
(x,y).
\]
\end{proposition}

\subsection{R\'{e}duction}

La th\'{e}orie de la r\'{e}duction des formes quadratiques
binaires remonte \`{a} C. F. Gauss \cite{Gauss}. Elle concerne les
formes quadratiques ind\'{e}finies que l'on \'{e}crit avec des
coefficients r\'{e}els $\lambda \in \mathbb{R}\backslash \{0\}$ et
$\beta $, $\gamma \in \mathbb{R}$
\[
\lambda f(x,y)=\lambda (x^2+\beta xy+\gamma y^2).
\]
Chacune a un discriminant strictement positif $\Delta (\lambda f)=\lambda
^2(\beta ^2-4\gamma )=\lambda ^2\Delta (f)$.\ Elle poss\`{e}de un minimum
arithm\'{e}tique
\[
m(\lambda f)=\inf_{(x,y)\in \mathbf{Z}^2-\{(0,0)\}}\left| \lambda
f(x,y)\right| =\left| \lambda \right| m(f).
\]
Ceci donne sa constante de Markoff, ne d\'{e}pendant pas du coefficient $%
\lambda $
\[
C(\lambda f)=m(\lambda f)/\sqrt{\Delta (\lambda f)}=m(f)/\sqrt{\Delta (f)}%
=C(f).
\]
Le spectre de Markoff est d\'{e}fini comme \'{e}tant l'ensemble de
toutes les constantes de Markoff de formes quadratiques
r\'{e}elles ind\'{e}finies. Il poss\`{e}de un sous ensemble
particulier $Mark$ de constantes des formes quadratiques
ind\'{e}finies \`{a} coefficients entiers. C'est le spectre
quadratique. Le lien entre les deux spectres a fait l'objet de
diff\'{e}rents travaux \cite{Cusick}\cite{Tornheim}.

L'\'{e}quivalence de deux formes $\lambda f$ et $\lambda ^{\prime }f^{\prime
}$ est d\'{e}finie avec des entiers $v_{11}$, $v_{12}$, $v_{21}$, $v_{22}$
v\'{e}rifiant :
\[
\lambda ^{\prime }f^{\prime }(v_{11}x+v_{12}y,v_{21}x+v_{22}y)=\lambda
f(x,y),\;\;v_{11}v_{22}-v_{12}v_{21}=\pm 1.
\]
Elle donne avec des notations comparables \`{a} celles de A.\ A.\ Markoff
\cite{Markoff} le classique lemme de r\'{e}duction :

\begin{proposition}
Pour toute forme quadratique r\'{e}elle ind\'{e}finie $\lambda f(x,y)$ il
existe une forme r\'{e}duite \'{e}quivalente $\lambda _0f_0(x,y)$,
v\'{e}rifiant les conditions suivantes :
\[
\lambda _0f_0(x,y)=\lambda _0(x^2+\beta _0xy+\gamma _0y^2)=\lambda _0(x-\xi
_0y)(x-\xi _0^{\prime }y),
\]
\[
\xi _0=\frac{-\beta _0+\sqrt{\beta _0^2-4\gamma _0}}2=[\alpha _0,\alpha
_1,...,\alpha _j,...]>1,
\]
\[
-1<\xi _0^{\prime }=-(1/\eta _0)=\frac{-\beta _0-\sqrt{\beta _0^2-4\gamma _0}%
}2=-[0,\alpha _{-1},\alpha _{-2},...,\alpha _{-j},...]<0.
\]
La suite des nombres entiers strictement positifs $(\alpha _n)_{n\in \mathbb{Z}%
} $ est associ\'{e}e de fa\c {c}on unique (\`{a} la sym\'{e}trie pr\`{e}s $%
\alpha _j\rightarrow \alpha _{-j}$ et aux d\'{e}calages pr\`{e}s
$\alpha _j\rightarrow \alpha _{j+t}$ o\`{u} $t\in \mathbb{Z}$)
\`{a} $\lambda f(x,y)$. Si l'on consid\`{e}re les diff\'{e}rentes
valeurs
\[
\xi _j=[\alpha _j,\alpha _{j+1},...,\alpha _{2j},...]>1,
\]
\[
-1<\xi _j^{\prime }=-(1/\eta _j)=-[0,\alpha _{j-1},\alpha _{j-2},...,\alpha
_0,...]<0,
\]
\[
\frac 2{L_j}=\xi _j-\xi _j^{\prime }=\sqrt{\beta _j^2-4\gamma _j},
\]
d\'{e}finissant pour tout $j\in \mathbb{Z}$ une forme r\'{e}duite
\'{e}quivalente \`{a} $\lambda f(x,y)$ :
\[
\lambda _jf_j(x,y)=\lambda _j(x^2+\beta _jxy+\gamma _jy^2)=\lambda _j(x-\xi
_jy)(x-\xi _j^{\prime }y).
\]
Le nombre $\lambda _j=\lambda _jf_j(1,0)$ est repr\'{e}sent\'{e} par la
forme $\lambda f(x,y)$.\ Et on a :
\[
C(\lambda f)=C(f_0)=C(f_j)=\inf_{j\in \mathbb{Z}}(\frac{L_j}2).
\]
\end{proposition}

Depuis \cite{Markoff}, il est clair que travailler sur les formes de Markoff
est \'{e}quivalent \`{a} utiliser la th\'{e}orie classique de la
r\'{e}duction des formes quadratiques :

\begin{proposition}
Toute forme quadratique ind\'{e}finie $f(x,y)$ \`{a} coefficients entiers
d\'{e}finit un nombre fini de formes de Markoff $F_\theta (x,y)$
\'{e}quivalentes \`{a} $f(x,y)$, de nombres de Markoff $\theta _a(S)$
correspondant compris entre 0 et 1, et de suites associ\'{e}es $(S^{*},a)$.
De plus on a \'{e}quivalence des propri\'{e}t\'{e}s suivantes :

1/ $F_\theta (x,y)$ forme de Markoff

2/ $F_\theta (-x,y)$ forme r\'{e}duite
\end{proposition}

\subsection{Calcul des constantes et approximation diophantienne}

L'\'{e}tude du spectre quadratique $Mark$ dans le spectre de Markoff est
faisable de fa\c {c}on exhaustive en \'{e}tudiant \cite{Perrine} les
constantes des formes $F_\theta (x,y)$ :
\[
\Delta (F_\theta )=\left[ \frac{((a+1)m+K_1-K_2)^2-4\varepsilon
_1\varepsilon _2}{m^2}\right] =\frac{\Delta _a(S)}{m^2},
\]
\[
0<m(F_\theta )=\inf \{\mid F_\theta (x,y)\mid ;(x,y)\in \mathbb{Z}%
^2-\{(0,0)\}\}=\frac{m-s}m\leq F_\theta (1,0)=1.
\]
La th\'{e}orie du polygone de Klein \cite{Klein2} permet d'\'{e}crire
\[
0<C(F_\theta )=m(F_\theta )\frac m{\sqrt{\Delta _a(S)}}=\frac{m-s}{\sqrt{%
\Delta _a(S)}}\leq \frac m{\sqrt{\Delta _a(S)}}.
\]
Elle fournit un lien avec l'approximation diophantienne :

\begin{proposition}
Soit $\theta _a(S)$ un nombre de Markoff r\'{e}el alg\'{e}brique
de degr\'{e} 2 associ\'{e} \`{a} la forme $F_\theta (x,y)$,
l'ensemble des points d'accumulation de l'ensemble
\[
\{\mid q(q\theta _a(S)-p)\mid ;p,q\in \mathbb{Z}\},
\]
est fini et s'\'{e}crit sous la forme
\[
\{\frac{\mid m_j\mid }{\sqrt{\Delta _a(S)}};m_j\in
\mathbb{Z}^{*}\},
\]
o\`{u} $m_j$ est un entier repr\'{e}sent\'{e} par la forme $mF_\theta (x,y)$
sur une r\'{e}duite $(p_j/q_j)$ de $\theta _a(S)=[0,\underline{S^{*},a}]$:
\[
mF_\theta (p_j,q_j)=m_j.
\]
C'est aussi l'ensemble des points d'accumulation de l'ensemble
\[
\{\mid q(q\overline{\theta _a(S)}-p)\mid ;p,q\in \mathbb{Z}\}.
\]
Sa plus petite valeur n'est autre que la constante de Markoff $C(F_\theta
)=C(\theta _a(S))$. Sa plus grande valeur peut \^{e}tre tr\`{e}s
diff\'{e}rente de $C(\theta _a(S))$. Et si l'on note
\[
\theta _a(S)=[0,\underline{S^{*},a}]=[b_0,b_1,b_2,...],
\]
on peut aussi \'{e}crire avec les r\'{e}duites de ce nombre
\[
q_j(q_j\theta _a(S)-p_j)=\frac{(-1)^j}{%
(b_{j+1}+[0,b_{j+2},b_{j+3},...]+[0,b_j,b_{j-1},...,b_1])},
\]
\[
C(F_\theta )=C(\theta _a(S))=\frac 1{\lim \sup_{j\rightarrow \infty
}(b_{j+1}+[0,b_{j+2},b_{j+3},...]+[0,b_j,b_{j-1},...,b_1])}.
\]
\end{proposition}

\subsection{Extrema positif et n\'{e}gatif}

On est conduit \`{a} se demander si le minimum arithm\'{e}tique de $F_\theta
$ est atteint positivement ou n\'{e}gativement.\ On note $\nu _\theta $ la
plus grande valeur strictement n\'{e}gative repr\'{e}sent\'{e}e par $%
F_\theta $ et $\mu _\theta $ la plus petite valeur strictement positive
repr\'{e}sent\'{e}e par $F_\theta $.\ On pose :
\[
1\geq \mu _\theta =\frac{m-s_\mu }m>0,\;\;\nu _\theta =-\frac{m-s_\nu }m<0.
\]
La situation o\`{u} $-\nu _\theta =\mu _\theta $, comme dans la th\'{e}orie
de Markoff classique, est exceptionnelle.\ C'est pourquoi on ne doit plus
l'utiliser comme un argument d\'{e}terminant dans l'\'{e}tude des constantes
de Markoff, ainsi que cela est fait depuis les travaux de Remak \cite{Remak}%
, notamment dans \cite{Cassels}. Consid\'{e}rant la p\'{e}riode du nombre de
Markoff associ\'{e} \`{a} $(S^{*},a)=(\lhd X_2^{*},c,T,b,X_2,a)$, on a
\'{e}t\'{e} conduit \`{a} se demander si les nombres $b$ et $c$ ne
d\'{e}terminent pas la fa\c {c}on dont la forme $F_\theta $ atteint ses
valeurs $\mu _\theta $ ou $\nu _\theta $. En fait ceci d\'{e}pend de $%
\varepsilon _1$ et $\varepsilon _2$ car on a :
\[
\dfrac 1{c+\dfrac 1{[T,b,X_2,a,\lhd X_2^{*},c,...]}+\dfrac 1{[X_2\rhd
,a,X_2^{*},b,T^{*},c,...]}}=\varepsilon _2\frac{mF_\theta (k_2,m_2)}{\sqrt{%
\Delta _a(S)}}>0.
\]
\[
\dfrac 1{b+\dfrac 1{[X_2,a,\lhd X_2^{*},c,T,b,...]}+\dfrac
1{[T^{*},c,X_2\rhd ,a,X_2^{*},b,...]}}=-\varepsilon _1\frac{mF_\theta
(k_1,m_1)}{\sqrt{\Delta _a(S)}}>0.
\]
\\
Si l'on \'{e}crit le dernier nombre sous la forme
\[
\frac{m-s_b}{\sqrt{\Delta _a(S)}},
\]
on obtient le r\'{e}sultat essentiel suivant :

\begin{proposition}
Avec les expressions pr\'{e}c\'{e}dentes qui d\'{e}finissent
$s_b$, on a :
\[
s_b=(b-a)m_1m_2-u.
\]
\end{proposition}

Cette formule remarqu\'{e}e dans l'article \cite{Perrine7} a une
d\'{e}monstration directe :
\begin{eqnarray*}
mF_\theta (k_1,m_1) &=&mk_1^2+((a+1)m-K_2-K_1)k_1m_1-((a+1)K_1-l)m_1^2 \\
&=&k_1(mk_1-m_1K_1)+(a+1)m_1(mk_1-m_1K_1)+m_1(m_1l-K_2k_1) \\
&=&-\varepsilon _1(k_1+(a+1)m_1)m_2+\varepsilon _1m_1k_2 \\
&=&\varepsilon _1((b+1)m_1m_2-m-u)-\varepsilon _1(a+1)m_1m_2 \\
&=&-\varepsilon _1(m-((b-a)m_1m_2-u)).
\end{eqnarray*}
Elle donne un compl\'{e}ment \`{a} la proposition 1.2.2 :

\begin{proposition}
La forme de Markoff v\'{e}rifie avec les param\`{e}tres introduits
\[
\varepsilon _1F_\theta (k_1,m_1)=\varepsilon _2F_\theta
(k_2-(a+1)m_2,m_2)=-((m+(a-b)m_1m_2+u)/m)<0.
\]
\end{proposition}

On obtient maintenant en comparant les cas $\varepsilon _1=1$ et $%
\varepsilon _1=-1$ :

\begin{proposition}
Pour toute forme de Markoff $F_\theta $, on a le majorant suivant
pour son minimum arithm\'{e}tique :
\[
m(F_\theta )\leq \frac{m+u+(a-b)m_1m_2}m,
\]
avec les in\'{e}galit\'{e}s suivantes :
\[
(b-a)m_1m_2<m+u<(a+b+2)m_1m_2-(a+1)\partial Km_2^2,
\]
\[
\partial Km_2<m_1.
\]
\end{proposition}

Vouloir \'{e}tudier s\'{e}parement les deux extrema positif ou
n\'{e}gatif pourrait conduire \`{a} consid\'{e}rer chacune des
deux parties du polygone de Klein pour elle-m\^{e}me. En fait les
fractions continues adapt\'{e}es pour ce faire sont les fractions
continues r\'{e}guli\`{e}res r\'{e}duites, dites de
Jung-Hirzebruch, qui s'\'{e}crivent :
\[
\lbrack [a_0,a_1,...,a_n]]=a_0-\frac 1{a_1-\dfrac 1{...-\dfrac 1{a_n}}}.
\]
Ces nouvelles r\'{e}duites correspondent \cite{Finkel'shtein} \`{a} des
sommets du polygone de Klein sup\'{e}rieur si et seulement si on a $a_n\neq
2 $. Elles sont reli\'{e}es aux fractions continues ordinaires utilis\'{e}es
ci-dessus (\cite{Hirzebruch0} (p. 215) \cite{Myerson} \cite{Dimca}) par la
formule g\'{e}n\'{e}rale suivante :
\[
\lbrack a_0,a_1,z]=[[a_0+1,2_{a_1-1},z+1]].
\]

\subsection{L'\'{e}quation de Markoff g\'{e}n\'{e}ralis\'{e}e}

Dans le cas le plus g\'{e}n\'{e}ral, on peut mettre en
\'{e}vidence de plusieurs fa\c {c}ons l'existence d'une
\'{e}quation diophantienne g\'{e}n\'{e}ralisant celle de Markoff.
Comme dans \cite{Cassels} on peut utiliser une nouvelle forme
quadratique reli\'{e}e \`{a} $F_\theta (x,y)$ :
\[
\phi _\theta (z,y)=z^2+((a+1)m+K_1-K_2)zy-\varepsilon _Sy^2=m^2F_\theta
(x,y),\;\;z=mx-K_1y.
\]
Elle poss\`{e}de la propri\'{e}t\'{e} de multiplicativit\'{e} suivante :

\begin{proposition}
On a
\[
\phi _\theta (z_1,y_1)\phi _\theta (z_2,y_2)=\phi _\theta
(z_1z_2+\varepsilon _Sy_1y_2,y_1z_2+z_1y_2+((a+1)m+K_1-K_2)y_1y_2).
\]
\end{proposition}

Elle est invariante par diff\'{e}rentes transformations \cite{Cassels} :

\begin{proposition}
On a :
\begin{eqnarray*}
\phi _\theta (z,y)=\phi _\theta (-z,-y) \\
=-\varepsilon _S\phi _\theta (y,-\varepsilon _Sz) \\
=\phi _\theta (z+((a+1)m+K_1-K_2)y,-y) \\
=\phi _\theta (-z,y-((a+1)m+K_1-K_2)\varepsilon _Sz) \\
=-\varepsilon _S\phi _\theta (y-\varepsilon _S((a+1)m+K_1-K_2)z,\varepsilon
_Sz) \\
=-\varepsilon _S\phi _\theta (-y,-\varepsilon
_Sz-((a+1)m+K_1-K_2)\varepsilon _Sy).
\end{eqnarray*}
\end{proposition}

Cette derni\`{e}re proposition donne $\phi _\theta (-\varepsilon
_1m_2,m_1)=m^2F_\theta (k_1,m_1)$ et l'expression vue pour $s_b$ fait
apparaitre l'\'{e}quation $M^{s_1s_2}(b,\partial K,u)$ recherch\'{e}e, dont
les termes ne d\'{e}pendent que de la suite $S^{*}$ :

\begin{proposition}
Soit $S^{*}=(a_0,a_1,...,a_n)=(X_1,b,X_2)$ une suite d'entiers positifs
donnant les param\`{e}tres $m$, $m_1$, $m_2$, $\partial K$, $u$, $%
\varepsilon _1$, $\varepsilon _2$, le triplet d'entiers $(m,m_1,m_2)\in (%
\mathbb{N}\backslash \{0\})^3$ est solution de l'\'{e}quation diophantienne $%
M^{s_1s_2}(b,\partial K,u)$%
\[
m^2+\varepsilon _2m_1^2+\varepsilon _1m_2^2=(b+1)mm_1m_2+\varepsilon
_2\partial Km_1m_2-um.
\]
\end{proposition}

En notant $u_\theta =u+(a-b)m_1m_2=-s_b$ pour tout $a\in
\mathbb{N}\backslash
\{0\}$, le triplet d'entiers $(m,m_1,m_2)$ v\'{e}rifie aussi l'\'{e}quation $%
M^{s_1s_2}(a,\partial K,u_\theta )$
\[
m^2+\varepsilon _2m_1^2+\varepsilon _1m_2^2=(a+1)mm_1m_2+\varepsilon
_2\partial Km_1m_2-u_\theta m.
\]

\subsection{Autres d\'{e}montrations}

Trois autres d\'{e}monstrations de cette proposition ont \'{e}t\'{e}
d\'{e}couvertes. Elles sont d\'{e}taill\'{e}es dans l'ouvrage \cite{Perrine9}%
.

$\bullet $ Une premi\`{e}re g\'{e}n\'{e}ralise le calcul original de Markoff
\cite{Markoff}.

$\bullet $ Une seconde met en oeuvre les sommes de Dedekind
\cite{Perrine6}, dont le lien avec l'\'{e}quation de Markoff a
\'{e}t\'{e} reconnu depuis longtemps \cite{Hirzebruch}(pp.
158-165) au travers de leur classique formule de
r\'{e}ciprocit\'{e} \cite{Rademacher}. La somme de Dedekind est
d\'{e}finie pour $(\delta ,\gamma )\in \mathbb{Z}\times
\mathbb{Z}-\{0\}$ comme suit :
\[
s(\delta ,\gamma )=s(\delta ,\left| \gamma \right| )=\sum_{k=1}^{\left|
\gamma \right| }\left( \left( \frac{k\delta }{\left| \gamma \right| }\right)
\right) \left( \left( \frac k{\left| \gamma \right| }\right) \right) .
\]
La premi\`{e}re mention des sommes $s(\delta ,\gamma )$ se trouve
dans l'\'{e}tude de la fonction $\eta $ faite par R.\ Dedekind
dans son commentaire du fragment XXVIII de B.\ Riemann
\cite{Riemann} (p. 397).\ Cette fonction est issue des calculs
d'Eisenstein pour donner des produits infinis exprimant les
fonctions elliptiques \cite{Weil4}, et analogues \`{a} ceux
d\'{e}couverts par Euler pour les fonctions trigonom\'{e}triques
\cite {Euler2} (Tome1 ch.\ IX).\ La somme de Dedekind est
pr\'{e}sente dans l'exposant donnant $\varepsilon $, la racine
$24i\grave{e}me$ de l'unit\'{e}
de la formule de transformation de $\eta $ par un \'{e}l\'{e}ment de $PSL(2,%
\mathbb{Z})$ :
\[
\eta (\frac{\alpha \tau +\beta }{\gamma \tau +\delta })=\varepsilon (\gamma
\tau +\delta )^{\frac 12}\eta (\tau ).
\]

$\bullet $ Une troisi\`{e}me d\'{e}monstration interpr\`{e}te l'\'{e}quation
$M^{s_1s_2}(b,\partial K,u)$ comme une formule de trace utilisant les
matrices
\[
A_b=M_{(\lhd X_2^{*},b)}=\left[
\begin{array}{cc}
bm_2+k_{21} & m_2 \\
bk_2+l_2 & k_2
\end{array}
\right] ,\;\;\varepsilon _A=\det (A_b)
\]
\[
B_c=M_{(X_1^{*}\rhd ,c)}=\left[
\begin{array}{cc}
(c+1)m_1-k_1 & m_1 \\
(c+1)(m_1-k_{12})-(k_1-l_1) & m_1-k_{12}
\end{array}
\right] ,\;\;\varepsilon _B=\det (B_c)
\]
\[
A_bB_c=M_{(\lhd X_2^{*},b)}M_{(X_1^{*}\rhd ,c)}=M_{(\lhd S\rhd ,c)}=\left[
\begin{array}{cc}
(c+1)m-K_1 & m \\
(c+1)K_2-l & K_2
\end{array}
\right] .
\]
Ces matrices sont dans $GL(2,\mathbb{Z})$ et non seulement dans $SL(2,\mathbb{Z})$%
.\ Notre \'{e}quation se d\'{e}duit d'une formule de Fricke qui donne pour $%
tr(A_bB_cA_b^{-1}B_c^{-1})$ la valeur :
\[
\varepsilon _Atr(A_b)^2+\varepsilon _Btr(B_c)^2+\varepsilon _A\varepsilon
_Btr(A_bB_c)^2-\varepsilon _A\varepsilon _Btr(A_b)tr(B_c)tr(A_bB_c)-2.
\]
Il suffit de calculer par une autre m\'{e}thode la trace du commutateur $%
A_bB_cA_b^{-1}B_c^{-1}$ dans le cas o\`{u} $b=c$ pour retrouver notre
\'{e}quation diophantienne comme simple formule de trace \cite{Perrine9}.

\subsection{Compl\'{e}ment}

Dans le cas g\'{e}n\'{e}ral il n'y a pas d'hypoth\`{e}se \`{a} faire sur le
nombre $\delta =$ pgcd$(m_1,m_2)$. Il s'agit d'un nombre qui peut \^{e}tre
diff\'{e}rent de $1$ et divise $u$. Il v\'{e}rifie :

\begin{proposition}
On a les \'{e}galit\'{e}s
\[
\delta =\text{pgcd}(m_1,m_2)=\text{pgcd}(m_2,m)=\text{pgcd}(m,m_1)=\text{pgcd%
}(m,m_1,m_2).
\]
\end{proposition}

La situation g\'{e}n\'{e}rale se distingue donc clairement de la th\'{e}orie
de Markoff classique o\`{u} l'on a toujours $\delta =1$. Comme cette
derni\`{e}re condition est utilis\'{e}e de fa\c {c}on assez centrale dans
l'expos\'{e} \cite{Cassels}, notamment au travers de ses lemmes 5 et 6, on
comprend a posteriori pourquoi il a fallu changer de paradigme pour
d\'{e}gager notre g\'{e}n\'{e}ralisation de la th\'{e}orie de Markoff.

\section{Perspectives}

Les calculs qui pr\'{e}c\`{e}dent s'appliquent \`{a} toutes les formes
quadratiques binaires ind\'{e}finies.\ Ceci explique pourquoi les
\'{e}quations diophantiennes mises en \'{e}vidence sont tr\`{e}s
g\'{e}n\'{e}rales. On a indiqu\'{e} qu'elles sont aussi donn\'{e}es par une
formule de trace, ainsi que par une propri\'{e}t\'{e} de la fonction $\eta $
de Dedekind.\ Il s'agit l\`{a} de r\'{e}sultats tout \`{a} fait nouveaux qui
ouvrent un domaine de r\'{e}flexion tr\`{e}s important.\ On peut chercher
\`{a} g\'{e}n\'{e}raliser ce qui pr\'{e}c\`{e}de \`{a} des formes
homog\`{e}nes de plus grand degr\'{e} ou poss\'{e}dant plus de variables. Il
est possible qu'il faille privil\'{e}gier dans ce contexte un algorithme
\cite{Gomory} \cite{Moussafir} \cite{Lachaud} \cite{Dimca}
g\'{e}n\'{e}ralisant les fractions continues r\'{e}guli\`{e}res r\'{e}duites
$[[a_0,a_1,...,a_n]]$ de Jung-Hirzebruch, dont on peut syst\'{e}matiser
l'utilisation dans ce qui pr\'{e}c\`{e}de.\

La fonction $\eta $ de Dedekind vient des calculs d'Eisenstein pour la
d\'{e}composition des fonctions elliptiques en produits infinis \cite{Weil4}%
. Une question qui se pose est de savoir s'il existe une fonction
g\'{e}n\'{e}ralisant $\eta $ pour d'autres fonctions trigonom\'{e}triques.\
Un projet est de d\'{e}duire de l\`{a} des sommes plus g\'{e}n\'{e}rales que
celles de Dedekind, et de comprendre ce que pourrait \^{e}tre une formule de
r\'{e}ciprocit\'{e} correspondante, ainsi qu'une \'{e}quation diophantienne
associ\'{e}e. Ce projet est accessible \`{a} partir de la th\'{e}orie des
groupes de Lie \cite{Baker}. Chercher \`{a} partir de l\`{a} des formules de
trace plus g\'{e}n\'{e}rales semble \^{e}tre un sujet d'une grande
importance.

En liaison avec des travaux de C. Procesi \cite{Procesi} une autre piste
concerne l'\'{e}tude d'une formule plus g\'{e}n\'{e}rale que celle de Fricke
pour la trace du commutateur de deux matrices $2\times 2$.

Egalement, en liaison avec ce qui a \'{e}t\'{e} vu pour les extrema positif
et n\'{e}gatif, il est int\'{e}ressant d'examiner les cons\'{e}quences pour
les approximations asym\'{e}triques des nombres irrationnels et le
r\'{e}sultat classique de B.\ Segre \cite{Alzer}.

\chapter{R\'{e}solution compl\`{e}te de nos \'{e}quations}

\section{Introduction}

Ayant identifi\'{e} une bonne g\'{e}n\'{e}ralisation de l'\'{e}quation de
Markoff classique, on a \'{e}tudi\'{e} ensuite la r\'{e}solution directe de
l'\'{e}quation diophantienne $M^{s_1s_2}(a,\partial K,u_\theta )$, o\`{u} $%
s_1$ et $s_2$ signes respectifs de $\varepsilon _1$ et $\varepsilon _2\in
\{-1,+1\}$, $a\in \mathbb{N}\backslash \{0\}$, $\partial K\in \mathbb{Z}$, $%
u_\theta \in \mathbb{Z}$ :
\[
x^2+\varepsilon _2y^2+\varepsilon _1z^2=(a+1)xyz+(\varepsilon _2\partial
K)yz-u_\theta x,
\]
\[
x,y,z\in \mathbb{N}\backslash \{0\}.
\]
Il s'agissait de comprendre comment s'organisent les triplets de solutions
que l'on note $(m,m_1,m_2)$.\ Une m\'{e}thode de r\'{e}solution a
\'{e}t\'{e} mise au point sur des cas particuliers $M^{++}(2,0,0)$, $%
M^{++}(2,0,-2)$, $M^{++}(3,0,1)$. Elle est essentiellement d\'{e}crite dans
\cite{Perrine4}. D\'{e}sormais cette m\'{e}thode est compl\`{e}te et permet
la r\'{e}solution de toutes les \'{e}quations $M^{s_1s_2}(a,\partial
K,u_\theta )$.

\section{M\'{e}thode de r\'{e}solution et cons\'{e}quences}

\subsection{Invariance par le groupe du triangle}

La m\'{e}thode de r\'{e}solution classique de l'\'{e}quation de Markoff
pr\'{e}sent\'{e}e dans \cite{Cassels}, en \'{e}vitant les redondances entre
des triplets de solutions pouvant se d\'{e}duire les uns des autres, casse
en r\'{e}alit\'{e} la structure de l'ensemble des solutions. Pour
l'\'{e}tendre \`{a} une \'{e}quation $M^{s_1s_2}(a,\partial K,u_\theta )$
mieux vaut consid\'{e}rer toutes les solutions, sans restriction.\ Pour
simplifier le probl\`{e}me il est aussi utile de consid\'{e}rer les
solutions dans $\mathbb{Z}^3$. Pour tout ensemble de solutions dans $\mathbb{Z}^3$%
, on dit que son intersection avec l'ensemble
$(\mathbb{N}\backslash \{0\})^3$ est son empreinte dans
$(\mathbb{N}\backslash \{0\})^3$.

Il existe diff\'{e}rentes possibilit\'{e}s pour d\'{e}duire une
solution dans $\mathbb{Z}^3$ d'une autre. L'\'{e}quation
$M^{s_1s_2}(a,\partial K,u_\theta )$ est invariante par les
involutions suivantes :
\[
N:(x,y,z)\longrightarrow (x,-y,-z).
\]
\[
X:(m,m_1,m_2)\longmapsto ((a+1)m_1m_2-m-u_\theta ,m_1,m_2)=(m^{\prime
},m_1,m_2),
\]
\[
Y:(m,m_1,m_2)\longmapsto (m,\varepsilon _2((a+1)mm_2+\varepsilon _2\partial
Km_2)-m_1,m_2)=(m,m_1^{\prime },m_2),
\]
\[
Z:(m,m_1,m_2)\longmapsto (m,m_1,\varepsilon _1((a+1)mm_1+\varepsilon
_2\partial Km_1)-m_2)=(m,m_1,m_2^{\prime }),
\]
On a les conditions
\[
N^2=X^2=Y^2=Z^2=Id.
\]
\[
XN=NX,\;YN=NY,\;ZN=NZ.
\]
Pour $\varepsilon _1=\varepsilon _2$, il existe une autre involution qui
laisse invariante l'\'{e}quation :
\[
P:(x,y,z)\longrightarrow (x,z,y).
\]
Elle v\'{e}rifie :
\[
P^2=Id,\;XP=PX,\;ZP=PY,\;YP=PZ,\;NP=PN.
\]

Modifiant $X$, remarquons que si on utilise $m_{\bullet }=(a+1)m_1m_2-m$ au
lieu de $m^{\prime }$, l'\'{e}quation $M^{s_1s_2}(a,\partial K,u_\theta )$
se transforme en une \'{e}quation de m\^{e}me forme qui s'\'{e}crit $%
M^{s_1s_2}(a,\partial K-\varepsilon _2u_\theta (a+1),-u_\theta )$. Cette
observation permet \'{e}ventuellement de concentrer l'attention sur les
\'{e}quations telles que $u_\theta =0$ ou $s=-u_\theta >0$.

Avec les involutions $X$, $Y$ et $Z$, s'introduit $\mathbf{T}_3$, le groupe
du triangle aussi not\'{e} $\mathbf{T}^{*}(\infty ,\infty ,\infty )$.\ C'est
le produit libre de trois groupes cycliques \`{a} deux \'{e}l\'{e}ments $%
\mathbf{C}_2$ :
\[
\mathbf{T}_3=\mathbf{C}_2*\mathbf{C}_2*\mathbf{C}_2.
\]
Par le th\'{e}or\`{e}me de la forme normale pour un tel produit libre \cite
{Cohen} (p. 26), tout \'{e}l\'{e}ment de $\mathbf{T}_3$ peut \^{e}tre
\'{e}crit comme un mot $ch=ch(X,Y,Z)$, produit d'involutions formelles $X$, $%
Y$, $Z$, dont deux lettres cons\'{e}cutives sont toujours diff\'{e}rentes.
Notre \'{e}quation est invariante par l'action du groupe $\mathbf{C}_2\times
\mathbf{T}_3$ construite avec $N$, $X$, $Y$, $Z$. Et comme sa partie la
moins \'{e}vidente vient de l'action induite de $\mathbf{T}_3$, c'est sur
cette derni\`{e}re que l'on met l'accent.

\subsection{Diff\'{e}rentes structures d'arbres sur le groupe du triangle}

Dans le cas particulier d'une action transitive et libre du groupe $\mathbf{T%
}_3$ sur un ensemble $\Omega $, on dit avec John H. Conway \cite{Conway} que
le $\mathbf{T}_3$-espace $\Omega $ est un topographe. Le groupe $\mathbf{T}%
_3 $ lui m\^{e}me peut \^{e}tre structur\'{e} en topographe. Il
poss\`{e}de donc une structure de graphe en forme d'arbre,
c'est-\`{a}-dire avec les
d\'{e}finitions de \cite{Serre} de graphe sans aucun circuit de forme $Cir_n$%
, o\`{u} $n\geq 1$. Ses sommets sont les \'{e}l\'{e}ments de $\mathbf{T}_3$,
la racine de l'arbre \'{e}tant l'unit\'{e} du groupe, et ses ar\^{e}tes sont
\'{e}tiquet\'{e}es avec $X$, $Y$, $Z$. Les chemins (ou g\'{e}od\'{e}siques)
de l'arbre sont aussi d\'{e}crits \`{a} partir de la racine par des mots $%
ch\in \mathbf{T}_3$, de sorte que les \'{e}l\'{e}ments de $\mathbf{T}_3$ se
repr\'{e}sentent de deux fa\c {c}ons, soit par les sommets du topographe
soit par ses chemins ayant pour origine sa racine. De chaque sommet sont
issues trois ar\^{e}tes qui correspondent \`{a} chaque lettre $X$, $Y$, ou $%
Z $.

Avec \cite{Perrine3} on a pu d\'{e}finir sur $\mathbf{T}_3$ une nouvelle
structure d'arbre sur l'ensemble des mots r\'{e}duits de $\mathbf{T}_3$ qui
commencent par $XY$ (suivi donc d'un mot commen\c {c}ant par $X$ ou $Z$,
\'{e}ventuellement vide).\ On dit qu'il s'agit des mots de Cohn.\ Ils sont
classables par longueur croissante avec les transformations $G$ et $D$
suivantes de $\mathbf{T}_3$ dans $\mathbf{T}_3$ :

$\bullet $ A gauche, on \'{e}crit le mot de d\'{e}part sous la forme $XW$,
et on fabrique $W^{\prime }$ \`{a} partir de $W$ en permutant $Y$ et $Z$. On
d\'{e}finit ensuite le transform\'{e} \`{a} gauche de $XW$ comme \'{e}tant
le mot $XYW^{\prime }$.\ Il est clair que pour $XW$ de longueur $n$ et
commen\c {c}ant par $XY$, son transform\'{e} est de longueur $n+1$ et
commence par $XYZ$. La transformation $G:XW\rightarrow XYW^{\prime }$ est
injective.

$\bullet $ A droite, on \'{e}crit le mot de d\'{e}part sous la forme $VW$,
o\`{u} $V$ ne contient que des lettres $X$ et $Y$ (au moins $2$), et $W$
commence par $Z$ ou est \'{e}ventuellement vide. On fabrique alors $%
V^{\prime }$ en permutant $X$ et $Y$ dans $V$. On d\'{e}finit ensuite $%
XV^{\prime }W$ comme \'{e}tant le transform\'{e} \`{a} droite de $VW$. Il
est \'{e}vident que le terme $XV^{\prime }W$ commence par $XYX$ et est de
longueur $n+1$ lorsque $VW$ commence par $XY$ et est de longueur $n$. La
transformation $D:VW\rightarrow XV^{\prime }W$ est injective.

On a obtenu ainsi une propri\'{e}t\'{e} qui a pu \^{e}tre utilis\'{e}e pour
montrer que dans la plupart des cas l'\'{e}quation $M^{s_1s_2}(a,\partial
K,u_\theta )$ poss\`{e}de une infinit\'{e} de solutions :

\begin{proposition}
Dans le groupe $\mathbf{T}_3$ engendr\'{e} par $X$, $Y$ et $Z$, pour toute
longueur $n\geq 2$ il existe $2^{n-2}$ mots de Cohn de longueur $n$. Ils
sont naturellement organis\'{e}s en arbre par les transformations $G$ et $D$
d\'{e}finies de $\mathbf{T}_3$ dans $\mathbf{T}_3$.
\end{proposition}

Egalement, on peut consid\'{e}rer dans $\mathbf{T}_3$ l'ensemble des mots
r\'{e}duits qui commencent par $X$ (suivi donc d'un mot commen\c {c}ant par $%
Y$ ou $Z$, \'{e}ventuellement vide).\ On dit qu'il s'agit des mots de
Cassels. En changeant $Y$ en $Z$ dans la proposition pr\'{e}c\'{e}dente, on
a facilement :

\begin{proposition}
Dans le groupe $\mathbf{T}_3$ engendr\'{e} par $X$, $Y$ et $Z$, pour toute
longueur $n\geq 1$ il existe $2^{n-1}$ mots de Cassels de longueur $n$. Ils
sont naturellement organis\'{e}s en arbre.
\end{proposition}

\subsection{Le groupe du triangle dans $GL(2,\mathbb{Z})$}

Dans \cite{Perrine1b}, et en tirant les cons\'{e}quences de la
th\'{e}orie de Markoff classique, on a montr\'{e} comment le
groupe $\mathbf{T}_3$ est \'{e}troitement li\'{e} au groupe
$GL(2,\mathbb{Z})$. On consid\`{e}re pour
cela, avec le morphisme d'ab\'{e}lianisation $\pi ^{\prime }$ du groupe $Aut(%
\mathbf{F}_2)$ \`{a} valeurs dans $GL(2,\mathbb{Z})$, deux
matrices engendrant dans $GL(2,\mathbb{Z})$ un groupe di\'{e}dral
$\mathbf{D}_6$ \`{a} $12$ \'{e}l\'{e}ments :
\[
\pi ^{\prime }(t)=\left[
\begin{array}{cc}
1 & 1 \\
-1 & 0
\end{array}
\right] ,\;\;\pi ^{\prime }(o)=\left[
\begin{array}{cc}
0 & -1 \\
-1 & 0
\end{array}
\right] .
\]
On compl\`{e}te en consid\'{e}rant trois matrices d'ordre 2 :
\[
\pi ^{\prime }(X_0)=\left[
\begin{array}{cc}
1 & 0 \\
-2 & -1
\end{array}
\right] ,\;\;\pi ^{\prime }(Y_0)=\left[
\begin{array}{cc}
-1 & -2 \\
0 & 1
\end{array}
\right] ,\;\;\pi ^{\prime }(Z_0)=\left[
\begin{array}{cc}
1 & 0 \\
0 & -1
\end{array}
\right] .
\]
Elles permettent de faire agir le groupe $\mathbf{T}_3$ dans
$GL(2,\mathbb{Z})$ en d\'{e}finissant le produit suivant o\`{u}
$ch\in \mathbf{T}_3$ et $\pi _0^{\prime }(\mathbf{T}_3)$ de fa\c
{c}on \'{e}vidente
\[
ch(\pi ^{\prime }(X_0),\pi ^{\prime }(Y_0),\pi ^{\prime }(Z_0))=\pi
_0^{\prime }(ch(X,Y,Z))\in \pi _0^{\prime }(\mathbf{T}_3).
\]
On en a d\'{e}duit la d\'{e}composition ternaire repr\'{e}sentant le groupe $%
\mathbf{T}_3$ dans $GL(2,\mathbb{Z})$ :

\begin{proposition}
Tout \'{e}l\'{e}ment $V\in GL(2,\mathbb{Z})$ se d\'{e}compose
d'une et d'une seule fa\c {c}on sous la forme
\[
\pi ^{\prime }(o)^h\pi ^{\prime }(t)^kch(\pi ^{\prime }(X_0),\pi ^{\prime
}(Y_0),\pi ^{\prime }(Z_0)),\text{ }
\]
\[
\text{o\`{u} }h=0,1;\;\;k=0,1,...,5;\text{\ \ }ch\in \mathbf{T}_3.
\]
Les \'{e}l\'{e}ments de $\pi _0^{\prime }(\mathbf{T}_3)$, sont
caract\'{e}ris\'{e}s par les conditions $h=0$ et $k=0$. Le groupe $\pi
_0^{\prime }(\mathbf{T}_3)$ n'est pas normal dans le groupe $GL(2,\mathbb{Z})$%
.\ Il est isomorphe par $\pi _0^{\prime }$ au groupe
$\mathbf{T}_3$. Les \'{e}l\'{e}ments du groupe $\mathbf{D}_6$ non
normal dans $GL(2,\mathbb{Z})$ sont caract\'{e}ris\'{e}s par la
condition
\[
ch(\pi ^{\prime }(X_0),\pi ^{\prime }(Y_0),\pi ^{\prime }(Z_0))=\mathbf{1}%
_2.
\]
\end{proposition}

Le groupe $\mathbf{D}_6$ introduit deux relations
d'\'{e}quivalence entre \'{e}l\'{e}ments de $GL(2,\mathbb{Z})$
\[
V_1\;\Re _{\mathbf{D}_6}\;V_2\;\;\Leftrightarrow \;\;V_1V_2^{-1}\in \mathbf{D%
}_6\;\;\Leftrightarrow \;\;V_2\in \mathbf{D}_6V_1,
\]
\[
V_1\;_{\mathbf{D}_6}\Re \;V_2\;\;\Leftrightarrow \;\;V_1^{-1}V_2\in \mathbf{D%
}_6\;\;\Leftrightarrow \;\;V_2\in V_1\mathbf{D}_6.
\]
Le quotient \`{a} droite $GL(2,\mathbb{Z})/\Re _{\mathbf{D}_6}=(GL(2,\mathbb{Z})/%
\mathbf{D}_6)_d$ des classes $\mathbf{D}_6V_1$ et le quotient \`{a} gauche $%
GL(2,\mathbb{Z})/_{\mathbf{D}_6}\Re
=(GL(2,\mathbb{Z})/\mathbf{D}_6)_g$ des classes $V_1\mathbf{D}_6$
o\`{u} $V_1\in GL(2,\mathbb{Z})$ sont \'{e}quipotents. Ces deux
ensembles sont diff\'{e}rents car $\mathbf{D}_6$ n'est pas normal
dans le groupe $GL(2,\mathbb{Z})$. L'\'{e}criture de $V\in
GL(2,\mathbb{Z})$ dans le dernier r\'{e}sultat \'{e}nonc\'{e}
donne
\[
Vch(\pi ^{\prime }(X_0),\pi ^{\prime }(Y_0),\pi ^{\prime }(Z_0))\text{ }%
^{-1}=\pi ^{\prime }(o)^h\pi ^{\prime }(t)^k\in \mathbf{D}_6.
\]
Elle d\'{e}termine un unique \'{e}l\'{e}ment $ch(\pi ^{\prime }(X_0),\pi
^{\prime }(Y_0),\pi ^{\prime }(Z_0))\in \pi _0^{\prime }(\mathbf{T}_3)$ tel
que
\[
V\;\Re _{\mathbf{D}_6}\;ch(\pi ^{\prime }(X_0),\pi ^{\prime }(Y_0),\pi
^{\prime }(Z_0)).\;
\]
D'o\`{u} une autre interpr\'{e}tation du topographe qui est identifiable
\`{a} l'arbre complet de la th\'{e}orie de Markoff ou encore au groupe du
triangle $\mathbf{T}_3$ :

\begin{proposition}
Le groupe $\mathbf{T}_3$ est \'{e}quipotent au quotient (\`{a} droite ou
\`{a} gauche) du groupe $GL(2,\mathbb{Z})$ par son sous-groupe non normal $%
\mathbf{D}_6$. C'est en particulier un $GL(2,\mathbb{Z})$-espace
homog\`{e}ne.
\end{proposition}

On a pu en d\'{e}duire une proposition pr\'{e}alable \`{a} des r\'{e}sultats
connus de la $K$-th\'{e}orie (\cite{Rotman} (p. 193), \cite{Rosenberg} (p.
218 et p. 75), \cite{Soule} (p. 261), \cite{Swinnerton}).

\begin{proposition}
On a pour $GL(2,\mathbb{Z})$ les groupes d'homologie suivants
\[
H_1(GL(2,\mathbb{Z}),\mathbb{Z})=GL(2,\mathbb{Z})/[GL(2,\mathbb{Z}),GL(2,\mathbb{Z})]\simeq
\mathbf{D}_6/[\mathbf{D}_6,\mathbf{D}_6]\simeq \mathbf{C}_2\times \mathbf{C}%
_2,
\]
\[
H_2(GL(2,\mathbb{Z}),\mathbb{Z})\simeq \mathbf{C}_2.
\]
\end{proposition}

En utilisant le groupe libre \`{a} deux \'{e}l\'{e}ments
$\mathbf{F}_2\simeq [SL(2,\mathbb{Z}),SL(2,\mathbb{Z})]$, dont on
a montr\'{e} dans \cite{Perrine1b} qu'il est reli\'{e} \`{a}
l'\'{e}quation de Markoff classique, on a obtenu :

\begin{proposition}
Tout \'{e}l\'{e}ment $V\in GL(2,\mathbb{Z})$ se d\'{e}compose
d'une et d'une seule fa\c {c}on sous la forme
\[
\pm W(A_0,B_0)O^hW_k(S,T),\text{ }
\]
\[
h\in \{0,1\},
\]
\[
W(A_0,B_0)\in \mathbf{F}_2=[SL(2,\mathbb{Z}),SL(2,\mathbb{Z})],
\]
\[
W_k(S,T)\in \{\mathbf{1}_2,S,ST,STS,STST,STSTS\}\;\text{avec }k=0,1,...,5.
\]
Les \'{e}l\'{e}ments du sous-groupe $SL(2,\mathbb{Z})$ normal dans $GL(2,\mathbb{Z}%
)$ sont caract\'{e}ris\'{e}s par la condition $h=0$.
\end{proposition}

Les matrices cit\'{e}es dans cette proposition sont les trois
g\'{e}n\'{e}rateurs de $GL(2,\mathbb{Z})$ :
\[
S=\left[
\begin{array}{cc}
0 & -1 \\
1 & 0
\end{array}
\right] ,\;\;T=\left[
\begin{array}{cc}
1 & 1 \\
0 & 1
\end{array}
\right] ,\;\;O=\left[
\begin{array}{cc}
-1 & 0 \\
0 & 1
\end{array}
\right] ,
\]
ainsi que des mots $W(A_0,B_0)$ \'{e}crits multiplicativement en fonction
des deux commutateurs qui engendrent $\mathbf{F}_2$ d'apr\`{e}s \cite
{MagnusKarassSolitar} (p. 97-98) :
\[
A_0=[(TS)^{-1},S^{-1}]=\left[
\begin{array}{cc}
1 & 1 \\
1 & 2
\end{array}
\right] ,\;B_0=[(TS)^{-2},S^{-1}]^{-1}=\left[
\begin{array}{cc}
1 & -1 \\
-1 & 2
\end{array}
\right] .
\]
On a explicit\'{e} tous les passages entre les deux
repr\'{e}sentations ternaires des matrices du groupe
$GL(2,\mathbb{Z})$, groupe dont on a pu
\'{e}galement retrouver une pr\'{e}sentation \`{a} deux g\'{e}n\'{e}rateurs $%
T$ et $I=OS$ qui est minimale \cite{BeylRosenberger} :
\[
GL(2,\mathbb{Z})=<I,T^{-1}\mid I^2=([T^{-1},I]T^{-1})^4=([T^{-1},I]T^{-1}I)^2=%
\mathbf{1}_2>.
\]
Le sous-groupe $\pi _0^{\prime }(\mathbf{T}_3)$ est engendr\'{e} par trois
matrices calculables en $I$ et $T^{-1}$ :
\[
\pi ^{\prime }(X_0)=T^{-1}IOT^{-1}IOIT^{-1}B_0^{-1},\;\;\pi ^{\prime
}(Y_0)=IOIOA_0^{-1}TS,\;\;\pi ^{\prime }(Z_0)=IS.
\]
De plus \cite{BeylRosenberger} le groupe du triangle
$\mathbf{T}_3$ est isomorphe \`{a} $PGL(2,\mathbb{Z})$ avec :
\[
PGL(2,\mathbb{Z})=<\overline{I},\overline{T}^{-1}\mid \overline{I}^2=([%
\overline{T}^{-1},\overline{I}]\overline{T}^{-1})^2=([\overline{T}^{-1},%
\overline{I}]\overline{T}^{-1}\overline{I})^2=\mathbf{1}>.
\]
On peut v\'{e}rifier que $\mathbf{F}_2\simeq
[PSL(2,\mathbb{Z}),PSL(2,\mathbb{Z})]$ est d'indice $2$ dans ce
groupe, et que l'on a aussi :
\[
\lbrack PGL(2,\mathbb{Z}),PGL(2,\mathbb{Z})]=<[\overline{I},\overline{T}^{-1}],[%
\overline{I},\overline{T}]\mid [\overline{I},\overline{T}^{-1}]^3=[\overline{%
I},\overline{T}]^3=\mathbf{1}>\simeq \mathbf{C}_3\star \mathbf{C}_3.
\]

\subsection{For\^{e}t et bouquets de solutions}

R\'{e}soudre l'\'{e}quation $M^{s_1s_2}(a,\partial K,u_\theta )$ dans $\mathbb{Z%
}^3$ consiste \`{a} d\'{e}terminer la structure du $\mathbf{T}_3$-espace de
ses triplets de solutions. C'est une union de $\mathbf{T}_3$-espaces
connexes (des $\mathbf{T}_3$-orbites).\ On dit alors que chaque $\mathbf{T}_3$%
-espace connexe de solutions dans $\mathbb{Z}^3$ est un bouquet.\ On le note $%
Bq\subset \mathbb{Z}^3$. L'union des bouquets possibles $Bq_1$, $Bq_2$, ...., $%
Bq_n$, ..., est la for\^{e}t des solutions dans $\mathbb{Z}^3$ de
l'\'{e}quation $M^{s_1s_2}(a,\partial K,u_\theta )$. Bouquets et
for\^{e}t \'{e}tant des $\mathbf{T}_3$-espaces, ils peuvent
\^{e}tre structur\'{e}s comme un graphe dont les sommets sont les
triplets de solutions et dont les ar\^{e}tes sont non
orient\'{e}es. De chaque sommet partent trois ar\^{e}tes. Chaque
ar\^{e}te est \'{e}tiquet\'{e}e par l'involution $X$, $Y$ ou $Z$
permettant de passer d'une extremit\'{e} de l'ar\^{e}te \`{a}
l'autre. Les d\'{e}finitions de \cite{Serre} s'appliquent encore,
permettant de consid\'{e}rer aussi des arbres de solutions, ce
sont des graphes sans aucun circuit de forme $Cir_n$, o\`{u}
$n\geq 1$. L'\'{e}tude d'exemples montre que tous les bouquets de
solutions que l'on rencontre ne sont pas des arbres.

\subsection{Hauteur et r\'{e}duction des triplets de solutions}

Pour tout triplet $(m,m_1,m_2)\in \mathbb{Z}^3$ de solutions de l'\'{e}quation $%
M^{s_1s_2}(a,\partial K,u_\theta )$, on d\'{e}finit sa hauteur
\[
h=\max (\mid m\mid ,\mid m_1\mid ,\mid m_2\mid )\geq 0.
\]
On peut consid\'{e}rer trois autres valeurs construites avec les involutions
$X$, $Y$, $Z$ :
\[
h_X=\max (\mid m^{\prime }\mid ,\mid m_1\mid ,\mid m_2\mid ),
\]
\[
h_Y=\max (\mid m\mid ,\mid m_1^{\prime }\mid ,\mid m_2\mid ),
\]
\[
h_Z=\max (\mid m\mid ,\mid m_1\mid ,\mid m_2^{\prime }\mid ).
\]
On dit qu'un triplet $(m,m_1,m_2)$ n'est pas fondamental si et seulement si
l'un des nombres $h_X$, $h_Y$, $h_Z$ est strictement plus petit que $h$.
Dans le cas contraire, un triplet $(m,m_1,m_2)$ qui ne v\'{e}rifie pas cette
derni\`{e}re condition est appel\'{e} fondamental.\ Les in\'{e}galit\'{e}s
qui caract\'{e}risent cette situation permettent d'identifier les triplets
fondamentaux, chacun d'entre eux d\'{e}finissant un bouquet de solutions par
l'action du groupe $\mathbf{T}_3$.

Consid\'{e}rons un triplet quelconque d'un bouquet de solutions de
l'\'{e}quation $M^{s_1s_2}(a,\partial K,u_\theta )$. Si $h_X<h$ on applique $%
X$ et on change de triplet, si $h_Y<h$ on applique $Y$ et on change de
triplet, si $h_Z<h$ on applique $Z$ et on change de triplet. Ceci donne un
algorithme dont l'avancement dans le bouquet que l'on consid\`{e}re est
contr\^{o}l\'{e} par la r\'{e}duction de la hauteur qui d\'{e}croit en
restant positive.\ Losque la hauteur est minimale, on identifie un triplet
fondamental dans le bouquet consid\'{e}r\'{e} pour l'\'{e}quation $%
M^{s_1s_2}(a,\partial K,u_\theta )$. On dispose ainsi d'une
m\'{e}thode analogue \`{a} la descente infinie de Fermat pour
calculer toutes les solutions dans $\mathbb{Z}^3$ de cette
\'{e}quation, et les classer en bouquets.

Si l'on travaille dans $(\mathbb{N}\backslash \{0\})^3$ la hauteur
est d\'{e}finie sans valeur absolue.\ Il se peut que pour un
triplet donn\'{e} l'algorithme pr\'{e}c\'{e}dent ne permette plus
par application de $X$, $Y$ ou $Z$, de trouver un nouveau triplet
dans l'ensemble $(\mathbb{N}\backslash \{0\})^3$.\ Un tel triplet
sur lequel l'algorithme s'arr\^{e}te est dit minimal.

\subsection{Solutions fondamentales dans $(\mathbb{N}\backslash \{0\})^3$}

On a un r\'{e}sultat de finitude g\'{e}n\'{e}ral \cite{Perrine9}
pour les solutions fondamentales d'une \'{e}quation
$M^{s_1s_2}(a,\partial K,u_\theta )$ :

\begin{proposition}
Consid\'{e}rons les solutions dans $(\mathbb{N}\backslash
\{0\})^3$ d'une \'{e}quation diophantienne $M^{s_1s_2}(a,\partial
K,u_\theta )$.\ Elles ne sont fondamentales que dans un nombre
fini de cas, hors le cas des \'{e}quations $M^{--}(a,-2-u_\theta
(a+1),u)$ o\`{u} $u_\theta <0$ :
\[
x^2-y^2-z^2=(a+1)xyz+(u_\theta (a+1)-2)yz-u_\theta x.
\]
\ Ces derni\`{e}res ont une infinit\'{e} de solutions fondamentales valant $%
(-u_\theta ,m_1,m_1)$, avec $m_1\in \mathbb{N}\backslash \{0\}$
quelconque, et les bouquets correspondants, en nombre infini, sont
finis et s'\'{e}crivent
\[
\{(-u_\theta ,m_1,m_1),((a+1)m_1^2,m_1,m_1)\}.
\]
En dehors de ces cas particuliers, on ne trouve ainsi qu'un nombre
fini de bouquets pour l'action du groupe $\mathbf{T}_3$ ayant une
empreinte non vide dans $(\mathbb{N}\backslash \{0\})^3$.
\end{proposition}

Ce r\'{e}sultat a donn\'{e} une proposition garantissant qu'on ne trouve
dans l'essentiel des cas qu'un nombre fini de solutions fondamentales.

\begin{proposition}
Consid\'{e}rons les solutions dans $(\mathbb{N}\backslash
\{0\})^3$ d'une \'{e}quation diophantienne $M^{s_1s_2}(a,\partial
K,u_\theta )$.\ Si elle poss\`{e}de une empreinte de bouquet
contenant une infinit\'{e} de solutions distinctes, alors elle n'a
qu'un nombre fini de bouquets pour l'action du groupe
$\mathbf{T}_3$ ayant une empreinte non vide dans
$(\mathbb{N}\backslash \{0\})^3$.
\end{proposition}

\subsection{Solutions minimales dans $(\mathbb{N}\backslash \{0\})^3$}

Certaines empreintes de bouquet ne sont identifiables que gr\^{a}ce \`{a}
des solutions minimales.\ Pour ces derni\`{e}res, on a la
caract\'{e}risation suivante \cite{Perrine9} :

\begin{proposition}
Soit une solution $(m,m_1,m_2)\in (\mathbb{N}\backslash \{0\})^3$
d'une \'{e}quation diophantienne $M^{s_1s_2}(a,\partial K,u_\theta
)$ v\'{e}rifiant \`{a} une inversion pr\`{e}s des indices la
condition $m_1\geq m_2\geq 1$. Elle est minimale si et seulement
si on a l'une des conditions suivantes :
\[
\varepsilon _2m_1^2+\varepsilon _1m_2^2-\varepsilon _2\partial Km_1m_2\leq
0,\;\;\varepsilon _2m^2+\varepsilon _1\varepsilon _2m_2^2+\varepsilon
_2u_\theta m\leq 0.
\]
\end{proposition}

Il se peut qu'une \'{e}quation $M^{s_1s_2}(a,\partial K,u_\theta
)$ ait un nombre fini de solutions minimales, et aucune solution
fondamentale.\ C'est le cas de l'\'{e}quation $M^{++}(2,0,-2)$.\
Pour $\varepsilon _1=\varepsilon _2=1$, les deux conditions
$\partial K\leq 2$ et $u_\theta \leq 0$ ne donnent qu'un nombre
fini de solutions minimales et de solutions fondamentales. Dans ce
cas, on a \'{e}tabli l'existence d'un nombre fini d'empreintes de
bouquets de solutions dans $(\mathbb{N}\backslash \{0\})^3$ pour
l'\'{e}quation $M^{++}(a,\partial K,u_\theta )$. Pour les autres
cas, la situation est assez diverse en fonction des param\`{e}tres
$a$, $\partial K$, $u_\theta $, mais dans l'essentiel des cas le
nombre d'empreintes de bouquet reste fini.

\subsection{Les triplets de Cohn et leur utilisation}

On dit qu'une solution $(m,m_1,m_2)\in (\mathbb{N}\backslash
\{0\})^3$ d'une \'{e}quation $M^{s_1s_2}(a,\partial K,u_\theta )$
est un triplet de Cohn \cite{Cohn} si et seulement si on a
$m>m_1>m_2$. Toutes les solutions possibles dans
$(\mathbb{N}\backslash \{0\})^3$ ne sont pas de ce type, comme le
montre le cas o\`{u} $\varepsilon _1=\varepsilon _2$ et une
permutation de $y$ et $z$ dans l'\'{e}quation \'{e}tudi\'{e}e.
Mais de telles solutions apparaissent naturellement \`{a} l'issue
des calculs du chapitre pr\'{e}c\'{e}dent. En effet toute paire de
suites $X_2$ et $T$ d\'{e}termine des fractions continues de plus
en plus longues expliquant a posteriori les in\'{e}galit\'{e}s
d\'{e}finissant les triplets de Cohn :
\[
m_2/k_2=[\lhd X_2^{*}],\;\;m_1/k_1=[\lhd X_2^{*},c,T],\;\;m/K_1=[\lhd
X_2^{*},c,T,b,X_2].
\]
On a pu v\'{e}rifier que les triplets de Cohn d'une m\^{e}me empreinte de
bouquet sont donn\'{e}s par des chemins de $\mathbf{T}_3$ commen\c {c}ant
par $XY$. A partir de telles suites, on a mis au point un proc\'{e}d\'{e} de
construction d'un arbre de triplets de Cohn pour nos \'{e}quations \cite
{Perrine3}. On a utilis\'{e} pour cela les combinaisons $G$, $DD$, $GD$, des
transformations $G$ et $D$ mises en \'{e}vidence dans le groupe $\mathbf{T}%
_3 $, ceci donne des triplets de Cohn lorsque les suites
associ\'{e}es sont bien d\'{e}finies, c'est-\`{a}-dire \`{a}
coefficients entiers positifs (comme vu dans \cite{Perrine7} les
op\'{e}rateurs $\lhd $ et $\rhd $ peuvent cr\'{e}er des
probl\`{e}mes correspondant au fait que le bouquet concern\'{e}
n'est pas un arbre). Pour cela on change d'abord l'\'{e}quation
$M^{s_1s_2}(a,\partial K,u_\theta )$ en une \'{e}quation \'{e}quilibr\'{e}e $%
M^{s_1s_2}(c,\partial K_c,u)$ assurant la condition $b=c$ et ne modifiant
pas les suites $X_2$ et $T$.

\subsection{La construction algorithmique \`{a} droite et \`{a} gauche}

Les formules pour des transformations $G$, $DD$, $GD$, donnant un triplet de
Cohn \`{a} partir d'un autre sont les suivantes pour l'\'{e}quation $%
M^{s_1s_2}(c,\partial K_c,u)$ :

$\bullet $ La construction \`{a} gauche est d\'{e}finie sur les suites par :
\[
X_2^G=(\lhd T^{*},c,X_2),\;\;T^G=T.
\]
On en d\'{e}duit
\[
X_1^G=(\lhd X_2^{*},c,T\rhd ,c,T),
\]
\[
(S^G\rhd )=(X_2^{*},c,T\rhd ,c,T^{*},c,\lhd T^{*},c,X_2).
\]
L'\'{e}quation diophantienne correspondant aux nouvelles suites et dont le
triplet de Cohn $(m^G,m_1^G,m_2^G)$ est une solution, s'\'{e}crit:
\[
M^{s_2,s_1}(c,\partial K_c,\varepsilon _1\varepsilon _2u):x^2+\varepsilon
_1y^2+\varepsilon _2z^2=(c+1)xyz+\varepsilon _1\partial K_cyz-\varepsilon
_1\varepsilon _2ux.
\]

$\bullet $ La construction \`{a} droite est plus complexe.\ Ceci a
\'{e}t\'{e} d\'{e}couvert dans \cite{Perrine1}. On doit en r\'{e}alit\'{e}
distinguer deux cas. En partant deux fois \`{a} droite, on d\'{e}finit
\[
X_2^{DD}=X_2^{*},\;\;T^{DD}=(\lhd X_2^{*},c,T,c,X_2\rhd ).
\]
Ceci donne :
\[
X_1^{DD}=(\lhd X_2,c,\lhd X_2^{*},c,T,c,X_2\rhd ),
\]
\[
(S^{DD}\rhd )=(X_2,c,\lhd X_2^{*},c,T^{*},c,X_2\rhd ,c,X_2^{*}).
\]
L'\'{e}quation diophantienne correspondant aux nouvelles suites et dont le
triplet de Cohn $(m^{DD},m_1^{DD},m_2^{DD})$ est solution, s'\'{e}crit :
\[
M^{s_1,s_2}(c,\partial K_c,u):x^2+\varepsilon _2y^2+\varepsilon
_1z^2=(c+1)xyz+\partial K_cyz-\varepsilon _2ux.
\]

$\bullet $ La construction \`{a} gauche une fois apr\`{e}s un passage \`{a}
droite est d\'{e}finie avec :
\[
X_2^{DG}=(\lhd X_2^{*},c,T),\;\;T^{DG}=(X_2^{*},c,T^{*},c,X_2).
\]
Ceci donne pour les autres suites que l'on consid\`{e}re
\[
X_1^{DG}=(\lhd T^{*},c,X_2\rhd ,c,X_2^{*},c,T^{*},c,X_2),
\]
\[
(S^{DG}\rhd )=(T^{*},c,X_2\rhd ,c,X_2^{*},c,T,c,X_2,c,\lhd X_2^{*},c,T).
\]
On trouve encore une \'{e}quation diophantienne correspondant aux nouvelles
suites, dont le triplet de Cohn $(m^{DG},m_1^{DG},m_2^{DG})$ est une
solution :
\[
M^{s_2,s_1}(c,\varepsilon _2\partial K_c,\varepsilon _1u):x^2+\varepsilon
_1y^2+\varepsilon _2z^2=(c+1)xyz+\varepsilon _1\varepsilon _2\partial
K_cyz-\varepsilon _1ux.
\]

\subsection{Cons\'{e}quence pour la r\'{e}solution de nos \'{e}quations}

Les transformations $G$, $DD$, $GD$, ont donn\'{e} le r\'{e}sultat
suivant pour l'\'{e}quation $M^{s_1s_2}(c,\partial K_c,u)$ :

\begin{proposition}
Consid\'{e}rons un triplet de Cohn $(m,m_1,m_2)$ associ\'{e} \`{a} deux
suites $X_2$ et $T$, solution de l'\'{e}quation diophantienne
\'{e}quilibr\'{e}e
\[
M^{s_1s_2}(c,\partial K_c,u):x^2+\varepsilon _2y^2+\varepsilon
_1z^2=(c+1)xyz+\varepsilon _2\partial K_cyz-ux.
\]
On obtient pour les \'{e}quations diophantiennes transform\'{e}es \`{a}
droite et \`{a} gauche de la pr\'{e}c\'{e}dente les expressions
\[
G:M^{s_1s_2}(c,\partial K_c,u)\longmapsto M^{s_2s_1}(c,\partial
K_c,\varepsilon _1\varepsilon _2u),
\]
\[
DD:M^{s_1s_2}(c,\partial K_c,u)\longmapsto M^{s_1s_2}(c,\varepsilon
_2\partial K_c,\varepsilon _2u),
\]
\[
GD:M^{s_1s_2}(c,\partial K_c,u)\longmapsto M^{s_2s_1}(c,\varepsilon
_2\partial K_c,\varepsilon _1u).
\]
De plus le processus de construction donn\'{e} sur les suites fournit,
lorsque les suites sont bien d\'{e}finies, un triplet de Cohn solution de
l'\'{e}quation correspondante, de taille strictement plus grande que celle
du triplet $(m,m_1,m_2)$. Il existe alors une infinit\'{e} de solutions pour
l'\'{e}quation $M^{s_1s_2}(c,\partial K_c,u)$ et un nombre fini d'empreintes
de bouquets correspondantes.
\end{proposition}

La transposition \`{a} des valeurs $a$ ou $b$ diff\'{e}rentes de $c$ ne pose
pas de probl\`{e}me, donnant un r\'{e}sultat analogue pour $%
M^{s_1s_2}(a,\partial K,u_\theta )$ ou $M^{s_1s_2}(b,\partial K,u)$.

\subsection{Construction des suites de d\'{e}part $X_2$ et $T$}

Les nombres $\varepsilon _1$, $\varepsilon _2$, $a$, $\partial K$,
$u_\theta $ sont donn\'{e}s par l'\'{e}quation
$M^{s_1s_2}(a,\partial K,u_\theta )$ que l'on consid\`{e}re.
Disposant par la m\'{e}thode de r\'{e}solution de cette
\'{e}quation d'un triplet $(m,m_1,m_2)\in (\mathbb{N}\backslash
\{0\})^3$ de solutions, on peut construire deux suites
associ\'{e}es $X_1$ et $X_2$ en r\'{e}solvant les \'{e}quations de
Bezout en $(K_1,k_1)$ et $(K_2,k_2)$. On se ram\`{e}ne alors \`{a}
une \'{e}quation $M^{s_1s_2}(b,\partial K,u)$.

\subsubsection{ Cas particulier o\`{u} $\varepsilon _1=\varepsilon _2$}

Dans tous les exemples \'{e}tudi\'{e}s o\`{u} $\varepsilon _1=\varepsilon _2$%
, on a trouv\'{e} un cas o\`{u} $T=\emptyset $.\ On a pu d\'{e}montrer que
cette remarque est g\'{e}n\'{e}rale.\

\begin{proposition}
Consid\'{e}rons une \'{e}quation $M^{s_1s_2}(b,\partial K,u)$ o\`{u} $%
\varepsilon _1=\varepsilon _2$%
\[
x^2+\varepsilon _2y^2+\varepsilon _2z^2=(b+1)xyz+\varepsilon _2\partial
Kyz-ux,
\]
telle que l'on puisse trouver $m_1$ et $m_2$ dans
$\mathbb{N}\backslash \{0\}$ v\'{e}rifiant
\[
m_1^2-(b+\partial K+1)m_1m_2+m_2^2=-u-\varepsilon _2.
\]
Alors elle poss\`{e}de un triplet de solutions $(m,m_1,m_2)$ tel que
\[
m=m_1^2-\partial Km_1m_2+m_2^2.
\]
En notant $c=b+\partial K$ et dans le cas o\`{u} l'on a $m_1-cm_2\in \mathbb{N}%
\backslash \{0\}$, condition assur\'{e}e si $u<0$, on peut construire une
infinit\'{e} de solutions de l'\'{e}quation \'{e}quilibr\'{e}e associ\'{e}e
gr\^{a}ce aux transformations $G$, $DD$, $GD$, avec $T=\emptyset $ et $X_2$
suite d\'{e}finie avec $k_{21}=m_1-cm_2>0$ par
\[
\frac{m_2}{m_1-cm_2}=[X_2],\;\;\det (M_{X_2})=\varepsilon _2.
\]
\end{proposition}

Dans tous ces cas on a la solution $(\varepsilon _2,m_1,m_2)$ pour
l'\'{e}quation $M^{s_1s_2}(b,\partial K,u)$ :
\[
m_1^2-(b+1+\partial K)m_1m_2+m_2^2=-u-\varepsilon _2.
\]

La valeur de $\varepsilon _1\varepsilon _2$ est une forte contrainte, elle
impose $\varepsilon _S=-1$.\ En r\'{e}alit\'{e}, si l'on \'{e}tudie des
nombres $\theta _a(S)$ on peut toujours changer la suite $S$ en $S^{\prime
}=(S,a,S)$, et se ramener avec cette derni\`{e}re suite \`{a} $\varepsilon
_{S^{\prime }}=-1$. Moyennant cette transformation d'allongement de la suite
$S$, on peut par exemple dans l'\'{e}tude des constantes de Markoff faire en
sorte que la contrainte $\varepsilon _1=\varepsilon _2$ soit toujours
v\'{e}rifi\'{e}e. On peut appliquer alors l'involution $P$ de fa\c {c}on
\`{a} ce que la longueur de la suite $\lhd X_1$ soit plus grande ou
\'{e}gale \`{a} la longueur de la suite $X_2$. Cette normalisation ne change
pas l'\'{e}quation \'{e}tudi\'{e}e mais donne naturellement un triplet de
Cohn.

\subsubsection{Cas g\'{e}n\'{e}ral pour $\varepsilon _1$ et $\varepsilon _2$}

La proposition qui pr\'{e}c\`{e}de a \'{e}t\'{e} g\'{e}n\'{e}ralis\'{e}e au
cas o\`{u} l'on n'a plus n\'{e}cessairement la condition $\varepsilon
_1=\varepsilon _2$ ni a fortiori la normalisation introduite avant. On a
trouv\'{e} par exemple pour $T=(1)$ :

\begin{proposition}
On consid\`{e}re un triplet $(m,m_1,m_2)\in \mathbb{Z}^3$
v\'{e}rifiant les deux relations
\[
-u-\varepsilon _2=m_1^2-(b+\partial K+1)m_1m_2+\varepsilon _1\varepsilon
_2m_2^2,
\]
\[
m=m_1^2-\partial Km_1m_2+\varepsilon _1\varepsilon _2m_2^2.
\]
Il est solution de l'\'{e}quation $M^{s_1s_2}(b,\partial K,u)$. Si ce
triplet correspond \`{a} une suite $T=(1)$ avec laquelle on peut \'{e}crire $%
X_1=(\lhd X_2^{*},c,1)$, on a
\[
\varepsilon _1=-\varepsilon _2,\;\partial K=(c-b),\;m_1=(c+1)m_2+k_{21},
\]
\[
u+\varepsilon _2=m_2^2-(c+1)m_2k_{21}-k_{21}^2=\Psi _{(c,1)}(m_2,k_{21}).
\]
Avec $c=b+\partial K$ et dans le cas o\`{u} $m_1-(c+1)m_2\in \mathbb{N}%
\backslash \{0\}$, condition assur\'{e}e si $u<0$, on peut construire une
infinit\'{e} de solutions de l'\'{e}quation \'{e}quilibr\'{e}e associ\'{e}e
gr\^{a}ce aux transformations $G$, $DD$, $GD$, avec $T=(1)$ et $X_2$ suite
d\'{e}finie avec $k_{21}=m_1-(c+1)m_2>0$ par
\[
\frac{m_2}{m_1-(c+1)m_2}=[X_2],\;\;\det (M_{X_2})=\varepsilon _2.
\]
\end{proposition}

Les premi\`{e}res \'{e}galit\'{e}s de cette proposition proviennent des
relations suivantes du cas g\'{e}n\'{e}ral, sp\'{e}cialis\'{e}es compte tenu
de la suite $T$ choisie :
\[
-u-\varepsilon _2\mu =m-(b+1)m_1m_2,\;\;\mu m=m_1^2-\partial
Km_1m_2+\varepsilon _1\varepsilon _2m_2^2.
\]

\subsection{Remarques compl\'{e}mentaires}

On a dans le cas g\'{e}n\'{e}ral une forme quadratique $\Psi _{(c,T)}$
\[
u+\varepsilon _2\mu =\Psi _{(c,T)}(m_2,k_{21})=(c\kappa _2+\lambda
)m_2^2-(c\mu +\kappa _1-\kappa _2)m_2k_{21}-\mu k_{21}^2.
\]
Le discriminant de $\Psi _{(c,T)}$ est positif dans l'essentiel des cas,
assurant que la forme $\Psi _{(c,T)}$ est ind\'{e}finie. Pour une valeur $u$
donn\'{e}e et sachant que $\varepsilon _2=\pm 1$, l'\'{e}quation que l'on
consid\`{e}re poss\`{e}de alors une infinit\'{e} de solutions en $%
(m_2,k_{21})$ d\`{e}s qu'elle en poss\`{e}de une. D'o\`{u} une infinit\'{e}
de possibilit\'{e}s pour la suite $X_2$ lorsque la suite $T$ est donn\'{e}e.
Un calcul comparable est faisable d\'{e}terminant une infinit\'{e} des
possibilit\'{e}s pour $T$ lorsque $X_2$ est donn\'{e}e. Ceci permet de
comprendre autrement l'existence de l'arbres des triplets de Cohn mis en
\'{e}vidence ci-dessus.

On a pu \'{e}tablir :

\begin{proposition}
Dans les cas o\`{u} $\varepsilon _1=\varepsilon _2=1$, on a :
\[
G=XYPX,\;\;GD=XYP,\;\;DD=XY.
\]
\end{proposition}

Ces expressions expliquent autrement pourquoi, dans le cas correspondant, on
trouve des triplets de Cohn avec les trois transformations $G$, $GD$, $DD$.
En effet on a d\'{e}j\`{a} indiqu\'{e} que ces triplets sont
caract\'{e}ris\'{e}s par le fait qu'ils correspondent \`{a} des mots
r\'{e}duits qui commencent par $XY$.

\subsection{Un exemple d'application}

Tous les exemples peuvent \^{e}tre trait\'{e}s gr\^{a}ce aux m\'{e}thodes
qui pr\'{e}c\`{e}dent.\ On illustre ici sur un cas, celui des \'{e}quations $%
M^{++}(2,0,u)$.\ Pour $\partial K=0$, soit $c=b$. Avec $\varepsilon
_1=\varepsilon _2=1$ on obtient :
\[
m_1=bm_2+k_{21},
\]
\[
m=(b^2+1)m_2^2+2bm_2k_{21}+k_{21}^2=m_2^2+m_1^2,
\]
\[
u=(b-1)m_2^2-(b-1)m_2k_{21}-k_{21}^2-1=\Psi _{(c,T)}(m_2,k_{21})_{.}
\]
Ceci donne un triplet de Cohn $((bm_2+k_{21}),m_2,1)$ pour l'\'{e}quation $%
M^{++}(b,0,u)$. Pour $b=2$ et une infinit\'{e} de valeurs $u=-s<0$,
l'\'{e}quation $M^{++}(2,0,u)$ a des solutions $(m,m_1,m_2)\in (\mathbb{N}%
\backslash \{0\})^3$, notamment si on a avec $(p,q)\in
(\mathbb{N}\backslash
\{0\})^2$ $:$%
\[
s=p^2+q^2+1-3pq>0.
\]
On trouve une infinit\'{e} de telles expressions avec les nombres de
Fibonacci :
\[
s=(1+4F_{2t+1}^2-2F_{2t+1}F_{2t}-F_{2t}^2)=F_{2t+3}^2+F_{2t}^2+1-3F_{2t+3}F_{2t}>0.
\]
Dans d'autres cas, il n'y a aucune solution dans
$(\mathbb{N}\backslash \{0\})^3 $. On a en effet \'{e}tabli :

\begin{proposition}
Consid\'{e}rons une \'{e}quation $M^{++}(2,0,u)$ avec $u<0$%
\[
x^2+y^2+z^2=3xyz-ux.
\]
Elle poss\`{e}de des solutions $(m,m_1,m_2)\in
(\mathbb{N}\backslash \{0\})^3$ si et seulement si on peut en
trouver une v\'{e}rifiant
\[
0<m<s=-u,\;\;0<m_2<\sqrt{(s-m)m}.
\]
Dans ce cas qui arrive pour une infinit\'{e} de valeurs $s>0$, elle
poss\`{e}de une infinit\'{e} de solutions.\ De plus pour $0<s\leq 50$
l'\'{e}quation $M^{++}(2,0,u)$ n'admet aucune solution lorsque l'on a
\[
-u=s\in \{1,3,7,9,11,19,23,27,31,43,47\}.
\]
\end{proposition}

Dans l'essentiel des cas on peut \'{e}crire :
\[
0<s=p_k^2-3p_kp_{k-1}+p_{k-1}^2+1<\;m=p_k^2+p_{k-1}^2,\;\;\;m_2=p_{k-1}.
\]
Les nombres $p_k$ et $p_{k-1}$ se d\'{e}duisent de nombres de Fibonacci et
donnent des constantes de Markoff s'\'{e}crivant :
\[
C(\theta _2(S))=\frac{3p_kp_{k-1}-1}{\sqrt{9(p_k^2+p_{k-1}^2)^2-4}}<\frac
13.
\]
Lorsque $p_{k-1}$ augmente ind\'{e}finiment, ces constantes convergent vers
la valeur $(1/3)$.\ Ceci a donn\'{e} :

\begin{proposition}
Le spectre de Markoff quadratique $Mark$ a pour plus grande valeur
d'accumulation $(1/3)$, par valeurs inf\'{e}rieures et par valeurs
sup\'{e}rieures.
\end{proposition}

La derni\`{e}re proposition peut se d\'{e}duire d'une autre expression :
\[
-u=-(F_{2t}^2+6F_{2t+1}F_{2t}-F_{4t+3})=F_{2t+1}^2+F_{2t}^2+1-3F_{2t+1}F_{2t}<0.
\]
Pour une infinit\'{e} des valeurs $u>0$ l'\'{e}quation
$M^{++}(2,0,u)$ a des solutions dans $(\mathbb{N}\backslash
\{0\})^3$.\

\subsection{La condition de divisibilit\'{e} \'{e}quivalente et ses
cons\'{e}quences}

Toute \'{e}quation diophantienne $M^{s_1s_2}(a,\partial K,u_\theta )$ se
d\'{e}duit en r\'{e}alit\'{e} d'une simple condition de divisibilit\'{e} :
\[
m\mid m_1^2-\partial Km_1m_2+\varepsilon _2\varepsilon _1m_2^2.
\]
Supposons que l'on note $m_1^2-\partial Km_1m_2+\varepsilon _2\varepsilon
_1m_2^2=\mu m$, en rempla\c {c}ant dans l'\'{e}quation et simplifiant par $%
m\neq 0$ il reste
\[
m+\varepsilon _2\mu =(a+1)m_1m_2-u_\theta .
\]
Cette expression d\'{e}termine $u_\theta $. En la combinant avec la
pr\'{e}c\'{e}dente de fa\c {c}on \`{a} \`{a} \'{e}liminer le terme $\mu $,
on retrouve l'\'{e}quation $M^{s_1s_2}(a,\partial K,u_\theta )$ dont les
propri\'{e}t\'{e}s essentielles sont donc contenues dans la seule condition
de divisibilit\'{e}. Sans \'{e}liminer $\mu $, on a aussi l'\'{e}quation $%
M^{-s_1,-s_2}(a,\partial K,u_\theta +2\varepsilon _2\mu )$. Ceci illustre le
ph\'{e}nom\`{e}ne des \'{e}quations \`{a} solutions communes \'{e}voqu\'{e}
dans \cite{Perrine5}. Si l'on note maintenant
\[
\partial ^{a+1}K=\varepsilon _2(a+1)m+\partial K=\varepsilon
_2((a+1)m+K_1-K_2),
\]
on a la condition de divisibilit\'{e} \'{e}quivalente
\[
m\mid (m_1^2-(\partial ^{a+1}K)m_1m_2+\varepsilon _1\varepsilon
_2m_2^2)=\phi _\theta (m_1,-\varepsilon _2m_2).
\]
Le discriminant $\Delta _0=(\partial K)^2-4\varepsilon _1\varepsilon _2$
commun aux pr\'{e}c\'{e}dentes conditions de divisibilit\'{e} permet de
classifier les \'{e}quations singuli\`{e}res, c'est-\`{a}-dire telles que $%
\Delta _0\leq 0$ ou $\Delta _0$ carr\'{e} parfait, comme suit :

$\bullet $ Pour $\varepsilon _2=1$, une \'{e}quation $M^{s_1s_2}(a,\partial
K,u_\theta )$ est dite pointue si elle est de forme :
\[
x^2+y^2+z^2=(a+1)xyz-u_\theta x,
\]
\[
x^2+y^2+z^2=(a+1)xyz\pm yz-u_\theta x.
\]
On dit qu'il s'agit d'une \'{e}quation d\'{e}g\'{e}n\'{e}r\'{e}e lorsqu'elle
s'\'{e}crit :
\[
x^2+y^2+z^2=(a+1)xyz\pm 2yz-u_\theta x,
\]
\[
x^2+y^2-z^2=(a+1)xyz-u_\theta x.
\]

$\bullet $ Pour $\varepsilon _2=-1$, une \'{e}quation est dite pointue si
elle s'\'{e}crit :
\[
x^2-y^2-z^2=(a+1)xyz-u_\theta x,
\]
\[
x^2-y^2-z^2=(a+1)xyz\pm yz-u_\theta x,
\]
On dit qu'on a affaire \`{a} une \'{e}quation d\'{e}g\'{e}n\'{e}r\'{e}e
lorsqu'elle est de forme :
\[
x^2-y^2-z^2=(a+1)xyz\pm 2yz-u_\theta x,
\]
\[
x^2-y^2+z^2=(a+1)xyz-u_\theta x.
\]

\subsection{Le cas des \'{e}quations o\`{u} $u=0$}

Consid\'{e}rons un nombre de Markoff $\theta _a(S)$ d\'{e}finissant la
constante $C(\theta _a(S))$. L'application du lemme de Dickson \cite{Dickson}
(ch.8, vol.2, p. 408-409) permet de faire l'hypoth\`{e}se que l'on a :
\[
S^{*}=(a_n,a_{n-1},...,a_0),\;\;\forall i=0,...,n,\;\;a_i\leq a,
\]
\[
C(\theta _a(S))=\frac 1{\xi _0-\xi _0^{\prime }}=\frac 1{a+[0,\underline{S,a}%
]+[0,\underline{S^{*},a}]}=\frac m{\sqrt{\Delta _a(S)}}.
\]
Dans le cas o\`{u} le minimum donnant la constante est obtenu pour un unique
indice $j\in \{0,1,...,(n+1)\}$, on dit que la constante est uniquement
atteinte.\ Mais il peut \^{e}tre obtenu sur plusieurs indices diff\'{e}rents
$j\in \{0,1,...,(n+1)\}$, on dit dans ce cas que la constante est
multiplement atteinte. Si le minimum est atteint pour $j=0$, on dit que l'on
est dans le cas super-r\'{e}duit.

Le cas super-r\'{e}duit de constante multiplement atteinte a donn\'{e} :

\begin{proposition}
Dans le cas super-r\'{e}duit o\`{u} la constante de Markoff de $\theta _a(S)$
est obtenue pour deux indices diff\'{e}rents $0$ et $j\in \{1,...,(n+1)\}$,
on a une d\'{e}composition naturelle
\[
S^{*}=(X_1,a,X_2),
\]
Avec les param\`{e}tres associ\'{e}s \`{a} la suite $S^{*}$, l'\'{e}quation
de Markoff associ\'{e}e s'\'{e}crit $M^{s_1s_2}(a,\partial K,0)$%
\[
x^2+\varepsilon _2y^2+\varepsilon _1z^2=(a+1)xyz+\varepsilon _2\partial Kyz.
\]
\end{proposition}

La situation d\'{e}crite par cette proposition g\'{e}n\'{e}ralise celle de
la th\'{e}orie de Markoff classique. Pour $\varepsilon _1=\varepsilon _2=1$,
la condition $u=0$ n'est d'ailleurs conciliable avec la condition $\partial
K=0$ que lorsqu'on a $a=2$.\ C'est le sens du r\'{e}sultat d\'{e}montr\'{e}
par G.\ Frobenius \cite{Frobenius}. Pour g\'{e}n\'{e}raliser l'\'{e}quation
de Markoff classique \`{a} d'autres cas identifi\'{e}s par la derni\`{e}re
proposition, on doit supposer $\partial K\neq 0$. Et une r\'{e}ciproque de
cette proposition est facile. Ces r\'{e}sultats ont permis d'\'{e}tudier
\cite{Perrine7} des \'{e}quations comme $M^{++}(2,2,0)$ de solution $(3,1,1)$%
, $M^{++}(2,-2,0)$ de solution $(3,2,1)$, $M^{++}(3,-1,0)$ de solution $%
(3,1,1)$, ainsi que les constantes associ\'{e}es.

\subsection{Application \`{a} l'\'{e}tude du spectre de Markoff}

La m\'{e}thode d'analyse du spectre de Markoff d\'{e}velopp\'{e}e
par l'auteur \cite{Perrine4} a \'{e}t\'{e} illustr\'{e}e ci-dessus
au voisinage de $(1/3)$.\ Elle consiste \`{a} utiliser une
\'{e}quation donn\'{e}e $M^{s_1s_2}(a,\partial K,u_\theta )$ pour
d\'{e}crire un endroit particulier du spectre. Chaque solution
d'une telle \'{e}quation fournit des
suites $X_2$ et $T$, et permet la construction d'une constante de forme $%
C(\theta _a(S))=C(F_\theta )$ dans le spectre quadratique. Par ailleurs, les
branches infinies donn\'{e}es par tout bouquet de solutions de
l'\'{e}quation fournissent des points d'accumulation du spectre
alg\'{e}brique $Mark$.\ Ces points peuvent correspondre, comme dans la
th\'{e}orie de Markoff classique \`{a} des constantes de formes quadratiques
\`{a} coefficients r\'{e}els.\ Ce sont alors des constantes du spectre de
Markoff complet. L'op\'{e}ration de passage de $Mark$ au spectre complet (%
\cite{Cusick} Chapitre 3, \cite{Cusick1}) correspond \`{a} une op\'{e}ration
de fermeture topologique. Le spectre de Markoff est ainsi analys\'{e} comme
superposition de sous-ensembles de constantes de nombres quadratiques $%
\theta _a(S)$ et de leurs points d'accumulation. On a trouv\'{e} ainsi de
nouveaux trous du spectre et \'{e}valu\'{e} sa complexit\'{e} au voisinage
de $(1/3)$.

On peut montrer avec l'expression de $C(\theta _a(S))$ que cette constante
est situ\'{e}e dans le segment
\[
U_a=[\frac 1{\sqrt{a^2+4a}},\frac 1{\sqrt{a^2+4}}].
\]
Le segment $U_1$ est r\'{e}duit \`{a} l'ensemble $\{1/\sqrt{5}\}$ qui
contient la plus grande constante du spectre de Markoff. Le segment $U_2$
donne dans sa partie sup\'{e}rieure, entre $(1/3)$ et $(1/\sqrt{8})$ les
constantes fournies par la th\'{e}orie de Markoff classique. Ce sont des
nombres isol\'{e}s \`{a} l'exception du plus petit $(1/3)$ qui est un point
d'accumulation par valeurs sup\'{e}rieures de constantes de Markoff. Il est
connu qu'au dessus de la valeur $(1/3,334367...)$ de R.\ T.\ Bumby le
spectre des constantes de Markoff est de mesure nulle (\cite{Cusick} p. 76).
Comme l'a montr\'{e} Mary E.\ Gbur Flahive \cite{Gbur}, cette partie du
spectre contient cependant une infinit\'{e} de points d'accumulation dont la
valeur $(1/(\sqrt{5}+1))$ d\'{e}couverte par C.\ J.\ Hightower \cite
{Hightower}. J.\ R.\ Kinney et T. S.\ Pitcher ont affich\'{e} l'existence
d'une infinit\'{e} de trous dans le spectre de Markoff au dessus de $(1/%
\sqrt{12})$, aussi pr\`{e}s que souhait\'{e} de cette valeur qui
est \'{e}galement un point d'accumulation de valeurs du spectre,
mais l'existence de ces trous reste \`{a} confirmer
(\cite{Perrine} IV 143). L'ensemble $U_2$ ne rencontre pas
l'ensemble $U_3$, ce qui met en \'{e}vidence un trou bien connu du
spectre de Markoff
\[
]\frac 1{\sqrt{13}},\frac 1{\sqrt{12}}[.
\]
La valeur $(1/\sqrt{13})$ est la plus grande valeur de $U_3$.\ Elle est
isol\'{e}e comme l'a montr\'{e} O.\ Perron (\cite{Cusick} p.\ 15) en
exhibant le trou maximal
\[
]\frac{22}{65+9\sqrt{3}},\frac 1{\sqrt{13}}[.
\]
La plus petite valeur de $U_3$ est $(1/\sqrt{21})$, elle est donc
aussi comprise dans $U_4$ dont la plus grande valeur vaut
$(1/\sqrt{20})$. Entre les deux derni\`{e}res bornes cit\'{e}es se
trouve la valeur $\textbf{F}$ de G.\ A.\ Freiman (\cite{Cusick}
p.\ 55) situ\'{e}e au bord d'un trou du spectre, et telle que
toute valeur r\'{e}elle comprise entre $0$ et $\textbf{F}$ soit
une constante de Markoff :
\[
\textbf{F}^{-1}=4+\frac{253589820+283748\sqrt{462}}{491993569}.
\]
C'est dans la partie basse de $U_2$ et dans la partie haute de $U_3$ que la
distribution des constantes de Markoff est la plus mal connue et que l'on
travaille donc.

Lorsque la valeur de $a$ augmente, le nombre de possibilit\'{e}s
pour les suites $T$ et $X_2$ s'accro\^{i}t. La distribution des
constantes dans le
segment $U_{a+1}$ est ainsi plus compliqu\'{e}e que celle existant dans $U_a$%
. Toute constante $C(\theta _a(S))$ de $U_a$ dans cet ensemble donne de plus
gr\^{a}ce au lemme de Dickson (\cite{Cassels} p.\ 408) une valeur de $%
U_{a+1} $ elle-m\^{e}me point d'accumulation du spectre. Ainsi la
plus grande constante du spectre de Markoff $(1/\sqrt{5})\in U_1$
donne le point d'accumulation $(1/(1+\sqrt{5}))$ de C.\ J.\
Hightower dans $U_2$. L'article de W.\ R.\ Lawrence
\cite{Lawrence} montre un ph\'{e}nom\`{e}ne comparable mais de
plus grande complexit\'{e}, en \'{e}tablissant que la distribution
des constantes de Markoff dans la partie basse de l'ensemble $U_a$
est plus compliqu\'{e}e que celle que l'on trouve dans sa partie
haute.

D\'{e}crivant le spectre par valeurs d\'{e}croissantes, plus on se
rapproche de $0$ plus sa complexit\'{e} cro\^{i}t. Apr\`{e}s une
partie discr\`{e}te, puis une autre cantorienne, l'aspect
chaotique du spectre dispara\^{i}t d'un coup lorsqu'il devient
continu sous la valeur de Freiman $\textbf{F}$. Une telle
structure ressemble \`{a} celle du spectre d'un op\'{e}rateur.

\section{Perspectives}

Une m\'{e}thode de r\'{e}solution des \'{e}quations $M^{s_1s_2}(a,\partial
K,u_\theta )$ a \'{e}t\'{e} mise au point. On a donn\'{e} de nombreux
exemples d'\'{e}quations dont toutes les solutions sont connues et entrent
dans notre formalisme g\'{e}n\'{e}ral. Un projet important est de
r\'{e}soudre le maximum d'\'{e}quations de ce type pour approfondir la
connaissance du spectre de Markoff. On peut automatiser cette
r\'{e}solution. Une des difficult\'{e}s pour fournir des r\'{e}sultats
g\'{e}n\'{e}raux concerne le calcul du maximum qui d\'{e}finit toute
constante de Markoff.\ Sur tous les cas pratiques ce n'est pas un
probl\`{e}me gr\^{a}ce \`{a} la th\'{e}orie du polygone de Klein \cite
{Klein2}.

La m\'{e}thode que l'on a d\'{e}velopp\'{e}e pour \'{e}tudier nos
\'{e}quations rend moins cruciale une d\'{e}monstration de la conjecture de
Frobenius, Cassels et Zagier \cite{Zagier} \cite{Button} pour l'arbre de la
th\'{e}orie de Markoff classique. On a d'ailleurs pu montrer dans \cite
{Perrine7} que cette conjecture est bien sp\'{e}cifique \`{a} la th\'{e}orie
classique.\ On n'a pas de r\'{e}sultat analogue pour les triplets d'autres
\'{e}quations $M^{s_1s_2}(a,\partial K,u_\theta )$. La conjecture reste
cependant ouverte, et on peut l'aborder avec les proc\'{e}d\'{e}s qui ont
\'{e}t\'{e} r\'{e}sum\'{e}s dans ce qui pr\'{e}c\`{e}de. Cependant, cette
approche n'a pas encore permis de conclure.

La notion de hauteur est essentielle pour faire fonctionner l'algorithme que
l'on a mis au point pour r\'{e}soudre nos \'{e}quations. En fait il s'agit
simplement d'une m\'{e}thode de descente infinie adapt\'{e}e de celle
tr\`{e}s classique de Pierre De Fermat.\ On dispose donc maintenant d'un
ensemble d'exemples concrets d'\'{e}quations diophantiennes non
compl\`{e}tement triviales sur lesquelles tester un certain nombre de
conjectures classiques sur les hauteurs (\cite{Lang} chapitre 2).

On a vu dans le chapitre pr\'{e}c\'{e}dent que nos \'{e}quations \'{e}taient
aussi donn\'{e}es par une formule de trace (voir \cite{Perrine9}).\ La
question se pose de savoir si toutes le sont. Ceci revient \`{a} approfondir
la fa\c {c}on dont le groupe du triangle $\mathbf{T}_3$ se plonge dans $GL(2,%
\mathbb{Z})$, et \`{a} g\'{e}n\'{e}raliser l'approche de
\cite{Perrine1b} par la trace \`{a} toutes nos \'{e}quations. Un
point particulier sur lequel
l'auteur voudrait se pencher est le fait que tout groupe d\'{e}nombrable $%
\mathbf{G}$ puisse \^{e}tre plong\'{e} en tant que sous groupe de $GL(2,\mathbb{%
Z})$.\ On pourrait ainsi d\'{e}finir une trace pour ses \'{e}l\'{e}ments
\cite{Lehrer}, et la question se pose de savoir si cette trace d\'{e}pend du
plongement que l'on consid\`{e}re.\ Ceci donnerait aussi un d\'{e}but de
r\'{e}ponse \`{a} la probl\'{e}matique \'{e}voqu\'{e}e dans \cite{Alperin}
et explicable par le fait que tout groupe de matrices ferm\'{e} dans $GL(n,%
\mathbb{R})$ est un groupe de Lie \cite{Baker}. On pourrait aussi
pour un tel
groupe $\mathbf{G}$ consid\'{e}rer les relations $\Re _{\mathbf{G}}$ et $_{%
\mathbf{G}}\Re $ qui s'en d\'{e}duisent \`{a} droite et \`{a}
gauche. On trouverait au quotient une structure arborescente.\
Pour $\mathbf{G}$ d'indice fini dans $GL(2,\mathbb{Z})$ ceci fait
un lien avec la th\'{e}orie des dessins d'enfants
(\cite{Waldschmidt2} p.\ 99).\ Et lorsque $\mathbf{G}$ est fini,
ceci fait un lien avec l'interpr\'{e}tation de nos equations. Ce
d\'{e}veloppement conduit \`{a} g\'{e}n\'{e}raliser notre article
\cite {Perrine1b} avec une v\'{e}ritable correspondance de Galois
entre groupes
finis ou d\'{e}nombrables et structures arborescentes d\'{e}finies dans $%
GL(2,\mathbb{Z})$, ainsi que sur une approche de la th\'{e}orie de
Galois inverse \cite{Serre5}. Les cons\'{e}quences pour les
groupes de tresses et les groupes de classes d'applications
(mapping class groups au sens de \cite {Birman}) pourraient se
r\'{e}v\'{e}ler tr\`{e}s importantes. Ceux-ci sont
en effet d\'{e}nombrables, et seraient donc aussi plongeables dans $GL(2,%
\mathbb{Z})$, tout comme les groupes $GL(a+1,\mathbb{Z})$ dont les
propri\'{e}t\'{e}s seraient donc accessibles par
$GL(2,\mathbb{Z})$, groupe dont on voudrait aussi d\'{e}velopper
l'arithm\'{e}tique.

Une perspective connexe est d'\'{e}tendre ce qui pr\'{e}c\`{e}de \`{a} $%
GL(a+1,\mathbb{Z})$ et des \'{e}quations poss\'{e}dant un nombre
plus grand de termes, comme par exemple celle d\'{e}j\`{a}
\'{e}tudi\'{e}e par A. Hurwitz qui g\'{e}n\'{e}ralise
l'\'{e}quation de Markoff classique \cite{Baragar} :

\[
\sum_{i=0}^{i=a}x_i^2=(a+1)\prod_{i=0}^{i=a}x_i.
\]
Les r\'{e}sultats sous-jacents relatifs \`{a} des arbres $\mathbf{T}_{a+1}$
\`{a} $a+1$ branches en chaque noeud, et g\'{e}n\'{e}ralisant $\mathbf{T}_3$%
, pourraient s'av\'{e}rer tr\`{e}s importants. Le lien entrevu dans \cite
{Perrine1b} avec le th\'{e}or\`{e}me de Dyer et Formanek \cite{LyndonSchupp}
laisse penser que des r\'{e}sultats profonds entre $\mathbf{T}_{a+1}$ et $%
GL(a+1,\mathbb{Z})$ sont ainsi accessibles.

L'auteur envisage aussi d'\'{e}tudier la fa\c {c}on dont
$GL(2,\mathbb{Z})$ est utilisable pour coder de l'information.\
Des id\'{e}es de ce genre ont d\'{e}j\`{a} \'{e}t\'{e}
pr\'{e}sent\'{e}es par W.\ Magnus qui a travaill\'{e} pour la
soci\'{e}t\'{e} Telefunken apr\`{e}s 1930 (voir \cite {Magnus2} p.
186).

\chapter{Approche alg\'{e}brique}

\section{Introduction}

La question \'{e}tudi\'{e}e ensuite concerne la signification
alg\'{e}brique de nos \'{e}quations diophantiennes
$M^{s_1s_2}(a,\partial K,u_\theta )$. On a pu en donner une
interpr\'{e}tation gr\^{a}ce aux r\'{e}seaux de rang 2 sur
$\mathbb{Z}$. Ceci a permis de poursuivre le classement de ces
\'{e}quations diophantiennes avec ce qui est connu pour les corps
quadratiques, et de r\'{e}interpr\'{e}ter certains des
r\'{e}sultats d\'{e}j\`{a} obtenus. Une observation essentielle a
\'{e}t\'{e} que tout r\'{e}seau complet d'un corps quadratique
donne en fait naissance \`{a} une \'{e}quation de Markoff
g\'{e}n\'{e}ralis\'{e}e, permettant d'envisager ses bouquets de
solutions comme d\'{e}crivant des relations entre des id\'{e}aux
d'ordres quadratiques. On a aussi montr\'{e} comment nos
\'{e}quations donnent des indications sur les points entiers et
rationnels des courbes elliptiques en les plongeant dans des
surfaces cubiques qui sont rationnelles. Ce point fait
appara\^{i}tre un ph\'{e}nom\`{e}ne quantique de changement brutal
des caract\'{e}ristiques d'une courbe elliptique r\'{e}elle
lorsque le plan qui lui donne naissance \`{a} l'intersection avec
la surface cubique se d\'{e}place. Toute courbe elliptique
r\'{e}elle peut \^{e}tre obtenue ainsi, ceci ouvre une perspective
int\'{e}ressante. Le contenu de ce chapitre a \'{e}t\'{e}
pr\'{e}sent\'{e} aux Journ\'{e}es Arithm\'{e}tiques de Lille
\cite{Perrine8}.

\section{Lien de nos \'{e}quations avec des corps quadratiques r\'{e}els}

Dans l'essentiel des cas le nombre $\Delta _\phi
=((a+1)m+K_1-K_2)^2-4\varepsilon _1\varepsilon _2$ est positif. La condition
de divisibilit\'{e} condensant l'\'{e}quation $M^{s_1s_2}(a,\partial
K,u_\theta )$ s'\'{e}crit :
\[
4m\mid (2m_1-\partial ^{a+1}Km_2)^2-\Delta _\phi (m_2)^2=4\phi _\theta
(m_1,-\varepsilon _2m_2).
\]
Deux cas apparaissent selon la parit\'{e} de $\partial ^{a+1}K=\varepsilon
_2((a+1)m+K_1-K_2)$, que l'on regroupe en posant
\[
\tau =0\text{ et }d=\frac{\Delta _\phi }4\;\text{si \ }\Delta
_\phi \equiv 0\;(\mod\,4),\;\;\tau =1\text{ et }d=\Delta _\phi
\;\text{si \ }\Delta _\phi \equiv 1\;(\mod\,4),
\]
\[
k=\frac{\partial ^{a+1}K-\tau }2\in \mathbb{Z},\;\;\varpi =\frac{\tau +\sqrt{%
\Delta _\phi }}2=\frac{\tau +\sqrt{d}}{2^\tau },
\]
\[
P_\varpi (x)=\frac{(2x-\tau )^2-\Delta _\phi }4=x^2-\tau x-\frac{\Delta
_\phi -\tau }4.
\]
Avec ces notations, la condition de divisibilit\'{e} s'\'{e}crit simplement
\[
m\mid m_2^2P_\varpi (\frac{m_1-m_2k}{m_2}).
\]
Dans le cas d'\'{e}quations d\'{e}g\'{e}n\'{e}r\'{e}es,
$\mathbb{Q}(\sqrt{d})$
n'est pas un corps quadratique. Dans le cas d'\'{e}quations pointues, $\mathbb{Q%
}(\sqrt{d})$ est un corps quadratique imaginaire, $\mathbb{Q}(i)$
pour les cas pointus $n^{\circ }1$ o\`{u} l'on retrouve la
th\'{e}orie de Markoff classique, $\mathbb{Q}(j)$ pour les cas
pointus $n^{\circ }2$. Dans les autres cas $\mathbb{Q}(\sqrt{d})$
est un corps quadratique r\'{e}el li\'{e} \`{a} l'\'{e}quation
$M^{s_1s_2}(a,\partial K,u_\theta )$.

\subsection{Construction de $\mathbb{Z}$-modules complets}

L'\'{e}tude de la condition de divisibilit\'{e} mise en
\'{e}vidence est un probl\`{e}me tr\`{e}s classique de th\'{e}orie
des nombres (voir par exemple \cite{Legendre} Tome\ 1 p. 200).\
Elle s'interpr\`{e}te dans le corps quadratique
$\mathbb{Q}(\sqrt{d})$ en posant avec $\delta
=$pgcd$(m,m_1,m_2)>0$ :
\[
\mathbf{c}_2=m/\delta ,\;\;\mathbf{e}_2\equiv (m_2k-m_1)/\delta \;\;(%
\mod\,\mathbf{c}_2)\;\text{avec }0\leq \mathbf{e}_2<\mathbf{c}_2,\;\;\mathbf{f}%
_2=m_2/\delta >0,
\]
Avec par exemple \cite{Faisant} (p. 11) ou \cite{Borevitch} (pp. 144-169),
elle signifie qu'il existe un $\mathbb{Z}$-module complet de $\mathbb{Q}(\sqrt{d})$%
, dit aussi r\'{e}seau de rang $2$ sur $\mathbb{Z}$. Il s'agit
d'un id\'{e}al
de l'ordre $\mathcal{O}_{m_2}=\mathbb{Z}[m_2\varpi ]$ du corps quadratique $%
\mathbb{Q}(\sqrt{d})$ not\'{e}
\[
\mathbb{M}_2^{\diamond }=(\delta
)(\mathbf{c}_2;\mathbf{e}_2+\mathbf{f}_2\varpi
)=\{xm+y(m_2(k+\varpi )-m_1)\mid x,y\in \mathbb{Z}\}.
\]
L'anneau des stabilisateurs du r\'{e}seau $\mathbb{M}_2^{\diamond
}$ est un
ordre $\mathcal{O}_{\mathbf{c}_2}=\mathbb{Z}[(m_2/\delta )\varpi ]$ de $\mathbb{Q}(%
\sqrt{d})$. En tant que module sur $\mathbb{Z}$ le r\'{e}seau $\mathbb{M}%
_2^{\diamond }$ a pour norme $N(\mathbb{M}_2^{\diamond })=m\delta
$. La forme
quadratique associ\'{e}e \`{a} cette base est \`{a} coefficients dans $\mathbb{Z%
}$ et s'\'{e}crit :
\[
f_{\mathbb{M}_2^{\diamond }}(x,y)=\frac 1\delta (mx^2+(m_2\partial
^{a+1}K-2m_1)xy+(\mu -\varepsilon _2(a+1)m_1m_2)y^2).
\]
Le lien avec les formes quadratiques $\phi _\theta (z,y)$ et
$F_\theta (x,y)$ appara\^{i}t alors en posant $\mathbf{z}=mx-m_1y$
et $\mathbf{y}=\varepsilon
_2m_2y$ dans la forme $f_{\mathbb{M}_2^{\diamond }}$ associ\'{e}e \`{a} $\mathbb{M}%
_2^{\diamond }$ :
\[
m\delta f_{\mathbb{M}_2^{\diamond }}(x,y)=\phi _\theta (\mathbf{z},\mathbf{y}%
)=N(\mathbf{z-y}(m\theta _a(S)-K_1)).
\]
La forme $\phi _\theta $ est donc une norme du corps quadratique $\mathbb{Q}(%
\sqrt{d})$, ce qui explique sa propri\'{e}t\'{e} de multiplicativit\'{e}.
Les calculs pr\'{e}c\'{e}dents mettent l'accent sur le r\'{e}seau $\mathbb{M}%
_\theta =\{\mathbf{x}m-\mathbf{y}m\theta _a(S)\mid
\mathbf{x},\mathbf{y}\in \mathbb{Z}\}$, avec lequel on a obtenu :

\begin{proposition}
La forme quadratique associ\'{e}e \`{a} $[1,-(m\theta _a(S)-K_1)]$ base de
l'ordre maximal $\mathcal{O}_\theta =\mathbb{Z}[\varpi ]=\mathbb{Z}[\mathbf{-}%
(m\theta _a(S)-K_1)]$ du corps quadratique $\mathbb{Q}(\sqrt{d})$ vaut, avec $N(%
\mathcal{O}_\theta )=1,$
\[
\phi _\theta (\mathbf{z},\mathbf{y})=f_{\mathcal{O}_\theta }(\mathbf{z},%
\mathbf{y})=N(\mathbf{z-y}(m\theta _a(S)-K_1)).
\]
Cet ordre contient un id\'{e}al entier $\mathbb{M}_\theta =\{\mathbf{x}m+%
\mathbf{y}m\theta _a(S)\mid \mathbf{x},\mathbf{y}\in \mathbb{Z}\}$, de norme $m$%
, et dont la forme quadratique associ\'{e}e \`{a} la base $[m,-m\theta
_a(S)] $ vaut
\[
mF_\theta (\mathbf{x},\mathbf{y})=f_{\mathbb{M}_\theta }(\mathbf{x},\mathbf{y})=%
\frac{N(\mathbf{x-y}m\theta _a(S))}{N(\mathbb{M}_\theta )}.
\]
\end{proposition}

\subsection{D'autres $\mathbb{Z}$-modules complets}

L'ordre $\mathcal{O}_{m_2}=$ $\mathbb{Z}[m_2\varpi ]$ est un
sous-anneau de
l'ordre maximal $\mathcal{O}_\theta $. On peut poser avec son id\'{e}al $%
\mathbb{M}_2^{\diamond }$ :

$\bullet $ Pour $\varepsilon _2=1$ :
\[
\mathbb{M}_2=\mathbb{M}_2^{\diamond
}=\{(x+y((a+1)m_2-k_2))m+(ym_2)m\theta _a(S)\mid x,y\in
\mathbb{Z}\}\subset \mathbb{M}_\theta .
\]

$\bullet $ Pour $\varepsilon _2=-1$ :
\[
\overline{\mathbb{M}_2}=\mathbb{M}_2^{\diamond }=\{(x-y((a+1)m_2-k_2))m-(ym_2)m%
\overline{\theta _a(S)}\mid x,y\in \mathbb{Z}\}\subset
\overline{\mathbb{M}_\theta }.
\]

Avec le r\'{e}seau $\mathbb{M}_{\delta \theta
}=\{\mathbf{x}m-\mathbf{y}\delta
m\theta _a(S)\mid \mathbf{x},\mathbf{y}\in \mathbb{Z}\}$ de $\mathbb{Q}(\sqrt{d})$%
, on a alors :

\begin{proposition}
Avec les notations pr\'{e}c\'{e}dentes et les r\'{e}seaux introduits, la
condition de divisibilit\'{e} donne les inclusions
\[
\mathbb{M}_2\subset \mathbb{M}_{\delta \theta }\subset \mathbb{M}_\theta ,\;\overline{%
\mathbb{M}_2}\subset \overline{\mathbb{M}_{\delta \theta }}\subset \overline{\mathbb{M%
}_\theta }.
\]
\end{proposition}

\subsection{Une d\'{e}composition en produit}

Dans ce que l'on vient de voir, on aurait pu permuter $m_1$ et $m_2$.
D'o\`{u} un calcul comparable \`{a} ce qui pr\'{e}c\`{e}de, dans l'ordre $%
\mathcal{O}_{m_1}=$ $\mathbb{Z}[m_1\varpi ]$ du m\^{e}me corps quadratique $%
\mathbb{Q}(\sqrt{d})$. Ceci d\'{e}finit un r\'{e}seau
$\mathbb{M}_1^{\diamond
}=(\delta )(\mathbf{c}_1;\mathbf{e}_1+\mathbf{f}_1\varpi )$, sa norme $%
m\delta $, sa forme quadratique associ\'{e}e de discriminant
$(m_1^2\Delta _\phi /\delta ^2)$, son anneau de stabilisateurs
$\mathcal{O}_{(m_1/\delta )}=\mathbb{Z}[(m_1/\delta )\varpi ].$ La
forme quadratique associ\'{e}e se calcule facilement. L'ordre
$\mathcal{O}_{m_1}=$ $\mathbb{Z}[m_1\varpi ]$ est un autre
sous-anneau de l'ordre maximal $\mathcal{O}_\theta $ qui permet de
poser :

$\bullet $ Pour $\varepsilon _1=-1$ :
\[
\mathbb{M}_1=\mathbb{M}_1^{\diamond }=\{(x-yk_1)m+(ym_1)m\theta
_a(S)\mid x,y\in \mathbb{Z}\}\subset \mathbb{M}_{\delta \theta
}\subset \mathbb{M}_\theta .
\]

$\bullet $ Pour $\varepsilon _1=1$ :
\[
\overline{\mathbb{M}_1}=\mathbb{M}_1^{\diamond }=\{(x+yk_1)m-(ym_1)m\overline{%
\theta _a(S)}\mid x,y\in \mathbb{Z}\}\subset \overline{\mathbb{M}_{\delta \theta }}%
\subset \overline{\mathbb{M}_\theta }.
\]

Il devient alors int\'{e}ressant de consid\'{e}rer le produit $\mathbb{M}_1\mathbb{%
M}_2$, ce qui a bien un sens (\cite{Faisant} p.20). En compl\'{e}tant avec
les classes de similitude \cite{Faisant} (p. 22), on a ainsi obtenu :

\begin{proposition}
Dans l'id\'{e}al $\mathbb{M}_{\delta \theta }\mathbb{=}\{\mathbf{x}m-\mathbf{y}%
\delta m\theta _a(S)\mid \mathbf{x},\mathbf{y}\in \mathbb{Z}\}$ de l'ordre $%
\mathcal{O}_\theta =\mathbb{Z}[\varpi ]$ existent deux r\'{e}seaux
\[
\mathbb{M}_1=\{(x-yk_1)m+(ym_1)m\theta _a(S)\mid x,y\in
\mathbb{Z}\},
\]
\[
\mathbb{M}_2=\{(x+y((a+1)m_2-k_2))m+(ym_2)m\theta _a(S)\mid x,y\in
\mathbb{Z}\}.
\]
Le premier est un id\'{e}al de l'anneau $\mathcal{O}_{m_1}=$ $\mathbb{Z}%
[m_1\varpi ]$. Il poss\`{e}de pour anneau de stabilisateurs $\mathcal{O}%
_{(m_1/\delta )}=\mathbb{Z}[(m_1/\delta )\varpi ]$ et a pour norme
$m\delta $.
Le second est un id\'{e}al de l'anneau $\mathcal{O}_{m_2}=$ $\mathbb{Z}%
[m_2\varpi ]$. Il poss\`{e}de en tant qu'anneau de stabilisateurs l'ordre $%
\mathcal{O}_{(m_2/\delta )}=\mathbb{Z}[(m_2/\delta )\varpi ]$ et a
aussi pour norme $m\delta $. Enfin on a
\[
\mathbb{M}_1\mathbb{M}_2=m\mathbb{M}_{\delta \theta }=\{\mathbf{x}m^2-\mathbf{y}%
\delta m^2\theta _a(S)\mid \mathbf{x},\mathbf{y}\in \mathbb{Z}\},
\]
ou avec les classes de similitudes des r\'{e}seaux $[\mathbb{M}_1][\mathbb{M}_2]=[%
\mathbb{M}_{\delta \theta }]$. On a des conditions comparables
pour les r\'{e}seaux conjugu\'{e}s.
\end{proposition}

\subsection{Equation d'un $\mathbb{Z}$-module complet quelconque}

La donn\'{e}e d'un id\'{e}al $\mathbf{I}=(\delta )(\mathbf{c};\mathbf{e}+%
\mathbf{f}\varpi )$ quelconque dans un ordre $\mathcal{O}_{m_2}$
d'un corps quadratique $\mathbb{Q}(\sqrt{d})$, o\`{u} $d$ sans
facteur carr\'{e}, conduit inversement \`{a} une condition de
divisibilit\'{e} et \`{a} une \'{e}quation diophantienne, et ceci
pour toute valeur $m_2$.\ Pour le voir, on g\'{e}n\'{e}ralise les
calculs pr\'{e}c\'{e}dents en les prenant \`{a} l'envers. Ceci a
donn\'{e} :

\begin{proposition}
Tout id\'{e}al d'un ordre $\mathcal{O}_{m_2}$ d'un corps
quadratique quelconque $\mathbb{Q}(\sqrt{d})$ d\'{e}finit une
relation diophantienne. Avec
les conditions $\varepsilon _2^{\prime }\in \mathbb{Z}\backslash \{0\}$ et $%
\varepsilon _1^{\prime }=\varepsilon _2^{\prime }\varepsilon
^{\prime }\in \mathbb{Z}$ elle s'\'{e}crit
\[
m^2+\varepsilon _2^{\prime }m_1^2+\varepsilon _1^{\prime
}m_2^2=(a+1)mm_1m_2-\varepsilon _2^{\prime }\mathbf{\partial }%
^{a+1}m_1m_2-u^{\prime }m.
\]
Elle correspond avec $(m,m_1,m_2)\in (\mathbb{N}\setminus
\{0\})^3$ aux
conditions suivantes o\`{u} $\mathbf{\partial }^{a+1}\in \mathbb{Z}$ et $%
\varepsilon ^{\prime }\in \mathbb{Z}$
\[
m\mid (m_1^2-\mathbf{\partial }^{a+1}m_1m_2+\varepsilon ^{\prime
}m_2^2),\;\;\delta =\text{pgcd}(m,m_1,m_2).
\]
\end{proposition}

Une telle \'{e}quation en $(m,m_1,m_2)$ g\'{e}n\'{e}ralise nos \'{e}quations
$M^{s_1s_2}(a,\partial K,u_\theta )$. Elle est diff\'{e}rente de celles
\'{e}tudi\'{e}es dans \cite{Mordell} ou \cite{Rosenberger2}. Elle correspond
seulement \`{a} la donn\'{e}e d'un r\'{e}seau d'un corps quadratique.\ Le
fait que toute forme quadratique enti\`{e}re binaire ind\'{e}finie peut
\^{e}tre r\'{e}duite, et donne donc une forme de Markoff, montre que l'on
peut traiter la r\'{e}solution des nouvelles \'{e}quations ici mises en
\'{e}vidence par les m\^{e}mes moyens que ceux d\'{e}velopp\'{e}s ci
dessus.\ De telles \'{e}quations ont par exemple \'{e}t\'{e}
\'{e}tudi\'{e}es par G.\ Rosenberger \cite{Rosenberger2}. Remarquons que la
proposition que l'on vient de faire s'applique pour tout id\'{e}al d'un
ordre de corps quadratique quelconque, m\^{e}me avec $d$ n\'{e}gatif. La
situation ici d\'{e}crite est donc beaucoup plus g\'{e}n\'{e}rale que celle
que l'on envisageait ci-dessus. La diff\'{e}rence est que l'on a\ $%
\varepsilon _1^{\prime }\in \mathbb{Z},\;\varepsilon _2^{\prime }\in \mathbb{Z}%
\backslash \{0\}$. La d\'{e}composition en produit de deux
r\'{e}seaux appara\^{i}t maintenant li\'{e}e au fait que l'on a
$\varepsilon ^{\prime }=\pm
1 $, et donc que $\Delta _\phi $ est de forme $(\mathbf{\partial }%
^{a+1})^2\pm 4$.\ Cette propri\'{e}t\'{e} permet d'\'{e}changer les
r\^{o}les de $m_1$ et $m_2$ dans la condition de divisibilit\'{e}, donc de
construire un autre id\'{e}al avec lequel le produit d'id\'{e}aux peut
\^{e}tre fait.

En r\'{e}alit\'{e}, pour parvenir \`{a} la derni\`{e}re proposition on a
impos\'{e} la contrainte suppl\'{e}mentaire que $d$ soit sans facteur
carr\'{e}. Si l'on admet au contraire de poser $\Delta _\phi =(\mathbf{%
\partial }^{a+1})^2\pm 4=\lambda ^2d$, avec $\lambda \in \mathbb{Z}$, ce qui ne
change pas le corps quadratique que l'on consid\`{e}re et conduit \`{a}
r\'{e}soudre un \'{e}quation de Pell-Fermat pour identifier $\lambda $, on
peut d\'{e}velopper le calcul pr\'{e}c\'{e}dent en imposant $\varepsilon
_1^{\prime }$, $\varepsilon _2^{\prime }\in \{-1,+1\}$. Ceci montre que nos
\'{e}quations sont en fait aussi g\'{e}n\'{e}rales que les
pr\'{e}c\'{e}dentes.\ En choisir une revient lorsqu'elle est non
singuli\`{e}re \`{a} consid\'{e}rer un r\'{e}seau complet dans un corps
quadratique, et non un r\'{e}seau quelconque d'un tel corps.

On a pu d\'{e}velopper cette approche en examinant la signification pour nos
\'{e}quations du fait que les r\'{e}seaux correspondants sont strictement
semblables, ainsi que la traduction pour les r\'{e}seaux de l'action du
groupe du triangle $\mathbf{T}_3$ sur les solutions et de l'existence d'un
nombre fini de bouquets de solutions. On trouve dans \cite{Hirzebruch0} des
indications sur l'interpr\'{e}tation g\'{e}om\'{e}trique qui peut \^{e}tre
donn\'{e}e de tels r\'{e}sultats. Le formalisme qui en d\'{e}coule permet de
syst\'{e}matiser les r\'{e}sultats disponibles sur le lien entre arbres,
ordres maximaux et formes quadratiques, tels que cit\'{e}s dans \cite{Pays}
ou \cite{Vigneras} (p.\ 41). Le point essentiel en vue est un lien entre le
nombre de classes d'un corps quadratique et le nombre de bouquets de
solutions pour certaines de nos \'{e}quations.

\section{Lien de nos \'{e}quations avec les courbes elliptiques}

L'id\'{e}e approfondie maintenant peut \^{e}tre comprise tr\`{e}s simplement
de fa\c {c}on g\'{e}om\'{e}trique. Avec des variables $(x,y,z)\in \mathbb{R}^3$%
, on consid\`{e}re une surface cubique r\'{e}elle d'\'{e}quation $%
M^{s_1s_2}(b,\partial K,u)$. Coup\'{e}e par un plan, elle donne une courbe
cubique dont on \'{e}tablit dans diff\'{e}rents cas qu'elle est elliptique.
Disposant alors, gr\^{a}ce \`{a} l'action du groupe $\mathbf{T}_3$,
d'informations sur les points entiers de la surface, on esp\`{e}re en
d\'{e}duire des cons\'{e}quences pour les points entiers de la courbe
elliptique. Diff\'{e}rentes tentatives faites pour concr\'{e}tiser cette
id\'{e}e sur l'\'{e}quation de Markoff classique se sont
r\'{e}v\'{e}l\'{e}es infructueuses. Mais on a pu la d\'{e}velopper sur nos
\'{e}quations g\'{e}n\'{e}ralis\'{e}es, on va expliquer comment et pourquoi.
On donne d'abord un exemple pour montrer comment cette approche fonctionne.

\subsection{Un exemple}

On consid\`{e}re l'\'{e}quation $M^{++}(2,0,-2)$.\ On conna\^{i}t
un triplet de solutions $(m,m_1,m_2)=(73,8,3)$.\ Il correspond aux
param\`{e}tres
\[
K_1=K_2=46,\;k_1=k_{12}=5,\;k_2=k_{21}=2.
\]
Ces valeurs v\'{e}rifient par exemple la relation $2m_1=5m_2+1$. En la
combinant avec la relation $M^{++}(2,0,-2)$ liant $m$, $m_1$, $m_2$, on
obtient :

\begin{proposition}
Consid\'{e}rons la courbe r\'{e}elle $E$ d'\'{e}quation cubique
\[
30xz^2-4x^2+6xz-29z^2+8x-10z-1=0,
\]
Il s'agit d'une courbe elliptique o\`{u} existe un point entier $%
(x,z)=(m,m_2)=(73,3)$. Inversement tout point entier $(x,z)=(m,m_2)\in \mathbb{Z%
}^2$ de cette courbe elliptique $E$ est de plus tel qu'il existe
un point entier $(x,y,z)=(m,m_1,m_2)\in \mathbb{Z}^3$ situ\'{e}
sur la surface cubique r\'{e}elle $M^{++}(2,0,-2)$ d'\'{e}quation
\[
x^2+y^2+z^2=3xyz+2x.
\]
\end{proposition}

La partie d\'{e}licate consiste \`{a} d\'{e}montrer que $E$ est
bien elliptique. On utilise pour cela l'algorithme de
r\'{e}duction de Nagell \cite{Nagell}, tel qu'il est
pr\'{e}sent\'{e} dans \cite{HCohen} ou \cite {Connell}. On renvoie
\`{a} \cite{Perrine9} pour la d\'{e}monstration effective.

\subsection{Cas singuliers}

On d\'{e}signe par $M^{s_1s_2}(b,\partial K,u)$ la surface cubique que l'on
consid\`{e}re, not\'{e}e comme l'\'{e}quation la d\'{e}finissant. On la
coupe par un plan $\Pi _{(t_{1,\rho },t_{2,\rho })}$ d'\'{e}quation $%
u=t_{1,\rho }z-t_{2,\rho }y$. Cette \'{e}quation d\'{e}rive de
l'expression de $u$ d\'{e}j\`{a} vue, sachant que l'on note avec
$\rho \in \mathbb{Z}$
\[
t_{1,\rho }=k_1+k_{12}-\rho m_1=t_1-(\rho -1)m_1,\;\;t_{2,\rho
}=k_2+k_{21}-\rho m_2=t_2-(\rho -1)m_2.
\]
L'intersection est une courbe que l'on note $E_{(t_{1,\rho },t_{2,\rho })}$.

Le calcul pr\'{e}c\'{e}dent ne peut absolument pas fonctionner pour
l'\'{e}quation de Markoff classique $M^{++}(2,0,0)$ car elle donne $%
t_1=t_2=u=0$. Dans un tel cas dit totalement singulier, le plan $\Pi
_{(t_{1,\rho },t_{2,\rho })}$ avec lequel couper notre surface cubique n'est
pas d\'{e}fini.\ A fortiori, on n'obtient pas une courbe elliptique,
m\^{e}me en changeant la valeur de $\rho $. De nombreux cas totalement
singuliers ont pu \^{e}tre fabriqu\'{e}s. Hors ces cas qu'on laisse
maintenant de c\^{o}t\'{e}, on voit que d'autres situations dites
partiellement singuli\`{e}res se pr\'{e}sentent.\ Le plan $\Pi _{(t_{1,\rho
},t_{2,\rho })}$ est calculable, mais son intersection avec la surface
cubique $M^{s_1s_2}(b,\partial ,u)$ est une courbe de degr\'{e}
inf\'{e}rieur ou \'{e}gal \`{a} 2. On a donn\'{e} des exemples dans \cite
{Perrine9}.

\subsection{Cas g\'{e}n\'{e}ral}

On consid\`{e}re maintenant les cas non singuliers o\`{u} l'on a
n\'{e}cessairement $t_{1,\rho }t_{2,\rho }\neq 0$. Pour la courbe cubique $%
E_{(t_{1,\rho },t_{2,\rho })}$ on trouve une \'{e}quation \`{a} coefficients
entiers. L'algorithme de Nagell peut lui \^{e}tre appliqu\'{e}. Hors
quelques cas particuliers que l'on peut expliciter, la courbe fabriqu\'{e}e
par cet algorithme est elliptique. Les cas qui \'{e}chappent peuvent
\^{e}tre \'{e}tudi\'{e}s de fa\c {c}on s\'{e}par\'{e}e. De sorte qu'on a mis
en \'{e}vidence pour toute surface cubique r\'{e}elle $M^{s_1s_2}(b,\partial
K,u)$ un ensemble de courbes elliptiques $E_{(t_{1,\rho },t_{2,\rho })}$ qui
lui sont attach\'{e}es, et de points entiers en nombre fini sur la courbe $%
E_{(t_{1,\rho },t_{2,\rho })}$ qui sont \'{e}galement sur la surface. En se
limitant \`{a} $\rho =0$, tout point entier de la surface cubique $%
M^{s_1s_2}(b,\partial K,u)$ appara\^{i}t sur une courbe elliptique $%
E_{(t_{1,\rho },t_{2,\rho })}$ contenue dans la surface.

Inversement, si l'on consid\'{e}re un point entier $(x,z)=(m,m_2)\in \mathbb{Z}%
^2$ d'une courbe elliptique $E_{(t_{1,\rho },t_{2,\rho })}$, son
\'{e}quation fournit dans $\mathbb{Z}$ une condition qui impose
que $m_1$ soit
rationnel.\ La forme particuli\`{e}re de l'\'{e}quation de degr\'{e} 2 en $%
m_1$ d\'{e}duite de l'\'{e}quation $M^{s_1s_2}(b,\partial K,u)$ montre alors
qu'en r\'{e}alit\'{e} $m_1$ est entier.

En d'autres termes les points entiers de la courbe $E_{(t_{1,\rho
},t_{2,\rho })}$ sont exactement les points entiers de la surface $%
M^{s_1s_2}(b,\partial K,u)$ qui sont situ\'{e}s dans le plan $\Pi
_{(t_{1,\rho },t_{2,\rho })}$.

Par le th\'{e}or\`{e}me de Mordell (\cite{Mordell} chapter 27), on ne trouve
qu'un nombre fini de points entiers sur la courbe $E_{(t_{1,\rho },t_{2,\rho
})}$. Cependant, en g\'{e}n\'{e}ral la surface $M^{s_1s_2}(b,\partial K,u)$
poss\`{e}de une infinit\'{e} de points entiers comme on l'a vu avec les
contructions arborescentes faites au moyen des triplets de Cohn. Ils se
classent d'ailleurs, dans le cas le plus g\'{e}n\'{e}ral, en un nombre fini
d'orbites pour l'action du groupe $\mathbf{T}_3$.\ Ceci permet de classer
les points de la courbe $E_{(t_{1,\rho },t_{2,\rho })}$. Pour des
compl\'{e}ments sur les points entiers des courbes elliptiques et leur
calcul effectif, on renvoie \`{a} \cite{Smart} (XIII.3.). Une \'{e}tude plus
globale de cette situation reste \`{a} faire, sachant que le contexte des
surfaces elliptiques (\cite{Silverman} chapter 3) fournit des
\'{e}l\'{e}ments de compr\'{e}hension int\'{e}ressants et que l'on peut
consid\'{e}rer des plans plus g\'{e}n\'{e}raux avec lesquels couper la
surface.

\subsection{Description g\'{e}om\'{e}trique de la surface cubique}

La surface r\'{e}elle cubique $M^{s_1s_2}(b,\partial K,u)$ peut \^{e}tre
\'{e}tudi\'{e}e avec des m\'{e}thodes classiques de g\'{e}om\'{e}trie
alg\'{e}brique (voir par exemple \cite{Hartshorne}). On complexifie les
variables pour simplifier les \'{e}nonc\'{e}s lorsque c'est n\'{e}cessaire.

\subsubsection{Points singuliers}

L'\'{e}quation d\'{e}finissant la surface est d'ordre 3 est :
\[
F(x,y,z)=(b+1)xyz-x^2-\varepsilon _2y^2-\varepsilon _1z^2+\varepsilon
_2\partial Kyz-ux=0.
\]
Les points singuliers non \`{a} l'infini, points doubles lorsqu'ils
existent, sont calculables :
\[
x=0,\;\;\partial K=\pm 2,\;\;2z=\partial Ky,\;\;u=(b+1)yz,
\]
\[
x=u=(\varepsilon _2\varepsilon ^{\prime }(b+1)y^2/3),\;\;\varepsilon
_2\partial K=2\varepsilon ^{\prime }-u(b+1),\;\;z=\varepsilon _2\varepsilon
^{\prime }y.
\]
En dehors de tous ces cas qui sont assez nombreux et contiennent par exemple
la th\'{e}orie de Markoff classique, la surface ne poss\`{e}de pas de point
singulier, et est donc non singuli\`{e}re.

\subsubsection{G\'{e}n\'{e}ratrices}

La surface a des points doubles \`{a} l'infini, les points \`{a}
l'infini des axes du rep\`{e}re. Il s'agit des sommets \textbf{A,
B, C,} d'un triangle dont les c\^{o}t\'{e}s sont des
g\'{e}n\'{e}ratrices, c'est-\`{a}-dire des droites contenues dans
la surface, mais dans ce cas situ\'{e}es \`{a} l'infini sur la
surface.\ Par construction, les autres g\'{e}n\'{e}ratrices de la
surface sont \`{a} distance finie et parall\`{e}les \`{a} l'un des
plans de coordonn\'{e}es. Elles peuvent toutes \^{e}tre
calcul\'{e}es \cite{Perrine9}.\ Au total il existe huit
g\'{e}n\'{e}ratrices parall\`{e}les au plan $yOz$. Par le m\^{e}me
proc\'{e}d\'{e} on obtient huit g\'{e}n\'{e}ratrices parall\`{e}les au plan $%
xOy$ et huit g\'{e}n\'{e}ratrices parall\`{e}les au plan $xOz$.

Au total, on trouve ainsi les $(3\times 8)+3=27$
g\'{e}n\'{e}ratrices r\'{e}elles ou complexes de Cayley et Salmon
pour la surface cubique \'{e}tudi\'{e}e \cite{Henderson}. En
utilisant une m\'{e}thode classique (par exemple \cite{Bouligand}
p. 466) on en d\'{e}duit une repr\'{e}sentation rationnelle de la
surface qui ne fait que traduire dans ce cas particulier le fait
que toute surface du troisi\`{e}me ordre est rationnelle
(unicursale). Il est int\'{e}ressant d'expliciter une telle
repr\'{e}sentation rationnelle de $M^{s_1s_2}(b,\partial K,u)$
pour comprendre, \`{a} l'intersection avec des plans comme ceux
utilis\'{e}s dans ce qui pr\'{e}c\`{e}de, les cons\'{e}quences
pour les courbes elliptiques que l'on a mises en \'{e}vidence
ci-dessus. Dans le cas o\`{u} un point double existe \`{a}
distance finie sur la surface, toute droite passant par ce point
d\'{e}finit aussi une telle repr\'{e}sentation rationnelle de la
surface cubique. Dans les autres cas, on peut \'{e}galement
appliquer la m\'{e}thode de la tangente due \`{a} B.\ Segre
\cite{Segre} pour construire une repr\'{e}sentation rationnelle de
la surface.

\subsubsection{Repr\'{e}sentation rationnelle de la surface cubique
r\'{e}elle}

On a d\'{e}crit dans \cite{Perrine9} la construction d'une telle
repr\'{e}sentation.\ On consid\`{e}re la trace de la surface d'\'{e}quation $%
M^{s_1s_2}(b,\partial K,u)$ dans le plan $(b+1)x+\varepsilon _2\partial K=0$%
. C'est en dehors de cas limites ou impossibles, une conique. Ceci permet de
consid\'{e}rer un point $\Omega (\mathcal{X},\mathcal{Y},\mathcal{Z})$ sur
cette conique dont les coordonn\'{e}es sont \'{e}crites avec un premier
param\`{e}tre $\mu $. On passe alors dans un rep\`{e}re d'origine $\Omega $
avec $x=\mathcal{X}+x_0,\;\;y=\mathcal{Y}+y_0,\;\;z=\mathcal{Z}+z_0$.
L'\'{e}quation de la surface s'\'{e}crit alors avec des polyn\^{o}mes
homog\`{e}nes $\Phi _i$ de degr\'{e} $i$ en $x_0$, $y_0$, $z_0$ :
\[
\Phi _3(x_0,y_0,z_0)+\Phi _2(x_0,y_0,z_0)+\Phi _1(x_0,y_0,z_0)=0.
\]
Le plan tangent en $\Omega $ \`{a} la surface a pour \'{e}quation $\Phi
_3(x_0,y_0,z_0)=(b+1)x_0y_0z_0=0$. On change \`{a} nouveau de rep\`{e}re en
l'utilisant pour poser
\[
x_1=x_0,\;\;y_1=y_0,\;\;z_1=((b+1)\mathcal{YZ}-2\mathcal{X}%
-u)x_0-2\varepsilon _2\mathcal{Y}y_0-2\varepsilon _1\mathcal{Z}z_0.
\]
L'\'{e}quation de la surface s'\'{e}crit avec des polyn\^{o}mes $\Psi _i$ de
degr\'{e} $i$ en $x_1$, $y_1$, $z_1$ :
\[
\Psi _3(x_1,y_1,z_1)+\Psi _2(x_1,y_1,z_1)+\Psi _1(x_1,y_1,z_1)=0.
\]
Avec une droite d'\'{e}quation $z_1=0$ et $x_1=\lambda y_1$ passant par le
point double $\Omega $ du plan tangent, et coupant donc la surface en un
troisi\`{e}me point dont les coordonn\'{e}es sont calculables, on obtient
une repr\'{e}sentation en $\lambda $ et $\mu $ en rempla\c {c}ant $\mathcal{X%
}$, $\mathcal{Y}$, $\mathcal{Z}$, par leurs expressions en fonction de $\mu $
et en r\'{e}duisant les formules qui en r\'{e}sultent. Ceci donne une
repr\'{e}sentation birationnelle \`{a} deux param\`{e}tres $\lambda $ et $%
\mu $ de la surface $M^{s_1s_2}(b,\partial K,u)$ qui est donc
(\cite {Hartshorne} p.\ 422) une surface r\'{e}elle rationnelle de
dimension de Kodaira $\kappa =-1$. Il en d\'{e}coule la
possibilit\'{e} de la comparer \`{a} un plan projectif r\'{e}el
construit sur les deux variables $\lambda $ et $\mu $. Cette
repr\'{e}sentation d\'{e}g\'{e}n\`{e}re en celle utilis\'{e}e par
H.\ Cohn dans l'article \cite{Cohn4} et due \`{a} R.\ Fricke
\cite{Fricke} pour le cas de la th\'{e}orie de Markoff classique.
On trouve dans \cite{Bajaj} des r\'{e}f\'{e}rences pour obtenir
d'autres repr\'{e}sentations rationnelles des surfaces
$M^{s_1s_2}(b,\partial K,u)$.\ Elles donnent la possibilit\'{e} de
d\'{e}crire l'ensemble des points rationnels $E(\mathbb{Q})$ des
courbes elliptiques $E$ que l'on introduit \`{a} l'intersection de
la surface cubique avec un plan d'\'{e}quation
rationnelle. Ces points sont param\'{e}tr\'{e}s au moyen de $\lambda $ et $%
\mu $ v\'{e}rifiant une contrainte alg\'{e}brique suppl\'{e}mentaire en
rempla\c {c}ant $y$ et $z$ par leurs expressions dans la relation
d\'{e}finissant le plan.

\section{Perspectives}

Le dernier sujet \'{e}voqu\'{e}, o\`{u} changer de plan revient \`{a}
d\'{e}former la courbe elliptique r\'{e}elle $E$ avec de temps en temps des
sauts quantiques pour les structures alg\'{e}briques qu'elle porte, reste
enti\`{e}rement \`{a} explorer. On a pens\'{e} \`{a} l'utiliser par pour
construire des courbes elliptiques de grand rang. La surface $%
M^{s_1s_2}(b,\partial K,u)$ est utilis\'{e}e pour contr\^{o}ler la
g\'{e}om\'{e}trie des courbes elliptiques r\'{e}elle $E$ qu'elle contient.
Ces courbes ne sont d'ailleurs pas rationnelles.\ Elles donnent un bon
exemple de la remarque bien connue (\cite{LevyBruhl} p.171) que les sections
planes d'une surface rationnelle ne sont pas n\'{e}cessairement des courbes
rationnelles. La m\'{e}thode suivie a consist\'{e} \`{a} utiliser la plus
petite vari\'{e}t\'{e} rationnelle contenant une vari\'{e}t\'{e}
alg\'{e}brique donn\'{e}e pour \'{e}tudier cette derni\`{e}re. Remarquons
que l'on peut adapter \`{a} la surface $M^{s_1s_2}(b,\partial K,u)$ la
construction de la structure de groupe d'une courbe elliptique.

On trouve dans \cite{Kollar} (chapter 1) une approche moderne des
surfaces cubiques $\mathfrak{X}$ non singuli\`{e}res montrant
comment elles permettent de construire un r\'{e}seau
$\mathbb{Z}^7$ \'{e}quip\'{e} d'un produit scalaire de signature
$(1,-6)$. Ce r\'{e}seau peut \^{e}tre d\'{e}crit en terme
d'homologie ou de cohomologie.\ Il est \'{e}gal \`{a} son groupe
de classes de diviseurs $Pic(\mathfrak{X})$. Sur de telles
surfaces, on peut d\'{e}velopper une th\'{e}orie de Galois avec le
groupe de Weyl $W(E_6)$, qui correspond aux permutations de leurs
27 droites dans 45 plans tritangents \cite {Hartshorne} (p. 405).
On met ainsi en \'{e}vidence pour une telle cubique sur
$\mathbb{C}$ un groupe simple \`{a} $29520$ \'{e}l\'{e}ments que
l'on peut repr\'{e}senter comme groupe unitaire $U_4(2)$ sur le
corps $F_4$, comme
groupe symplectique $PSp_4(3)$ sur le corps $F_3$, comme groupe orthogonal $%
O_6^{-}(2)$ sur le corps $F_2$ \cite{Conway3}. Les surfaces cubiques sont en
particulier des exemples bien connus de surfaces Del Pezzo \cite{Hartshorne}
(p.401).\ En se limitant au cas r\'{e}el, la th\'{e}orie de Galois que l'on
vient d'\'{e}voquer donne des indications sur les configurations que l'on
peut trouver. On trouve dans \cite{Hunt} (chapitres 5 et 6) de magnifiques
d\'{e}veloppements autour de $W(E_6)$. On a un lien \'{e}vident avec un
syst\`{e}me de Steiner particulier, le plan projectif d'ordre 2 dit plan de
Fano \cite{Assmus} (p.4), certains syst\`{e}mes r\'{e}guliers de poids \cite
{Saito3} (p.\ 522), et les alg\`{e}bres de Lie \cite{Leung}. Les surfaces
r\'{e}elles $M^{s_1s_2}(b,\partial K,u)$ rel\`{e}vent de cette approche.

Il est aussi possible d'envisager la transposition de l'article de
M.\ H.\ \`{E}l'-Huti \cite{ElHuti}.\ Des d\'{e}veloppements
comparables \`{a} ceux de \cite{Manin} \cite{Manin1} (p. 89)
permettent de calculer le
groupe de tous les automorphismes birationnels de la surface cubique $%
M^{s_1s_2}(b,\partial K,u)$, et de v\'{e}rifier que son action sur
l'ensemble des solutions enti\`{e}res de l'\'{e}quation diophantienne
correspondante est transitive. Le r\'{e}sultat obtenu est essentiellement le
m\^{e}me que celui de \`{E}l'-Huti.\ Il donne une repr\'{e}sentation
g\'{e}om\'{e}trique du groupe $\mathbf{T}_3$ par le groupe des
transformations de la surface engendr\'{e} par des r\'{e}flexions par
rapport aux points doubles \`{a} l'infini \textbf{A, B, C}. Ce groupe agit
transitivement dans l'ensemble des solutions enti\`{e}res de l'\'{e}quation
diophantienne $M^{s_1s_2}(b,\partial K,u)$.\ Ceci permet de disposer d'une
interpr\'{e}tation g\'{e}om\'{e}trique expliquant avec le groupe du triangle
$\mathbf{T}_3$ les structures arborescentes que l'on a construites avec les
triplets de Cohn.

On peut \'{e}galement caract\'{e}riser en tant que groupe
d'automorphismes birationnels de la surface le groupe engendr\'{e}
par $\mathbf{T}_3$ et le groupe $W$ de tous les automorphismes
projectifs de la surface qui sont bir\'{e}guliers en dehors de
l'ensemble des points des c\^{o}t\'{e}s du triangle \textbf{A, B,
C}. Ceci permet de d\'{e}crire le groupe de Brauer de la surface
$M^{s_1s_2}(b,\partial K,u)$ et d'\'{e}tudier sur des exemples non
triviaux des probl\`{e}mes comtemporains de g\'{e}om\'{e}trie
arithm\'{e}tique \cite{Lang} \cite{Cornell} \cite{Hulsbergen}.\ On
renvoie \`{a} \cite{Manin1} \cite{Colliot1} \cite{Swinnerton}
\cite{Serre3} \cite {Colliot} \cite{Jahnel} \cite{Bajaj}
\cite{Silverman1} pour la perspective d\'{e}j\`{a} envisag\'{e}e
dans \cite{Perrine3} indiquant qu'il n'y a pas de contre exemple
au principe de Hasse sur nos \'{e}quations. Une piste d'\'{e}tude
qui para\^{i}t aussi prometteuse \cite{Colliot} (p.\ 397) est de
faire un lien avec les surfaces de Severi-Brauer construites avec
la norme d'un corps cubique. Cette construction de F.\
Ch\^{a}telet fait jouer un r\^{o}le particulier au groupe des
permutations de trois \'{e}l\'{e}ments, groupe que l'on
repr\'{e}sente sur nos surfaces par des transformations
g\'{e}om\'{e}triques permutant les points doubles \textbf{A},\textbf{\ B }et%
\textbf{\ C. }L'\'{e}tude du lien avec les surfaces elliptiques (\cite
{Hartshorne} (chapitre V) \cite{Shioda} \cite{Friedman}) est \'{e}galement
une piste que l'on voudrait approfondir, en recherchant quel type d'ensemble
on doit extraire pour passer d'un type de surface \`{a} un autre.

Les autres r\'{e}sultats obtenus ont montr\'{e} que nos surfaces ont un lien
\'{e}troit avec des r\'{e}seaux complets des corps quadratiques, raison pour
laquelle on pense qu'elles ne donnent pas de contre-exemple au principe de
Hasse. Diff\'{e}rentes perspectives ont \'{e}t\'{e} identifi\'{e}es, dont
celle de relier arbres et ordres.\ L'interpr\'{e}tation locale sur nos
surfaces cubiques de tous ces r\'{e}sultats est possible. Une autre id\'{e}e
est d'\'{e}clairer la r\'{e}flexion sur les grandes conjectures non encore
r\'{e}solues sur les courbes elliptiques \cite{Wiles2}. On peut passer des
corps quadratiques \`{a} des corps plus g\'{e}n\'{e}raux et chercher \`{a}
transposer ce qui pr\'{e}c\`{e}de.

\chapter{Approche analytique}

\section{Introduction}

La th\'{e}orie de Markoff classique, notamment dans la
pr\'{e}sentation de Harvey Cohn \cite{Cohn2}, est li\'{e}e \`{a}
la g\'{e}om\'{e}trie de certains tores perc\'{e}s conformes et
\`{a} leurs g\'{e}od\'{e}siques.\ La question qui s'est pos\'{e}e
a \'{e}t\'{e} de savoir s'il en est de m\^{e}me de la
g\'{e}n\'{e}ralisation que l'on a mise au point
pr\'{e}c\'{e}demment.\ Ce probl\`{e}me a \'{e}t\'{e} r\'{e}solu.\
Pour le faire on a caract\'{e}ris\'{e} d'abord les tores
perc\'{e}s, puis on a fait le lien avec les matrices mises en
\'{e}vidence dans les calculs des chapitres pr\'{e}c\'{e}dents.\
Ceci est possible gr\^{a}ce \`{a} une \'{e}quation
g\'{e}n\'{e}ralisant celle de Markoff \`{a} tout tore perc\'{e}
conforme.\ Elle justifie a posteriori le bien fond\'{e} du choix
des \'{e}quations $M^{s_1s_2}(b,\partial K,u)$ que l'on a mises en
avant. Les d\'{e}finitions utilis\'{e}es pour la g\'{e}om\'{e}trie
hyperbolique sont classiques et issues de \cite{Perrine9}. On a pu
\`{a} partir de l\`{a} effectuer une classification des tores
perc\'{e}s conformes construits sur un m\^{e}me tore perc\'{e}
topologique. L'originalit\'{e} de ce qui suit r\'{e}side
essentiellement dans le traitement rigoureux des tores perc\'{e}s
paraboliques.\ Il confirme que ces tores sont donn\'{e}s par
l'\'{e}quation de la th\'{e}orie de Markoff classique. Le fait que
l'on caract\'{e}rise r\'{e}ellement ainsi tous les groupes de
Fricke a \'{e}t\'{e} \'{e}nonc\'{e} il y a longtemps
(\cite{FrickeKlein} \cite {Rosenberger} \cite{Keen4}), mais les
nombreuses d\'{e}monstrations qui existent dans la litt\'{e}rature
pr\'{e}sentent des lacunes (\cite{Gilman} p. 3), ce qui ne semble
pas le cas de notre approche. On donne dans la suite un exemple
d'\'{e}nonc\'{e} que l'on est oblig\'{e} de prendre avec une
grande prudence.\ Le contre-exemple que l'on a donn\'{e} dans le
cas d'un tore perc\'{e} hyperbolique a montr\'{e} qu'il est
associ\'{e} \`{a} un groupe non libre semble compl\`{e}tement
nouveau. Et le lien d\'{e}couvert avec une probl\'{e}matique de
g\'{e}om\'{e}trie alg\'{e}brique donne une perspective de
compr\'{e}hension commune pour les deux cas pr\'{e}c\'{e}dents.\
Elle relie le groupe de matrices que l'on consid\`{e}re \`{a} un
groupe de diviseurs d'une surface. Ceci a permis d'\'{e}laborer le
point de vue analytique de la th\'{e}orie dont le point de vue
alg\'{e}brique a \'{e}t\'{e} esquiss\'{e} au chapitre
pr\'{e}c\'{e}dent \cite {Serre4}.\ Le texte qui suit d\'{e}veloppe
l'approche qui a conduit \`{a} ces r\'{e}sultats.\ Ils ont
\'{e}t\'{e} pr\'{e}sent\'{e}s lors de conf\'{e}rences faites en
1996-1997 \`{a} une \'{e}cole th\'{e}matique du CNRS
\cite{Perrine2a} et \`{a} l'Institut des Mat\'{e}riaux du Mans
\cite {LeMehaute}. \

\section{Construction de tores perc\'{e}s conformes}

Les tores perc\'{e}s \'{e}tudi\'{e}s sont construits \`{a} partir du
demi-plan de Poincar\'{e} $\mathcal{H}$. On indexe de fa\c {c}on naturelle
chacun d'eux par des $n$-uplets de nombres r\'{e}els. Ces nombres sont
li\'{e}s par des relations qui les organisent en un nouvel objet
g\'{e}om\'{e}trique $\mathcal{V}$. On construit donc un ensemble de surfaces
de Riemann $(\mathcal{H}/\Gamma _s)_{s\in \mathcal{V}}$, des tores
perc\'{e}s dont le support topologique est le m\^{e}me, mais dont la
g\'{e}om\'{e}trie est d\'{e}crite d'une fa\c {c}on particuli\`{e}re en
chaque point de l'objet $s\in \mathcal{V}$. Cette approche, qui revient
\`{a} param\'{e}trer des structures de surfaces de Riemann diff\'{e}rentes
existantes sur un m\^{e}me objet topologique, est celle de la th\'{e}orie de
Teichm\"{u}ller \cite{Keen1} \cite{Imayoshi} \cite{Seppala} \cite{Nag}.\ On
l'a d\'{e}velopp\'{e}e sur les tores perc\'{e}s en \'{e}voquant le
probl\`{e}me du choix de l'objet $\mathcal{V}$ le plus pertinent et des
variables que l'on peut cacher en raisonnant \`{a} \'{e}quivalence conforme
ou isom\'{e}trique pr\`{e}s de $\mathcal{H}$.

\subsection{Les deux matrices d'un tore perc\'{e} conforme}

Pour construire un tore perc\'{e} $\mathcal{T}^{\bullet }$ par extraction
d'un point, on utilise quatre g\'{e}od\'{e}siques de $\mathcal{H}$
not\'{e}es $\alpha s$, $s\beta $, $\beta p$, $p\alpha $, ne se coupant pas,
et dont les extr\'{e}mit\'{e}s $\alpha $, $s$, $\beta $, $p$, sont
situ\'{e}es sur la droite r\'{e}elle qui constitue le bord de $\mathcal{H}$.
Elles d\'{e}limitent un domaine quadrangulaire de $\mathcal{H}$.\ On
convient que les sommets $\alpha $, $s$, $\beta $, $p$, apparaissent dans
cet ordre lorsque l'on d\'{e}crit ce bord de $-\infty $ \`{a} $+\infty $. Il
s'agit de nombres r\'{e}els. Mais on suppose que $p$ peut \'{e}ventuellement
prendre une valeur infinie. En effet les points $-\infty $ et $+\infty $ du
bord de $\mathcal{H}$ sont confondus au seul point \`{a} l'infini $\infty $,
compactifiant ce bord en une droite projective $\mathcal{S}^1=\mathbf{P}^1(%
\mathbb{R})$. Ce bord compactifie $\mathcal{H}$ lui-m\^{e}me d'une
certaine fa\c {c}on, sous forme d'une demi-sph\`{e}re ferm\'{e}e
(ou d'un disque ferm\'{e}). Pour retrouver le tore perc\'{e} \`{a}
partir de l\`{a}, on identifie deux \`{a} deux les
g\'{e}od\'{e}siques pr\'{e}c\'{e}dentes par des transformations
\[
t_A:\alpha p\rightarrow s\beta ,\;\;t_B:\alpha s\rightarrow p\beta .
\]
Ceci revient \`{a} construire le tore en collant gr\^{a}ce \`{a} $t_A$ et $%
t_B$ les bords du domaine quadrangulaire d\'{e}fini ci dessus. Dans cette
op\'{e}ration, le point extrait du tore correspond aux quatre points $\alpha
$, $s$, $\beta $, $p$, qui sont identifi\'{e}s par $t_A$ ou $t_B$. Ils n'ont
pas d'image dans l'objet construit car ils sont situ\'{e}s au bord de $%
\mathcal{H}$ et non dans $\mathcal{H}$.

Pour conserver un maximum de propri\'{e}t\'{e}s g\'{e}om\'{e}triques, et pas
seulement les propri\'{e}t\'{e}s topologiques sous jacentes, les
transformations $t_A$ et $t_B$ doivent \^{e}tre des isom\'{e}tries de $%
\mathcal{H}$ pour sa m\'{e}trique habituelle.\ Si on veut qu'elles
conservent aussi l'orientation et les angles, elles doivent
\^{e}tre des transformations conformes $t_A$ et $t_B$ donn\'{e}es
par des matrices $A$ et $B$ de $SL(2,\mathbb{R})$. Avec les
extr\'{e}mit\'{e}s des g\'{e}od\'{e}siques, les matrices $A$ et
$B$ remplissent des conditions qui permettent de les calculer en
fonction des nombres $\alpha $, $s$, $\beta $, $p$ :

\begin{proposition}
A une conjugaison pr\`{e}s par une matrice $M$ de
$SL(2,\mathbb{R})$, on a la repr\'{e}sentation param\'{e}trique
suivante pour les matrices $A$ et $B$ d\'{e}finissant un tore
perc\'{e} conforme, construites dans $SL(2,\mathbb{R})$ avec
$\alpha <0$ et $\beta >0$:
\[
A=\left[
\begin{array}{cc}
c\beta & -c\alpha \beta \\
c & (1/c\beta )-c\alpha
\end{array}
\right] \text{ o\`{u} }c\neq 0,
\]
\[
B=\left[
\begin{array}{cc}
c^{\prime }\alpha & -c^{\prime }\alpha \beta \\
c^{\prime } & (1/c^{\prime }\alpha )-c^{\prime }\beta
\end{array}
\right] \text{ o\`{u} }c^{\prime }\neq 0.
\]
De telles matrices sont associ\'{e}es aux valeurs $\alpha <0$, $s=0$, $\beta
>0$, et $p=\infty $, du bord de $\mathcal{H}$, qu'elles transforment comme
suit:
\[
A(\alpha )=s,\;\;A(p)=\beta ,\;\;B(\beta )=s,\;\;B(p)=\alpha .
\]
Elles donnent pour les g\'{e}od\'{e}siques associ\'{e}es de $\mathcal{H}$
\[
A(\alpha p)=s\beta ,\;\;B(\alpha s)=p\beta .
\]
\end{proposition}

Les expressions donn\'{e}es pour $A$ et $B$ dans cette proposition
r\'{e}sultent du calcul de leur d\'{e}terminant qui vaut $1$. Raisonner
\`{a} \'{e}quivalence conforme pr\`{e}s de $\mathcal{H}$ a permis de cacher
deux param\`{e}tres. Ceux qui restent d\'{e}finissent un objet
g\'{e}om\'{e}trique r\'{e}el $\mathcal{V}$ de dimension $4$ gr\^{a}ce auquel
on indexe toutes les possibilit\'{e}s de couples $(A,B).$ A \'{e}quivalence
conforme pr\`{e}s de $\mathcal{H}$, on indexe toutes les possibilit\'{e}s de
tores perc\'{e}s conformes avec les param\`{e}tres conserv\'{e}s $(\alpha
,\beta ,c,c^{\prime })\in \mathcal{V}$. L'objet g\'{e}om\'{e}trique $%
\mathcal{V}$ est d\'{e}fini par les contraintes
\[
\alpha <0,\;\;\beta >0,\;\;c\neq 0,\;\;c^{\prime }\neq 0.
\]

\subsection{Le groupe fuchsien d'un tore perc\'{e} conforme}

Ayant identifi\'{e} deux matrices $A$ et $B$ par le r\'{e}sultat
pr\'{e}c\'{e}dent, on consid\'{e}re dans $SL(2,\mathbb{R})$ le
groupe qu'elles
engendrent $G=gp(A,B)$. Son image par le morphisme canonique $\psi $ de $%
SL(2,\mathbb{R})$ dans $PSL(2,\mathbb{R})$ est not\'{e}e
\[
\Gamma =PG=Pgp(A,B)=G/G\cap \{\pm \mathbf{1}_2\}=gp(\psi (A),\psi
(B))=gp(a,b).
\]
Ce groupe de transformations conformes agit sur le demi-plan de Poincar\'{e}
$\mathcal{H}$.\ Au quotient, on trouve un tore perc\'{e} par extraction d'un
point $\mathcal{T}_\Gamma ^{\bullet }=\mathcal{H}/\Gamma $. En transportant
la m\'{e}trique de $\mathcal{H}$ sur ce quotient, la projection $\mathcal{H}%
\rightarrow \mathcal{T}_\Gamma ^{\bullet }$ devient une application
conforme.\ On dit que $A$ et $B$ sont les matrices du tore perc\'{e}
conforme $\mathcal{T}_\Gamma ^{\bullet }$ et que le groupe $\Gamma =gp(A,B)$
est un groupe fuchsien d\'{e}finissant $\mathcal{T}_\Gamma ^{\bullet }$.
Evidemment, un m\^{e}me tore perc\'{e} $\mathcal{T}_\Gamma ^{\bullet }$ peut
correspondre \`{a} d'autres couples de g\'{e}n\'{e}rateurs $(A,B)$ de $G$ et
\`{a} d'autres couples de g\'{e}n\'{e}rateurs $(a,b)$ du groupe $\Gamma $.
\\
\subsubsection{Notion de groupe de Fricke}

La th\'{e}orie de Markoff classique \cite{Cohn2} entre dans le cadre
g\'{e}om\'{e}trique que l'on vient de pr\'{e}senter avec
\[
c=\beta =-c^{\prime }=-\alpha =1,
\]
\[
A=A_0=\left[
\begin{array}{cc}
1 & 1 \\
1 & 2
\end{array}
\right] ,\;\;B=B_0=\left[
\begin{array}{cc}
1 & -1 \\
-1 & 2
\end{array}
\right] .
\]
Ces deux matrices engendrent \cite{MagnusKarassSolitar} le
sous-groupe normal d\'{e}riv\'{e} du groupe discret
$SL(2,\mathbb{Z})$, d'o\`{u} un groupe fuchsien de
$PSL(2,\mathbb{Z})$ isomorphe \`{a} $\mathbf{F}_2$, le groupe
libre
de rang $2$. G\'{e}n\'{e}ralisant cet exemple, on dit qu'un groupe fuchsien $%
\Gamma =PG$ est un groupe de Fricke si et seulement s'il
v\'{e}rifie les deux conditions \cite{Rosenberger} \cite{Schmidt}
: \\
(1):\ Le groupe $\Gamma $ est isomorphe \`{a} un groupe libre
\`{a} deux g\'{e}n\'{e}rateurs
$\mathbf{F}_2=\mathbb{Z}*\mathbb{Z}$. \\
(2):\ La surface de
Riemann $\mathcal{H}/\Gamma $ poss\`{e}de un espace topologique
support qui est hom\'{e}omorphe \`{a} un tore
topologique perc\'{e} par extraction d'un point. \\
Dans le cas g\'{e}n\'{e}ral, il n'est pas toujours simple de
d\'{e}montrer que $\Gamma $ est un groupe fuchsien \cite{Gilman}.
Il n'est pas non plus n\'{e}cessairement facile de montrer que
l'on a affaire \`{a} un groupe libre \cite{Newman}. Pour cela, on
a besoin de connaitre un minimum des propri\'{e}t\'{e}s des
matrices $A$ et $B$. Dans la suite on donne un exemple qui montre
que certains r\'{e}sultats classiques dans ce domaine
\cite{LyndonUllman} \cite{Purzitsky} sont \`{a} appliquer avec
prudence. Notre d\'{e}finition m\^{e}me des groupes de Fricke
n'est pas la plus commun\'{e}ment admise.\ On trouve par exemple
dans \cite{Bers} une d\'{e}finition des groupes modulaires de
Fricke qui englobe celle qui pr\'{e}c\`{e}de. Ces d\'{e}finitions
trouvent leur origine dans l'ouvrage \cite{FrickeKlein}.
\\
\subsubsection{Image inverse}

Notons $a$ et $b$ les deux g\'{e}n\'{e}rateurs du sous-groupe fuchsien $%
\Gamma $ de $PSL(2,\mathbb{R})$, et soient $A$ et $B$ deux images
inverses
respectives de $a$ et $b$. On peut consid\'{e}rer en remontant \`{a} $SL(2,%
\mathbb{R})$ quatre sous-groupes \`{a} deux g\'{e}n\'{e}rateurs
d'image $\Gamma $ dans $PSL(2,\mathbb{R})$ par la projection
canonique $\psi $
\[
gp(A,B),\;\;gp(-A,B),\;\;gp(A,-B),\;\;gp(-A,-B).
\]
Les points correspondants $\alpha $, $s=0$, $\beta $, $p=\infty $,
d\'{e}finis par chacun des groupes pr\'{e}c\'{e}dents sont identiques. En
consid\'{e}rant les quatre possibilit\'{e}s pr\'{e}c\'{e}dentes, on dit que $%
gp(A,B)$ est le groupe principal d\'{e}fini par $\Gamma $ si et seulement si
on a
\[
tr(A)\geq 0,\;\;tr(B)\geq 0.
\]
On dit que les trois autres groupes $gp(-A,B)$,$\;gp(A,-B)$,$\;gp(-A,-B)$,
sont les groupes conjugu\'{e}s de $gp(A,B)$. La remont\'{e}e d'un groupe $%
\Gamma \subset PSL(2,\mathbb{R})$ en un groupe $G\subset SL(2,\mathbb{R})$ dont $%
\Gamma $ est l'image est \'{e}tudi\'{e}e dans \cite{Kra}. On a :

\begin{proposition}
Le groupe principal $gp(A,B)$ d\'{e}fini par un groupe de Fricke $\Gamma =$ $%
gp(a,b)$ est libre. La projection canonique $\psi $ telle que $\psi (A)=a$
et $\psi (B)=b$ est un isomorphisme de $gp(A,B)$ sur $gp(a,b)$. Pour les
groupes conjugu\'{e}s on a aussi des isomorphismes
\[
\psi :gp(A,-B)\simeq \Gamma ,\;\;\psi :gp(-A,B)\simeq \Gamma ,\;\;\psi
:gp(-A,-B)\simeq \Gamma .
\]
Enfin, on a pour l'oppos\'{e} de la matrice unit\'{e}
\[
-\mathbf{1}_2\notin gp(A,B),\;\;-\mathbf{1}_2\notin gp(A,-B),\;\;-\mathbf{1}%
_2\notin gp(-A,B),\;\;-\mathbf{1}_2\notin gp(-A,-B).
\]
\end{proposition}

\subsection{Hyperbolicit\'{e} des deux matrices d'un tore perc\'{e}}

Avec l'expression calcul\'{e}e la matrice $A$ et de $B$, on a facilement
\cite{Katok1} :

\begin{proposition}
Les matrices $A$ et $B$ d'un tore perc\'{e} $\mathcal{T}_\Gamma
^{\bullet }$ sont hyperboliques, c'est-\`{a}-dire telles que :
\[
\left| tr(A)\right| >2,\;\;\left| tr(B)\right| >2.
\]
Elles poss\`{e}dent chacune deux points fixes r\'{e}els non \`{a} l'infini
sur le bord de $\mathcal{H}$, et une g\'{e}od\'{e}sique invariante qui les
relie, son axe. En particulier, pour le groupe principal $gp(A,B)$ d'un tore
perc\'{e} conforme, on a $tr(A)>2,\;\;tr(B)>2$.
\end{proposition}

La position respective des extr\'{e}mit\'{e}s des axes, les points fixes $%
a^{+}$, $a^{-}$, $b^{+}$, $b^{-}$, de $A$ et $B,$ n'est pas
indiff\'{e}rente, ou ce qui revient au m\^{e}me le fait que les axes de $A$
et $B$ se coupent dans $\mathcal{H}$. Ces deux axes ne peuvent d'ailleurs
\^{e}tre identiques que si l'on a $c^{\prime 2}\alpha =c^2\beta $. Or les
signes de $\alpha $ et $\beta $ garantissent que cette \'{e}galit\'{e} n'est
jamais assur\'{e}e.\ L'introduction un birapport permet de retrouver un
r\'{e}sultat connu :

\begin{proposition}
Avec les deux conditions $\alpha <0$ et $\beta >0$ et les expressions
donn\'{e}es pour $A$ et $B$, les axes de ces deux matrices hyperboliques
sont toujours distincts.\ Ils ne se coupent que si et seulement si on a la
condition
\[
0>[a^{+},a^{-};b^{+},b^{-}]=\frac{(b^{+}-a^{+})}{(b^{+}-a^{-})}\times \frac{%
(b^{-}-a^{-})}{(b^{-}-a^{+})}.
\]
Celle-ci est \'{e}quivalente au fait que tout intervalle du bord de $%
\mathcal{H}$ contenant deux points fixes de l'une des transformations $A$ ou
$B$ contient aussi un point fixe de l'autre.
\end{proposition}

Les d\'{e}finitions du birapport (''rapport de rapport'' plut\^{o}t que
''cross product'') sont diverses selon les auteurs. Notre d\'{e}finition est
celle de \cite{Toubiana} \cite{Sidler}.

\subsection{Intervention des commutateurs}

On consid\`{e}re le commutateur de $A$ et $B$, que l'on d\'{e}finit comme
suit :
\[
L=[A,B]=ABA^{-1}B^{-1}.
\]
Il s'agit ici de la d\'{e}finition classique du commutateur donn\'{e}e par
exemple dans \cite{Beardon} \cite{Katok1} et non de celle que l'on trouve
dans \cite{Bourbaki}. On peut le calculer. Il permet de consid\'{e}rer avec
\cite{Cohn} une autre matrice $C^{\circ }$ de $G$ telle que
\[
C^{\circ }BA=1,\;\;ABC^{\circ }=L.
\]
Le commutateur s'introduit naturellement dans notre contexte parce que l'on
a
\[
L(s)=ABA^{-1}B^{-1}(s)=ABA^{-1}(\beta )=AB(p)=A(\alpha )=s.
\]
En d'autres termes il contient toute l'information n\'{e}cessaire
\`{a} la d\'{e}finition du tore perc\'{e} conforme d\'{e}fini par
$A$ et $B$. Si $A$ et $B$ commutent, toute possibilit\'{e} de
d\'{e}finir le tore dispara\^{i}t.\ Dans le cas contraire tout
point fixe de $L$ permet de d\'{e}finir les points possibles $s$,
$\beta $, $p$, $\alpha $. Dans le cas g\'{e}n\'{e}ral,
on trouve deux possibilit\'{e}s pour $s$, donc pour $\beta $, $p$, $\alpha $%
. Remarquons aussi que dans le cas encore plus g\'{e}n\'{e}ral pour $A$ et $%
B $ il n'y a pas de raisons que $s$, $\beta $, $p$, $\alpha $, soient
r\'{e}els, le proc\'{e}d\'{e} peut alors donner des tores complets. Mais on
laisse ces cas de c\^{o}t\'{e}, concentrant l'attention sur les tores
perc\'{e}s construits, o\`{u} sont r\'{e}els les nombres $s$, $\beta $, $p$,
$\alpha $. Ceci donne \cite{Katok1} :

\begin{proposition}
Avec les expressions des matrices $A$ et $B$ du tore perc\'{e} $\mathcal{T}%
_\Gamma ^{\bullet }$, le commutateur $L=[A,B]$ est tel que
\[
tr(L)=tr([A,B])\leq -2.
\]
\end{proposition}

On dit que $[A,B]$ est une matrice parabolique lorsque $tr([A,B])=-2$ a
lieu.\ Lorsque l'on a l'in\'{e}galit\'{e} stricte $tr([A,B])<-2$, on dit
qu'elle est hyperbolique.\ La matrice inverse $L^{-1}$ permet d'introduire
une matrice $C$ v\'{e}rifiant
\[
CAB=1,\;\;BAC=L^{-1}=[B,A]=[A,B]^{-1},\;\;tr(L^{-1})=tr(L).
\]
On a aussi un autre commutateur $K$ qui d\'{e}finit le m\^{e}me tore
perc\'{e} que $L$ avec
\[
ABC=1,\;\;CBA=K=[B^{-1},A^{-1}],\;\;tr(K)=tr(L),
\]
\[
BAC^{\circ }=1,\;\;C^{\circ }AB=K^{-1}=[A^{-1},B^{-1}],\;\;tr(K^{-1})=tr(K),
\]
\[
K(p)=B^{-1}A^{-1}BA(p)=B^{-1}A^{-1}B(\beta )=B^{-1}A^{-1}(s)=B^{-1}(\alpha
)=p.
\]
Pour les traces des matrices que l'on vient de consid\'{e}rer, il est facile
d'\'{e}tablir :

\begin{proposition}
On a
\[
tr(C)=tr(C^{\circ }),
\]
\[
tr(L)=tr(L^{-1})=tr(K)=tr(K^{-1})\leq -2,
\]
\[
tr(L)+2=tr(A)^2+tr(B)^2+tr(AB)^2-tr(A)tr(B)tr(AB)\leq 0.
\]
\end{proposition}

La derni\`{e}re \'{e}galit\'{e} de cette proposition est due \`{a} Fricke.
Elle introduit un nombre qui est utilis\'{e} dans la suite
\[
\sigma =tr(A)^2+tr(B)^2+tr(AB)^2-tr(A)tr(B)tr(AB).
\]

\subsection{Tores perc\'{e}s paraboliques et hyperboliques}

La derni\`{e}re proposition identifie deux cas pour $K$ et $L$ (comparer
\`{a} \cite{Wolpert}).\ Illustrons avec $K$

$\bullet $ Si $tr(K)=-2$, on a $c^2\beta =-c^{\prime 2}\alpha $, et les
matrices $K$ et $L$ sont paraboliques. La matrice $K$ se simplifie
\[
K=\left[
\begin{array}{cc}
-1 & 2(1-c^2\alpha \beta -c^{\prime 2}\alpha \beta )/(c^{\prime 2}\alpha )
\\
0 & -1
\end{array}
\right] .
\]
Elle donne une transformation parabolique du demi-plan de Poincar\'{e} $%
\mathcal{H}$. Son unique point fixe est $p=\infty $.\ Il permet de
d\'{e}finir un tore associ\'{e} unique $\mathcal{T}_\Gamma ^{\bullet }$
gr\^{a}ce aux matrices $A$ et $B$. Cette transformation ne laisse aucune
g\'{e}od\'{e}sique de $\mathcal{H}$ invariante. Elle correspond \`{a} une
translation parall\`{e}lement \`{a} l'axe r\'{e}el. On dit que $\mathcal{T}%
_\Gamma ^{\bullet }$ est un tore perc\'{e} conforme parabolique.

$\bullet $ Si $tr(K)<-2$, les matrices $K$ et $L$ sont hyperboliques.\ $K$
laisse invariante une g\'{e}od\'{e}sique de $\mathcal{H}$, l'axe de $K$, qui
avec $c^2\beta +c^{\prime 2}\alpha \neq 0$ est la g\'{e}od\'{e}sique des
points $z=x+iy$ de $\mathcal{H}$ v\'{e}rifiant
\[
x=\frac{(c^2\alpha \beta +c^{\prime 2}\alpha \beta -1)}{(c^2\beta +c^{\prime
2}\alpha )}.
\]
Elle poss\`{e}de deux points fixes sur le bord de $\mathcal{H}$ : le point
\`{a} l'infini $p=\infty $ et l'intersection $p^{\prime }$ de cette
g\'{e}od\'{e}sique avec le bord de $\mathcal{H}$. Le point \`{a} l'infini $p$
permet de d\'{e}finir un tore associ\'{e} $\mathcal{T}_\Gamma ^{\bullet }$
avec les points $B(p)=\alpha $, $A(\alpha )=B(\beta )=s$,$\;A(p)=\beta $. On
dit que $\mathcal{T}_\Gamma ^{\bullet }$ est un tore perc\'{e} conforme
hyperbolique. Dans ce cas, il est possible de s'assurer que la
g\'{e}od\'{e}sique $pp^{\prime }$ invariante dans $\mathcal{H}$ par $K$
donne dans $\mathcal{T}_\Gamma ^{\bullet }$ une g\'{e}od\'{e}sique
ferm\'{e}e entourant la piq\^{u}re.\ En extrayant alors le disque piqu\'{e}
ayant cette g\'{e}od\'{e}sique pour bord, on fabrique une nouvelle surface
trou\'{e}e $\mathcal{T}_\Gamma ^{\circ }$. Dans $\mathcal{H}$, on peut
repr\'{e}senter le domaine fondamental correspondant \`{a} l'image
r\'{e}ciproque de $\mathcal{T}_\Gamma ^{\circ }$. On peut v\'{e}rifier qu'il
est stable par le groupe $gp(A,B)$. Tout se passe comme si la surface $%
\mathcal{T}_\Gamma ^{\bullet }$ prolongeait $\mathcal{T}_\Gamma ^{\circ }$
de fa\c {c}on \`{a} r\'{e}duire le trou \`{a} une piq\^{u}re.\ Les deux
objets $\mathcal{T}_\Gamma ^{\bullet }$ et $\mathcal{T}_\Gamma ^{\circ }$
ont m\^{e}me support topologique, mais pas m\^{e}me support conforme.\

Le fait remarquable dans ce cas est que le tore perc\'{e} hyperbolique est
d\'{e}doubl\'{e} gr\^{a}ce \`{a} l'autre extr\'{e}mit\'{e} $p^{\prime }$ de
la g\'{e}od\'{e}sique invariante par $K$ et aux points qui s'en
d\'{e}duisent avec $B(p^{\prime })=\alpha ^{\prime }$, $A(\alpha ^{\prime
})=B(\beta ^{\prime })=s^{\prime }$,$\;A(p^{\prime })=\beta ^{\prime }$. Ce
second tore est distinct du tore pr\'{e}c\'{e}dent. Lorsque $c^2\beta
+c^{\prime 2}\alpha $ tend vers $0$, on constate que $p^{\prime }$ tend vers
$p=\infty $, $s^{\prime }$ vers $s=0$, $\alpha ^{\prime }$ vers $\alpha $, $%
\beta ^{\prime }$ vers $\beta $. Le tore perc\'{e} devient parabolique mais
double \`{a} la limite. Ceci illustre le ph\'{e}nom\`{e}ne du double de
Schottky d'une surface de Riemann non compacte (\cite{Cohn5} p.235).

\subsection{Une repr\'{e}sentation \`{a} trois param\`{e}tres}

Ayant param\'{e}tr\'{e} tous les tores perc\'{e}s conformes avec un objet
g\'{e}om\'{e}trique $\mathcal{V}$ de dimension $4$, on voit maintenant
comment trouver d'autres param\'{e}trisations de tous ces tores par un objet
g\'{e}om\'{e}trique diff\'{e}rent de $\mathcal{V}$. On privil\'{e}gie les
nombres :
\[
\lambda =c^{\prime }\alpha ,\;\;\mu =c\beta ,
\]
\[
\theta _\alpha =-c^{\prime 2}\alpha =-c^{\prime }\lambda >0,\;\;\theta
_\beta =c^2\beta =c\mu >0,\;\;\Theta =(\theta _\alpha /\theta _\beta )>0,
\]
\[
M=tr(AB)^2-tr([A,B])-2=tr(AB)^2-\sigma ,
\]
\[
M_2=tr(A)tr(AB)-tr(B)+\Theta tr(B),
\]
\[
M_1=tr(B)tr(AB)-tr(A)+\Theta ^{-1}tr(A).
\]
On obtient :
\[
\lambda =(M_2/M),\;\;\mu =(M_1/M).
\]
Les expressions de $tr(A)$, $tr(B)$, $tr(AB)$ donnent maintenant :
\[
M^2+M_1^2+\Theta ^{-1}M_2^2=tr(A)MM_1,
\]
\[
M^2+\Theta M_1^2+M_2^2=tr(B)MM_2,
\]
\[
M^2+\Theta M_1^2+\Theta ^{-1}M_2^2=tr(AB)M_1M_2.
\]
Les trois relations pr\'{e}c\'{e}dentes ont une solution commune en $\Theta $
pourvu que l'on ait :
\[
M^2+M_1^2+M_2^2=tr(A)MM_1+tr(B)MM_2-tr(AB)M_1M_2.
\]
Lorsque la valeur de $\Theta $ est diff\'{e}rente de 1, on trouve, avec $%
\varepsilon =\pm 1$ :
\[
\mu =\frac{-(2tr(B)tr(AB)-tr(A)\sigma )+\varepsilon tr(A)\sqrt{\sigma
^2-4\sigma }}{2(\sigma -tr(AB)^2)},
\]
\[
\lambda =\frac{-(2tr(A)tr(AB)-tr(B)\sigma )-\varepsilon tr(B)\sqrt{\sigma
^2-4\sigma }}{2(\sigma -tr(AB)^2)}.
\]
Ces expressions n'ont un sens qu'\`{a} condition d'avoir un argument positif
dans les radicaux. Comme par construction $\lambda $ et $\mu $ sont
r\'{e}els et existent bien, cette condition est assur\'{e}e. Le cas
parabolique o\`{u} $tr([A,B])=-2$ se singularise en annulant le terme $%
\sigma ^2-4\sigma $. Ceci simplifie les expressions de $\lambda $ et $\mu $.

Si l'on revient aux expressions des matrices $A$ et $B$, on observe qu'elles
sont totalement d\'{e}termin\'{e}es par les trois nombres r\'{e}els $tr(A)$,
$tr(B)$, $tr(AB)$, \`{a} un param\`{e}tre r\'{e}el pr\`{e}s cependant, que
l'on peut supposer ici \^{e}tre $\theta _\alpha $. La question naturelle qui
se pose est donc de savoir ce qui lie des couples de matrices $(A,B)$
correspondant aux m\^{e}mes traces, mais \`{a} des valeurs $\theta _\alpha $
distinctes. Consid\'{e}rons donc de tels couples $(A,B)$ et $(A^{\prime
},B^{\prime })$. Avec $s=0$ et $p=\infty $, on a par construction :
\[
\alpha =-\frac{\lambda ^2}{\theta _\alpha },\;\;\beta =\frac{\mu ^2\Theta }{%
\theta _\alpha }.
\]
Ceci donne le birapport suivant
\[
\lbrack \alpha ,\beta ;s,p]=-\frac 1\Theta (\frac \lambda \mu )^2.
\]
Le m\^{e}me raisonnement fait pour $(A^{\prime },B^{\prime })$ conduit au
m\^{e}me birapport. Sur la droite projective constituant le bord de $%
\mathcal{H}$, on met ainsi en \'{e}vidence deux quadruplets de
points ayant m\^{e}me birapport.\ Il en d\'{e}coule selon un
r\'{e}sultat connu(\cite{Frenkel} p. 248, \cite{Rees} p. 76)
l'existence d'une homographie $h$ de $PGL(2,\mathbb{R})=GL(2,\mathbb{R})/(\mathbb{R}%
\backslash \{0\})$ les \'{e}changeant. Elle permet la construction d'une
transformation conforme de $\mathcal{H}$ autorisant \`{a} se limiter \`{a} $%
\theta _\alpha =1$ et \`{a} \'{e}noncer :

\begin{proposition}
A une conjugaison pr\`{e}s par une matrice de $SL(2,\mathbb{R})$,
on a la repr\'{e}sentation param\'{e}trique suivante \`{a} trois
param\`{e}tres pour les matrices $A$ et $B$ du tore perc\'{e}
$\mathcal{T}_\Gamma ^{\bullet }$
\[
A=\left[
\begin{array}{cc}
\mu & (\mu \lambda ^2) \\
(1/\Theta \mu ) & ((1+(\lambda ^2/\Theta ))/\mu )
\end{array}
\right] ,\;\;B=\left[
\begin{array}{cc}
\lambda & -(\lambda \mu ^2\Theta ) \\
-(1/\lambda ) & ((1+\Theta \mu ^2)/\lambda
\end{array}
\right] .
\]
La donn\'{e}e des trois param\`{e}tres $\lambda \neq 0$, $\mu \neq 0$, $%
\Theta >0$, d\'{e}termine les matrices $A$, $B$, et $AB$, et donc leurs
traces selon les expressions
\[
tr(A)=((1+(\lambda ^2/\Theta )+\mu ^2)/\mu ),
\]
\[
tr(B)=((1+\lambda ^2+\Theta \mu ^2)/\lambda ),
\]
\[
tr(AB)=((1+(\lambda ^2/\Theta )+\Theta \mu ^2)/\lambda \mu ).
\]
Ces valeurs v\'{e}rifient les conditions suppl\'{e}mentaires
\[
1+\lambda ^2+\mu ^2=tr(A)\mu +tr(B)\lambda -tr(AB)\lambda \mu ,
\]
\[
\alpha =-\lambda ^2,\;\;s=0,\;\;\beta =\mu ^2\Theta ,\;\;p=\infty .
\]
Inversement, les trois param\`{e}tres intervenant dans ces matrices ne
d\'{e}pendent que des trois valeurs $tr(A)$, $tr(B)$, $tr(AB)$, et d'un
signe, avec les expressions
\[
\lambda =\frac{-(2tr(A)tr(AB)-tr(B)\sigma )-\varepsilon tr(B)\sqrt{\sigma
^2-4\sigma }}{2(\sigma -tr(AB)^2)}\neq 0,
\]
\[
\mu =\frac{-(2tr(B)tr(AB)-tr(A)\sigma )+\varepsilon tr(A)\sqrt{\sigma
^2-4\sigma }}{2(\sigma -tr(AB)^2)}\neq 0,
\]
\[
\Theta =\frac{2tr(A)^2+2tr(B)^2-tr(B)^2\sigma +\varepsilon tr(B)^2\sqrt{%
\sigma ^2-4\sigma }}{2tr(A)^2+2tr(B)^2-tr(A)^2\sigma -\varepsilon tr(A)^2%
\sqrt{\sigma ^2-4\sigma }}>0,
\]
o\`{u} l'on a
\[
\varepsilon =\pm 1,\;\;\sigma =tr(A)^2+tr(B)^2+tr(AB)^2-tr(A)tr(B)tr(AB)\leq
0.
\]
De plus on a \'{e}quivalence des trois propri\'{e}t\'{e}s suivantes :
\[
tr(L)=-2,\;\;\sigma =0,\;\;\Theta =1.
\]
\end{proposition}

Les expressions donn\'{e}es pour $A$ et $B$ dans cette proposition
n'utilisent que trois param\`{e}tres parce qu'on a cach\'{e} $\theta _\alpha
$ en raisonnant \`{a} une transformation conforme de $\mathcal{H}$ pr\`{e}s.
Les param\`{e}tres qui restent d\'{e}finissent un objet g\'{e}om\'{e}trique $%
\mathcal{V}^{\prime }$ qui est r\'{e}el de dimension $3$.\ Il indexe avec
des param\`{e}tres $(\lambda ,\mu ,\Theta )\in \mathcal{V}^{\prime }$ les
couples $(A,B)$ correspondants, et donc les diff\'{e}rentes possibilit\'{e}s
pour les classes de tores perc\'{e}s conformes. L'espace $\mathcal{V}%
^{\prime }$ est d\'{e}fini par les contraintes $\lambda \neq 0$, $\mu \neq 0
$, $\Theta >0$. Raisonnant sur $\Gamma =Pgp(A,B)$ on peut se limiter \`{a} $%
\lambda >0,\;\mu >0,\;\Theta >0$.

\subsection{Autre repr\'{e}sentation \`{a} quatre param\`{e}tres}

Dans le r\'{e}sultat qui pr\'{e}c\`{e}de, on a introduit une
dissym\'{e}trie dans les r\^{o}les jou\'{e}s par $\theta _\alpha $
et $\theta _\beta $.\ En r\'{e}tablissant la sym\'{e}trie entre
$\theta _\alpha $ et $\theta _\beta $ on a obtenu :

\begin{proposition}
A une conjugaison pr\`{e}s par une matrice de $SL(2,\mathbb{R})$,
on a la repr\'{e}sentation param\'{e}trique suivante \`{a} quatre
param\`{e}tres pour les matrices $A$ et $B$ du tore perc\'{e}
conforme $\mathcal{T}_\Gamma ^{\bullet }$
\[
A=\left[
\begin{array}{cc}
\mu & (\mu \lambda ^2/\Theta _\alpha ) \\
(\Theta _\beta /\mu ) & ((1+(\Theta _\beta /\Theta _\alpha )\lambda ^2)/\mu )
\end{array}
\right] ,
\]
\[
B=\left[
\begin{array}{cc}
\lambda & -(\lambda \mu ^2/\Theta _\beta ) \\
-(\Theta _\alpha /\lambda ) & ((1+(\Theta _\alpha /\Theta _\beta )\mu
^2)/\lambda )
\end{array}
\right] .
\]
Les param\`{e}tres intervenant dans ces expressions ne d\'{e}pendent que des
trois valeurs $tr(A)$, $tr(B)$, $tr(AB)$, et d'une valeur $\varepsilon =\pm
1 $ :
\[
\sigma =tr(A)^2+tr(B)^2+tr(AB)^2-tr(A)tr(B)tr(AB)\leq 0,
\]
\[
\lambda =\frac{-(2tr(A)tr(AB)-tr(B)\sigma )-\varepsilon tr(B)\sqrt{\sigma
^2-4\sigma }}{2(\sigma -tr(AB)^2)}\neq 0,
\]
\[
\mu =\frac{-(2tr(B)tr(AB)-tr(A)\sigma )+\varepsilon tr(A)\sqrt{\sigma
^2-4\sigma }}{2(\sigma -tr(AB)^2)}\neq 0,
\]
\[
\Theta _\alpha =2tr(A)^2+2tr(B)^2-tr(B)^2\sigma +\varepsilon tr(B)^2\sqrt{%
\sigma ^2-4\sigma }>0,
\]
\[
\Theta _\beta =2tr(A)^2+2tr(B)^2-tr(A)^2\sigma -\varepsilon tr(A)^2\sqrt{%
\sigma ^2-4\sigma }>0.
\]
\[
\alpha =-(\lambda ^2/\Theta _\alpha ),\;\;s=0,\;\;\beta =(\mu ^2/\Theta
_\beta ),\;\;p=\infty .
\]
A une conjugaison pr\`{e}s d\'{e}finie par une dilatation d'amplitude $\tau
^2$ telle que
\[
\theta _\alpha =\Theta _\alpha \tau ^2,\;\;\theta _\beta =\Theta _\beta \tau
^2,
\]
on retrouve les expressions param\'{e}triques ant\'{e}rieures
\[
A=\left[
\begin{array}{cc}
\mu & (\mu \lambda ^2/\theta _\alpha ) \\
(\theta _\beta /\mu ) & ((1+(\theta _\beta /\theta _\alpha )\lambda ^2)/\mu )
\end{array}
\right] ,\;\;B=\left[
\begin{array}{cc}
\lambda & -(\lambda \mu ^2/\theta _\beta ) \\
-(\theta _\alpha /\lambda ) & ((1+(\theta _\alpha /\theta _\beta )\mu
^2)/\lambda )
\end{array}
\right] .
\]
A une conjugaison pr\`{e}s d\'{e}finie par une dilatation d'amplitude $%
\Theta _\alpha $, on retrouve aussi les expressions d\'{e}j\`{a} vues avec
le param\`{e}tre $\Theta =(\Theta _\alpha /\Theta _\beta )$.
\end{proposition}

Cette proposition peut \^{e}tre interpr\'{e}t\'{e}e avec un nouvel objet
g\'{e}om\'{e}trique $\mathcal{V}^{\prime \prime }$ de dimension $4$
permettant de param\'{e}trer tous les tores perc\'{e}s conformes d'une
nouvelle fa\c {c}on. On utilise ici des quadruplets $(tr(B),tr(A),tr(BA),%
\varepsilon )\in \mathcal{V}^{\prime \prime }$ l'objet $\mathcal{V}^{\prime
\prime }$ est d\'{e}fini par $\varepsilon =\pm 1$ et la condition
\[
tr(A)^2+tr(B)^2+tr(AB)^2-tr(A)tr(B)tr(AB)\leq 0.
\]
Le bord de $\mathcal{V}^{\prime \prime }$ correspondant \`{a} la condition $%
\sigma =0$ ne donne que des tores perc\'{e}s conformes paraboliques. Dans ce
cas d'ailleurs, les tores perc\'{e}s associ\'{e}s \`{a} $\varepsilon =1$ et $%
\varepsilon =-1$ sont identiques. Ce bord peut donc \^{e}tre
param\'{e}tr\'{e} en oubliant $\varepsilon $, uniquement par des triplets $%
(tr(B),tr(A),tr(BA))$ v\'{e}rifiant l'\'{e}quation de Markoff classique :
\[
tr(A)^2+tr(B)^2+tr(AB)^2-tr(A)tr(B)tr(AB)=0.
\]
On a montr\'{e} dans \cite{Perrine1b} que dans ce cas le groupe
$G=gp(A,B)$ est un groupe libre \`{a} deux g\'{e}n\'{e}rateurs
$\mathbf{F}_2$ s'il est contenu dans $GL(2,\mathbb{Z})$. Il
d\'{e}termine un groupe de Fricke $Pgp(A,B)$ engendr\'{e} par les
classes de $A$ et $B$. La suite montre comment l'\'{e}quation de
Markoff param\'{e}trise en fait tous les groupes de Fricke par des
points du bord de $\mathcal{V}^{\prime \prime }$. Ceci revient
\`{a} dire que pour un tore perc\'{e} conforme les
propi\'{e}t\'{e}s d'\^{e}tre de Fricke ou parabolique sont
\'{e}quivalentes \cite{FrickeKlein} \cite {Rosenberger}.

On conjecture que les groupes correspondant aux tores perc\'{e}s
conformes hyperboliques ont, en dehors du bord de
$\mathcal{V}^{\prime \prime }$, deux g\'{e}n\'{e}rateurs $A$ et
$B$ qui sont li\'{e}s. Construire les relations les liant est un
probl\`{e}me essentiel dont les cons\'{e}quences pourraient
\^{e}tre importantes. On donne dans la suite un exemple o\`{u}
l'on a r\'{e}ussi \`{a} le faire.\ Cet exemple illustre notre
conjecture.

\subsection{R\^{o}le des transformations anti-conformes}

Dans la proposition qui pr\'{e}c\`{e}de, on voudrait pouvoir se
limiter dans tous les cas \`{a} une param\'{e}trisation des tores
perc\'{e}s par des
triplets $(tr(B),tr(A),tr(BA))$, et donc se passer \'{e}galement du terme $%
\varepsilon $ pour les tores perc\'{e}s conformes hyperboliques.
C'est possible si on ne raisonne qu'\`{a} isom\'{e}trie pr\`{e}s
de $\mathcal{H}$, c'est-\`{a}-dire en faisant agir aussi ses
transformations anti-conformes.
Pour le voir il a suffi d'expliquer ce qui diff\'{e}rencie les deux cas $%
\varepsilon =+1$ et $\varepsilon =-1$ correspondant \`{a} un m\^{e}me
triplet $(tr(A),tr(B),tr(AB))$. Ceci a permis d'\'{e}noncer :

\begin{proposition}
Pour les deux couples de matrices $(A^{+},B^{+})$ et
$(B^{-},A^{-})$ correspondant \`{a} un m\^{e}me triplet de traces
tel que $\sigma <0$ ainsi que respectivement \`{a} $\varepsilon
=1$ et $\varepsilon =-1$, il existe une matrice $D\in
S^{*}L(2,\mathbb{R})$ telle que
\[
B^{-}=DA^{+}D^{-1},\;\;A^{-}=DB^{+}D^{-1},\;\;\det (D)=-1.
\]
La matrice $D$ d\'{e}finit une transformation anti-conforme $\psi
(D)=h_{+}^{-}$ du demi-plan de Poincar\'{e} $\mathcal{H}$ dans lui-m\^{e}me
qui transforme les g\'{e}od\'{e}siques comme suit (en inversant les sens de
parcours):
\[
\alpha ^{+}p\rightarrow p\beta ^{-},\;\;\alpha ^{+}s\rightarrow s\beta
^{-},\;\;s\beta ^{+}\rightarrow \alpha ^{-}s,\;\;p\beta ^{+}\rightarrow
\alpha ^{-}p;
\]
\[
h_{+}^{-}(\alpha ^{+})=\beta ^{-},\;\;h_{+}^{-}(\beta ^{+})=\alpha
^{-},\;\;h_{+}^{-}(s)=s,\;\;h_{+}^{-}(p)=p;
\]
\[
h_{+}^{-}(z)=(\frac{\alpha ^{-}}{\beta ^{+}})\overline{z}=(\frac{\beta ^{-}}{%
\alpha ^{+}})\overline{z}.
\]
Elle donne pour les divers param\`{e}tres intervenant
\[
(tr(A^{+}),tr(B^{+}),tr(A^{+}B^{+}))=(tr(B^{-}),tr(A^{-}),tr(A^{-}B^{-})),
\]
\[
(\lambda ^{+},\mu ^{+},\Theta ^{+})=(\mu ^{-},\lambda ^{-},(1/\Theta ^{-})),
\]
\[
[\alpha ^{+},\beta ^{+};s,p]=[\beta ^{-},\alpha ^{-};s,p],
\]
\[
\alpha ^{-}=-(\Theta _\beta ^{+}/\Theta _\alpha ^{-})\beta ^{+}=-(\Theta
_\alpha ^{+}/\Theta _\beta ^{-})\beta ^{+},
\]
\[
\beta ^{-}=-(\Theta _\alpha ^{+}/\Theta _\beta ^{-})\alpha ^{+}=-(\Theta
_\beta ^{+}/\Theta _\alpha ^{-})\alpha ^{+}.
\]
\end{proposition}

Ce r\'{e}sultat permet de se limiter au cas $\varepsilon =1$ dans les
calculs courants faits autour des tores perc\'{e}s, lorsque l'on raisonne
\`{a} isom\'{e}trie pr\`{e}s de $\mathcal{H}$. Il est int\'{e}ressant de se
demander ce que donne la proposition pr\'{e}c\'{e}dente lorsque $\sigma $
tend vers $0$. On trouve \`{a} la limite un tore perc\'{e} conforme
parabolique o\`{u} $s=0$ et $p=\infty $. Ceci explique comment tout tore
perc\'{e} conforme parabolique est anti-conform\'{e}ment \'{e}quivalent
\`{a} lui-m\^{e}me. Dans les autres cas, la derni\`{e}re proposition
correspond aux observations qui ont \'{e}t\'{e} faites pr\'{e}c\'{e}demment
sur le d\'{e}doublement des tores perc\'{e}s hyperboliques (et les doubles
de Schottky d'une surface de Riemann non compacte \cite{Cohn5} p.235). Une
transformation anti-conforme lie les deux tores perc\'{e}s obtenus.

\section{Signification g\'{e}om\'{e}trique de nos \'{e}quations}

\subsection{C\^{o}ne attach\'{e} \`{a} un tore perc\'{e}}

Revenant sur les nombres $M$, $M_1$, $M_2$, qui ont \'{e}t\'{e} introduits
pr\'{e}c\'{e}demment, on a obtenu :

\begin{proposition}
Soient $A$ et $B$ les matrices d'un tore perc\'{e} conforme $\mathcal{T}%
_\Gamma ^{\bullet }$ quelconque.\ Avec les expressions connues o\`{u} $%
\varepsilon =\pm 1$
\[
\sigma =tr(A)^2+tr(B)^2+tr(AB)^2-tr(A)tr(B)tr(AB),
\]
\[
\Theta =\frac{2tr(A)^2+2tr(B)^2-tr(B)^2\sigma +\varepsilon tr(B)^2\sqrt{%
\sigma ^2-4\sigma }}{2tr(A)^2+2tr(B)^2-tr(A)^2\sigma -\varepsilon tr(A)^2%
\sqrt{\sigma ^2-4\sigma }},
\]
\[
M_1=tr(B)tr(AB)-tr(A)+\Theta ^{-1}tr(A),
\]
\[
M_2=tr(A)tr(AB)-tr(B)+\Theta tr(B),
\]
\[
M=tr(AB)^2-\sigma ,
\]
on a la relation $(FR^{*})$ suivante :
\[
M^2+M_1^2+M_2^2=tr(A)MM_1+tr(B)MM_2-tr(AB)M_1M_2.
\]
\end{proposition}

L'\'{e}quation $(FR^{*})$ d\'{e}finit une quadrique en $M$, $M_1$, $M_2$,
qui est un c\^{o}ne en ces param\`{e}tres directement donn\'{e} par la
matrice
\[
\left[
\begin{array}{ccc}
1 & -\dfrac{tr(A)}2 & -\dfrac{tr(B)}2 \\
-\dfrac{tr(A)}2 & 1 & \dfrac{tr(AB)}2 \\
-\dfrac{tr(B)}2 & \dfrac{tr(AB)}2 & 1
\end{array}
\right] .
\]
On dit que c'est le c\^{o}ne $(FR^{*})$ associ\'{e} au couple de
g\'{e}n\'{e}rateurs $(A,B)$ du groupe $gp(A,B)$ du tore perc\'{e} $\mathcal{T%
}_\Gamma ^{\bullet }$ que l'on consid\`{e}re. Le d\'{e}terminant de la
matrice qui le d\'{e}finit vaut
\[
1-\frac 14(tr(A)^2+tr(B)^2+tr(AB)^2-tr(A)tr(B)tr(AB))=\frac{4-\sigma }4\geq
1.
\]

Pour le tore perc\'{e} conforme associ\'{e}, on peut consid\'{e}rer que la
relation $(FR^{*})$ est une bonne g\'{e}n\'{e}ralisation de l'\'{e}quation
de Markoff classique \cite{Markoff}. En effet, si $\Theta =1$, soit $\sigma
=0$, elle se simplifie par un facteur $tr(AB)^2$ en
\[
tr(A)^2+tr(B)^2+tr(AB)^2=tr(A)tr(B)tr(AB).
\]

\subsection{Lien avec nos \'{e}quations $M^{s_1s_2}(b,\partial K,u)$}

Il est apparu que l'\'{e}quation $(FR^{*})$ correspond aux \'{e}quations qui
ont \'{e}t\'{e} \'{e}tudi\'{e}es dans ce qui pr\'{e}c\`{e}de.
\\
\subsubsection{Une \'{e}quation \'{e}quivalente}

On a fait appara\^{i}tre dans \cite{Perrine6} une \'{e}quation
\'{e}quivalente \`{a} $M^{s_1s_2}(b,\partial K,u)$.\ On appelle
$M(b,r,s,t)$ cette nouvelle \'{e}quation :
\[
x^2+y^2+z^2=(b+1)xyz+ryz+szx+txy,
\]
o\`{u}
\[
r=\varepsilon _1K_1-\varepsilon _2K_2,\;\;s=-(\varepsilon
_1k_1+k_{12}),\;\;t=\varepsilon _2k_2+k_{21}.
\]
Le lien s'effectue avec l'\'{e}quation $M^{s_1s_2}(b,\partial K,u)$
gr\^{a}ce aux deux \'{e}galit\'{e}s suivantes
\[
\varepsilon _1m_2=K_1m_1-k_1m,\;\;\varepsilon _2m_1=k_2m-K_2m_2.
\]
\\
\subsubsection{Mise en \'{e}vidence du tore perc\'{e} et du c\^{o}ne}

Dans le cas le plus g\'{e}n\'{e}ral pour \'{e}tablir l'\'{e}quation $%
M^{s_1s_2}(b,\partial K,u)$ on a vu que l'on pouvait utiliser une formule de
Fricke pour calculer la trace du commutateur $%
[A_b,B_c]=A_bB_cA_b^{-1}B_c^{-1}$ o\`{u}
\[
A_b=M_{(\lhd X_2^{*},b)}=\left[
\begin{array}{cc}
bm_2+k_{21} & m_2 \\
bk_2+l_2 & k_2
\end{array}
\right] ,
\]
\[
B_c=M_{(X_1^{*}\rhd ,c)}=\left[
\begin{array}{cc}
(c+1)m_1-k_1 & m_1 \\
(c+1)(m_1-k_{12})-(k_1-l_1) & m_1-k_{12}
\end{array}
\right] ,
\]
Ceci d\'{e}finit $t$, $s$, $r$, par un simple calcul de traces. De
fa\c {c}on \`{a} disposer de matrices appartenant \`{a}
$SL(2,\mathbb{R})$ on fait l'hypoth\`{e}se que l'on a
\[
\det (A_b)=\det (B_c)=\det (A_bB_c)=1=\varepsilon _1=\varepsilon _2.
\]
L'\'{e}quation \'{e}quivalente $M(b,r,s,t)$ prend alors la forme :
\[
m^2+m_1^2+m_2^2=tr(A_b)mm_1+tr(B_c)mm_2-tr(A_bB_c)m_1m_2.
\]
On reconna\^{i}t l'\'{e}quation $(FR^{*})$ du c\^{o}ne qui a
\'{e}t\'{e} associ\'{e}e \`{a} un tore perc\'{e} conforme. Ce tore
est d\'{e}duit du groupe $gp(A_b,B_c)$ avec :
\[
L(s)=A_bB_cA_b^{-1}B_c^{-1}(s)=s,\;\;\alpha
=A_b^{-1}(s),\;\;p=B_c^{-1}(\alpha ),\;\;\beta =A_b(p).
\]
Dans le cas parabolique, on trouve avec ces conditions un tore perc\'{e}
unique.\ C'est le cas de la th\'{e}orie de Markoff classique. Dans le cas
hyperbolique qui est le cas le plus fr\'{e}quent, on identifie ainsi deux
tores perc\'{e}s.
\\
\subsubsection{Un exemple de tore perc\'{e} hyperbolique}

On a d\'{e}taill\'{e} un exemple hyperbolique qui correspond \`{a}
l'\'{e}quation $M^{++}(3,0,1)$.\ Les matrices \`{a} consid\'{e}rer
sont dans $SL(2,\mathbb{Z})$ :
\[
A=\left[
\begin{array}{cc}
11 & 3 \\
7 & 2
\end{array}
\right] =M_{(1,1,1,3)},\;\;B=\left[
\begin{array}{cc}
37 & 11 \\
10 & 3
\end{array}
\right] =M_{(3,1,2,3)}.
\]
On peut calculer les deux tores perc\'{e}s conformes. L'un est donn\'{e} par
les points
\[
s_{+}=\frac{4363+\sqrt{3122285}}{1658}=[3,\underline{1,2,3,3,3,3,2,1}%
]\approx 3,697225,
\]
\[
\beta _{+}=\frac{1477+\sqrt{3122285}}{982}=[\underline{3,3,3,2,1,1,2,3}%
]\approx 3,303461,
\]
\[
p_{+}=\frac{-44517-\sqrt{3122285}}{155578}=[-1,1,2,2,1,\underline{%
3,3,2,1,1,2,3,3}]\approx -2,297497,
\]
\[
\alpha _{+}=\frac{1477-\sqrt{3122285}}{982}=[-1,1,2,\underline{%
2,1,1,2,3,3,3,3}]\approx -0,295315.
\]
Le second tore est donn\'{e} par les points
\[
s_{-}=\frac{4363-\sqrt{3122285}}{1658}=[1,1,1,\underline{3,3,3,3,2,1,1,2}%
]\approx 1,565743,
\]
\[
\beta _{-}=\frac{1477-\sqrt{3122285}}{982}=[-1,1,2,\underline{2,1,1,2,3,3,3,3%
}]\approx -0,295315,
\]
\[
p_{-}=\frac{-44517+\sqrt{3122285}}{155578}=[-1,1,2,1,1,1,\underline{%
3,3,3,2,1,1,2,3}]\approx -0,274782,
\]
\[
\alpha _{-}=\frac{1477+\sqrt{3122285}}{982}=[\underline{3,3,3,2,1,1,2,3}%
]\approx 3,303461.
\]
Un point remarquable est que dans ce cas on a
\[
\beta _{+}=\alpha _{-},\;\;\beta _{-}=\alpha _{+},
\]
d'o\`{u} deux matrices $U=B^{-1}A=-A^{-1}B$ et $V=BA^{-1}=-AB^{-1}$ telles
que
\[
U^2=V^2=-\mathbf{1}_2,\;\;A=-VB=BU,\;\;B=VA=-AU,
\]
\[
\beta _{+}=U(\;\beta _{-})=V(\;\beta _{-}).
\]
Dans le groupe $\Gamma =gp(\psi (A),\psi (B))$ on trouve ainsi des relations
entre $\psi (A)$ et $\psi (B).$ Elles \'{e}tablissent que ce groupe n'est
pas libre.\ Ce n'est donc pas un groupe de Fricke, m\^{e}me si par
construction la surface de Riemann $\mathcal{H}/\Gamma $ est hom\'{e}omorphe
\`{a} un tore perc\'{e}. Les points fixes de $A$ et $B$ sont respectivement
\[
a^{+}=-[\underline{\lhd X_2^{*},a}]=-[\underline{1,1,1,3}]=\frac{9+\sqrt{165}%
}{14}\approx 1,5604,
\]
\[
\;\;a^{-}=-[0,\underline{X_2\rhd ,a}]=-[0,\underline{1,1,1,3}]=\frac{9-\sqrt{%
165}}6\approx -0,6409,
\]
\[
b^{+}=-[\underline{X_1^{*}\rhd ,a}]=-[\underline{3,1,2,3}]=\frac{34+\sqrt{%
1586}}{20}\approx 3,6912,
\]
\[
b^{-}=-[0,\underline{\lhd X_1,a}]=-[0,\underline{2,1,3,3}]=\frac{32-\sqrt{%
1586}}{22}\approx -0,3557.
\]
Leurs axes respectifs se coupent donc. D'autre part, un calcul simple montre
que $s_{+}$ et $s_{-}$ sont des points fixes r\'{e}els de la matrice de
trace $\sigma -2=1767$
\[
L=ABA^{-1}B^{-1}=\left[
\begin{array}{cc}
-1298 & 4799 \\
-829 & 3065
\end{array}
\right] \in SL(2,\mathbb{Z}).
\]
Cet exemple est int\'{e}ressant car il est en contradiction avec
un th\'{e}or\`{e}me \'{e}tabli par R.C.\ Lyndon et J.L.\ Ullman
\cite{LyndonUllman} (p.\ 164) qui permettrait dans ce cas de
conclure que le groupe $gp(\psi (A),\psi (B))$ est libre. Le
constat que cet article pr\'{e}sente au moins deux difficult\'{e}s
a d\'{e}j\`{a} \'{e}t\'{e} fait dans l'article \cite {Purzitsky}
(pp.\ 213-214). Il est confirm\'{e}.

L'\'{e}quation $(FR^{*})$ du c\^{o}ne est locale et change pour chaque point
$(m,m_1,m_2)$ de la surface cubique $M^{++}(3,0,1)$. Au point $(130,11,3)$
elle s'\'{e}crit :
\[
x^2+y^2+z^2=tr(A)xy+tr(B)zx-tr(AB)yz=13xy+40xz-520yz.
\]
Equivalente avec $15m_2-4m_1=1$ \`{a} $M^{++}(3,0,1)$, elle donne aussi
l'\'{e}quation $M(b,r,s,t)$ qui s'\'{e}crit :
\[
x^2+y^2+z^2=4xyz-15xz+4xy.
\]
\\
\subsubsection{Une piste d'approfondissement}

L'exemple pr\'{e}c\'{e}dent permet de comprendre le lien de la
surface cubique avec le groupe fuchsien $\Gamma =gp(\psi (A),\psi
(B))$. Pour prolonger la r\'{e}flexion on a trouv\'{e} des
indications dans \cite {Shafarevich} (Tome 1, chapitre III 1.6 p.\
164). Avec la repr\'{e}sentation param\'{e}trique
g\'{e}n\'{e}ralisant celle de Fricke qui a \'{e}t\'{e} construite
au chapitre pr\'{e}c\'{e}dent, on d\'{e}duit une application
r\'{e}guli\`{e}re de la surface dans le plan projectif et surtout
un pinceau non d\'{e}g\'{e}n\'{e}r\'{e} de coniques. On peut alors
faire appara\^{i}tre dans cette situation un groupe ab\'{e}lien
libre \cite{Shafarevich} (Tome 1, chapitre III 1.6,
th\'{e}or\`{e}me 4) \`{a} partir duquel on peut esp\'{e}rer
reconstruire les matrices $2\times 2$ que l'on consid\`{e}re. Dans
une telle interpr\'{e}tation qui reste \`{a} d\'{e}tailler
compl\`{e}tement, on mat\'{e}rialise le groupe des classes de
diviseurs de la surface en chaque point entier $(m,m_1,m_2)$ en
utilisant une application $s$ du plan projectif $\mathbf{P}^1$
dans la surface d\'{e}finissant la courbe $S=s(\mathbf{P}^1)$ et
une fibre non singuli\`{e}re $F$ dont on d\'{e}duit le groupe
$gp(A,B)$. Ceci donne une nouvelle piste pour approfondir la
situation que l'on consid\`{e}re, en la rattachant \`{a} une
probl\'{e}matique importante de g\'{e}om\'{e}trie alg\'{e}brique.

\section{Th\'{e}orie compl\`{e}te pour les tores perc\'{e}s paraboliques}

Dans le cas des tores perc\'{e}s paraboliques on peut r\'{e}duire encore le
nombre des param\`{e}tres.\ On a vu pr\'{e}c\'{e}demment que ce cas est
celui des groupes de Fricke et qu'il existe un lien direct avec
l'\'{e}quation de Markoff classique. Ceci permet de d\'{e}velopper une
th\'{e}orie compl\`{e}te de la r\'{e}duction pour ces tores perc\'{e}s \cite
{Rosenberger}.\ Elle g\'{e}n\'{e}ralise ce qui a \'{e}t\'{e} construit dans
\cite{Perrine1b} pour la th\'{e}orie de Markoff classique, ou dans le
chapitre 2 pour la r\'{e}solution de nos \'{e}quations par descente infinie.

\subsection{Repr\'{e}sentations \`{a} deux param\`{e}tres}

En supposant que $gp(A,B)$ est un groupe principal, on peut supposer $%
\lambda $ et $\mu $ positifs. Seules deux valeurs suffisent alors \`{a}
d\'{e}finir les matrices $A$ et $B$ dans le cas parabolique.\ On a ainsi
\'{e}nonc\'{e} :

\begin{proposition}
Pour un tore perc\'{e} conforme parabolique $\mathcal{T}_\Gamma ^{\bullet }$%
, on a \`{a} une conjugaison pr\`{e}s par une matrice de
$SL(2,\mathbb{R})$, la repr\'{e}sentation param\'{e}trique
suivante pour les matrices $A$ et $B$ du groupe principal de
$\mathcal{T}_\Gamma ^{\bullet }$
\[
A=\left[
\begin{array}{cc}
\mu & (\mu \lambda ^2/\Theta _\alpha ) \\
(\Theta _\alpha /\mu ) & ((1+\lambda ^2)/\mu )
\end{array}
\right] ,\;\;B=\left[
\begin{array}{cc}
\lambda & -(\lambda \mu ^2/\Theta _\alpha ) \\
-(\Theta _\alpha /\lambda ) & ((1+\mu ^2)/\lambda )
\end{array}
\right] ,
\]
avec
\[
\Theta _\alpha =2(tr(A)^2+tr(B)^2),\;\;\alpha =-(\lambda ^2/\Theta _\alpha
),\;\;s=0,\;\;\beta =(\mu ^2/\Theta _\alpha ),\;\;p=\infty .
\]
Ceci donne la repr\'{e}sentation param\'{e}trique suivante des traces
\[
tr(A)=\frac{1+\lambda ^2+\mu ^2}\mu ,\;\;tr(B)=\frac{1+\lambda ^2+\mu ^2}%
\lambda ,\;\;tr(AB)=\frac{1+\lambda ^2+\mu ^2}{\lambda \mu },
\]
o\`{u}
\[
\lambda =(tr(A)/tr(AB))>0,\;\;\mu =(tr(B)/tr(AB))>0.
\]
Ce cas est caract\'{e}ris\'{e} par la relation de Fricke
\[
tr(A)^2+tr(B)^2+tr(AB)^2=tr(A)tr(B)tr(AB).
\]
Cette relation signifie que la repr\'{e}sentation param\'{e}trique
pr\'{e}c\'{e}dente de $\mathcal{T}_\Gamma ^{\bullet }$ est \`{a} deux
param\`{e}tres $\lambda $ et $\mu $. A une dilatation d'amplitude $\tau
=\Theta _\alpha ^{-1}$ pr\`{e}s, on peut faire dispara\^{i}tre le param\`{e}tre $%
\Theta _\alpha $ des \'{e}critures pr\'{e}c\'{e}dentes en raisonnant \`{a}
une transformation conforme pr\`{e}s. Les deux matrices \`{a} consid\'{e}rer
prennent alors la forme
\[
A=\left[
\begin{array}{cc}
\mu & \mu \lambda ^2 \\
(1/\mu ) & ((1+\lambda ^2)/\mu )
\end{array}
\right] ,\;\;B=\left[
\begin{array}{cc}
\lambda & -\lambda \mu ^2 \\
-(1/\lambda ) & ((1+\mu ^2)/\lambda )
\end{array}
\right] .
\]
Le groupe $Pgp(A,B)$ qu'elles d\'{e}finissent est un groupe de Fricke. Et $%
gp(A,B)$ est un groupe libre \`{a} deux g\'{e}n\'{e}rateurs isomorphe \`{a} $%
\mathbf{F}_2$.
\end{proposition}

Cette proposition peut \^{e}tre interpr\'{e}t\'{e}e avec un objet
g\'{e}om\'{e}trique $\mathcal{V}^{\prime \prime \prime }$ qui est une
surface de Riemann d'\'{e}quation
\[
x^2+y^2+z^2=xyz\text{.}
\]
Chaque point $(x,y,z)=(tr(B),tr(A),tr(AB))$ de
$\mathcal{V}^{\prime \prime \prime }$ d\'{e}finit un couple
$(\lambda ,\mu )$ permettant la donn\'{e}e d'un tore perc\'{e}
conforme parabolique $\mathcal{T}_{gp(\psi (A),\psi (B))}^{\bullet
}$. La param\'{e}trisation des matrices en $\lambda $ et $\mu $
est due \`{a} Fricke \cite{Fricke} \cite{Cohn4}. De plus tous les
tores perc\'{e}s paraboliques sont ainsi obtenus avec les couples
$(\lambda ,\mu )\in \mathbb{R}^2\backslash \{(0,0)\}$.

L'\'{e}nonc\'{e} v\'{e}ritablement nouveau de cette proposition est celui
qui affiche que le groupe fuchsien $gp(\psi (A),\psi (B))=Pgp(A,B)$ est
toujours un groupe de Fricke. On utilise pour le d\'{e}montrer le
th\'{e}or\`{e}me 8 (p.\ 221) de \cite{Purzitsky} avec
\[
tr(A)>2,\;\;tr(B)>2,\;\;tr(L)=tr(ABA^{-1}B^{-1})=-2.
\]
On peut calculer les points fixes $a^{+}$, $a^{-}$, $b^{+}$, $b^{-}$, de $A$
et $B$ en fonction de $\lambda $, $\mu $, et s'assurer du signe n\'{e}gatif
de $[a^{+},a^{-};b^{+},b^{-}]$. Ayant ainsi v\'{e}rifi\'{e} toutes les
conditions de th\'{e}or\`{e}me cit\'{e} on l'applique pour conclure que le
groupe $gp(A,B)$ est discret et libre \`{a} deux g\'{e}n\'{e}rateurs tout
comme $Pgp(A,B)$. Comme par construction la surface de Riemann $\mathcal{H}%
/Pgp(A,B)$ est hom\'{e}omorphe \`{a} un tore perc\'{e} en un point, il en
r\'{e}sulte que $Pgp(A,B)$ est un groupe de Fricke. Comme la r\'{e}ciproque
se d\'{e}duit ais\'{e}ment de \cite{Perrine1b} en montrant que les traces
sont li\'{e}es par une \'{e}quation de Markoff classique, cette
propri\'{e}t\'{e} est bien caract\'{e}ristique du cas parabolique. De plus
on a donn\'{e} pr\'{e}c\'{e}demment un exemple hyperbolique o\`{u} cette
propri\'{e}t\'{e} n'est pas assur\'{e}e. En d'autres termes on a obtenu une
\'{e}quivalence entre la cat\'{e}gorie des groupes de Fricke et celle des
tores perc\'{e}s conformes paraboliques.

\subsection{Des exemples de tores perc\'{e}s paraboliques}

Diff\'{e}rents exemples de groupes de Fricke associ\'{e}s \`{a} des tores
perc\'{e}s conformes paraboliques sont bien connus.

$\bullet $ Le lien avec les travaux de A.\ Schmidt \cite{Schmidt} introduit
\[
\mathbf{A}_0=tr(AB),\;\;\mathbf{B}_0=tr(A),\;\;\mathbf{C}_0=tr(B),\;\;k=(1+%
\lambda ^2+\mu ^2)/\theta ,
\]
\[
T_0=BA,\;\;U_0=A,\;\;V_0=B^{-1}.
\]
Ceci donne une nouvelle repr\'{e}sentation param\'{e}trique (voir \cite
{Perrine9}) pr\'{e}cisant comment $A$ et $B$ agissent sur les bords du
domaine fondamental $p\alpha s\beta $.
\[
A=\pm \left[
\begin{array}{cc}
\sqrt{\beta \theta } & -\alpha \sqrt{\beta \theta } \\
\sqrt{\theta /\beta } & ((1-\alpha \theta )/\sqrt{\beta \theta })
\end{array}
\right] ,\;B=\pm \left[
\begin{array}{cc}
\sqrt{-\alpha \theta } & -\beta \sqrt{-\alpha \theta } \\
-\sqrt{(\theta /-\alpha )} & ((1-\beta \theta )/\sqrt{-\alpha \theta })
\end{array}
\right] .
\]
Les travaux de A.\ Schmidt \cite{Schmidt} introduisent une notion
de groupe de Fricke \'{e}tendu dont un groupe de Fricke est un
sous groupe d'indice 2. Un tel groupe \'{e}tendu n'est autre qu'un
groupe isomorphe au groupe du triangle $\mathbf{T}_3$ dans lequel
l'indice 2 d\'{e}finit de fa\c {c}on unique le groupe de Fricke.
Le groupe \'{e}tendu correspond \`{a} une sph\`{e}re triplement
perc\'{e}e dont le tore perc\'{e} est un rev\^{e}tement \`{a} deux
feuilles. On peut prolonger ces rep\'{e}sentations
de $\mathbf{F}_2$ et $\mathbf{T}_3$ en une repr\'{e}sentation de $GL(2,\mathbb{Z%
})$.

$\bullet $ Presque toutes les matrices $A\in SL(2,\mathbb{R})$
permettent de trouver une matrice $B$ avec laquelle $gp(A,B)$
d\'{e}termine un tore perc\'{e} conforme parabolique :

\begin{proposition}
Consid\'{e}rons une matrice \`{a} coefficients r\'{e}els
\[
A=\pm \left[
\begin{array}{cc}
\mathbf{a} & \mathbf{b} \\
\mathbf{c} & \mathbf{d}
\end{array}
\right] \in SL(2,\mathbb{R}),\;\text{o\`{u} }\mathbf{bc}>0,\;\;\mathbf{ba}%
>0,\;\;\mathbf{ac}>0,
\]
alors $A$ d\'{e}termine un tore perc\'{e} conforme parabolique avec
\[
B=\pm \left[
\begin{array}{cc}
\sqrt{\mathbf{bc}} & -\mathbf{a}\sqrt{\dfrac{\mathbf{b}}{\mathbf{c}}} \\
-\mathbf{a}\sqrt{\dfrac{\mathbf{c}}{\mathbf{b}}} & \dfrac{(1+\mathbf{a}^2)}{%
\sqrt{\mathbf{bc}}}
\end{array}
\right] \in SL(2,\mathbb{R}).
\]
Le groupe $gp(A,B)$ est libre \`{a} deux g\'{e}n\'{e}rateurs.
\end{proposition}

Cette proposition donne des exemples classiques \cite{Cohn1} \cite{Schmidt}
\cite{SeriesHaas} :

$\bullet $ Le groupe de Klein est d\'{e}fini avec $\lambda =1$, $\mu =\theta
=2$. Il est d\'{e}termin\'{e} par $A$ :
\[
A=\left[
\begin{array}{cc}
2 & 1 \\
1 & 1
\end{array}
\right] ,\;\;B=\left[
\begin{array}{cc}
1 & -2 \\
-2 & 5
\end{array}
\right] \text{.}
\]

$\bullet $ Le groupe de la th\'{e}orie de Markoff, qui est en
r\'{e}alit\'{e} le groupe libre $\mathbf{F}_2$, est d\'{e}fini avec $\lambda
=$ $\mu =\theta =1$. Il est d\'{e}termin\'{e} par la seule donn\'{e}e de la
matrice $A_0$ :
\[
A_0=\left[
\begin{array}{cc}
1 & 1 \\
1 & 2
\end{array}
\right] ,\;\;B_0=\left[
\begin{array}{cc}
1 & -1 \\
-1 & 2
\end{array}
\right] \text{.}
\]
Il est possible de voir que ce cas se ram\`{e}ne au pr\'{e}c\'{e}dent.

$\bullet $ Le groupe de Hecke est d\'{e}fini avec $\lambda =$ $\mu =\sqrt{2}$%
, $\theta =1$. Il est d\'{e}termin\'{e} par :
\[
A=\left[
\begin{array}{cc}
\sqrt{2}/2 & \sqrt{2}/4 \\
\sqrt{2} & 3\sqrt{2}/2
\end{array}
\right] ,\;\;B=\left[
\begin{array}{cc}
\sqrt{2}/2 & -\sqrt{2}/4 \\
-\sqrt{2} & 3\sqrt{2}/2
\end{array}
\right] \text{.}
\]

$\bullet $ Le groupe $G_\theta $ est engendr\'{e} par les matrices suivantes
o\`{u} $\theta >0$
\[
A_\theta =\left[
\begin{array}{cc}
\mu & (\mu \lambda ^2/\theta ) \\
(\theta /\mu ) & ((1+\lambda ^2)/\mu )
\end{array}
\right] ,\;\;B_\theta =\left[
\begin{array}{cc}
\lambda & -(\lambda \mu ^2/\theta ) \\
-(\theta /\lambda ) & ((1+\mu ^2)/\lambda )
\end{array}
\right] .
\]
Il est conform\'{e}ment \'{e}quivalent au groupe $G_1$ donn\'{e} avec $%
\theta =1$ par :
\[
D_\theta =\frac 1{\sqrt{\theta }}\left[
\begin{array}{cc}
\theta & 0 \\
0 & 1
\end{array}
\right] ,\;\;A_1=D_\theta A_\theta D_\theta ^{-1},\;\;B_1=D_\theta B_\theta
D_\theta ^{-1}.
\]
On en d\'{e}duit l'expression de la matrice de passage d'un groupe $G_\theta
$ \`{a} tout autre groupe $G_{\theta ^{\prime }}$. Si l'on note
respectivement $\mathcal{T}_{\Gamma _\theta }^{\bullet }$ et $\mathcal{T}%
_{\Gamma _{\theta ^{\prime }}}^{\bullet }$ les tores perc\'{e}s conformes
associ\'{e}s, ils sont conform\'{e}ment \'{e}quivalents lorsque $\theta $ et
$\theta ^{\prime }$ sont de m\^{e}me signe, et anti-conform\'{e}ment
\'{e}quivalents dans le cas contraire.

\subsection{Classification des groupes de Fricke par les triplets de traces}

Avec un formalisme sur les traces analogue \`{a} celui de \cite{Perrine1b}
on a trouv\'{e} :

\begin{proposition}
Soient $(A,B)$ et $(A^{\prime },B^{\prime })$ les syst\`{e}mes de
g\'{e}n\'{e}rateurs respectifs de deux groupes principaux de
groupes de Fricke $\Gamma $ et $\Gamma ^{\prime }$ associ\'{e}s
\`{a} des tores perc\'{e}s conformes paraboliques,\ on a
\'{e}quivalence des propri\'{e}t\'{e}s suivantes :

1/\ Les couples $(A,B)$ et $(A^{\prime },B^{\prime })$ sont
\'{e}quivalents par un m\^{e}me automorphisme int\'{e}rieur de
$GL(2,\mathbb{R})$ :
\[
A=DA^{\prime }D^{-1},\;\;B=DB^{\prime }D^{-1},\;\;\text{o\`{u} }D\in GL(2,%
\mathbb{R}).
\]

2/\ On a \'{e}galit\'{e} des deux triplets suivants
\[
\Pi (A,B)=(tr(B^{-1}),tr(A),tr(B^{-1}A^{-1})),
\]
\[
\Pi (A^{\prime },B^{\prime })=(tr(B^{\prime -1}),tr(A^{\prime
}),tr(B^{\prime -1}A^{\prime -1})).
\]

3/ Les couples $(A,B)$ et $(A^{\prime },B^{\prime })$ d\'{e}finissent les
m\^{e}mes param\`{e}tres $\lambda $, $\mu \in \mathbb{R}^{+}$%
\[
\lambda =(tr(A)/tr(AB))=(tr(A^{\prime })/tr(A^{\prime }B^{\prime })),
\]
\[
\mu =(tr(B)/tr(AB))=(tr(B^{\prime })/tr(A^{\prime }B^{\prime })).
\]
\end{proposition}

De fa\c {c}on \'{e}vidente, on a $1/\Rightarrow 2/\Rightarrow 3/$.
Le plus d\'{e}licat est d'\'{e}tablir l'implication $3/\Rightarrow
1/$. On le fait avec une m\'{e}thode g\'{e}om\'{e}trique directe
bas\'{e}e sur la comparaison de birapports. Il en r\'{e}sulte
l'existence d'une homographie de la droite r\'{e}elle projective
sur le bord de $\mathcal{H}$, et donc d'une matrice $D\in
GL(2,\mathbb{R})$ associ\'{e}e \`{a} cette homographie. La matrice
$D$ est explicitement calculable, et on v\'{e}rifie qu'elle
satisfait \`{a} la condition $1/$.\ Ceci termine la
d\'{e}monstration en explicitant la transformation de M\"{o}bius
de $\mathcal{H}$ recherch\'{e}e. De plus on a pu s'assurer que
l'on a :

\begin{proposition}
Toute \'{e}quivalence conforme d'un tore perc\'{e} parabolique $\mathcal{T}%
_\Gamma ^{\bullet }$ dans lui-m\^{e}me donn\'{e}e par une conjugaison de $%
GL(2,\mathbb{R})$ est \'{e}gale \`{a} l'identit\'{e}.
\end{proposition}

\subsection{R\'{e}duction des tores perc\'{e}s paraboliques}

Ayant class\'{e} les tores perc\'{e}s paraboliques au moyen des
transformations conformes de $\mathcal{H}$, on a examin\'{e} ce que l'on
peut faire sans changer de groupe, mais en changeant seulement de
syst\`{e}me de g\'{e}n\'{e}rateurs $(A,B)$. Ceci permet de contruire une
th\'{e}orie de la r\'{e}duction dans tout groupe de Fricke.
\\
\subsubsection{Les involutions}

Le groupe $\Gamma =Pgp(A,B)$ est un groupe de Fricke pour tout tore
perc\'{e} parabolique $\mathcal{T}_\Gamma ^{\bullet }$ , et le goupe $%
gp(A,B) $ est libre \`{a} deux g\'{e}n\'{e}rateurs $A$ et $B$.\ On applique
les automorphismes involutifs du groupe $gp(A,B)$ dont les expressions sont
issues de la th\'{e}orie de Markoff classique \cite{Perrine1b} :
\[
X_\phi :(A,B)\longrightarrow (A^{-1},ABA),
\]
\[
Y_\phi :(A,B)\longrightarrow (BAB,B^{-1}),
\]
\[
Z_\phi :(A,B)\longrightarrow (A^{-1},B).
\]
Leur action sur le triplet des traces $%
(x,y,z)=(tr(B^{-1}),tr(A),tr(B^{-1}A^{-1}))$ est :
\[
\widetilde{X_\phi }:(x,y,z)\longrightarrow (yz-x,y,z),
\]
\[
\widetilde{Y_\phi }:(x,y,z)\longrightarrow (x,xz-y,z),
\]
\[
\widetilde{Z_\phi }:(x,y,z)\longrightarrow (x,y,xy-z).
\]
Ces transformations laissent invariante la relation $x^2+y^2+z^2=xyz$.
\\
\subsubsection{Nappe principale et groupe principal}

La derni\`{e}re \'{e}quation cit\'{e}e est celle d'une surface r\'{e}elle
poss\'{e}dant un point double $(0,0,0)$ et quatre nappes se d\'{e}duisant de
la nappe principale d\'{e}finie par les conditions $x>0,\;\;y>0,\;\;z>0$. La
nappe principale est invariante par les transformations $\widetilde{X_\phi }$%
, $\widetilde{Y_\phi }$, $\widetilde{Z_\phi }$. On passe d'une nappe aux
autres par des transformations \'{e}videntes. Elles peuvent ne pas laisser
le groupe $gp(A,B)$ invariant.\ Comme on raisonne sur un tore perc\'{e}
parabolique, on a recours aux deux param\`{e}tres $\lambda =(tr(A)/tr(AB))$
et $\mu =(tr(B)/tr(AB))$. Pour la nappe principale, on a $\lambda >0$ et $%
\mu >0$, c'est \`{a} dire des valeurs dans le premier quart de plan
r\'{e}el. Pour les autres couples matrices dont les param\`{e}tres sont dans
un des autres quarts de plan, on note les param\`{e}tres correspondants $%
\varepsilon _\lambda \lambda $ et $\varepsilon _\mu \mu $, avec $\lambda >0$
et $\mu >0$, $\varepsilon _\lambda $ et $\varepsilon _\mu $ dans $\{+1,-1\}$%
. Ceux-ci d\'{e}terminent des couples de matrices que l'on peut \'{e}crire $%
(\varepsilon _\mu A,\varepsilon _\lambda B)$. Les groupes $gp(\varepsilon
_\mu A,\varepsilon _\lambda B)$ et $gp(A,B)$ peuvent \^{e}tre
diff\'{e}rents, mais les groupes de transformations associ\'{e}s sont
identiques et d\'{e}terminent m\^{e}me groupe de Fricke. Tous donnent les
m\^{e}mes points $\alpha $, $s=0$, $\beta $, $p=\infty $. On peut donc se
limiter \`{a} consid\'{e}rer le groupe principal $gp(A,B)$, avec les
conditions $\lambda >0$ et $\mu >0$ caract\'{e}risant la nappe principale.
Les autres sont ses groupes conjugu\'{e}s.

La remont\'{e}e d'un groupe $\Gamma \subset PSL(2,\mathbb{R})$ \`{a} un groupe $%
G\subset SL(2,\mathbb{R})$ dont $\Gamma $ est l'image est
\'{e}tudi\'{e}e dans \cite{Kra}. Le groupe $\Gamma $ se remonte en
$G$ si et seulement s'il n'a pas d'\'{e}l\'{e}ment d'ordre 2. Dans
le cas parabolique, il n'y pas de difficult\'{e}.
\\
\subsubsection{La r\'{e}duction}

Le processus de r\'{e}duction peut \^{e}tre transpos\'{e} facilement du
groupe principal \`{a} tout groupe conjugu\'{e}. Sur le groupe principal $%
gp(A,B)$ on construit algorithmiquement une suite des transformations $%
\widetilde{X_\phi }$, $\widetilde{Y_\phi }$, $\widetilde{Z_\phi }$, de fa\c
{c}on \`{a} r\'{e}duire tout syst\`{e}me de g\'{e}n\'{e}rateurs $(A,B)$.
Consid\'{e}rons que ce syst\`{e}me d\'{e}finisse avec le triplet de traces
associ\'{e} sur la nappe principale les quatre nombres
\[
m=\max (x,y,z)>0,
\]
\[
m_x=\max (yz-x,y,z)>0,
\]
\[
\ m_y=\max (x,xz-y,z)>0,
\]
\[
m_z=\max (x,y,xy-z)>0.
\]
On dit que le triplet $(x,y,z)$ n'est pas r\'{e}duit si et seulement si l'un
des nombres $m_x$, $m_y$, $m_z$, est strictement plus petit que $m$. On
s'assure facilement que pour tout triplet non r\'{e}duit, deux des nombres $%
m_x$, $m_y$, $m_z$, sont plus grands que $m$, le troisi\`{e}me \'{e}tant
plus petit que $m$. Ceci permet de choisir une unique involution parmi $%
\widetilde{X_\phi }$, $\widetilde{Y_\phi }$, $\widetilde{Z_\phi }$, avec
laquelle on construit un nouveau triplet $(x_1,y_1,z_1)$ tel que la valeur $%
m_1=\max (x_1,y_1,z_1)$ soit strictement plus petite que $m$.\ On poursuit
en renouvelant le proc\'{e}d\'{e} \`{a} partir de ce dernier triplet,
d\'{e}veloppant un processus de descente infinie analogue \`{a} celui que
l'on a utilis\'{e} pour r\'{e}soudre nos \'{e}quations. Il est facile de
v\'{e}rifier que l'algorithme s'arr\^{e}te sur un triplet r\'{e}duit. Ceci
donne :

\begin{proposition}
Tout groupe principal du groupe de Fricke $\Gamma $ associ\'{e} \`{a} un
tore perc\'{e} conforme parabolique $\mathcal{T}_\Gamma ^{\bullet }$
poss\`{e}de un syst\`{e}me de g\'{e}n\'{e}rateurs r\'{e}duit.
\end{proposition}

L'action des involutions $\widetilde{X_\phi }$, $\widetilde{Y_\phi }$, $%
\widetilde{Z_\phi }$, se traduit sur les param\`{e}tres $\lambda $ et $\mu $
gr\^{a}ce \`{a} des involutions d\'{e}finissant une action de $\mathbf{T}_3$
sur le quart de plan :
\[
\mathbf{X}_\phi :(\lambda ,\mu )\longrightarrow (\lambda ,\frac{1+\lambda ^2}%
\mu ).
\]
\[
\mathbf{Y}_\phi :(\lambda ,\mu )\longrightarrow (\frac{1+\mu ^2}\lambda ,\mu
),
\]
\[
\mathbf{Z}_\phi :(\lambda ,\mu )\longrightarrow (\frac \lambda {\lambda
^2+\mu ^2},\frac \mu {\lambda ^2+\mu ^2}).
\]
On fait alors appara\^{i}tre un int\'{e}ressant pavage d'un quart
de plan r\'{e}el par un triangle curviligne, pavage du \`{a} une
action du groupe du triangle $\mathbf{T}_3$. Les points invariants
par $\mathbf{X}_\phi $ dans le quart de plan qui correspond \`{a}
la nappe principale sont port\'{e}s par une hyperbole $H_X$
d'\'{e}quation $\mu ^2-\lambda ^2=1$.\ Ceux qui sont invariants
par $\mathbf{Y}_\phi $ sont port\'{e}s par l'hyperbole $H_Y$
d'\'{e}quation $\lambda ^2-\mu ^2=1$.\ Les points invariants par $\mathbf{Z}%
_\phi $ sont port\'{e}s par le cercle $H_Z$ d'\'{e}quation $\lambda ^2+\mu
^2=1$. Ces trois courbes d\'{e}terminent un triangle curviligne de sommets $%
\mathbf{L}(1,0)$, $\mathbf{M}(0,1)$, $\mathbf{N}(\infty ,\infty )$ qui
constitue un domaine fondamental pour l'action sur le quart de plan du
groupe $\mathbf{T}_3$.
\\
\subsubsection{La super-r\'{e}duction}

Dans le triangle curviligne $\mathbf{LMN}$ lui-m\^{e}me, on a la condition
\[
\mu ^2\leq 1+\lambda ^2.
\]
Mais on peut \'{e}changer le r\^{o}le des matrices $A$ et $B$ sans changer
de groupe, c'est-\`{a} dire permuter $\lambda $ et $\mu $.\ On obtient $%
\lambda \leq \mu $ avec cette transformation
\[
P_1:(A,B)\longrightarrow (B,A).
\]
On obtient aussi $1\leq \lambda $ avec la transformation suivante
\[
P_2:(A,B)\longrightarrow (A,B^{-1}A^{-1}).
\]
On dit qu'un syst\`{e}me de g\'{e}n\'{e}rateurs $(A,B)$ du groupe principal
associ\'{e} \`{a} un tore perc\'{e} parabolique $\mathcal{T}_\Gamma
^{\bullet }$ est super-r\'{e}duit si et seulement si on a les conditions
\[
1\leq \lambda \leq \mu ,\;\;\mu ^2\leq 1+\lambda ^2.
\]
Ce qui pr\'{e}c\`{e}de permet d'\'{e}noncer :

\begin{proposition}
Tout groupe principal du groupe de Fricke $\Gamma $ associ\'{e}
\`{a} un tore perc\'{e} conforme parabolique $\mathcal{T}_{\Gamma}
^{\bullet }$ poss\`{e}de un syst\`{e}me de g\'{e}n\'{e}rateurs
super-r\'{e}duit.
\end{proposition}

\subsubsection{Exemple des tores perc\'{e}s de Klein et de Hecke}

On peut illustrer ce que donne l'algorithme sur les exemples connus de
groupes de Fricke \cite{Cohn1}.

$\bullet $ Tore de Klein : Ce cas a \'{e}t\'{e} donn\'{e} avec $\lambda =1$,
$\mu =\theta =2$, qui ne respectent pas la condition de
super-r\'{e}duction.\ Le triplet correspondant est $(x,y,z)=(6,3,3)$.\ Il
donne $m=6$, $m_x=3$, $m_y=15$, $m_z=15$.\ On identifie ainsi la
transformation $\mathbf{X}_\phi $ qui conduit \`{a} calculer les matrices
suivantes :
\[
A=\left[
\begin{array}{cc}
1 & -1 \\
-1 & 2
\end{array}
\right] ,\;\;B=\left[
\begin{array}{cc}
1 & 1 \\
1 & 2
\end{array}
\right] .
\]
On a alors $(x,y,z)=(3,3,3)$ et $m=3<$ $m_x=$ $m_y=$ $m_z=6$. On est cette
fois dans le triangle curviligne $\mathbf{LMN}$ avec les valeurs $\lambda
=\mu =1$. On voit alors que l'on se ram\`{e}ne simplement au groupe de la
th\'{e}orie de Markoff, o\`{u} $B=A_0$ et $A=B_0$.\ Avec $\lambda =\mu =1$
on est alors dans le cas d'un syst\`{e}me super-r\'{e}duit de
g\'{e}n\'{e}rateurs du groupe principal consid\'{e}r\'{e}.

$\bullet $ Tore de Hecke : Ce cas a \'{e}t\'{e} \'{e}voqu\'{e} avec les
valeurs $\lambda =\mu =\sqrt{2}/2$ et $\theta =1$. Ces valeurs ne respectent
pas la condition de super-r\'{e}duction. On se trouve cette fois dans le
triangle curviligne\textbf{\ }$\mathbf{LMN}$. Le triplet correspondant est
maintenant $(x,y,z)=(2\sqrt{2},2\sqrt{2},4)$.\ Il correspond aux valeurs $%
m=4 $, $m_x=m_y=2\sqrt{2}$, $m_z=4$.\ On n'identifie ainsi aucune
transformation applicable $\mathbf{X}_\phi $, $\mathbf{Y}_\phi $, $\mathbf{Z}%
_\phi $.\ Par contre on peut appliquer $P_2$ qui donne les matrices
suivantes correspondant aux valeurs $\lambda =1$ et $\mu =\sqrt{2}$ :
\[
A=\left[
\begin{array}{cc}
\sqrt{2}/2 & \sqrt{2}/4 \\
\sqrt{2} & 3\sqrt{2}/2
\end{array}
\right] ,\;\;B=\left[
\begin{array}{cc}
4 & -(1/2) \\
2 & 0
\end{array}
\right] .
\]

\subsection{Module d'un tore perc\'{e} conforme parabolique}

On dit que deux tores perc\'{e}s conformes paraboliques sont de m\^{e}me
type si et seulement s'il existe une \'{e}quivalence conforme transformant
l'un en l'autre. L'algorithme de r\'{e}duction permet de remplacer le couple
de g\'{e}n\'{e}rateurs $(A,B)$ d'un groupe de Fricke gr\^{a}ce aux
involutions $X_\phi $, $Y_\phi $, $Z_\phi $.\ Il ne change pas le tore
perc\'{e} conforme sur lequel on travaille. En combinant les deux
m\'{e}thodes, on associe aux diff\'{e}rents types de tores perc\'{e}s
conformes paraboliques avec les calculs qui pr\'{e}c\`{e}dent un nombre
r\'{e}el $(\mu ^2/\lambda ^2)$, le module du tore perc\'{e} que l'on
consid\`{e}re. Les conditions de super-r\'{e}duction garantissent que l'on
peut se ramener \`{a}
\[
1\leq \frac{\mu ^2}{\lambda ^2}\leq 2.
\]
Ceci a permis d'\'{e}noncer :

\begin{proposition}
Tous les types de tores perc\'{e}s conformes paraboliques sont associ\'{e}s
\`{a} un nombre r\'{e}el positif $(\mu ^2/\lambda ^2)$ compris entre 1 et 2,
le module du tore perc\'{e} conforme parabolique consid\'{e}r\'{e}.\ La
valeur 1 correspond \`{a} un tore perc\'{e} de Klein.\ La valeur 2
correspond \`{a} un tore perc\'{e} de Hecke. Toute valeur comprise entre 1
et 2 correspond \`{a} un tore perc\'{e} conforme.
\end{proposition}

Cette proposition classe \`{a} l'aide de leur module les tores perc\'{e}s
conformes paraboliques que l'on peut construire sur un m\^{e}me tore
perc\'{e} topologique. Deux tores perc\'{e}s conformes correspondants \`{a}
des modules $(\mu ^2/\lambda ^2)$ diff\'{e}rents ne peuvent \^{e}tre de
m\^{e}me type. A l'inverse, deux tores correspondants \`{a} un m\^{e}me
module $(\mu ^2/\lambda ^2)$ peuvent ne pas \^{e}tre de m\^{e}me type.
Consid\'{e}rons pour le voir
\[
\mu ^{\prime }=\kappa \mu ,\;\lambda ^{\prime }=\kappa \lambda ,\;\kappa
\neq 0.
\]
Sauf le cas o\`{u} $\kappa =1$, les quadruplets $(\alpha ,s,\beta ,p)$ et $%
(\alpha ^{\prime },s^{\prime },\beta ^{\prime },p^{\prime })$ se
d\'{e}duisent par une homographie sans que celle-ci permette de
conclure \`{a} des relations convenables entre les matrices
associ\'{e}es $A$, $B$ et $A^{\prime }$, $B^{\prime }$. Les traces
seules permettent de garantir que l'on a affaire \`{a} des tores
paraboliques conform\'{e}ment \'{e}quivalents. Par exemple le tore
de la th\'{e}orie de Markoff est donn\'{e} avec $\lambda =\mu =1$,
mais il n'est pas conform\'{e}ment \'{e}quivalent \`{a} celui que
l'on obtient avec $\kappa =2$ et les matrices
\[
A^{\prime }=\left[
\begin{array}{cc}
2 & 8 \\
(1/2) & (5/2)
\end{array}
\right] ,\;B^{\prime }=\left[
\begin{array}{cc}
2 & -8 \\
-(1/2) & (5/2)
\end{array}
\right] ,
\]
car les triplets de traces associ\'{e}s comprennent un rationnel
non entier qui rend impossible de trouver $M\in SL(2,\mathbb{Z})$
telle que $A^{\prime }=M^{-1}A_0M$ et $B^{\prime }=M^{-1}B_0M$.\
Les exemples de ce genre ont \'{e}t\'{e} \'{e}tudi\'{e}s dans
\cite{Cohn2} o\`{u} des indications sont donn\'{e}es sur la valeur
des constantes de Markoff correspondantes, mais la th\'{e}orie
d\'{e}velopp\'{e}e par cet auteur est moins compl\`{e}te que la
n\^{o}tre. Les deux tores perc\'{e}s de l'exemple que l'on vient
de donner ne sont pas conform\'{e}ment \'{e}quivalents, alors
qu'ils sont de m\^{e}me module \'{e}gal \`{a} $1$ par
hypoth\`{e}se. Cette situation ne se
reproduit pas dans le cas particulier du tore de Hecke qui est de module $2$%
.\ Consid\'{e}rons en effet les in\'{e}galit\'{e}s qui le
d\'{e}finissent, elles imposent $\lambda =1$ et $\mu =\sqrt{2}$.
Elles caract\'{e}risent de fa\c {c}on unique le type conforme du
tore de Hecke. En fait, se donner un couple $(\lambda ,\mu )$ avec
les contraintes trouv\'{e}es ant\'{e}rieurement est \'{e}quivalent
\`{a} se donner un couple $((\mu ^2/\lambda ^2),\lambda )$,
c'est-\`{a}-dire plus que le seul module $(\mu ^2/\lambda ^2)$,
avec cette fois les contraintes
\[
1\leq \lambda \leq \frac 1{\sqrt{(\mu ^2/\lambda ^2)-1}}.
\]
La fixation du module $(\mu ^2/\lambda ^2)$ permet de se ramener \`{a} un
m\^{e}me domaine fondamental $p=\infty $, $\alpha =-1$, $s=0$, $\beta =(\mu
^2/\lambda ^2)$.\ Mais le facteur suppl\'{e}mentaire $\lambda $ est en plus
n\'{e}cessaire pour d\'{e}crire alors la fa\c {c}on dont les bords de ce
domaine sont identifi\'{e}s par $A$ et $B$, ce que l'on a pu d\'{e}crire
g\'{e}om\'{e}triquement dans \cite{Perrine9}. On retrouve ainsi le fait que
le type conforme d'un tore perc\'{e} conforme parabolique n\'{e}cessite deux
param\`{e}tres pour \^{e}tre bien d\'{e}fini, ainsi que le fait que les
tores de Hecke sont d\'{e}finis \`{a} transformation conforme pr\`{e}s de $%
\mathcal{H}$ par leur seul module.

La th\'{e}orie de la r\'{e}duction que l'on a d\'{e}velopp\'{e}e n'est pas
celle de \cite{Katok2} qui correspond plut\^{o}t \`{a} un codage des
g\'{e}od\'{e}siques ferm\'{e}es d'un quotient $\mathcal{H}/\Gamma $ o\`{u} $%
\Gamma $ groupe fuchsien.

\subsection{Apparition des quaternions}

On consid\`{e}re une matrice $B\in SL(2,\mathbb{R})$ telle que $%
tr(B)=((1+\lambda ^2+\mu ^2)/\lambda )$. Avec le groupe $G_1$ introduit
pr\'{e}c\'{e}demment, $B_1\in G_1$ et la condition $BD=DB_1$o\`{u}
\[
B_1=\left[
\begin{array}{cc}
\lambda & -(\lambda \mu ^2) \\
-(1/\lambda ) & ((1+\mu ^2)/\lambda )
\end{array}
\right] ,\;\;D=\left[
\begin{array}{cc}
\mathbf{x} & \mathbf{y} \\
\mathbf{z} & \mathbf{t}
\end{array}
\right] ,\;\;\det (D)=\pm 1,
\]
on a obtenu :

\begin{proposition}
Si $B\in SL(2,\mathbb{R})$, on a \'{e}quivalence des deux
propri\'{e}t\'{e}s :

1/ $tr(B)=((1+\lambda ^2+\mu ^2)/\lambda )$.

2/ Il existe $D\in GL(2,\mathbb{R})$ telle que $B=DB_1D^{-1}$
o\`{u}
\[
B_1=\left[
\begin{array}{cc}
\lambda & -\lambda \mu ^2 \\
-(1/\lambda ) & ((1+\mu ^2)/\lambda )
\end{array}
\right] .
\]
\end{proposition}

Si on combine maintenant cette proposition avec la recherche d'une matrice $%
D^{\prime }$ telle que $A=D^{\prime }A_1D^{\prime -1}$ et $tr(A)=((1+\lambda
^2+\mu ^2)/\mu )$, on est conduit \`{a} \'{e}crire
\[
B^{-1}A^{-1}=DB_1^{-1}D^{-1}D^{\prime }A_1^{-1}D^{\prime -1},
\]
\[
tr(B^{-1}A^{-1})=tr(B_1^{-1}(D^{-1}D^{\prime })A_1^{-1}(D^{-1}D^{\prime
})^{-1})=((1+\lambda ^2+\mu ^2)/\lambda \mu ).
\]
Ceci introduit une matrice
\[
W=D^{-1}D^{\prime }=\left[
\begin{array}{cc}
\varpi _1 & \varpi _4 \\
\varpi _3 & \varpi _2
\end{array}
\right] ,
\]
et le calcul effectif de sa trace donne un \'{e}quation quadratique que l'on
peut interpr\'{e}ter comme la norme d'un quaternion. Une solution de cette
\'{e}quation est donn\'{e}e par $\varpi _1=\varpi _2=\pm 1$, $\varpi
_3=\varpi _4=0.$ Les autres solutions sont calculables et fournissent des
quaternions non triviaux que l'on peut utiliser pour donner une
caract\'{e}risation du couple $(A,B)$ par le triplet de traces
\[
\Pi (A,B)=(tr(B^{-1}),tr(A),tr(B^{-1}A^{-1})).
\]

\section{Perspectives}

Dans ce qui pr\'{e}c\`{e}de on a donn\'{e} une nouvelle interpr\'{e}tation
g\'{e}om\'{e}trique des \'{e}quations diophantiennes $M^{s_1s_2}(a,\partial
K,u_\theta )$ de notre th\'{e}orie de Markoff g\'{e}n\'{e}ralis\'{e}e. Le
lien a \'{e}t\'{e} fait avec la th\'{e}orie des tores perc\'{e}s conformes,
et on a vu une diff\'{e}rence entre le cas parabolique o\`{u} la
g\'{e}n\'{e}ralisation est compl\`{e}te et le cas hyperbolique o\`{u} les
r\'{e}sultats sont plus lacunaires. Pour les tores paraboliques, on dispose
d'une th\'{e}orie de la r\'{e}duction compl\`{e}te qui classe les triplets
de traces sous l'action du groupe du triangle $\mathbf{T}_3$.\ Tous sont
issus d'une \'{e}quation de Markoff classique gr\^{a}ce \`{a} la condition
\[
tr(A)^2+tr(B)^2+tr(AB)^2-tr(A)tr(B)tr(AB)=\sigma =0.
\]
Ils s'interpr\`{e}tent avec les couples de g\'{e}n\'{e}rateurs du
groupe libre non commutatif \`{a} deux g\'{e}n\'{e}rateurs
$\mathbf{F}_2$ auquel tout groupe de Fricke est isomorphe
\cite{Rosenberger}. Ce cas g\'{e}n\'{e}ralise de fa\c {c}on
compl\`{e}te aux tores perc\'{e}s paraboliques la th\'{e}orie de
Markoff classique. Cette pr\'{e}sentation explique les
d\'{e}veloppements que l'on trouve dans \cite{Cohn1} et \cite
{Cohn2} auxquels elle donne une interpr\'{e}tation.\ Elle
explicite les travaux de \cite{Schmidt}. On peut compl\'{e}ter ce
qui pr\'{e}c\`{e}de par un calcul des constantes de Markoff
associ\'{e}es, sachant que les fractions continues ne
s'\'{e}crivent plus dans ce cas avec des $1$ et des $2$, mais
contiennent d'autres valeurs (voir \cite{Cohn2}). Trouver un
exemple o\`{u} la seule valeur suppl\'{e}mentaire est \'{e}gale
\`{a} $3$ ne para\^{i}t pas un d\'{e}fi insurmontable.

Les tores perc\'{e}s conformes hyperboliques sont eux-m\^{e}mes donn\'{e}s
par une \'{e}quation $M^{s_1s_2}(a,\partial K,u_\theta )$ dont on a
donn\'{e} une interpr\'{e}tation g\'{e}om\'{e}trique et pour lesquels on a
une action du groupe de triangle $\mathbf{T}_3$ et une r\'{e}duction
associ\'{e}e.\ On les a \'{e}galement class\'{e}s \`{a} isom\'{e}trie
pr\`{e}s de $\mathcal{H}$ avec la condition diff\'{e}rente (comparer \`{a}
\cite{Jin}) :
\[
tr(A)^2+tr(B)^2+tr(AB)^2-tr(A)tr(B)tr(AB)=\sigma <0.
\]
Pour que ce classement soit \`{a} \'{e}quivalence conforme pr\`{e}s de $%
\mathcal{H}$ il faut ajouter une valeur $\varepsilon =\pm 1$ correspondant
\`{a} l'orientation du tore perc\'{e}. On a une autre r\'{e}duction pour les
tores $\sigma $-hyperboliques ainsi d\'{e}finis.\ On a vu que le groupe
correspondant n'est plus de Fricke.\ On a donn\'{e} un exemple o\`{u} une
relation entre g\'{e}n\'{e}rateurs a \'{e}t\'{e} calcul\'{e}e, il faut voir
si la m\'{e}thode utilis\'{e}e est g\'{e}n\'{e}ralisable, et ce que devient
la super-r\'{e}duction. Ceci est examin\'{e} au chapitre suivant.\ Les
g\'{e}od\'{e}siques invariantes ont permis de comprendre comment les
surfaces de Riemann qui en r\'{e}sultent sont prolong\'{e}es d'une surface
\`{a} un trou vers une surface \`{a} une piq\^{u}re. On a vu comment ce cas
recouvre une situation o\`{u} apparaissent deux tores trou\'{e}s conformes
li\'{e}s entre eux (doubles de Schottky), qui se recouvrent dans le cas
parabolique limite. On peut d\'{e}velopper une th\'{e}orie de la
r\'{e}duction pour les tores hyperboliques en prenant soin de travailler
simultan\'{e}ment sur les deux tores. Il faudrait aussi voir ce que devient
dans le cas hyperbolique le lien avec les quaternions.

L'exemple hyperbolique identifi\'{e} avec $\sigma =1769$ concerne les
matrices
\[
A=\left[
\begin{array}{cc}
11 & 3 \\
7 & 2
\end{array}
\right] ,\;\;B=\left[
\begin{array}{cc}
37 & 11 \\
10 & 3
\end{array}
\right] ,\;\;H=\left[
\begin{array}{cc}
3065 & -4799 \\
829 & -1298
\end{array}
\right] .
\]
Le groupe associ\'{e} $G=<A,B,H\mid [A,B]H=\mathbf{1}_2>$ n'est pas libre
car il contient des \'{e}l\'{e}ments particuliers $U=B^{-1}A$ et $V=BA^{-1}$
tels que $U^2=V^2=-\mathbf{1}_2$. La piste d'approfondissement \`{a} la
g\'{e}om\'{e}trie alg\'{e}brique qui a \'{e}t\'{e} signal\'{e}e dans ce cas
autour de \cite{Shafarevich} (Tome 1, chapitre III 1.6 p.\ 164) constitue un
sujet important \`{a} creuser. On a esquiss\'{e} l'interpr\'{e}tation qui
reste \`{a} d\'{e}tailler mat\'{e}rialisant le groupe des classes de
diviseurs de nos surfaces cubiques en chaque point entier en utilisant une
courbe $S=s(\mathbf{P}^1)$ repr\'{e}sentant le plan projectif dans la
surface $M^{s_1s_2}(a,\partial K,u_\theta )$ et une fibre non singuli\`{e}re
$F$. Cette approche construit un groupe libre \`{a} deux g\'{e}n\'{e}rateurs
dont il faut comprendre \`{a} quoi il correspond dans le cas hyperbolique.
Ce cas pourrait avoir une grande importance pour la classification des
fibr\'{e}s vectoriels \cite{Drezet} \cite{Drezet1} \cite{Drezet2} \cite
{LePotier} \cite{Klyachko} \cite{Klyachko1}.
\[
\]

\chapter{G\'{e}n\'{e}ralisation aux surfaces de Riemann}

\section{Introduction}

Les r\'{e}flexions pour g\'{e}n\'{e}raliser la th\'{e}orie de
Markoff classique \`{a} des situations plus vastes ont conduit
\`{a} \'{e}tudier de plus pr\`{e}s la g\'{e}om\'{e}trie conforme
des surfaces de Riemann.\ Un expos\'{e} sur cette question est
donn\'{e} dans \cite{Perrine9} o\`{u} l'on d\'{e}crit la vision
que l'auteur a de ce dernier sujet, et les liens qu'il a
formalis\'{e}s avec des th\`{e}mes d'actualit\'{e} en
math\'{e}matiques ou en physique. Le r\'{e}sum\'{e} qui suit
s'appuie sur les expos\'{e}s classiques sur le sujet
(\cite{Siegel} \cite{Lehner} \cite{Farkas} \cite {Beardon}
\cite{Katok1} \cite{Miranda} \cite{Serre1} \cite{Cohn6} \cite
{Douady} \cite{Stillwell} \cite{Zieschang} \cite{Berger1}
\cite{Lehmann}
\cite{Doubrovine} \cite{Reyssat} \cite{Nakahara} \cite{Nash1} \cite{Nash2}%
...). Il indique quelques perspectives de recherches que l'auteur a choisi
d'explorer. Le point de d\'{e}part a \'{e}t\'{e} d'\'{e}tendre les travaux
pr\'{e}sent\'{e}s pr\'{e}c\'{e}demment sur les tores perc\'{e}s au cadre
plus g\'{e}n\'{e}ral des surfaces de Riemann. On a concr\'{e}tis\'{e} cette
id\'{e}e en tentant d'\'{e}clairer des probl\'{e}matiques contemporaines. Le
chaos quantique est ainsi devenu progressivement une pr\'{e}occupation
essentielle.\ On a recherch\'{e} les liens qu'il pouvait avoir avec la
th\'{e}orie de Markoff. En d'autres termes, il s'agissait de savoir si le
spectre de Markoff peut \^{e}tre obtenu comme spectre d'un op\'{e}rateur, ce
qui pourrait expliquer son apparition dans des objets physiques tels que des
oscillateurs \cite{Planat1}.

Les principaux r\'{e}sultats auxquels on est parvenu sont les suivants :

$\bullet $ On a reformul\'{e} l'approche sur les tores perc\'{e}s
conformes en la pla\c {c}ant dans le cadre plus vaste des groupes
fuchsiens.\ Ceci a permis de comprendre comment \'{e}tudier le cas
hyperbolique non compl\`{e}tement trait\'{e} dans ce qui
pr\'{e}c\`{e}de. Ceci a aussi fait le lien complet avec la
th\'{e}orie de Teichm\"{u}ller interpr\'{e}t\'{e}e ici
comme th\'{e}orie des repr\'{e}sentations d'un groupe de Poincar\'{e} dans $%
PSL(2,\mathbb{R})$.\ Cette approche g\'{e}n\'{e}ralise la
th\'{e}orie de Markoff de fa\c {c}on tr\`{e}s satisfaisante. On
dispose ainsi d'un espace de modules qui joue le r\^{o}le de
l'ensemble des triplets de traces pour les tores perc\'{e}s. On a
aussi un groupe qui agit sur cet espace, g\'{e}n\'{e}ralisant
l'action du groupe du triangle du cas des tores perc\'{e}s. On a
pu en d\'{e}duire des r\'{e}sultats nouveaux sur des \'{e}quations
diophantiennes \`{a} plus grand nombre de variables que l'on peut
traiter comme l'\'{e}quation de Markoff. Il existe d\'{e}j\`{a} un
tel exemple \cite{Baragar}. Pour d\'{e}velopper l'approche
g\'{e}om\'{e}trique correspondante, on a recherch\'{e} les objets
\`{a} consid\'{e}rer \`{a} la place des surfaces de Riemann qui
semblent de ce point de vue un peu limit\'{e}es. C'est ainsi que
les espaces de Stein ont \'{e}t\'{e} abord\'{e}s, mais ils ne sont
certainement pas la bonne notion pour de nouvelles
g\'{e}n\'{e}ralisations, et on a indiqu\'{e} pourquoi il faut
privil\'{e}gier les domaines de Riemann.

$\bullet $ On a ensuite \'{e}tudi\'{e} le lien avec le codage des
g\'{e}od\'{e}siques sur une surface de Riemann.\ Ceci constitue un sujet
o\`{u} les r\'{e}sultats g\'{e}n\'{e}raux disponibles restent limit\'{e}s
\cite{Series1} \cite{Series2}, mais qui a un lien tr\`{e}s important avec
l'\'{e}tude des syst\`{e}mes dynamiques et la th\'{e}orie ergodique,
notamment les g\'{e}od\'{e}siques ferm\'{e}es.\ La r\'{e}flexion a
\'{e}t\'{e} faite dans la perspective de sortir de la seule th\'{e}orie de
Markoff qui semble pourtant la seule o\`{u} on dispose de r\'{e}sultats
signicatifs \cite{Schmutz2}.

$\bullet $ La fonction \^{e}ta de Dedekind est sous-jacente \`{a}
tous nos travaux.\ On a montr\'{e} qu'elle donne un certain nombre
des fonctions transcendantes classiques sur lesquelles sont
construits les plus beaux r\'{e}sultats de la th\'{e}orie de
nombres.\ On a donc cherch\'{e} \`{a} rep\'{e}rer un certain
nombre d'expressions o\`{u} cette fonction $\eta $ appara\^{i}t,
donnant la fonction modulaire, les fonctions elliptiques, les
fonctions th\^{e}ta, etc... On a montr\'{e} comment ces
derni\`{e}res sont importantes pour la th\'{e}orie de
l'information, et notamment la dualit\'{e} codage/quantification.\
Une id\'{e}e que l'on a ensuite cherch\'{e} \`{a} approfondir est
que ces expressions expliquent \`{a} partir de la
d\'{e}composition en produit infini de $\eta $ beaucoup des
produits infinis classiques des autres fonctions. Un point a
\'{e}t\'{e} laiss\'{e} de c\^{o}t\'{e}, concernant les fonctions
$L$. Mais elles ont \'{e}galement une propri\'{e}t\'{e} en
relation avec ce sch\'{e}ma \cite {Ericksson}.

$\bullet $ On a ensuite cherch\'{e} \`{a} comprendre certaines des
expressions diff\'{e}rentielles mises en avant dans les premiers travaux de
Harvey Cohn relatifs \`{a} l'interpr\'{e}tation g\'{e}om\'{e}trique de la
th\'{e}orie de Markoff (\cite{Cohn1}, \cite{Cohn7}, \cite{Cohn8}, \cite
{Cohn1}). Ceci permet de faire le lien avec la sph\`{e}re \`{a} trois trous,
et de comprendre par l\`{a} m\^{e}me certains travaux de Asmus Schmidt \cite
{Schmidt}. Une approche hyperg\'{e}om\'{e}trique est \'{e}galement possible
\`{a} partir de l\`{a}. Mais le plus important est que ce d\'{e}veloppement
met l'accent sur la double uniformisation \`{a} l'oeuvre sur les tores
perc\'{e}s.

$\bullet $ Cette double uniformisation vient de ce que le tore perc\'{e} a
pour rev\^{e}tement conforme le demi-plan de Poincar\'{e}, mais qu'il peut
\^{e}tre compl\'{e}t\'{e} en un tore qui a pour rev\^{e}tement conforme $%
\mathbb{C}$. Le tore donne naturellement naissance aux courbes
elliptiques et donc \`{a} des \'{e}quations cubiques analogues
\`{a} celles consid\'{e}r\'{e}es avant, cependant que le tore
perc\'{e} donne pour ce qui le concerne naissance aux
\'{e}quations analogues \`{a} celle de Markoff. Comme l'auteur a
d\'{e}crit dans les chapitres pr\'{e}c\'{e}dents les relations que
ces deux types d'\'{e}quations alg\'{e}briques entretiennent, il
convenait de mieux comprendre cette situation qui a pour
cons\'{e}quence l'existence d'une riche structure de double
rev\^{e}tement conforme \cite {Mazur1} sur les tores perc\'{e}s.
Cette observation a des implications arithm\'{e}tiques profondes
comparables \`{a} celles du cas de la conjecture de Shimura
Taniyama Weil \cite{Wiles} o\`{u} l'un des rev\^{e}tements est
euclidien et l'autre hyperbolique. On en a d\'{e}duit une
construction g\'{e}n\'{e}rale permettant de param\'{e}trer presque
tous les points d'un tore perc\'{e} avec des fonctions
elliptiques. Toutes les cons\'{e}quences d'une telle id\'{e}e ne
sont pas tir\'{e}es, notamment ce que l'on pourrait en tirer pour
une conjecture de Selberg.\ Le r\'{e}sultat essentiel auquel on
est parvenu est que la fonction \^{e}ta de Dedekind peut \^{e}tre
interpr\'{e}t\'{e}e avec le laplacien d'un tore.\ Cette approche
d\'{e}bouche sur la mise en avant d'un nouvel invariant
fondamental dont on a rep\'{e}r\'{e} l'utilisation en physique.

$\bullet $ C'est \`{a} partir de ces observations que les r\'{e}flexions sur
le chaos quantique apparaissent naturellement.\ Toute la r\'{e}flexion
s'organise autour d'une \'{e}quation de Schr\"{o}dinger dont l'espace des
phases est un tore ou un tore perc\'{e}. Dans le premier cas, on a mis en
\'{e}vidence un lien profond avec la loi de r\'{e}ciprocit\'{e} quadratique,
et donc la fonction $\eta $ de Dedekind et les sommes de Gauss, mais il est
curieux de constater que le temps devient discret. Le cas o\`{u} l'espace
des phases est un tore perc\'{e} reste \`{a} \'{e}tudier, et l'auteur
conjecture que c'est celui qui donne l'interpr\'{e}tation recherch\'{e}e du
spectre de Markoff comme spectre d'un op\'{e}rateur li\'{e} \`{a} une
\'{e}quation de Schr\"{o}dinger.

$\bullet $ En approfondissant ce sujet, on est parvenu \`{a} r\'{e}soudre le
probl\`{e}me de Riemann-Hilbert associ\'{e} \`{a} la th\'{e}orie de Markoff
classique. En d'autres termes on a pu \'{e}crire une \'{e}quation
diff\'{e}rentielle fuchsienne dont le groupe de monodromie est engendr\'{e}
par les deux matrices $A_0$ et $B_0$ qui engendrent le groupe $[SL(2,\mathbb{Z}%
),SL(2,\mathbb{Z})]\simeq \mathbf{F}_2$. Il reste \`{a}
approfondir l'analyse spectrale de l'op\'{e}rateur
diff\'{e}rentiel associ\'{e}.

$\bullet $ On a enfin esquiss\'{e} le lien avec la th\'{e}orie de fibr\'{e}s
vectoriels et montr\'{e} \`{a} partir de l\`{a} comment se d\'{e}veloppe une
approche par la $K$-th\'{e}orie. La th\'{e}orie de Markoff correspond dans
ce contexte \`{a} des fibr\'{e}s exceptionnels du plan projectif $P_2(\mathbb{C}%
)$ et aux h\'{e}lices de D. Yu. Nogin \cite{Nogin} \cite{Nogin1}.\ Cette
approche est particuli\`{e}rement importante car elle donne un cadre
permettant de comprendre un certain nombre de conjectures tr\`{e}s
importantes encore non r\'{e}solues comme les conjectures de Lichtenbaum ou
celle de Birch et Swinnerton-Dyer, gr\^{a}ce \`{a} une interpr\'{e}tation
automorphe de la $K$-th\'{e}orie. Au passage, des liens avec la conjecture
de Riemann ont \'{e}t\'{e} approfondis.

\section{Rappels succincts sur les surfaces de Riemann}

Le pr\'{e}sent paragraphe est un simple rappel sur les surfaces de Riemann
destin\'{e} \`{a} fixer les notations pour la suite. Il peut \^{e}tre omis
par le lecteur averti et ne d\'{e}veloppe rien que l'on ne trouve dans \cite
{Perrine9}.

Les objets g\'{e}om\'{e}triques que l'on a consid\'{e}r\'{e}s
pr\'{e}c\'{e}demment, les tores perc\'{e}s conformes, sont des quotients du
demi-plan de Poincar\'{e} $\mathcal{H}$ par l'action d'un groupe fuchsien $%
\Gamma $, c'est-\`{a}-dire d'un sous-groupe discret de
$PSL(2,\mathbb{R})$. Pour la plupart des surfaces de Riemann on
peut g\'{e}n\'{e}raliser la th\'{e}orie d\'{e}velopp\'{e}e pour
les tores perc\'{e}s en utilisant un groupe fuchsien. On peut en
effet construire par quotient de $\mathcal{H}$ pour l'action d'un
tel groupe toutes les surfaces de Riemann \`{a} l'exception de
celles dont le support topologique est hom\'{e}omorphe (\cite
{Farkas} p. 208) \`{a} la sph\`{e}re de Riemann $\mathcal{S}^2$,
\`{a} la sph\`{e}re perc\'{e}e par extraction d'un point
$\mathbb{C}$, \`{a} la
sph\`{e}re perc\'{e}e par extraction de deux points $\mathbb{C}^{*}$, au tore $%
\mathcal{T}$. Pour toute autre surface de Riemann $\mathcal{M}$ le
groupe de Poincar\'{e} $\pi _1(\mathcal{M},*)$ peut \^{e}tre
repr\'{e}sent\'{e} comme sous-groupe $\Gamma $ du groupe des
automorphismes $PSL(2,\mathbb{R})$ de son rev\^{e}tement conforme
$\mathcal{H}$. La surface $\mathcal{M}$ est de forme
$\mathcal{H}/\Gamma $, et une repr\'{e}sentation de groupes $\overline{\rho }%
:\pi _1(\mathcal{M},*)\rightarrow \Gamma $ porte les donn\'{e}es
g\'{e}om\'{e}triques de $\mathcal{M}$. Dans la suite les surfaces $\mathcal{M%
}$ sont connexes, et donc connexes par arcs, de sorte que leur groupe de
Poincar\'{e} $\pi _1(\mathcal{M},*)$ ne d\'{e}pend pas du point de base
servant \`{a} le d\'{e}finir.

\subsection{Uniformisation des surfaces de Riemann}

Un rev\^{e}tement conforme d'une surface de Riemann $\mathcal{M}$ est dit
universel si et seulement si son groupe de Poincar\'{e} $\pi _1(\mathcal{M}%
,*)$ est r\'{e}duit \`{a} un \'{e}l\'{e}ment neutre. La surface de Riemann
est alors simplement connexe. Toutes les surfaces de Riemann simplement
connexes sont connues \`{a} \'{e}quivalence conforme pr\`{e}s (\cite{Farkas}
p. 206).\ Elles correspondent aux trois mod\`{e}les de g\'{e}om\'{e}trie
classique (\cite{Wolf} \cite{Nakahara} p.\ 486) dont la structure conforme
est unique sur le support topologique que l'on consid\`{e}re, \`{a} courbure
constante positive (cas sph\'{e}rique), \`{a} courbure nulle (cas
euclidien), et \`{a} courbure n\'{e}gative (cas hyperbolique). Il s'agit des
suivantes :

$\bullet $ La sph\`{e}re de Riemann $\mathcal{S}^2\mathbb{=}\mathbf{P}_1(\mathbb{C}%
)$ de type conforme $(0,0,0)$.

$\bullet $ Le plan complexe $\mathbb{C}=\mathcal{S}^2\backslash
\{\infty \}$ qui est du type conforme $(0,1,0)$.

$\bullet $ Le demi-plan de Poincar\'{e} $\mathcal{H}$ qui est du type
conforme $(0,0,1)$.

Le th\'{e}or\`{e}me de Killing-Hopf (\cite{Stillwell} p.\ 135)
garantit que toute surface de Riemann peut ainsi \^{e}tre obtenue
comme quotient de l'une des trois surfaces $\mathcal{S}^2$,
$\mathbb{C}$, $\mathcal{H}$, par l'action d'un sous-groupe de leur
groupe d'automorphismes conformes.\ Il constitue la base de trois
th\'{e}ories de Galois s'appliquant respectivement pour les
surfaces de Riemann. Si $\mathcal{M}$ est une surface de Riemann
de rev\^{e}tement
simplement conforme $\mathcal{M}^{sc}$ et de groupe de rev\^{e}tement $%
\Gamma $, sous-groupe de $Aut(\mathcal{M}^{sc})$, on a \'{e}quivalence
conforme des surfaces $\mathcal{M}$ et $\mathcal{M}^{sc}/\Gamma $. D'o\`{u}
l'importance de connaitre les groupes d'automorphismes des surfaces
simplement connexes (\cite{Farkas} p. 206) :
\[
Aut(\mathcal{S}^2)\simeq PSL(2,\mathbb{C})\;\text{groupe des
transformations de M\"{o}bius complexes},
\]
\[
Aut(\mathbb{C})\simeq P\Delta L(2,\mathbb{C})\;\text{groupe
donn\'{e} par les matrices triangulaires sup\'{e}rieures},
\]
\[
Aut(\mathcal{H})\simeq PSL(2,\mathbb{R})\simeq
Isom^{+}(\mathcal{H})\text{ groupe des matrices r\'{e}elles}.
\]
Le groupe $PSL(2,\mathbb{C})=\{M\in GL(2,\mathbb{C})\mid \det
(M)=1\}/\{\pm 1\}$ de la sph\`{e}re contient les deux autres
groupes cit\'{e}s, ce qui r\'{e}unit les trois th\'{e}ories de
Galois que l'on vient d'\'{e}voquer en une seule.\ Les sous
groupes de $PSL(2,\mathbb{C})$ sont les groupes klein\'{e}ens
\cite{Series3}.

On conna\^{i}t tous les types conformes de surfaces de Riemann qui
ont les surfaces $\mathcal{S}^2$ ou $\mathbb{C}$ pour
rev\^{e}tement universel (\cite {Farkas} p. 208).\ Ce sont les cas
suivants :

$\bullet $ La sph\`{e}re de Riemann $\mathcal{S}^2$ est la seule surface de
Riemann qui poss\`{e}de $\mathcal{S}^2$ pour rev\^{e}tement universel
conforme. Son groupe de Poincar\'{e} est trivial.

$\bullet $ $\mathbb{C}$, $\mathbb{C}^{*}\simeq \mathbb{C}/\omega
\mathbb{Z}$, et les tores compacts
$\mathcal{T}=\mathcal{T}_\Lambda =\mathbb{C}/\Lambda $ sont les
seules surfaces de Riemann qui ont
$\mathbb{C}=\mathcal{S}^2\backslash \{\infty \}$ pour
rev\^{e}tement universel conforme.

- La sph\`{e}re perc\'{e}e d'une piq\^{u}re
$\mathbb{C}=\mathcal{T}_0^{\bullet }=\mathcal{S}^2\backslash
\{\infty \}$ est du type conforme $(0,1,0)$. Son groupe de
Poincar\'{e} est trivial.

- La sph\`{e}re perc\'{e}e de deux piq\^{u}res $\mathbb{C}^{*}=\mathcal{T}%
_0^{\bullet \bullet }=\mathcal{S}^2\backslash \{0,\infty \}$ est du type
conforme $(0,2,0)$.\ On peut la repr\'{e}senter par un cylindre $\mathbb{C}%
/\omega \mathbb{Z}$ d\'{e}fini avec $\omega \in \mathbb{C}^{*}$.\
Elle a pour groupe de Poincar\'{e} $\pi _1(\mathbb{C}^{*},*)\simeq
\mathbb{Z}$. Dans $\mathbb{C}$
le domaine fondamental est une bande permettant de paver avec le groupe $%
\mathbb{Z}$ tout l'espace $\mathbb{C}$.

- Les tores compacts $\mathcal{T}_\Lambda =\mathbb{C}/\Lambda $
sont du type conforme $(1,0,0)$, conform\'{e}ment \'{e}quivalents
\`{a} des courbes elliptiques. La projection canonique donne un
r\'{e}v\^{e}tement universel d'un tel tore et est d\'{e}finissable
avec des fonctions elliptiques. Son
groupe de Poincar\'{e} est $\pi _1(\mathcal{T},*)=\mathbb{Z\oplus Z}\simeq \mathbb{%
Z}^2$. On peut montrer que $Aut(\mathcal{M})$ est une extension de $\mathbb{C}%
/\Lambda $ par un groupe fini (\cite{Farkas} p. 296,
\cite{Reyssat} p.\ 48), en g\'{e}n\'{e}ral $\{\pm 1\}$.\ Mais deux
cas se distinguent correspondant \`{a} la sym\'{e}trie carr\'{e}e
$\Lambda \simeq \mathbb{Z}\oplus i\mathbb{Z}$ et \`{a} la
sym\'{e}trie hexagonale $\Lambda \simeq \mathbb{Z}\oplus
j\mathbb{Z}$.

$\bullet $ Trois types conformes de surfaces de Riemann compl\'{e}mentaires
aux pr\'{e}c\'{e}dents sont caract\'{e}ris\'{e}s par le fait que $\mathcal{H}
$ est cette fois leur rev\^{e}tement universel conforme (\cite{Farkas} p.
210) mais que leur groupe de Poincar\'{e} est commutatif. En dehors de $%
\mathcal{H}$ lui-m\^{e}me, on trouve les suivants :

- La sph\`{e}re perc\'{e}e d'une piq\^{u}re et d'un trou $\mathcal{D}%
^{\bullet }=\{z\in \mathbb{C};\;0<\mid z\mid <1\}$ qui est du type conforme $%
(0,1,1)$ et v\'{e}rifie $\pi _1(\mathcal{D}^{\bullet },*)\simeq
\mathbb{Z}$.

- La sph\`{e}re perc\'{e}e de deux trous $\mathcal{D}_\alpha ^{\circ
}=\{z\in \mathbb{C};\;0<\alpha <\mid z\mid <1\}$ qui est du type conforme $%
(0,0,2)$ et v\'{e}rifie $\pi _1(\mathcal{D}_\alpha ^{\circ
},*)\simeq \mathbb{Z} $.

$\bullet $ Dans tous les autres cas qui sont en nombre infini, le
rev\^{e}tement universel conforme de $\mathcal{M}$ est le
demi-plan de Poincar\'{e} $\mathcal{H}$ sur lequel agit son groupe
de Poincar\'{e} $\pi _1(\mathcal{M},*)$ qui est non commutatif. Ce
groupe est isomorphe \`{a} un groupe fuchsien $\Gamma \subset
PSL(2,\mathbb{R})\simeq Aut(\mathcal{H})$ agissant sur
$\mathcal{H}$ pour donner $\mathcal{M}\simeq \mathcal{H}/\Gamma $.
En pratique \cite{DeRham}, la surface $\mathcal{M}$ peut \^{e}tre
visualis\'{e}e globalement avec un domaine fondamental pour
l'action du groupe $\Gamma $ dans $\mathcal{H}$. Pour la
d\'{e}finition d'un domaine fondamental polygonal on peut utiliser
la m\'{e}thode de \cite{Keen}.

\subsection{Surfaces de Riemann d\'{e}finies par un groupe fuchsien}

Les groupes fuchsiens $\Gamma $ permettent de d\'{e}crire \`{a}
\'{e}quivalence conforme pr\`{e}s toutes les surfaces de Riemann
en dehors de celles qui ont $\mathcal{S}^2$ ou $\mathbb{C}$ pour
rev\^{e}tement universel. On a :

\begin{proposition}
Hors les types $(0,0,0)$, $(0,1,0)$, $(1,0,0)$, $(0,2,0)$, $(0,1,1)$, $%
(0,0,2)$, toute surface de Riemann est conform\'{e}ment \'{e}quivalente
\`{a} une surface qui peut \^{e}tre obtenue comme un espace quotient de
forme
\[
\mathcal{M}=\mathcal{H}/\Gamma ,
\]
o\`{u} $\Gamma \simeq \pi _1(\mathcal{M},*)$ groupe fuchsien non
commutatif isomorphe \`{a} un sous groupe de
$Aut(\mathcal{H})\simeq PSL(2,\mathbb{R}).$
\end{proposition}

Toutes les surfaces de type fini de genre $g\geq 2$ sont d\'{e}crites
ainsi.\ Mais c'est aussi le cas pour certaines surfaces de genre $0$ ou $1$.
Les tores perc\'{e}s conformes paraboliques qui sont de genre $1$ sont
donn\'{e}s par cette derni\`{e}re proposition.\ Avec une matrice parabolique
$P=L^{-1}$, le groupe fuchsien correspondant a pour pr\'{e}sentation $<%
\overline{A},\overline{B},\overline{P}\mid [\overline{A},\overline{B}]%
\overline{P}=\mathbf{1}_2>$. Les tores perc\'{e}s conformes
hyperboliques \'{e}galement
de genre $1$ sont donn\'{e}s par un groupe fuchsien dont on conna\^{i}t une pr\'{e}sentation $<%
\overline{A},\overline{B},\overline{H}\mid [\overline{A},\overline{B}]%
\overline{H}=\mathbf{1}_2>$ o\`{u} $H$ matrice hyperbolique. Les
sph\`{e}res \`{a} trois piq\^{u}res qui sont de genre $0$,
c'est-\`{a}-dire les pantalons, peuvent \^{e}tre obtenus de
m\^{e}me (\cite{Stillwell} p.114).\ Le groupe fuchsien
correspondant est isomorphe au groupe du triangle $\mathbf{T}_3$.

\subsection{Autre anti-\'{e}quivalence de cat\'{e}gories}

Le th\'{e}or\`{e}me fondamental de Riemann associe \`{a} toute surface de
Riemann compacte $\mathcal{M}$ une \'{e}quation polyn\^{o}miale $Q(x,y)=0$.
En normalisant $Q$ on se ram\`{e}ne \`{a} une relation alg\'{e}brique
irr\'{e}ductible entre les variables complexes $y$ et $x$ de forme suivante
\[
\Phi (y,x)=y^n+\phi _1(x)y^{n-1}+...+\phi _n(x)=0,
\]
o\`{u} $\phi _k(x)\;(1\leq k\leq n\leq m)$ sont des fonctions
rationnelles de $x$. Leurs d\'{e}nominateurs s'annulent en un
nombre fini de p\^{o}les o\`{u} l'on peut consid\'{e}rer que la
valeur prise par $y$ devient infinie. Ailleurs, la r\'{e}solution
en $y$ d'une telle \'{e}quation donne une fonction multivalente
$y(x)$, chaque valeur de $x$ permettant de d\'{e}finir $n$ valeurs
$y_i(x)$ dans $\mathbb{C}$ l\`{a} o\`{u} le discrimant de $\Phi $
n'est pas nul, c'est-\`{a}-dire dans un ouvert $\mathbb{C}_\Phi $
de $\mathbb{C}$
tel que $\mathbb{C}\backslash \mathbb{C}_\Phi $ ensemble fini. Pour tout point $%
x\in $ $\mathbb{C}_\Phi $, chaque uniformisation $y_i(x)$ donne
par le th\'{e}or\`{e}me des fonctions implicites une carte locale
holomorphe qui se prolonge analytiquement gr\^{a}ce au
th\'{e}or\`{e}me de Puiseux (\cite {Dieudonne3} p.\ 106,
\cite{Arnaudies2} th\'{e}or\`{e}me 7.7).\ L'ensemble de ces
prolongements d\'{e}finit une surface de Riemann
$\mathcal{N}\subset \mathcal{M}$ \`{a} $n$ feuilles au dessus de
$\mathbb{C}_\Phi $. On
compl\`{e}te avec la projection $\pi _x$ qui a chaque point $%
p_i=(x,y_i(x))\in \mathcal{N}$ associe $\pi _x((x,y_i(x)))=x\in $ $\mathbb{C}%
_\Phi $.\ Elle constitue un rev\^{e}tement \`{a} $n$ feuilles de
$\mathcal{N} $ au dessus de $\mathbb{C}_\Phi $. Les feuilles se
raccordent en des points singuliers de $\mathcal{M}$ qui sont ses
points de ramification. On voit comment les feuilles se raccordent
en observant les termes $y_i(x)$ pour $x$ tournant autour de
chaque point de $\mathbb{C}\backslash \mathbb{C}_\Phi $. En chacun
de ces points on d\'{e}termine ainsi une permutation sur les
feuilles se d\'{e}composant en cycles. Elle permet de prolonger le
rev\^{e}tement local induit par $\pi _x$ au-dessus d'un disque
perc\'{e} en $x$. On ajoute autant de points \`{a} $\mathcal{N}$
au-dessus de $x$ qu'il y a de cycles dans la permutation des
feuilles au voisinage. On fait de m\^{e}me au point
\`{a} l'infini $x=\infty $ en utilisant des coordonn\'{e}es homog\`{e}nes (%
\cite{Cartan} p.\ 205). On peut alors s'assurer que la surface $\mathcal{N}$
compl\'{e}t\'{e}e n'est autre que $\mathcal{M}$. Il en r\'{e}sulte en
recollant tous les morceaux un rev\^{e}tement global prolongeant $\pi _x$
\`{a} la surface de d\'{e}part et not\'{e} de m\^{e}me $\pi _x:\mathcal{M}%
\longrightarrow \mathbf{P}_1(\mathbb{C})$. Entre la courbe affine
d\'{e}finie dans $\mathbb{C}^2$ par $\Phi $ et $\mathcal{M}$ il
peut y avoir une diff\'{e}rence portant sur un nombre fini de
points. Mais en compl\'{e}tant par ces points dans
$\mathbf{P}^1(\mathbb{C})$ la surface de Riemann compacte devient
une courbe projective sur $\mathbb{C}$. On montre alors que $\pi
_x$ est m\'{e}romorphe sur $\mathcal{M}$ et qu'il en est de
m\^{e}me de la seconde projection $\pi _y$ d\'{e}finissable
gr\^{a}ce aux valeurs $y_i(x)$. Enfin il est facile de s'assurer
que le corps $\mathcal{K}(\mathcal{M})$ des fonctions
m\'{e}romorphes sur $\mathcal{M}$ s'identifie \`{a}
$\mathbb{C}(\pi _x,\pi _y)$ et a pour degr\'{e} $n$ sur le corps
$\mathbb{C}(\pi _x)$ qui est
de degr\'{e} de transcendance $1$ sur $\mathbb{C}$. De plus le corps $\mathbb{C}%
(\pi _x,\pi _y)$ s'identifie ais\'{e}ment au corps des fractions
de l'anneau $\mathbb{C}[X,Y]/Q(X,Y)\mathbb{C}[X,Y]\simeq
\mathcal{K}(\mathcal{M})$. Cette construction donne un foncteur
$\mathcal{K}$ de la cat\'{e}gorie des surfaces de Riemann compacte
connexe dans celle des corps de fonctions complexes,
c'est-\`{a}-dire des extensions de type fini et de degr\'{e} de
transcendance 1 de $\mathbb{C}$.\ Ce foncteur se prolonge en une
anti-\'{e}quivalence de cat\'{e}gories (une th\'{e}orie de Galois)
entre surfaces de Riemann compactes et corps de fonctions
complexes, elle-m\^{e}me prolongeable entre la cat\'{e}gorie des
surfaces de Riemann dee type fini et celle de certaines
$\mathbb{C}$-alg\`{e}bres (\cite{Douady} Tome 2, p.\ 138 et
\cite{Reyssat} p.\ 71). Des alg\`{e}bres vers les surfaces, on
proc\`{e}de en consid\'{e}rant l'ensemble des valuations de
l'alg\`{e}bre et en identifiant chacune d'elle \`{a} un point (une
place). La m\'{e}thode pour construire la structure de surface de
Riemann sur ces valuations est pr\'{e}cis\'{e}e dans
\cite{Chevalley}, \cite{Lang1} ou \cite{Arnaudies} (p. 92). On en
trouve un expos\'{e} simplifi\'{e} dans \cite{Edwards} qui permet
de bien comprendre l'analogie entre arithm\'{e}tique et corps de
fonctions ch\`{e}re \`{a} Andr\'{e} Weil \cite{Weil0}.\ Ceci donne
la signification de la notion de diviseur d'une surface, ainsi que
du th\'{e}or\`{e}me de Riemann-Roch (\cite{Reyssat} p.\ 94,
\cite{Edwards} p.\ 158, \cite{Arnaudies} p. 182). Le lien avec les
formes diff\'{e}rentielles en faisable avec le
quotient $\Omega _{\mathcal{K}(\mathcal{M})}$ du $\mathcal{K}(\mathcal{M})$%
-espace des symboles $df$ o\`{u} $f\in \mathcal{K}(\mathcal{M})$ par le
sous-espace engendr\'{e} par les relations $d(f+f^{\prime })-df-df^{\prime }$%
, $d(ff^{\prime })-fdf^{\prime }-(df)f^{\prime }$,$\;dc$ o\`{u}
$c\in \mathbb{C} $. On identifie dans cet espace un
$\mathbb{C}$-espace des formes diff\'{e}rentielles
m\'{e}romorphes, c'est-\`{a}-dire s'\'{e}crivant $fdz$ o\`{u}
$f\in \mathcal{K}(\mathcal{M})$ m\'{e}romorphe dans lequel on
trouve avec $f$ holomorphe un espace de cohomologie
$H^1(\mathcal{M},\mathbb{C})$.\

\subsection{$C^{*}$-alg\`{e}bres}

Il existe d'autres anti-\'{e}quivalences de cat\'{e}gories
concernant les surfaces de Riemann.\ Et par exemple
(\cite{MacLane} p.\ 93) le th\'{e}or\`{e}me de Gelfand-Naimark
(\cite{Guichardet0} p.\ 160) permet d'en construire une avec
l'anti-\'{e}quivalence qui existe entre la cat\'{e}gorie des
espaces topologiques s\'{e}par\'{e}s et celle des
$C^{*}$-alg\`{e}bres.\
Cette derni\`{e}re associe \`{a} tout espace topologique la $C^{*}$%
-alg\`{e}bre des fonctions complexes continues d\'{e}finies sur cet espace.\
En sens inverse la construction se fait en d\'{e}veloppant \cite{Connes1}
(th\'{e}or\`{e}me 6 p.\ 25). Il s'agit d'un cas particulier de la
transformation de Gelfand associant \`{a} toute alg\`{e}bre de Banach
commutative son spectre de caract\`{e}res localement compact, compact si
l'alg\`{e}bre est unitaire (\cite{Guichardet0} p. 108). Cette construction
donne un cadre tr\`{e}s naturel aux habituelles transformations de Fourier,
mais surtout elle permet de retrouver une d\'{e}monstration directe du fait
que tout espace compact peut \^{e}tre vu comme un espace alg\'{e}brique sur $%
\mathbb{C}$. On trouve dans \cite{SchwartzEnock} une tentative
d'extension de cette \'{e}quivalence aux alg\`{e}bres de Kac,
projet qui a fait l'objet d'intenses recherches autour de la
g\'{e}om\'{e}trie non commutative d'Alain Connes \cite{Connes}.\
Les $C^{*}$-alg\`{e}bres font quant \`{a} elles l'objet d'une
intense activit\'{e} de recherche car elles structurent les
ensembles d'observables de la m\'{e}canique quantique
(\cite{Waldschmidt2} p.548). La notion d'anti-\'{e}quivalence
signifie que diff\'{e}rentes th\'{e}ories parlent en
r\'{e}alit\'{e} des m\^{e}mes objets habill\'{e}s de
d\'{e}guisements diff\'{e}rents, ou consid\'{e}r\'{e}s de points
de vue diff\'{e}rents, notamment selon qu'ils sont \'{e}tudi\'{e}s
globalement ou localement. Il serait utile de comprendre quelles
propri\'{e}t\'{e}s suppl\'{e}mentaires sur les
$C^{*}$-alg\`{e}bres traduisent les propri\'{e}t\'{e}s certaines
surfaces de Riemann (\cite{Waldschmidt2} p. 548, \cite {Fischer}).
Parler d'anti-\'{e}quivalence de cat\'{e}gories ou de th\'{e}orie
de Galois revient essentiellement au m\^{e}me, la th\'{e}orie de
Galois classique ayant simplement donn\'{e} le premier exemple
historique d'une telle anti-\'{e}quivalence.

\subsection{Prolongement des surfaces et esp\`{e}ces de groupes fuchsiens}

On dit $\mathcal{M}$ est prolongeable en $\mathcal{M}^{\prime }$ ou que $%
\mathcal{M}^{\prime }$ prolonge $\mathcal{M}$ si et seulement s'il existe
une application holomorphe $f$ de $\mathcal{M}$ dans $\mathcal{M}^{\prime }$
telles que $\mathcal{M}^{\prime }\backslash f(\mathcal{M})$ ait un
int\'{e}rieur non vide. Un trou dans la surface $\mathcal{M}$ peut \^{e}tre
combl\'{e} avec un disque ferm\'{e} \`{a} une piq\^{u}re sans changer la
nature du support topologique de la surface. Les surfaces compactes
fournissent des exemples de surfaces non prolongeables. Les tores trou\'{e}s
conformes donnent au contraire des exemples de surfaces prolongeables en des
tores perc\'{e}s.

Le prolongement conduit \`{a} distinguer les groupes fuchsiens de
premi\`{e}re et ceux de seconde esp\`{e}ce (\cite{Beardon} p. 202). On
utilise pour cela l'ensemble $\Lambda (\Gamma )$ des points limites des
orbites $\Gamma z$ o\`{u} $z$ dans un domaine fondamental.\ Pour un groupe
fuchsien $\Gamma $ de seconde esp\`{e}ce, la surface de Riemann $\mathcal{H}%
/\Gamma $ est prolongeable. Et on trouve sur le bord de $\mathcal{H}$ pour $%
\Lambda (\Gamma )$ un ensemble vide, \`{a} un ou deux \'{e}l\'{e}ments, ou
un ensemble parfait et nulle part dense dans le bord de $\mathcal{H}$ (\cite
{Katok1} p. 67). Pour un groupe de premi\`{e}re esp\`{e}ce $\Gamma $, la
surface de Riemann $\mathcal{H}/\Gamma $ n'est pas prolongeable et
l'ensemble $\Lambda (\Gamma )$ est dense dans le bord de $\mathcal{H}$.

\subsection{Groupes fuchsiens \'{e}l\'{e}mentaires}

L'action d'un groupe fuchsien $\Gamma $ classe les points de $\mathcal{H}%
\cup \mathbb{R}\cup \{\infty \}$ en points paraboliques,
hyperboliques et elliptiques. Au del\`{a} des groupes de
premi\`{e}re ou de seconde esp\`{e}ce, il existe un autre type de
groupe fuchsien dit \'{e}l\'{e}mentaire caract\'{e}ris\'{e} par le
fait qu'il poss\`{e}de une
orbite finie pour son action dans la cl\^{o}ture euclidienne $\mathcal{H}%
\cup \mathbb{R}\cup \{\infty \}$ de $\mathcal{H}$. Un tel groupe est tel que $%
\Lambda (\Gamma )$ n'a pas plus de deux points (\cite{Katok1} 3.8
p. 78). Si un groupe fuchsien $\Gamma $ n'est pas
\'{e}l\'{e}mentaire il contient une infinit\'{e}
d'\'{e}l\'{e}ments hyperboliques, et tout \'{e}l\'{e}ment
elliptique est d'ordre fini (\cite{Katok1} p. 48). Si au contraire
un groupe fuchsien $\Gamma $ est \'{e}l\'{e}mentaire il est
cyclique (fini ou infini) ou conjugu\'{e} dans $PSL(2,\mathbb{R})$
\`{a} un groupe engendr\'{e} par les classes de $g(z)=-1/z$ et
$h(z)=kz$ o\`{u} $k>1$. Inversement, les sous groupes cycliques de
$PSL(2,\mathbb{R})$ engendr\'{e}s par un \'{e}l\'{e}ment
parabolique ou hyperbolique sont fuchsiens, et les sous-groupes
cycliques de $PSL(2,\mathbb{R})$ engendr\'{e}s par un
\'{e}l\'{e}ment elliptique sont fuchsiens si et seulement s'ils
sont finis.

\subsection{Signature d'un groupe fuchsien}

Avec \cite{Shimura} (ch. 1.3 \`{a} 1.5), \cite{KumarMurty} (ch.\ 10) ou \cite
{Knapp} (\S 9.5), on compl\`{e}te maintenant $\mathcal{H}$ en lui ajoutant
les pointes pour $\Gamma $.\ Elles sont situ\'{e}es sur son bord et donnent
un ensemble plus vaste $\mathcal{H}^{*}$. Ceci d\'{e}finit une nouvelle
surface de Riemann $\mathcal{H}^{*}/\Gamma $ qui est compacte si on suppose
que le groupe $\Gamma $ est de premi\`{e}re esp\`{e}ce. On la note $X(\Gamma
)$. L'ensemble des pointes pour $\Gamma $ ajout\'{e}es \`{a} $\mathcal{H}$
est fini et stable pour l'action de $\Gamma $.\ Il comble au quotient toutes
les piq\^{u}res de la surface de Riemann $\mathcal{H}/\Gamma $.

On dit de $X(\Gamma )$ qu'il s'agit d'une courbe modulaire lorsque $\Gamma
\subset PSL(2,\mathbb{Z})$ et $\Gamma $ contient le sous-groupe de congruence $%
\Gamma (n)=PG(n)$ de $PSL(2,\mathbb{Z})$ d\'{e}fini avec
\[
G(n)=\{\left[
\begin{array}{cc}
a & b \\
c & d
\end{array}
\right] \in SL(2,\mathbb{Z})\mid a\equiv d\equiv 1\;(\mod
\,n),\;b\equiv c\equiv 0\;(\mod\,n)\}.
\]
$G(2)$ est libre \`{a} deux g\'{e}n\'{e}rateurs (\cite{Iversen} p.154). On
note $X(\Gamma (n))=X(n)$. Ainsi la courbe modulaire$\;X(\Gamma
_0(n))=X_0(n) $ est d\'{e}finie avec $\Gamma =\Gamma _0(n)=PG_0(n)$ :
\[
G(n)\subset G_0(n)=\{\left[
\begin{array}{cc}
a & b \\
c & d
\end{array}
\right] \in SL(2,\mathbb{Z})\mid c\equiv 0\;(\mod\,n)\}\subset PSL(2,\mathbb{Z}%
).
\]
\\
\subsubsection{Cas o\`{u} le rev\^{e}tement universel d'une telle surface
n'est pas $\mathcal{H}$.}

Pour $\Gamma =\Gamma (1)=\Gamma _0(1)=PSL(2,\mathbb{Z})$, on trouve $\mathcal{H}%
^{*}=$ $\mathcal{H}\cup \mathbb{Q}\cup \{\infty \}$, et $\mathcal{H}^{*}/PSL(2,%
\mathbb{Z})$ est conform\'{e}ment \'{e}quivalent \`{a} la
sph\`{e}re de Riemann $\mathcal{S}^2$. Par construction on est
dans une situation relevant de ce qui est expliqu\'{e} dans
l'article de B.\ Mazur \cite{Mazur1} sur les doubles
rev\^{e}tements conformes.\ On a ici deux uniformisations qui
interagissent, une euclidienne et l'autre hyperbolique. Les points de $\mathbb{Q%
}$ sont tous paraboliques, d\'{e}ductibles de $\infty $ avec un
\'{e}l\'{e}ment de $PSL(2,\mathbb{Z})$, de sorte que $\mathcal{H}^{*}/PSL(2,%
\mathbb{Z})$ s'identifie \`{a} la surface modulaire
$\mathcal{H}/PSL(2,\mathbb{Z})$ compl\'{e}t\'{e}e de son point
\`{a} l'infini.\ Cette \'{e}quivalence conforme est donn\'{e}e par
l'invariant modulaire
\[
J:\tau \in \mathcal{H}/PSL(2,\mathbb{Z})\cup \{\infty \}\simeq \mathcal{H}%
^{*}/PSL(2,\mathbb{Z})\longmapsto J(\tau )\in \mathbb{C}\cup
\{\infty \}\simeq \mathcal{S}^2.
\]
Cet invariant d\'{e}finit $J(\tau )\in \mathbb{C}$ pour tout $\tau \in \mathcal{%
H}$, et $J(\tau )=\infty $ pour $\tau \in \mathbb{Q}\cup \{\infty
\}$. Le
demi-plan $\mathcal{H}$ devient ainsi un rev\^{e}tement ramifi\'{e} de $\mathbb{%
C}$ avec un point elliptique de ramification $2$ en $i$ et un point
elliptique de ramification $3$ en $(-1+i\sqrt{3})/2$. On retrouve ainsi (%
\cite{Katok1} p. 71) le domaine fondamental du groupe modulaire $PSL(2,\mathbb{Z%
})$ et ses deux classes de conjugaison de sous groupes cycliques
maximaux de $PSL(2,\mathbb{Z})$, l'une correspondant \`{a} des
groupes d'ordre $2$, l'autre \`{a} des groupes d'ordre $3$. Ceci
est li\'{e} \`{a} la pr\'{e}sentation de $PSL(2,\mathbb{Z})$ comme
produit libre d'un groupe cyclique d'ordre $2$ et d'un groupe
cyclique d'ordre $3$ tels que rappel\'{e}s par exemple dans \cite
{Serre1} (pp.\ 128-131).
\\
\subsubsection{Cas o\`{u} le rev\^{e}tement universel de $\mathcal{H}%
^{*}/\Gamma $ est $\mathcal{H}$}

Cette situation d\'{e}finit un nouveau groupe fuchsien $\Gamma ^{*}$ et une
\'{e}quivalence conforme
\[
\mathcal{H}^{*}/\Gamma \simeq \mathcal{H}/\Gamma ^{*}.
\]
La compacit\'{e} de $\mathcal{H}/\Gamma ^{*}$ a pour cons\'{e}quence que $%
\Gamma ^{*}$ ne construit pas de pointe, et que le nombre $r$ de classes de
points elliptiques pour l'action de $\Gamma ^{*}$ est fini. En revenant par
\'{e}quivalence conforme \`{a} $\mathcal{H}^{*}/\Gamma $ les
singularit\'{e}s demeurent. Par rev\^{e}tement, on peut \'{e}ventuellement
proc\'{e}der (\cite{Stillwell} ch. 8) \`{a} la d\'{e}singularisation de $%
\mathcal{H}/\Gamma ^{*}$. Les points elliptiques correspondent \`{a} des
points singuliers marqu\'{e}s sur la surface. La ramification (\cite{Douady}
ch. VI) d\'{e}crit les ph\'{e}nom\`{e}nes qui se manifestent pour
l'apparition de ces points. En consid\'{e}rant
\[
\mathcal{H}_{\Gamma ^{*}}=\mathcal{H}\backslash \{z\mid z\text{ elliptique
pour }\Gamma ^{*}\},
\]
on a les propri\'{e}t\'{e}s suivantes :

1- La surface $\mathcal{H}/\Gamma ^{*}$ prolonge $(\mathcal{H}_{\Gamma
^{*}}/\Gamma ^{*})$.

2- L'application canonique $\pi :\mathcal{H}\rightarrow \mathcal{H}/\Gamma
^{*}$ est localement bijective au voisinage de tout point de $\mathcal{H}%
_{\Gamma ^{*}}$.

3-\ Pour tout point elliptique $\mathbf{p}_i$ ($i=1$, ..., $r$) dans $%
\mathcal{H}/\Gamma ^{*}$ on peut d\'{e}finir un nombre $\upsilon _i$ tel que
$Card(\pi ^{-1}(\mathbf{p}_i))=\upsilon _i$.\ Ceci classe les points
elliptiques en classant les nombres $\upsilon _i$ par ordre croissant. On
dit que $\upsilon _i$ est l'indice de ramification du point $\mathbf{p}_i$
ou que les points $\mathbf{p}_1$,$...$, $\mathbf{p}_r$ sont marqu\'{e}s avec
les nombres $\upsilon _1$,$...$, $\upsilon _r$. Cette approche introduit
pour la surface de Riemann $\mathcal{M}\simeq \mathcal{H}/\Gamma $ une
signature, dite aussi signature de $\Gamma $ :
\[
(g;n:\upsilon _1,\upsilon _2,...,\upsilon _r,\upsilon _{r+1},...,\upsilon
_n;m),
\]
o\`{u} l'on note
\[
2\leq \upsilon _1\leq \upsilon _2\leq ...\leq \upsilon _r\leq \upsilon
_{r+1}=...=\upsilon _n=\infty .
\]
Cette signature indique que la surface $\mathcal{M}$ de genre $g$
poss\`{e}de $r$ points elliptiques $\mathbf{p}_1$,$...$, $\mathbf{p}_r$
d'indices de ramification $\upsilon _1$,$...$, $\upsilon _r$, des points
paraboliques $\mathbf{p}_{r+1}$,$...$, $\mathbf{p}_n$ en nombre $n-r$, ainsi
que $m$ trous. Son type conforme $(g,n,m)$ s'en d\'{e}duit.

\subsection{Invariant d'Euler-Poincar\'{e}}

La caract\'{e}ristique d'Euler-Poincar\'{e} d'un groupe fuchsien $\Gamma $
de signature $(g;n:\upsilon _1,\upsilon _2,...,\upsilon _r,\upsilon
_{r+1},...,\upsilon _n;m)$ est d\'{e}finie avec :
\[
-\chi (\Gamma )=2g-2+\sum_{i=1}^n(1-\frac 1{\upsilon
_i})+m=2g-2+\sum_{i=1}^r(1-\frac 1{\upsilon _i})+(n-r)+m.
\]
Ce nombre est positif lorsque le groupe fuchsien $\Gamma $ n'est
pas r\'{e}duit \`{a} l'unit\'{e}. Le covolume de $\Gamma $,
c'est-\`{a}-dire l'aire hyperbolique de $\mathcal{M}$, est
(\cite{Beardon} p. 269)
\[
Cov(\Gamma )=\mu (\mathcal{M})=2\pi (-\chi (\Gamma )).
\]
Dans le cas d'un groupe fuchsien $\Gamma $ de premi\`{e}re esp\`{e}ce, on a
n\'{e}cessairement $m=0$ et cette formule donne l'aire hyperbolique de tout
domaine fondamental convexe de $\Gamma $ dans le demi-plan de Poincar\'{e} $%
\mathcal{H}$. La caract\'{e}ristique d'Euler-Poincar\'{e} est aussi
l'invariant de la surface $\mathcal{M}$ d\'{e}fini classiquement comme somme
altern\'{e}e des nombres de Betti pour les $r$-simplexes construits par une
triangulation
\[
\chi (\Gamma )=\chi _{\mathcal{M}}=\sum_{j=0}^n(-1)^jb_j(\mathcal{M})=b_0(%
\mathcal{M})-b_1(\mathcal{M})+b_2(\mathcal{M}).
\]
Pour une surface compacte $\mathcal{M}$, $b_0(\mathcal{M})$ correspond au
nombre de composantes connexes, $b_2(\mathcal{M})$ est le nombre de
composantes connexes orientables (\cite{Lehmann} p.\ 257 et p.\ 260), et $%
b_1(\mathcal{M})$ est d\'{e}fini par le nombre de g\'{e}n\'{e}rateurs de $%
\pi _1(\mathcal{M},*)$ ou de son quotient commutatif $H_1(\mathcal{M},\mathbb{Z}%
)$, le premier groupe de l'homologie singuli\`{e}re de $\mathcal{M}$ :
\[
H_1(\mathcal{M},\mathbb{Z})\simeq \pi _1(\mathcal{M},*)/[\pi _1(\mathcal{M}%
,*),\pi _1(\mathcal{M},*)].
\]

\subsection{G\'{e}om\'{e}trie symplectique}

Dans chaque classe de $H_1(\mathcal{M},\mathbb{Z})$ on peut
trouver une courbe ferm\'{e}e $c(t)$ infiniment diff\'{e}rentiable
sur $\mathcal{M}$, m\^{e}me g\'{e}od\'{e}sique dans diff\'{e}rents
cas. Ceci permet en tout point $P\in \mathcal{M}$ d'intersection
de deux telles courbes $c_1(t_1)$ et $c_2(t_2)$
correspondant \`{a} deux \'{e}l\'{e}ments diff\'{e}rents $\gamma _1$ et $%
\gamma _2$ de $H_1(\mathcal{M},\mathbb{Z})$ de consid\'{e}rer la base $%
(\partial c_1/\partial t_1,\partial c_2/\partial t_2)$ de l'espace tangent
en $P$. Comme dans le cas que l'on privil\'{e}gie ici $\mathcal{M}$ est
orientable, avec un vecteur normal $n$ on peut d\'{e}finir $\varepsilon
(P)=1 $ ou $\varepsilon (P)=-1$ selon que le triplet $(n,\partial
c_1/\partial t_1,\partial c_2/\partial t_2)$ est direct ou non.\ Et en
sommant sur tous les points d'intersection des courbes $c_1(t_1)$ et $%
c_2(t_2)$ on obtient le nombre d'intersection $\gamma _1\sqcap
\gamma _2$. La g\'{e}om\'{e}trie symplectique s'introduit alors de
fa\c {c}on naturelle pour une telle surface de Riemann en
\'{e}tendant \`{a} tout le groupe d'homologie
$H_1(\mathcal{M},\mathbb{Z})$ ce nombre d'intersections qui
devient
une forme bilin\'{e}aire antisym\'{e}trique non d\'{e}g\'{e}n\'{e}r\'{e}e $%
H_1(\mathcal{M},\mathbb{Z})\times H_1(\mathcal{M},\mathbb{Z})\longrightarrow \mathbb{Z%
}$. On en d\'{e}duit l'existence de bases symplectiques et de dissections
canoniques associ\'{e}es (\cite{Waldschmidt2} p. 105) permettant de voir $%
\mathcal{M}$ au moyen d'un domaine fondamental de $\mathcal{H}$
sur le bord duquel peut \^{e}tre mat\'{e}rialis\'{e}e la
dissection canonique. On peut prolonger de fa\c {c}on naturelle de
$\mathbb{Z}$ \`{a} $\mathbb{R}$ cette forme
bilin\'{e}aire en $H_1(\mathcal{M},\mathbb{R})\times H_1(\mathcal{M},\mathbb{R}%
)\longrightarrow \mathbb{R}$. Ceci construit un espace vectoriel
symplectique \cite{Cannas}. Remarquons que le fait que
$\mathcal{M}$ est associ\'{e}e \`{a} un groupe fuchsien et donc
orientable est essentiel pour que la construction que l'on vient
de faire soit valide. On peut encore \'{e}tendre cette forme en
une forme hermitienne d\'{e}finie positive (\cite {Waldschmidt2}
p. 189) au travers de la notion de polarisation sur les
vari\'{e}t\'{e}s ab\'{e}liennes complexes qui caract\'{e}rise les
jacobiennes.\ Ce point est \'{e}voqu\'{e} plus loin.

\subsection{Approche topologique du groupe de Poincar\'{e}}

Pour toute surface de Riemann $\mathcal{M}$, la signature contient toutes
les donn\'{e}es topologiques essentielles, mais aucune donn\'{e}e conforme.
Elle donne une pr\'{e}sentation du groupe de Poincar\'{e} $\pi _1(\mathcal{M}%
,*)\simeq \Gamma $. On utilise pour le voir le th\'{e}or\`{e}me de Seifert
et Van Kampen (\cite{Gramain} p. 30) et un passage au quotient pour les
points elliptiques.\ Ceci donne :

\begin{proposition}
Le groupe de Poincar\'{e} $\pi _1(\mathcal{M},*\mathcal{)}$ de toute surface
$\mathcal{M}$ de type conforme $(g,n,m)$ admet une pr\'{e}sentation \`{a} $%
2g+n+m$ g\'{e}n\'{e}rateurs et $r+1$ relations o\`{u}

1/ Les g\'{e}n\'{e}rateurs sont $%
a_1,b_1,...,a_g,b_g,e_1,...,e_r,p_{r+1},...,p_n,h_1,...,h_m.$

2/\ Les relations sont
\[
\prod_{i=1}^g[a_i,b_i]e_1...e_rp_{r+1}...p_nh_1...h_m=1,\;\;\forall
i=1,...,r,\;\;e_i^{\upsilon _i}=1.
\]
Sa signature vaut
\[
(g;n:\upsilon _1,\upsilon _2,...,\upsilon _r,\infty _{n-r};m).
\]
\end{proposition}

Ce r\'{e}sultat permet le calcul du premier groupe d'homologie
\[
H_1(\mathcal{M},\mathbb{Z})\simeq \mathbb{Z}^{2g-r}\times \mathbb{Z}/\upsilon _1\mathbb{Z%
}\times ...\times \mathbb{Z}/\upsilon _r\mathbb{Z}.
\]
Dans le cas compact o\`{u} $n=r$ et $m=0$, on a $\pi _1(\mathcal{M},*)\simeq
\mathbf{F}_{2g-r}$, groupe libre \`{a} $2g-r$ g\'{e}n\'{e}rateurs.

\subsection{Approche conforme du groupe de Poincar\'{e}}

Les donn\'{e}es conformes d'une surface $\mathcal{M}$ de rev\^{e}tement
conforme $\mathcal{H}$ sont issues d'une repr\'{e}sentation injective du
groupe $\pi _1(\mathcal{M},*)$ dans le groupe $Aut(\mathcal{H})=PSL(2,\mathbb{R}%
)$.\ Ceci provient du r\'{e}sultat d\^{u} \`{a} Poincar\'{e} (\cite{Poincare}
\cite{Zieschang} (p. 114) \cite{Katok1} (p. 90)) :

\begin{proposition}
Soit $\Gamma $ un groupe fuchsien d\'{e}finissant une surface de Riemann de
type fini $\mathcal{M}=\mathcal{H}/\Gamma $ ayant pour signature $%
(g;n:\upsilon _1,\upsilon _2,...,\upsilon _n;m)$, $\Gamma $ admet une
pr\'{e}sentation \`{a} $2g+n+m$ g\'{e}n\'{e}rateurs et $r+1$ relations avec

1/ Les g\'{e}n\'{e}rateurs $\overline{A}_1,\overline{B}_1,...,\overline{A}_g,%
\overline{B}_g,\overline{E}_1,...,\overline{E}_r,\overline{P}_{r+1},...,%
\overline{P}_n,\overline{H}_1,...,\overline{H}_m$ dans
$PSL(2,\mathbb{R}).$

2/\ Les relations
\[
\prod_{i=1}^g[\overline{A}_i,\overline{B}_i]\overline{E}_1...\overline{E}_r%
\overline{P}_{r+1}...\overline{P}_n\overline{H}_1...\overline{H}%
_m=1,\;\;\forall i=1,...,r,\;\;\overline{E}_i^{\upsilon
_i}=\mathbf{1}_2.
\]
Les termes $\overline{H}_i$ sont hyperboliques et sont d\'{e}finis
\`{a} une permutation et \`{a} une conjugaison de
$PSL(2,\mathbb{R})$ pr\`{e}s. Il en est
de m\^{e}me des termes $\overline{P}_j$ qui sont paraboliques. Les termes $%
\overline{E}_k$ engendrent des sous-groupes finis maximaux et non
conjugu\'{e}s de $\Gamma $.\ Tout \'{e}l\'{e}ment elliptique de $\Gamma $
est conjugu\'{e} dans $PSL(2,\mathbb{R})$ d'une puissance d'un terme $\overline{%
E}_k$, et tout \'{e}l\'{e}ment parabolique de $\Gamma $ est de
m\^{e}me conjugu\'{e} d'une puissance d'un terme
$\overline{P}_j$.\ Tout \'{e}l\'{e}ment d'ordre fini dans $\Gamma
$ est elliptique. Si le groupe fuchsien $\Gamma $ est de
premi\`{e}re esp\`{e}ce, il n'y a pas de termes hyperboliques
$\overline{H}_i$. Dans ce cas le groupe $\Gamma $ est cocompact,
c'est-\`{a}-dire tel que $\mathcal{M}=\mathcal{H}/\Gamma $ soit
une surface de Riemann compacte, si et seulement s'il n'y a pas de
termes paraboliques.
\end{proposition}

\subsection{Remont\'{e}e \`{a} un groupe de matrices}

On peut maintenant revenir d'un groupe fuchsien $\Gamma $ \`{a} un groupe $G$
dans $SL(2,\mathbb{R})$ d\'{e}fini par image inverse de $PSL(2,\mathbb{R})$ dans $%
SL(2,\mathbb{R})$.\ Avec le morphisme canonique $P:$
$SL(2,\mathbb{R})\rightarrow PSL(2,\mathbb{R})$, on dit que le
sous-groupe $\Gamma $ de $PSL(2,\mathbb{R})$ est remont\'{e} en le
groupe $G$ dans $SL(2,\mathbb{R})$ si et seulement si la
restriction $P(G)$ est isomorphe \`{a} $\Gamma $. On a
d\'{e}j\`{a} vu pour les groupes de Fricke qu'il peut y avoir
plusieurs images r\'{e}ciproques $G$ de $\Gamma $ par $P$. En fait
(\cite{Seppala} p. 136) pour tout genre $g>1$ tout sous-groupe
fuchsien $\Gamma $ de $PSL(2,\mathbb{R})$ d\'{e}finit $2^{2g}$
groupes $G$ diff\'{e}rents remontant $\Gamma $ dans
$SL(2,\mathbb{R})$. Un r\'{e}sultat de Irwin Kra \cite{Kra}
indique aussi qu'un groupe $\Gamma \subset PSL(2,\mathbb{R})$ peut
\^{e}tre remont\'{e} dans $SL(2,\mathbb{R})$ si et seulement s'il
ne poss\`{e}de pas d'\'{e}l\'{e}ment d'ordre $2$. Ces derniers
peuvent en effet cr\'{e}er des probl\`{e}mes comme le montre
l'exemple de la transformation $f(z)=-(1/z)$ d'ordre $2$ dans $PSL(2,\mathbb{R}%
) $. La matrice qui correspond \`{a} $f$ dans $SL(2,\mathbb{R})$ est d'ordre $%
4. $ De tels \'{e}l\'{e}ments d'ordre $2$ appel\'{e}s ''casquettes
crois\'{e}es'', d\'{e}truisent l'orientabilit\'{e} de la surface
que l'on \'{e}tudie.\ Il faut faire appel comme dans
\cite{Seppala} (p. 70) aux notions plus vastes de surface de Klein
et de structure dianalytique pour trouver de tels \'{e}l\'{e}ments
dans le groupe correspondant que l'on peut alors consid\'{e}rer
comme un groupe klein\'{e}en (\cite{Zieschang} Theorem 3.2.8 p.
71, \cite{Zieschang1} Theorem 15.9 p. 35, \cite{Seppala} p. 89).
Mais ce cas ne peut se produire pour les surfaces de Riemann que
l'on \'{e}tudie ici o\`{u} le rev\^{e}tement est $\mathcal{H}$. De
plus on a :

\begin{proposition}
Soit $\Gamma $ un groupe fuchsien qui d\'{e}finit une surface de
Riemann de
type fini $\mathcal{M}=\mathcal{H}/\Gamma $ ayant pour signature $%
(g;n:\upsilon _1,\upsilon _2,...,\upsilon _n;m)$, le groupe
$\Gamma $ se remonte dans $SL(2,\mathbb{R})$.\ Il d\'{e}termine
m\^{e}me un unique groupe principal $G$ caract\'{e}ris\'{e} par le
fait que ses g\'{e}n\'{e}rateurs sont \`{a} trace positive. Le
groupe $\Gamma $ est isomorphe au groupe $G$ d\'{e}fini avec :

1/ Des g\'{e}n\'{e}rateurs $%
A_1,B_1,...,A_g,B_g,E_1,...,E_r,P_{r+1},...,P_n,H_1,...,H_m$.

2/\ Des relations
\[
\prod_{i=1}^g[A_i,B_i]E_1...E_rP_{r+1}...P_nH_1...H_m=1,\;\;\forall
i=1,...,r,\;\;E_i^{\upsilon _i}=\mathbf{1}_2.
\]
Les matrices $A_i$ et $B_i$ sont hyperboliques. Les \'{e}l\'{e}ments $%
E_1,...,E_r$, sont des \'{e}l\'{e}ments de torsion dans $G$.\ Ce sont des
matrices elliptiques ($0<tr(E_i)<2$) poss\'{e}dant un point fixe dans $%
\mathcal{H}$. Autour du point fixe de $E_i$ l'action se fait localement par
une matrice de rotation.\ Sur la surface de Riemann quotient, ceci donne un
point de ramification de multiplicit\'{e} $\upsilon _i$. La multiplicit\'{e}
du point c\^{o}ne correspondant est li\'{e}e \`{a} l'angle au sommet de ce
c\^{o}ne. Les \'{e}l\'{e}ments $P_{r+1},...,P_n$, sont paraboliques ($%
0<tr(P_i)=2$) poss\'{e}dant un point fixe sur le bord de $\mathcal{H}$. Sur
la surface de Riemann quotient, un tel point donne une piq\^{u}re. Les
\'{e}l\'{e}ments $H_1,...,H_m$, sont hyperboliques ($2<tr(H_i)$)
poss\'{e}dant une g\'{e}od\'{e}sique fixe dans $\mathcal{H}$. Sur la surface
de Riemann quotient, une telle g\'{e}od\'{e}sique permet de d\'{e}finir un
trou dont elle est le bord. On peut combler ce trou par un disque perc\'{e}
sans rien changer au support topologique, et en prolongeant seulement la
surface de Riemann que l'on consid\'{e}re. Si cette op\'{e}ration est faite,
la piq\^{u}re qui en r\'{e}sulte est l'image d'un point du bord de $\mathcal{%
H}$ dont on peut faire le tour avec une g\'{e}od\'{e}sique ferm\'{e}e
invariante par la matrice $H_i$ correspondante. Les deux derniers cas ne se
produisent pas si l'on a affaire \`{a} une surface compacte. Dans tous les
cas, on a $\mathcal{M}\simeq \mathcal{H}/G$.
\end{proposition}

Topologiquement, on voit bien que rien ne distingue les termes $p_i$ et $h_j$
dans la pr\'{e}sentation de $\pi _1(\mathcal{M},*)$, alors que dans $SL(2,%
\mathbb{R})$ la repr\'{e}sentation de ce groupe apporte du nouveau
qui correspond \`{a} la structure conforme et se mat\'{e}rialise
sur les valeurs des traces. On comprend aussi avec ces
observations pourquoi on n'a pas eu \`{a} consid\'{e}rer de tore
elliptique dans le chapitre pr\'{e}c\'{e}dent.

\subsection{Le th\'{e}or\`{e}me de Poincar\'{e}}

On trouve en \cite{Katok1} (Ch. 4) une r\'{e}ciproque partielle de
ce que l'on vient de voir. Il s'agit du th\'{e}or\`{e}me de
Poincar\'{e} qui indique que si $g\geq 0$, $r\geq 0$, $\upsilon
_i\geq 2$ (pour $i=1,...,r$) sont des nombres entiers tels que
l'on ait
\[
2g-2+\sum_{i=1}^r(1-\frac 1{\upsilon _i})>0,
\]
alors il existe un groupe fuchsien $\Gamma $ ayant pour signature $%
(g;r:\upsilon _1,\upsilon _2,...,\upsilon _r;0)$. On dispose d'une
construction explicite pour un tel groupe fuchsien dit
g\'{e}om\'{e}triquement fini, c'est-\`{a}-dire poss\'{e}dant un domaine
fondamental convexe polygonal \`{a} $4g+2r$ sommets d\'{e}limit\'{e} par un
nombre fini de c\^{o}t\'{e}s port\'{e}s par des g\'{e}od\'{e}siques. Ce
groupe admet une pr\'{e}sentation \`{a} $2g+r$ g\'{e}n\'{e}rateurs et $r+1$
relations avec

1/ Les g\'{e}n\'{e}rateurs $\overline{A}_1,\overline{B}_1,...,\overline{A}_g,%
\overline{B}_g,\overline{E}_1,...,\overline{E}_r.$

2/\ Les relations
\[
\prod_{i=1}^g[\overline{A}_i,\overline{B}_i]\overline{E}_1...\overline{E}%
_r=1,\;\;\forall i=1,...,r,\;\;\overline{E}_i^{\upsilon
_i}=\mathbf{1}_2.
\]
Le groupe $\Gamma $ ne contient pas d'\'{e}l\'{e}ment parabolique. Par
construction son covolume est fini, et ce groupe est cocompact. En sens
inverse pour tout groupe $\Gamma $ ayant ces propri\'{e}t\'{e}s, la
construction d'un domaine fondamental ayant les m\^{e}mes
caract\'{e}ristiques dans $\mathcal{H}$ et associ\'{e} \`{a} la surface $%
\mathcal{M}=\mathcal{H}/\Gamma $ est faisable par la m\'{e}thode
de \cite {Keen}. Selon la fa\c {c}on dont est donn\'{e} le groupe
fuchsien, il peut s'av\'{e}rer plus ou moins d\'{e}licat de
construire un domaine fondamental. Dans ce que l'on vient de voir
on conna\^{i}t des g\'{e}n\'{e}rateurs \`{a} partir desquels on
travaille. Si le groupe est plut\^{o}t donn\'{e} par des
congruences dans $PSL(2,\mathbb{Z})$, d'autres m\'{e}thodes
existent dont certaines sont automatis\'{e}es \cite{Verril}. Par
exemple le domaine fondamental d'un sous-groupe de congruence
$\Gamma $ de niveau $n$ est contenu dans un domaine fondamental du
groupe $\Gamma (n)$ identifi\'{e} dans \cite{Kulkarni}. Certains
groupes fuchsiens ne sont pas des groupes de congruence
(\cite{Lehner} p.\ 253).

Dans \cite{Katok1} la m\'{e}thode de construction de Poincar\'{e} est
\'{e}tendue \`{a} un groupe fuchsien de signature $(g;n:\upsilon _1,\upsilon
_2,...,\upsilon _r,\infty ,...,\infty ;m)$.\ Ceci donne la possibilit\'{e}
de construire un groupe fuchsien de premi\`{e}re esp\`{e}ce avec des
\'{e}l\'{e}ments paraboliques $P_{r+1},...,P_n$. Ce cas g\'{e}n\'{e}ralise
celui des tores perc\'{e}s conformes paraboliques qui sont de signature $%
(1;1:\infty ;1)$.\ Il existe une infinit\'{e} de tels tores perc\'{e}s non
conform\'{e}ment \'{e}quivalents, alors que la construction de \cite{Katok1}
n'en fournit qu'un.\ Pour les autres signatures le m\^{e}me constat peut
\^{e}tre fait, avec des domaines fondamentaux diff\'{e}rents donnant des
surfaces non conform\'{e}ment \'{e}quivalentes, mais qui sont
topologiquement identiques. Ceci signifie que la construction de
Poincar\'{e} peut \^{e}tre g\'{e}n\'{e}ralis\'{e}e, par exemple en ne pla\c
{c}ant plus le centre du polygone exhib\'{e} au centre du disque unit\'{e}.\

Le th\'{e}or\`{e}me de Poincar\'{e} est encore \'{e}tendu dans
l'\'{e}nonc\'{e} que l'on trouve dans \cite{Beardon} (p.\ 268) indiquant
qu'il existe un groupe fuchsien $\Gamma $ de type fini, non
\'{e}l\'{e}mentaire, et de signature $(g;n:\upsilon _1,\upsilon
_2,...,\upsilon _r,\upsilon _{r+1},...,\upsilon _n;m)$ si et seulement si on
a la condition pour la caract\'{e}ristique d'Euler-Poincar\'{e} :
\[
-\chi (\Gamma )=2g-2+\sum_{i=1}^r(1-\frac 1{\upsilon _i})+(n-r)+m>0.
\]
L'expression de cette caract\'{e}ristique d'Euler-Poincar\'{e} peut \^{e}tre
minor\'{e}e par la valeur positive $(1/42)$ qui correspond au groupe de
Hurwitz \`{a} trois g\'{e}n\'{e}rateurs elliptiques $\overline{E}_1,%
\overline{E}_2,\overline{E}_3,$ tels que $\overline{E}_1^2=\overline{E}_2^3=%
\overline{E}_3^7=1$. Cette observation permet d'\'{e}tablir le
th\'{e}or\`{e}me de Hurwitz (\cite{Farkas} p. 258) indiquant que le groupe $%
Aut(\mathcal{M})$ des automorphismes conformes d'une surface de Riemann
compacte $\mathcal{M}$ de genre $g\geq 2$ est fini et major\'{e} par $%
42\times (2g-2)=84(g-1)$. Ceci donne aussi le th\'{e}or\`{e}me de Schwarz (%
\cite{Farkas} p.\ 258), disant que si $\mathcal{M}$ est de genre $g\geq 2$,
alors $Aut(\mathcal{M})$ est un groupe fini.

\subsection{Groupes de Coxeter associ\'{e}s}

Pour un groupe fuchsien $\Gamma $ que l'on suppose ici pour simplifier de
signature $(g;r:\upsilon _1,\upsilon _2,...,\upsilon _r;0)$, on peut
introduire (\cite{Katok1} p.93) un polygone hyperbolique \`{a} $4g+2r$
c\^{o}t\'{e}s orient\'{e}s dont les sommets $s_j$ sont index\'{e}s
cycliquement, et dont les angles aux sommets sont tous calculables en
fonction de la signature de $\Gamma $. On peut alors consid\'{e}rer le
groupe $\Gamma ^{*}$ de toutes les isom\'{e}tries de $\mathcal{H}$ laissant
invariant les c\^{o}t\'{e}s de ce polygone. Il est engendr\'{e} par les
r\'{e}flexions $\sigma _j$ $(j=1,...,4g+2r)$ de $\mathcal{H}$ par rapport
aux c\^{o}t\'{e}s $s_js_{j+1}$ du domaine fondamental polygonal de $\Gamma $%
.\ En notant $(2\pi /2\nu _j)$ l'angle au sommet $s_j$, on obtient (\cite
{LaHarpe} p.135) les relations $\sigma _j^2=1$ et$\;(\sigma _{j-1}\sigma
_j)^{\nu _j}=1$.\ Tous les coefficients $\nu _j$ sont faciles \`{a}
expliciter en fonction de la signature du groupe $\Gamma $ ou de son groupe
principal $G$.\ Les reflexions retournant les angles, ceci introduit un
groupe d'isom\'{e}tries non toutes directes de $\mathcal{H}$ :
\[
\Gamma ^{*}=<\sigma _1,...,\sigma _{4g+2r}\mid \sigma _j^2=1,\;(\sigma
_{j-1}\sigma _j)^{\nu _j}=1>.
\]
Ce groupe agit proprement dans $\mathcal{H}$, ce qui signifie que pour tout
sous-ensemble compact $\mathcal{C}$ de $\mathcal{H}$ l'ensemble des
\'{e}l\'{e}ments $\gamma \in \Gamma ^{*}$ tels que $\gamma \mathcal{C}\cap
\mathcal{C}\neq \emptyset $ est fini. L'int\'{e}rieur du polygone de
d\'{e}part constitue un domaine fondamental pour cette action. Le groupe $%
\Gamma ^{*}$ est un groupe de Coxeter \cite{Bourbaki1} qui permet
d'expliciter le lien avec la th\'{e}orie de immeubles de Tits, ici des
immeubles hyperboliques (\cite{Ronan}). Dans $\Gamma ^{*}$ on retrouve $%
\Gamma $ comme sous-groupe d'indice 2 des transformations qui conservent
l'orientation (\cite{Katok1} th\'{e}or\`{e}me 3.5.4). Ceci donne comme
quotient $\mathcal{H}/\Gamma $ une surface de Riemann compacte de genre $g$
o\`{u} le polygone construit pr\'{e}c\'{e}demment se projette en un complexe
\`{a} $r+1$ sommets reli\'{e}s par $2g+r$ g\'{e}od\'{e}siques trac\'{e}es
sur la surface consid\'{e}r\'{e}e.\ Ce complexe permet de calculer
l'homologie singuli\`{e}re de la surface. Il correspond \`{a} une dissection
canonique de la surface de Riemann.

\subsection{Groupes de triangle hyperboliques}

Avec $g=0$ et $n=r=3$, ce qui pr\'{e}c\`{e}de garantit l'existence d'un
groupe de Coxeter $\mathbf{T}^{*}(\upsilon _1,\upsilon _2,\upsilon _3)$
d\'{e}finissable avec trois r\'{e}flexions $\overline{R}_1,\overline{R}_2,%
\overline{R}_3$ sur les cot\'{e}s d'un triangle g\'{e}od\'{e}sique de $%
\mathcal{H}$ pourvu que
\[
\sum_{i=1}^3\frac 1{\upsilon _i}<1.
\]
Ce groupe $\mathbf{T}^{*}(\upsilon _1,\upsilon _2,\upsilon _3)$ appel\'{e}
groupe de triangle hyperbolique a pour pr\'{e}sentation
\[
<\overline{R}_1,\overline{R}_2,\overline{R}_3\mid \overline{R}_1^2=\overline{%
R}_2^2=\overline{R}_3^2=(\overline{R}_1\overline{R}_2)^{\upsilon _3}=(%
\overline{R}_2\overline{R}_3)^{\upsilon _1}=(\overline{R}_3\overline{R}%
_1)^{\upsilon _2}=1>.
\]
Il poss\`{e}de un sous-groupe, le groupe de von Dyck, qui peut \^{e}tre vu (%
\cite{Katok1} p. 99-102) comme fuchsien d'indice 2 de signature $%
(0;3:\upsilon _1,\upsilon _2,\upsilon _3;0)$ :
\begin{eqnarray*}
\mathbf{T}(\upsilon _1,\upsilon _2,\upsilon _3) &=&<\overline{E}_1,\overline{%
E}_2\mid \overline{E}_1{}^{\upsilon _1}=\overline{E}_2{}^{\upsilon _2}=(%
\overline{E}_1\overline{E}_2)^{\upsilon _3}=1> \\
&=&<\overline{E}_1,\overline{E}_2,\overline{E}_3\mid \overline{E}%
_1{}^{\upsilon _1}=\overline{E}_2{}^{\upsilon _2}=\overline{E}_3{}^{\upsilon
_3}=\overline{E}_1\overline{E}_2\overline{E}_3=1>.
\end{eqnarray*}
On trouve $\mathbf{T}(2,3,\infty )=PSL(2,\mathbb{Z})=\Gamma (1)$
parmi les groupes de Dyck. Tout groupe de triangle est un quotient
du groupe utilis\'{e} dans la r\'{e}solution de nos \'{e}quations
et qui peut lui-m\^{e}me \^{e}tre consid\'{e}r\'{e} comme un
groupe de Coxeter \cite {Charney}
\[
\mathbf{T}_3\cong \mathbf{T}^{*}(\infty ,\infty ,\infty )=\mathbf{C}_2*%
\mathbf{C}_2*\mathbf{C}_2.
\]
Dans $\mathbf{T}_3$, le sous-groupe $\mathbf{F}_2\simeq [PSL(2,\mathbb{Z}%
),PSL(2,\mathbb{Z})]$ d'indice $2$ peut \^{e}tre vu comme le
groupe fuchsien qui construit la th\'{e}orie de Markoff classique.
Les groupes de Dyck sont sph\'{e}riques, euclidiens ou
hyperboliques, selon que le nombre suivant est plus grand,
\'{e}gal ou plus petit que 1 :
\[
\frac 1{\upsilon _1}+\frac 1{\upsilon _2}+\frac 1{\upsilon _3}.
\]
Les groupes de Dyck donnent concr\`{e}tement le passage entre les travaux
d\'{e}velopp\'{e}s dans le pr\'{e}sent ouvrage et la th\'{e}orie des
singularit\'{e}s \cite{Arnold2} \cite{Oka} \cite{Milnor3} \cite{BensonM}
\cite{Hirzebruch0} \cite{Dimca} \cite{Lamotke} \cite{Laufer}. La condition
de r\'{e}duction pr\'{e}sent\'{e}e dans \cite{Saito3} pour les syst\`{e}mes
de poids r\'{e}guliers et les singularit\'{e}s de surfaces associ\'{e}es est
comparable \`{a} celle vue au chapitre pr\'{e}c\'{e}dent pour les tores
paraboliques. Un syst\`{e}me r\'{e}gulier de poids est un quadruplet
d'entiers positifs $(t,x,y,z)$ tels que $t>m=\max (x,y,z)$ et
\[
\frac{(q^t-q^x)(q^t-q^y)(q^t-q^z)}{(1-q^x)(1-q^y)(1-q^z)}\;\text{%
polyn\^{o}me en }q\text{.}
\]
On peut lui associer une singularit\'{e} \`{a} l'origine d'une surface
d\'{e}finie par un polyn\^{o}me
\[
\sum_{xi+yj+zk=t}a_{ijk}X^iY^jZ^k.
\]
On associe \`{a} un tel syst\`{e}me un sous-groupe discret agissant sur $%
\mathcal{H}$, $\mathbb{C}$, $\mathcal{S}^2$. On trouve dans
\cite{Saito3} comment se fait le lien avec des groupes fuchsiens
et des surfaces $K_3$ lorsque l'on est dans le cas hyperbolique
o\`{u} $t-x-y-z=1>0$, et dans \cite {Nikulin} (p. 665) une
\'{e}vocation due \`{a} I.\ I. Piatetsky-Shapiro et I.\ R.\
Shafarevich du lien entre les surfaces r\'{e}elles $K_3$ et les
groupes de r\'{e}flexion hyperbolique. On indique aussi dans
\cite{Saito3} (p.\ 499 table 4) comment se fait le lien avec des
surfaces rationnelles, les groupes klein\'{e}ens et les
syst\`{e}mes de racines A-D-E lorsque l'on est dans le cas
sph\'{e}rique o\`{u} $t-x-y-z=-1<0$. On trouve ainsi les
sous-groupes finis du groupe $SU(2)$ des matrices unitaires de $SL(2,\mathbb{C}%
) $, rev\^{e}tement universel du groupe $SO(3)$ des rotations de l'espace
euclidien. Ceci donne les groupes de rotations $\mathbf{C}_{l+1}$, $\mathbf{D%
}_{l-2}$, $\mathbf{A}_4$, $\mathbf{S}_4$, $\mathbf{A}_5$, des poly\`{e}dres
platoniciens (pyramide, bipyramide, t\'{e}tra\`{e}dre, cube ou octa\`{e}dre,
icosa\`{e}dre ou dod\'{e}ca\`{e}dre), avec les groupes simples associ\'{e}s
par la correspondance de McKay (\cite{Conway2} p. 297, \cite{Baez}, \cite
{Slodowy}) et les polyn\^{o}mes correspondants \cite{Arnold2} (Tome 1 p.
139) :
\[
\]
\[
\begin{array}{ccc}
A(l)=\mathbf{T}(1,1,l+1) & \text{cyclique d'ordre }l+1\geq 2 & XZ+Y^{l+1} \\
D(l)=\mathbf{T}(2,2,l-2) & \text{di\'{e}dral d'ordre }4(l-2)\geq 8 &
X^2Y+Y^{l-1}+Z^2 \\
E(6)=\mathbf{T}(2,3,3) & \text{ t\'{e}tra\`{e}dral binaire }(\rightarrow
Fi_{24}) & X^2+Y^3+Z^4 \\
E(7)=\mathbf{T}(2,3,4) & \text{octa\`{e}dral binaire }(\rightarrow B=F_{2+})
& X^2+Y^3+YZ^3 \\
E(8)=\mathbf{T}(2,3,5) & \text{icosa\`{e}dral binaire }(\rightarrow M) &
X^2+Y^3+Z^5
\end{array}
\]
\[
\]
On obtient ainsi les groupes de Dyck sph\'{e}riques que l'on peut
voir comme sous-groupes finis $\Gamma $ de $PSL(2,\mathbb{C})$
agissant de mani\`{e}re discontinue sur $\mathcal{S}^2$,
d\'{e}finissant les cinq poly\`{e}dres r\'{e}guliers des pavages
classiques de la sph\`{e}re (voir \cite{Berger} tome 1 p. 44).\
Ils correspondent aux singularit\'{e}s simples ou de Klein
\cite{Klein1} \cite{Slodowy} \cite{Slodowy2} \cite{Arnold4}
(p.26), et aux diagrammes de Dynkin sans double liens
\cite{Bourbaki1} (Ch. VI \S\ 4 th. 3 p. 197) des groupes de
Coxeter associ\'{e}s \`{a} la r\'{e}solution de ces
singularit\'{e}s. D\'{e}j\`{a} identifi\'{e}s dans la scholie de
la proposition 18 du livre XIII des Elements d'Euclide,
\'{e}voqu\'{e}s dans le Tim\'{e}e de Platon, ils ont \'{e}t\'{e}
mis \`{a} contribution en 1621 dans le Secret du Monde par Jean\
Kepler pour justifier le syst\`{e}me h\'{e}liocentrique qui a
\'{e}t\'{e} propos\'{e} en 1543 par N.\ Copernic dans son livre
des R\'{e}volutions...

Bien que les surfaces $M^{s_1s_2}(b,\partial K,u)$ mises en \'{e}vidence
dans les chapitres ant\'{e}rieurs soient rationnelles et essentiellement
sans singularit\'{e}, il est int\'{e}ressant de constater que les types de
singularit\'{e}s isol\'{e}es possibles sur les surfaces cubiques sont
connus. On trouve notamment le point conique elliptique correspondant au
diagramme de Dynkin $E(6)$ et \`{a} la forme normale du cas euclidien o\`{u}
$t-x-y-z=1>0$ qui est l'\'{e}quation $XY^2-4Z^3+g_2X^2Y+g_3X^3$ d'une
surface elliptique (\cite{Friedman} p.182). Les autres possibilit\'{e}s \cite
{Fischer1} correspondent \`{a} des points doubles rationnels (des
singularit\'{e}s de Klein) et sont donn\'{e}es par $A(l)$ o\`{u} $%
l=1,...,5;\;D(l)$ o\`{u} $l=4,5;\;E(6).$

\subsection{Jacobienne et fonctions th\^{e}ta}

En se limitant encore \`{a} une surface $\mathcal{H}/\Gamma $ de signature $%
(g;r:\upsilon _1,\upsilon _2,...,\upsilon _r;0)$ le polygone bord
du domaine fondamental que l'on vient de mettre en \'{e}vidence
s'appelle une dissection canonique de la surface. Il permet de
reconstruire la surface par des transformations successives de son
domaine fondamental. Pour une surface $\mathcal{M}$ de genre $g$
suppos\'{e}e ici compacte, on fait appara\^{i}tre ainsi $2g$
cycles $\alpha _1$,..., $\alpha _{2g}$ avec lesquels les
diff\'{e}rentielles holomorphes $\omega _1$,..., $\omega _g$, de
la surface donnent une matrice de p\'{e}riodes
\[
\mathbf{\Omega }=\left[ \pi _{jk}\right] =\left[ \int_{\alpha _k}\omega
_j\right] ,\;j=1,...,g,\;k=1,...,2g.
\]
Ses $2g$ vecteurs colonnes $(\pi _{jk})_{k=1,...,2g}$ donnent un
sous-groupe discret de p\'{e}riodes $\Lambda $ de rang $g$ de
$\mathbb{C}^g$ d\'{e}finissant la surface jacobienne
$Jac(\mathcal{M})=\mathbb{C}^g/\Lambda $ de $\mathcal{M}$.\ On a
de plus un plongement canonique g\'{e}n\'{e}ralisant la situation
d\'{e}j\`{a} rencontr\'{e}e pour les courbes elliptiques :
\[
\kappa :u\in \mathcal{M}\longmapsto \kappa (u)=(\int_{u_0}^u\omega
_j)_{j=1,...g}\in Jac(\mathcal{M}).
\]
Chaque int\'{e}grale de cette application dite de Jacobi (ou Kodaira) est
mal d\'{e}finie car elle d\'{e}pend du chemin d'int\'{e}gration.\ Mais le $g$%
-uplet est quant \`{a} lui bien d\'{e}fini.\ On peut faire en sorte d'avoir $%
\alpha _1=a_1$,..., $\alpha _g=a_g$, $\alpha _{g+1}=b_1$, ..., $b_g$, et $%
\pi _{jk}=\delta _{jk}$ pour $k=1,...,g$, et poser avec une matrice $\mathbf{%
M}$ :
\[
\mathbf{M}=\left[ \int_{b_k}\omega _j\right] ,\;j=1,...,g,\;k=1,...,g,
\]
\[
\Lambda =\mathbb{Z}^g\oplus \mathbf{M}\mathbb{Z}^g.
\]
On v\'{e}rifie que l'on a $\mathbf{M=}^t\mathbf{M}$ et
Im$(\mathbf{M)}>0$, et ces deux conditions caract\'{e}risent le
demi-espace sup\'{e}rieur de Siegel $\mathcal{H}_g$ sur lequel on
peut faire agir de fa\c {c}on naturelle le groupe symplectique
$Sp(g,\mathbb{Z})$. L'int\'{e}r\^{e}t de cette construction est
que la surface jacobienne peut elle-m\^{e}me \^{e}tre
plong\'{e}e au moyen d'une fonction th\^{e}ta dans un espace projectif $%
\mathbf{P}^n(\mathbb{C})$ lorsqu'elle admet une polarisation. Il
s'agit d'une forme hermitienne $H$ d\'{e}finie sur
$\mathcal{M}\times \mathcal{M}$ telle
que $\Im(H)$ soit \`{a} valeurs enti\`{e}res sur le r\'{e}seau $%
\Lambda $. La surface jacobienne devient alors un groupe
alg\'{e}brique. Ceci est la cons\'{e}quence d'un r\'{e}sultat de
Lefschetz \cite{KumarMurty} \cite{Waldschmidt2} p. 192. Ce
r\'{e}sultat plonge de fa\c {c}on naturelle toute surface de
Riemann compacte dans un tel espace projectif, r\'{e}alisant de
fa\c {c}on concr\`{e}te le plongement donn\'{e} par les
th\'{e}or\`{e}mes de Chow ou Kodaira (\cite{Griffiths} p.\ 167 ou
p.\ 181 \cite{Serre4} p.29-30). Cette construction permet de
repr\'{e}senter le groupe des automorphismes $Aut(\mathcal{M})$
par un monomorphisme naturel dans le groupe $Sp(g,\mathbb{Z})$ qui
contient ainsi une information essentielle (\cite{Farkas} p. 287).
La fonction th\^{e}ta correspondante s'\'{e}crit pour $u\in
\mathbb{C}^g$ et $\mathbf{M\in }\mathcal{H}_g$
\[
\theta (u,\mathbf{M)=}\sum_{m\in \mathbb{Z}^g}\exp (\pi
i(^tm)\mathbf{M}m+2\pi i(^tm)u).
\]
La surface jacobienne $Jac(\mathcal{M})$ est une vari\'{e}t\'{e}
ab\'{e}lienne (\cite{Jost}, \cite{Arnaudies} tome 3 \S 24.2) sur laquelle on
dispose d'une polarisation canonique (\cite{Maurin} p. 311 \cite
{Waldschmidt2} p. 206). On peut caract\'{e}riser les vari\'{e}t\'{e}s
ab\'{e}liennes qui poss\`{e}dent une polarisation.\ Ce sont des
vari\'{e}t\'{e}s alg\'{e}briques projectives que l'on peut munir d'une loi
de groupe alg\'{e}brique d\'{e}finie avec des polyn\^{o}mes homog\`{e}nes et
deux applications $\mathcal{M}\times \mathcal{M}\rightarrow \mathcal{M}$ et $%
\mathcal{M}\rightarrow \mathcal{M}$ qui s'\'{e}crivent comme des fonctions
rationnelles \`{a} coefficients dans le corps $\mathcal{K}(\mathcal{M})$ des
fonctions m\'{e}romorphes d\'{e}finies sur $\mathcal{M}$. Un r\'{e}sultat
remarquable d\^{u} \`{a} T.\ Shioda est que les vari\'{e}t\'{e}s jacobiennes
sont caract\'{e}ris\'{e}es par des solitons, solutions de l'\'{e}quation de
Kadomtsev-Petviashvili de la th\'{e}orie des plasmas (\cite{Shioda}). Il
existe aussi des tores de dimension 2 sans plongement projectif (\cite
{Shafarevich} p.351-356), et qui ne sont donc pas des vari\'{e}t\'{e}s
ab\'{e}liennes.

\subsection{Fonctions automorphes}

On d\'{e}finit un facteur d'automorphie $\mu $ associ\'{e} au groupe
fuchsien $\Gamma \subset Aut(\mathcal{H})$ avec :
\[
\mu :\Gamma \times \mathcal{H}\longrightarrow
\mathcal{\,}\mathbb{C},
\]
\[
\forall \gamma _1,\gamma _2\in \Gamma ,\;\forall z\in \mathcal{H},\;\mu
(\gamma _1\gamma _2,z)=\mu (\gamma _1,\gamma _2z)\mu (\gamma _2,z).
\]
\[
\forall \gamma \in \Gamma ,\;\;\mu (\gamma ,.)\text{ holomorphe non nulle
sur }\mathcal{H}\text{.}
\]
Une fonction automorphe $f$ du groupe fuchsien $\Gamma $ et de facteur
d'automorphie $\mu $ est une fonction d\'{e}finie sur $\mathcal{H}$, souvent
suppos\'{e}e m\'{e}romorphe, telle que
\[
\forall \gamma \in \Gamma ,\;\forall z\in \mathcal{H},\;f(\gamma z)=\mu
(\gamma ,z)f(z).
\]
Une fonction automorphe est parfois dite modulaire. Mais l'auteur
pr\'{e}f\`{e}re r\'{e}server \`{a} ce dernier mot un sens plus pr\'{e}cis
destin\'{e} \`{a} \'{e}tudier des situations plus g\'{e}n\'{e}rales (\cite
{Kac} p.257).\ Si $\mu $ est constante et \'{e}gale \`{a} $1$, on dit
simplement ici que $f$ est une fonction $\Gamma $-automorphe. Il s'agit
d'une fonction d\'{e}finie sur $\mathcal{H}$ mais par une simple fonction
d\'{e}finie sur $\mathcal{M}=\mathcal{H}/\Gamma $ que l'on compose avec la
projection canonique de $\mathcal{H}$ sur $\mathcal{H}/\Gamma $. Tout
automorphisme $\gamma \in \Gamma \subset Aut(\mathcal{H})$ du demi-plan $%
\mathcal{H}$ peut \^{e}tre consid\'{e}r\'{e} localement comme une fonction
holomorphe permettant de d\'{e}finir
\[
\mu _\gamma =\frac{\partial \gamma }{\partial z},\;\;\mu _\gamma
(z)^{-1}=\mu (\gamma ,z).
\]
Si $F$ d\'{e}finie sur $\mathcal{H}$ est invariante par $\gamma
\in \Gamma \subset $ $Aut(\mathcal{H})\simeq PSL(2,\mathbb{R})$,
consid\'{e}rons l'expression associ\'{e}e
\[
F(\gamma z)=F(\frac{az+b}{cz+d})=F(z).
\]
Lorsque d\'{e}river $F$ en $z$ est possible, on obtient une fonction $%
f=F^{\prime }$ qui est $\Gamma $-automorphe et dont le facteur d'automorphie
est donn\'{e} par :
\[
f(\gamma z)=f(\frac{az+b}{cz+d})=(cz+d)^2f(z)=\mu (\gamma ,z)f(z).
\]
Avec des d\'{e}riv\'{e}es d'ordre sup\'{e}rieur, on doit introduire la
d\'{e}rivation de Schwarz (\cite{Ford} p.\ 99) pour trouver d'autres
formules de ce type. On peut montrer que les facteurs d'automorphie les plus
g\'{e}n\'{e}raux s'\'{e}crivent (\cite{Gunning} p. 19) avec $2k$ entier non
n\'{e}gatif $\;\mu (\gamma ,z)=(cz+d)^{2k}$. Ceci d\'{e}finit les fonctions $%
\Gamma $-automorphes de poids $2k$.\ Ces fonctions permettent la
d\'{e}finition des formes diff\'{e}rentielles de degr\'{e} $k$, dites encore
$k$-diff\'{e}rentielles ou formes automorphes \ $f\longrightarrow f(z)dz^k$.
Ces formes sont dites holomorphes si $f$ est holomorphe (\cite{Farkas} p.\
51 et 87).\ De telles formes permettent de consid\'{e}rer le $k$-i\`{e}me $%
\mathbb{C}$-espace de cohomologie $H^k(\mathcal{M},\mathbb{C})$
ainsi que l'alg\`{e}bre commutative gradu\'{e}e (\cite{Farkas} p.\
269)
\[
H^{*}(\mathcal{M},\mathbb{C})=\bigoplus_{k=0}^\infty
H^k(\mathcal{M},\mathbb{C}).
\]
Elles permettent l'\'{e}tude des aspects diff\'{e}rentiels de la surface $%
\mathcal{M}$ $=\mathcal{H}/\Gamma $ et de sa th\'{e}orie de Hodge \cite
{Lewis}. Il y a aussi un lien avec la th\'{e}orie des repr\'{e}sentations,
la th\'{e}orie du corps de classe et le programme de Langlands (\cite{Bump}
\cite{Benson} \cite{Hochschild} \cite{RamMurty} \cite{Gelbart}). Les
fonctions $\Gamma $-automorphes de poids $2k$ qui sont holomorphes sur $%
\mathcal{H}$ d\'{e}finissent de leur c\^{o}t\'{e} un
$\mathbb{C}$-espace vectoriel $\mathbf{M}_k(\Gamma )$ puis, en
d\'{e}signant l'espace $0$ par cette \'{e}criture si $k<0$, une
alg\`{e}bre gradu\'{e}e somme directe (\cite {Serre1} p.145)
\[
\mathbf{M}(\Gamma )=\bigoplus_{k=-\infty }^\infty \mathbf{M}_k(\Gamma ).
\]
Cette alg\`{e}bre a un lien avec l'alg\`{e}bre des fonctions
m\'{e}romorphes $\mathcal{K}(\mathcal{M})$ \'{e}voqu\'{e}e
ci-dessus sur la surface de Riemann
$\mathcal{M}=\mathcal{H}/\Gamma $. On trouve dans \cite{Dolgachev}
(p.75) une d\'{e}monstration du fait que si $\Gamma $ est un
sous-groupe d'indice fini de $PSL(2,\mathbb{Z})$ l'alg\`{e}bre
$\mathbf{M}(\Gamma )$ est de type fini sur $\mathbb{C}$, tous les
espaces $\mathbf{M}_k(\Gamma )$ \'{e}tant de dimension finie. Pour
le groupe $PSL(2,\mathbb{Z})$, on trouve dans la
m\^{e}me r\'{e}f\'{e}rence, ou dans \cite{Serre1} (p.145) l'isomorphisme de $%
\mathbf{M}(PSL(2,\mathbb{Z}))$ et de l'alg\`{e}bre de polyn\^{o}mes $\mathbb{C}%
[X,Y]$. Pour un groupe fuchsien plus g\'{e}n\'{e}ral $\Gamma $ tel que $%
\mathcal{M}$ $=\mathcal{H}/\Gamma $ soit une surface compacte le corps des
fractions de $\mathbf{M}(\Gamma )$ est une extension $\mathcal{K}(\mathcal{M)%
}$ de degr\'{e} fini du corps des fractions $\mathbb{C}(X,Y)$ de $\mathbb{C}[X,Y]$%
. En pratique ceci se traduit par le fait que deux fonctions $\Gamma $%
-automorphes ayant m\^{e}me domaine de d\'{e}finition sont
li\'{e}es par une relation alg\'{e}brique. Une cons\'{e}quence
importante (\cite{Ford} p.163) est que toute fonction automorphe
$f(\tau )$ d'un groupe fuchsien ayant pour domaine fondamental une
r\'{e}gion constitu\'{e}e de $k$ copies du domaine fondamental de
$PSL(2,\mathbb{Z})$ donne avec un polynome $\Phi $ de degr\'{e}
inf\'{e}rieur ou \'{e}gal \`{a} $k$ une relation alg\'{e}brique
$\Phi (f,J)=0 $ o\`{u} $J$ est l'invariant modulaire$.$

\section{La th\'{e}orie de Teichm\"{u}ller g\'{e}n\'{e}ralisant celle de
Markoff}

Le probl\`{e}me central de la th\'{e}orie de Teichm\"{u}ller
consiste \`{a} d\'{e}crire les diff\'{e}rentes structures
conformes qui existent sur un
m\^{e}me support topologique $\mathcal{M}_{top}$ d'une surface de Riemann $%
\mathcal{M}$ suppos\'{e}e ici connexe et de type fini. Le groupe de
Poincar\'{e} sur $\mathcal{M}_{top}$ point\'{e} est not\'{e} $\pi _1(%
\mathcal{M}_{top},*)$. Classiquement la th\'{e}orie de Teichmuller
est pr\'{e}sent\'{e}e avec tout un appareillage diff\'{e}rentiel.\
Or on peut la pr\'{e}senter de fa\c {c}on quasi alg\'{e}brique
lorsque $\mathcal{H}$ est le rev\^{e}tement conforme de la surface
$\mathcal{M}$.\ Ceci a permis d'expliciter comment elle
g\'{e}n\'{e}ralise la th\'{e}orie de Markoff. Le formalisme mis au
point sur les tores perc\'{e}s conformes a ainsi \'{e}t\'{e}
\'{e}tendu pour tout groupe fuchsien de signature $s$, permettant
une approche tr\`{e}s globale applicable \`{a} d'autres
\'{e}quations diophantiennes.\ Ceci a d\'{e}bouch\'{e} sur des
questions g\'{e}om\'{e}triques nouvelles dans la perspective de
sortir des surfaces pour appr\'{e}hender des objets plus
compliqu\'{e}s sur lesquels g\'{e}n\'{e}raliser les m\'{e}thodes
qui pr\'{e}c\`{e}dent. Les domaines de Riemann semblent
particuli\`{e}rement bien adapt\'{e}s \`{a} tel projet pour des
raisons que l'on explique.\ On d\'{e}crit maintenant comment ces
r\'{e}flexions ont \'{e}t\'{e} d\'{e}velopp\'{e}es, en renvoyant
au chapitre 7 de \cite{Perrine9} pour des compl\'{e}ments ainsi
qu'\`{a} \cite{Seppala} \cite{Harvey} \cite{Schneps}
\cite{Krushkal1}.

\subsection{Repr\'{e}sentations du groupe de Poincar\'{e}}

Les diff\'{e}rentes structures conformes sur $\mathcal{M}_{top}$
sont d\'{e}finies par les repr\'{e}sentations $\overline{\rho }$
du groupe $\pi _1(\mathcal{M}_{top},*)$ dans le groupe
$PSL(2,\mathbb{R})$, constituant l'espace des d\'{e}formations
\[
\mathcal{R}=\mathcal{R}(\pi
_1(\mathcal{M}_{top},*),PSL(2,\mathbb{R})).
\]
Au moyen de la notion de groupe principal, le calcul d'une d\'{e}formation $%
\overline{\rho }$ de signature $s=(g;n:\upsilon _1,\upsilon _2,...,\upsilon
_r,\upsilon _{r+1},...,\upsilon _n;m)$ est faisable analytiquement avec la
repr\'{e}sentation associ\'{e}e $\rho :$ $\pi _1(\mathcal{M}%
_{top},*)\rightarrow SL(2,\mathbb{R})$ telle que $\overline{\rho }=P\circ \rho $%
.\ Il suffit d'expliciter les coefficients des matrices images par $\rho $
des g\'{e}n\'{e}rateurs qui v\'{e}rifient les relations d'une
pr\'{e}sentation de $\pi _1(\mathcal{M}_{top},*)$. Le calcul des
g\'{e}n\'{e}rateurs $\rho (a_1)=A_1$, $\rho (b_1)=B_1$, $...$, $\rho
(a_g)=A_g$, $\rho (b_g)=B_g$, $\rho (e_1)=E_1$, $...$, $\rho (e_r)=E_r$, $%
\rho (p_{r+1})=P_{r+1}$, $...$, $\rho (p_n)=P_n$, $\rho (h_1)=H_1$, $...$, $%
\rho (h_m)=H_m$ n\'{e}cessite $3(2g+n+m)$ param\`{e}tres r\'{e}els
car on a quatre param\`{e}tres pour chacune des matrices et que
leurs d\'{e}terminants valent $1$. On doit prendre en compte entre
ces param\`{e}tres $3(r+1)$ \'{e}galit\'{e}s entre nombres
r\'{e}els issus des
relations qui lient les matrices, ainsi que les $n-r$ \'{e}galit\'{e}s $%
tr(P_i)=2$. Ceci repr\'{e}sente au total $n+2r+3$ contraintes liant ces
param\`{e}tres r\'{e}els. Trois param\`{e}tres suppl\'{e}mentaires peuvent
\^{e}tre \'{e}limin\'{e}s en raisonnant \`{a} un automorphisme int\'{e}rieur
pr\`{e}s, c'est-\`{a}-dire \`{a} une transformation conforme pr\`{e}s de $%
\mathcal{H}$. Ceci construit une vari\'{e}t\'{e} r\'{e}elle
$\mathcal{V}_s$ de dimension $6g-6-2r+2n+3m$ dont une partie
d\'{e}finie par les contraintes sur les traces sup\'{e}rieures
\`{a} $2$ param\'{e}trise les structures de Riemann possibles sur
le support topologique de $\mathcal{M}_{top}$. Chaque point
not\'{e} $\Pi (\overline{\rho })$ dans cette partie de
$\mathcal{V}_s$ correspond \`{a} une structure conforme sur
$\mathcal{M}_{top}$. Par exemple, pour les tores perc\'{e}s
paraboliques $g=1$, $r=0$, $n=1$, $m=0$, on a mis en \'{e}vidence
au chapitre pr\'{e}c\'{e}dent la nappe principale d'une
vari\'{e}t\'{e} $\mathcal{V}_{(1;1;0)}$ de dimension $2$
donn\'{e}e par l'\'{e}quation de Markoff. Et pour un tore
perc\'{e} hyperbolique $g=1$, $r=0$, $n=0$, $m=1$, on trouve de
m\^{e}me une partie d'une vari\'{e}t\'{e} de dimension $3$. On
trouve dans \cite{Keen1} d'autres exemples. Du groupe principal
$\rho (\pi _1(\mathcal{M}_{top},*))=G$ on d\'{e}duit alors la
repr\'{e}sentation associ\'{e}e $\overline{\rho }=P\circ
\rho $ et son image $\Gamma =PG=$ $\overline{\rho }(\pi _1(\mathcal{M}%
_{top},*))$ qui est un groupe fuchsien de signature $s=(g;n:\upsilon
_1,\upsilon _2,...,\upsilon _r,\upsilon _{r+1},...,\upsilon _n;m)$. On peut
pratiquement consid\'{e}rer qu'\`{a} chaque point $\Pi (\overline{\rho })$
de $\mathcal{V}_s$ est attach\'{e} le groupe fuchsien $\Gamma $, point que
l'on peut noter avec les matrices du groupe principal $G$ de $\Gamma $ et
par analogie avec ce que l'on a d\'{e}velopp\'{e} dans \cite{Perrine9} :
\[
\Pi (A_1,B_1,...,A_g,B_g,E_1,...,E_r,P_{r+1},...,P_n,H_1,...,H_m).
\]
Le calcul que l'on vient de faire ne d\'{e}termine pas tout l'espace des
d\'{e}formations de $\pi _1(\mathcal{M}_{top},*)$, mais uniquement son
sous-ensemble $\mathcal{R}_s=\mathcal{R}_s(\pi _1(\mathcal{M}_{top},*),PSL(2,%
\mathbb{R}))$. Il faut regrouper ces derniers espaces sur toutes
les signatures
correspondant au m\^{e}me type topologique $(g,n+m)$ de $\pi _1(\mathcal{M}%
_{top},*)$ pour retrouver $\mathcal{R}$.\ Comme le montre les tores
perc\'{e}s paraboliques et hyperboliques, on trouve des ph\'{e}nom\`{e}nes
de bord entre les espaces $\mathcal{R}_s$ correspondant aux sauts quantiques
que constituent pour la g\'{e}om\'{e}trie conforme le passage d'une
piq\^{u}re \`{a} un trou. Pour tout $\overline{\rho }\in \mathcal{R}$ une
structure conforme est donn\'{e}e par la consid\'{e}ration de :
\[
\mathcal{M}=\mathcal{H}/\overline{\rho }(\pi _1(\mathcal{M}_{top},*))=%
\mathcal{H}/\Gamma .
\]
Il s'agit d'une surface de Riemann qui selon les propri\'{e}t\'{e}s de $%
\overline{\rho }$ peut avoir pour support topologique $\mathcal{M}_{top}$ et
une signature ou une autre.\ On remarquera que des espaces topologiques
hom\'{e}otopes d\'{e}finissent des groupes de Poincar\'{e} isomorphes, mais
peuvent ne pas \^{e}tre hom\'{e}omorphes (\cite{Gramain} p.16) mais
qu'inversement la derni\`{e}re \'{e}galit\'{e} privil\'{e}gie un mod\`{e}le $%
\mathcal{M}_{top}$ parmi les classes d'hom\'{e}omorphie d'une m\^{e}me
classe d'hom\'{e}otopie.

\subsection{Equivalence des repr\'{e}sentations et r\'{e}duction}

Si $Int(PSL(2,\mathbb{R}))$ est le groupe des automorphismes int\'{e}rieurs de $%
PSL(2,\mathbb{R})$, on a avec la composition des morphismes une
action naturelle de ce groupe dans $\mathcal{R}_s$.\ Ceci permet
de cacher diff\'{e}rents param\`{e}tres li\'{e}s \`{a} ces
automorphismes int\'{e}rieurs et donc de raisonner \`{a}
\'{e}quivalence conforme pr\`{e}s de $\mathcal{H}$.\ On
d\'{e}finit (\cite{Seppala} p. 165) ainsi un quotient qui est
l'espace des modules de signature $s$ :
\[
\mathcal{M}od(s)=\mathcal{R}_s(\pi _1(\mathcal{M}_{top},*),PSL(2,\mathbb{R}%
))/Int(PSL(2,\mathbb{R})).
\]
Comme la vari\'{e}t\'{e} r\'{e}elle $\mathcal{V}_s$ \`{a} laquelle il
s'identifie par le calcul pr\'{e}c\'{e}dent, cet espace param\'{e}trise les
structures conformes de signature $s$ existant sur la surface topologique $%
\mathcal{M}_{top}$ support. Compte tenu de la fa\c {c}on dont on l'a
construit, remarquons que sur $\mathcal{V}_s$ il est naturel de
consid\'{e}rer les automorphismes int\'{e}rieurs d\'{e}finis par une matrice
$D\in GL(2,\mathbb{R})$.\ Ceci fait alors intervenir l'orientation de $\mathcal{%
M}$ et donne un r\'{e}sultat analogue \`{a} \cite{Perrine9} (prop.\ 5.5.3 et
prop.\ 6.5.3). De fa\c {c}on pr\'{e}cise, on a l'\'{e}quivalence entre
l'\'{e}galit\'{e} de
\[
\Pi (A_1,B_1,...,A_g,B_g,E_1,...,E_r,P_{r+1},...,P_n,H_1,...,H_m),
\]
\[
\Pi (A_1^{\prime },B_1^{\prime },...,A_g^{\prime },B_g^{\prime },E_1^{\prime
},...,E_r^{\prime },P_{r+1}^{\prime },...,P_n^{\prime },H_1^{\prime
},...,H_m^{\prime }),
\]
et l'existence de $D\in GL(2,\mathbb{R})$ telle que
\[
A_1^{\prime }=DA_1D^{-1},...,E_1^{\prime }=DE_1D^{-1},...,P_{r+1}^{\prime
}=DP_{r+1}D^{-1},...,H_m^{\prime }=DH_mD^{-1}.
\]
La partie de $\mathcal{V}_s$ mise en \'{e}vidence ci-dessus est
invariante pour l'action des automorphismes int\'{e}rieurs
d\'{e}finis par les matrices $D\in GL(2,\mathbb{R})$. Le calcul de
toutes les possibilit\'{e}s pour $D$ d\'{e}bouche sur des
consid\'{e}rations sur les quaternions ou les alg\`{e}bres de
Clifford qui les g\'{e}n\'{e}ralisent. Egalement, puisque
les calculs faits ont utilis\'{e} des repr\'{e}sentations $\rho :$ $\pi _1(%
\mathcal{M}_{top},*)\rightarrow SL(2,\mathbb{R})$,
c'est-\`{a}-dire des
repr\'{e}sentations de groupe sp\'{e}ciales $\rho :$ $\pi _1(\mathcal{M}%
_{top},*)\rightarrow GL(2,\mathbb{R})$, on peut utiliser les
r\'{e}sultats de cette th\'{e}orie dans l'\'{e}tude de la
situation que l'on consid\`{e}re, avec le fait qu'en
g\'{e}n\'{e}ral $\pi _1(\mathcal{M}_{top},*)$ est infini et non
commutatif. On fait ainsi le lien avec la notion de caract\`{e}re
qui, en sens inverse s'introduit dans la th\'{e}orie de
Teichmuller, par exemple avec la notion de caract\`{e}re de
Fricke. L'\'{e}quation alg\'{e}brique associ\'{e}e se retrouve par
les m\'{e}thodes de \cite {Horowitz}. Le lien avec les formes
quadratiques binaires peut \^{e}tre
retrouv\'{e} en g\'{e}n\'{e}ralisant le th\'{e}or\`{e}me de Frobenius Schur (%
\cite{Serre6} p.\ 121).

Il est aussi possible de faire agir de fa\c {c}on naturelle le groupe des
automorphismes $Aut(\pi _1(\mathcal{M}_{top},*))$ sur $\mathcal{R}_s$, ce
qui revient \`{a} changer de syst\`{e}me de g\'{e}n\'{e}rateurs du groupe $%
\pi _1(\mathcal{M}_{top},*)$. Comme l'action induite d'un \'{e}l\'{e}ment de
$Int(\pi _1(\mathcal{M}_{top},*))$ sur $\mathcal{R}_s$ donne l'identit\'{e},
on peut se contenter d'\'{e}tudier l'action sur $\mathcal{R}_s$ du groupe
des classes d'applications
\[
\Gamma _{\pi _1(\mathcal{M}_{top},*)}=Aut(\pi _1(\mathcal{M}%
_{top},*))/Int(\pi _1(\mathcal{M}_{top},*))=Out(\pi _1(\mathcal{M}%
_{top},*)).
\]
Cette d\'{e}marche g\'{e}n\'{e}ralise la th\'{e}orie de la r\'{e}duction qui
a \'{e}t\'{e} vue au chapitre pr\'{e}c\'{e}dent. Elle d\'{e}finit l'espace
de Teichmuller :
\[
\mathcal{T}eich(s)=\mathcal{M}od(s)/\Gamma _{\pi
_1(\mathcal{M}_{top},*)}.
\]
Il est identifiable \`{a} la partie de $\mathcal{V}_s$ mise en \'{e}vidence
ci-dessus que l'on peut maintenant interpr\'{e}ter comme domaine fondamental
pour l'action du groupe $\Gamma _{\pi _1(\mathcal{M}_{top},*)}$ dans toute
la vari\'{e}t\'{e} r\'{e}elle $\mathcal{V}_s$.

\subsection{Groupes fuchsiens arithm\'{e}tiques}

La question se pose de savoir si l'on peut remplacer dans ce que
l'on vient de voir le groupe $PSL(2,\mathbb{R})$ par
$PSL(2,\mathbb{Z})$. La r\'{e}ponse non
\'{e}vidente est partiellement donn\'{e}e dans le chapitre 5 de \cite{Katok1}%
. Elle conduit \`{a} se m\'{e}fier de la d\'{e}nomination de
groupe fuchsien arithm\'{e}tique utilis\'{e}e \'{e}galement dans
le contexte des groupes de Lie, et diff\'{e}rente de la notion
envisag\'{e}e ici qui se r\'{e}sume \`{a} la condition $\Gamma
\subset PSL(2,\mathbb{Z})$.

\subsection{Compl\'{e}ments sur les repr\'{e}sentations de groupes}

\subsubsection{Vari\'{e}t\'{e} de repr\'{e}sentations}

On peut aussi vouloir remplacer $PSL(2,\mathbb{R})$ par
$PSL(2,\mathbb{C})$, et raisonner sur des groupes klein\'{e}ens
plut\^{o}t que sur des groupes fuchsiens. Ceci conduit \`{a} la
notion de vari\'{e}t\'{e} de repr\'{e}sentations d'un groupe de
Poincar\'{e} \cite{Lubotzky} \cite {Brumfiel}
\[
\rho \in \mathcal{R}(\pi
_1(\mathcal{M}_{top},*),PSL(2,\mathbb{C}))\rightarrow (tr\rho
(g_1),tr\rho (g_2),...,tr\rho (g_p))\in \mathbb{C}^p,
\]
o\`{u} $\rho :\pi _1(\mathcal{M}_{top},*)\rightarrow
PSL(2,\mathbb{C})$ repr\'{e}sentation de $\pi
_1(\mathcal{M}_{top},*)$ dans $PSL(2,\mathbb{C})$,
avec $p$ nombre de g\'{e}n\'{e}rateurs choisis dans le groupe $\pi _1(%
\mathcal{M}_{top},*)$, $tr$ la trace dans $PSL(2,\mathbb{C})$
\`{a} distinguer de la trace de la matrice correspondante dans
$SL(2,\mathbb{C})$. L'ensemble
des repr\'{e}sentations complexes $\rho $ est $\mathcal{R}(\pi _1(\mathcal{M}%
_{top},*),PSL(2,\mathbb{C}))$. Ce nouveau sujet est lui-m\^{e}me
li\'{e} \`{a} l'\'{e}tude de l'espace de Teichm\"{u}ller qu'il
complexifie \cite{Seppala} (ch.\ 4). On construit de fa\c {c}on
naturelle des relations alg\'{e}briques entre les traces en
utilisant la m\'{e}thode de \cite{Horowitz} \cite {Procesi},
d'o\`{u} des vari\'{e}t\'{e}s autour des caract\`{e}res de Fricke
dans lesquelles on peut repr\'{e}senter l'espace de
Teichm\"{u}ller \cite {Bers}.\ La m\'{e}thode donne des structures
alg\'{e}briques qui g\'{e}n\'{e}ralisent \'{e}galement la
th\'{e}orie de Markoff \cite{Gonzales} \cite{Saito} \cite{Saito1}
\cite{Luo}. Le proc\'{e}d\'{e} peut d'ailleurs \^{e}tre
compar\'{e} \`{a} la m\'{e}thode utilis\'{e}e pour montrer que les
surfaces de Riemann compactes sont alg\'{e}briques (\cite{Cohn5}
(p.\ 120) \cite{Nag} (p.98) \cite{Serre4} \cite{Ly}).\ Le lien est
aussi faisable avec la classique th\'{e}orie des invariants
(\cite{Hilbert4} \cite{Dixmier} \cite {Liu}), puis les
alg\`{e}bres de Hopf, la th\'{e}orie de Galois et les groupes
quantiques (\cite{Bergman} \cite{Chase} (p. 52) \cite{Demidov}
\cite {Guichardet}).\ On a aussi un lien profond avec le calcul de
Heaviside (encore appel\'{e} ombral, symbolique, de Sylvester, de
Boole, de Leibnitz..., \cite{Humbert} \cite{Rota} \cite{Rota1})
qui a donn\'{e} naissance aux distributions par
g\'{e}n\'{e}ralisation de la fonction de Dirac (\cite{Heaviside}
\cite{Carson} \cite{Schwartz} \cite{Colombeau} \cite {Cartier}).
L'approfondissement de ce sujet conduit au calcul diff\'{e}rentiel
non commutatif \cite{Demidov}, aux D-modules (\cite {Coutinho}
\cite{Bertrand} (p.\ 14)), etc. Il constitue une perspective
essentielle pour des travaux \`{a} venir.

\subsubsection{Monodromie}

On appelle repr\'{e}sentation de monodromie d'un groupe $\Gamma =\pi _1(%
\mathcal{M}_{top},*)$ tout homomorphisme de groupes
\[
\rho :\pi _1(\mathcal{M}_{top},*)\longrightarrow GL(n,\mathbb{C}).
\]
L'image de $\rho $ est le groupe de monodromie. Ces repr\'{e}sentations
peuvent \^{e}tre class\'{e}es avec les automorphismes int\'{e}rieurs de $%
GL(n,\mathbb{C})$ et interviennent dans la r\'{e}solution des
\'{e}quations
diff\'{e}rentielles de Fuchs (\cite{Yosida1} p. 75, \cite{Gray}, \cite{Kuga}%
) qui sont de forme suivante o\`{u} les $a_i$ sont holomorphes, ou
encore m\'{e}romorphes dans le domaine consid\'{e}r\'{e} :
\[
\frac{d^nf}{dz^n}+a_1(z)\frac{d^{n-1}f}{dz^{n-1}}+...+a_n(z)f=0.
\]

$\bullet $ Pour le cas plus g\'{e}n\'{e}ral o\`{u} $n$ n'est pas
n\'{e}cessairement \'{e}gal \`{a} $2$, ce qui pr\'{e}c\`{e}de conduit \`{a}
l'\'{e}tude des groupes alg\'{e}briques et \`{a} la th\'{e}orie de Galois
diff\'{e}rentielle de Picard-Vessiot, Ritt, Kolchin, Pommaret, etc... On
renvoie pour l'approfondissement de ce sujet \`{a} \cite{Bertrand} \cite
{Yosida2}.

$\bullet $ Pour le cas $n=2$ et $\pi _1(\mathcal{M}_{top},*)\simeq \mathbf{F}%
_2$ engendr\'{e} par $A$ et $B$, les repr\'{e}sentations de monodromie sont
compl\`{e}tement d\'{e}crites dans \cite{Yosida1} (p.\ 80).\ Celles qui sont
irr\'{e}ductibles, c'est-\`{a}-dire sans sous espace propre invariant, sont
caract\'{e}ris\'{e}es \`{a} un automorphisme int\'{e}rieur pr\`{e}s de $GL(2,%
\mathbb{C})$ par des expressions
\[
\rho (A)=\left[
\begin{array}{cc}
\lambda _1 & 1 \\
0 & \lambda _2
\end{array}
\right] ,\;\rho (B)=\left[
\begin{array}{cc}
\mu _1 & 0 \\
(\nu _1+\nu _2)-(\lambda _1\mu _1+\lambda _2\mu _2) & \mu _2
\end{array}
\right] ,\;\lambda _i\mu _j\neq \nu _k.
\]
Elles sont d\'{e}termin\'{e}es de fa\c {c}on unique par les trois couples $%
(\lambda _1,\lambda _2)$, $(\mu _1,\mu _2)$, $(\nu _1,\nu _2)$ de valeurs
propres de $A$, $B$ et $AB$, pourvu qu'ils v\'{e}rifient les contraintes
cit\'{e}es. Par exemple en diagonalisant les matrices $A_0$ et $B_0$ de la
th\'{e}orie de Markoff classique, on v\'{e}rifie que les contraintes sont
v\'{e}rifi\'{e}es et que l'on a :
\[
\rho (A_0)=\left[
\begin{array}{cc}
\frac{3-\sqrt{5}}2 & 1 \\
0 & \frac{3+\sqrt{5}}2
\end{array}
\right] ,\;\rho (B_0)=\left[
\begin{array}{cc}
\frac{3-\sqrt{5}}2 & 0 \\
-4 & \frac{3+\sqrt{5}}2
\end{array}
\right] .
\]
On peut expliciter dans ce cas une solution du probl\`{e}me de
Riemann-Hilbert qui consiste \`{a} reconstruire \`{a} partir de la
repr\'{e}sentation de monodromie une \'{e}quation fuchsienne poss\'{e}dant $%
\rho $ pour repr\'{e}sentation de monodromie. Pour cela on utilise \cite
{Yosida} (th.4.3.2 p.85) pour calculer le sch\'{e}ma de Riemann associ\'{e}.
On reconstruit alors une \'{e}quation fuchsienne (une \'{e}quation
hyperg\'{e}om\'{e}trique perturb\'{e}e) qui est avec $\sigma _3+\tau _3=1$
et $\sigma _3+\sigma _3^{-1}=3$ :
\[
x(1-x)\frac{d^2u}{dx^2}+(1-2x)\frac{du}{dx}-(\sigma _3\tau _3)u=\frac 1{4\pi
^2x(1-x)}\log (\frac{3+\sqrt{5}}2)\log (\frac{3-\sqrt{5}}2)u.
\]
Elle identifie un op\'{e}rateur diff\'{e}rentiel dont l'analyse spectrale
reste \`{a} faire et \`{a} comparer avec le spectre de Markoff :
\[
L=D^2+\frac{(1-2x)}{x(1-x)}D-\frac{(\sigma _3\tau _3)4\pi ^2x(1-x)+\log (%
\frac{3+\sqrt{5}}2)\log (\frac{3-\sqrt{5}}2)}{4\pi ^2x^2(1-x)^2}.
\]

\subsection{La pr\'{e}sentation classique de la th\'{e}orie de Teichm\"{u}ller}

La th\'{e}orie de Teichm\"{u}ller a pour but de d\'{e}terminer
toutes les structures conformes sur une m\^{e}me surface
topologique. Comme une structure conforme d\'{e}finit une
structure diff\'{e}rentielle orient\'{e}e \`{a} deux dimensions,
on peut d\'{e}composer le probl\`{e}me en deux : construire
d'abord sur une structure topologique une structure
diff\'{e}rentielle ou la structure riemannienne unique qu'elle
d\'{e}finit, ensuite construire sur cette derni\`{e}re une
structure conforme.\ Le second probl\`{e}me a une solution
unique.\ Le premier est beaucoup plus d\'{e}licat, notamment si la
surface topologique n'est pas compacte. Au-dela de ce qui
pr\'{e}c\`{e}de une solution peut \^{e}tre obtenue par
diff\'{e}rents autres moyens comme la quasi-conformit\'{e}. On
donne ici quelques indications sur cette fa\c {c}on classique de
pr\'{e}senter la th\'{e}orie de Teichm\"{u}ller.

\subsubsection{Classes d'\'{e}quivalence conforme}

Deux m\'{e}triques $ds^2$ et $dt^2$ sont dites conform\'{e}ment
\'{e}quivalentes si l'application identique $Id:(\mathcal{M}%
,dt^2)\longrightarrow (\mathcal{M},ds^2)$ est une transformation conforme de
$\mathcal{M}$. En \'{e}crivant la m\'{e}trique $ds^2$ sous la forme $%
ds^2=\lambda \mid dz+\mu (z)d\overline{z}\mid ^2$ on voit que les classes
d'\'{e}quivalence conforme sont param\'{e}tr\'{e}es par $\mu (z)$. Leur
ensemble $Conf(\mathcal{M})$ peut \^{e}tre vu comme un ensemble quotient $%
Conf(\mathcal{M})=Met(\mathcal{M})/C_{+}^\infty (\mathcal{M})$, o\`{u} $Met(%
\mathcal{M})$ est l'ensemble de toutes les m\'{e}triques possibles sur $%
\mathcal{M}$, et $C_{+}^\infty (\mathcal{M})$ le groupe multiplicatif des
fonctions $\lambda $ r\'{e}elles positives non nulles, diff\'{e}rentiables,
d\'{e}finies sur $\mathcal{M}$.

\subsubsection{Espace des modules (ou des classes d'\'{e}quivalence
diff\'{e}omorphes)}

Deux m\'{e}triques $ds^2$ et $dt^2$ sur une surface topologique $\mathcal{M}$
sont dites diff\'{e}omorphiquement \'{e}quivalentes si et seulement si on a
un diff\'{e}omorphisme $f:(\mathcal{M},dt^2)\longrightarrow (\mathcal{M}%
,ds^2)$ pr\'{e}servant l'orientation et conforme.\ Avec $ds^2=\lambda \mid
dz+\mu (z)d\overline{z}\mid ^2$ on voit que les classes d'\'{e}quivalence
diff\'{e}omorphes correspondant \`{a} une m\^{e}me classe d'\'{e}quivalence
conforme d\'{e}termin\'{e}e par $\mu (z)$ sont param\'{e}tr\'{e}es par $%
\lambda $. Soit $Diff_{+}(\mathcal{M})$ le groupe des diff\'{e}omorphismes
de $\mathcal{M}$ dans $\mathcal{M}$. Les classes d'\'{e}quivalence
diff\'{e}omorphes d\'{e}finissent $\mathcal{M}od(\mathcal{M})=Met(\mathcal{M}%
)/Diff_{+}(\mathcal{M})$ l'espace des modules de la surface topologique $%
\mathcal{M}$. On a une surjection canonique de $\mathcal{M}od(\mathcal{M})$
dans $Conf(\mathcal{M})$.

\subsubsection{Espace de Teichm\"{u}ller}

Entre $C_{+}^\infty (\mathcal{M})$ et $Diff_{+}(\mathcal{M})$ existe le
groupe $Diff_0(\mathcal{M})$ de tous les diff\'{e}omorphismes isotopes \`{a}
l'identit\'{e}. On dit que les deux m\'{e}triques $ds^2$ et $dt^2$ sur la
surface topologique $\mathcal{M}$ sont fortement \'{e}quivalentes si et
seulement s'il existe $f\in Diff_0(\mathcal{M})$ de $\mathcal{M}$ tel que $%
f:(\mathcal{M},dt^2)\longrightarrow (\mathcal{M},ds^2)$ soit conforme. On
dit alors que $\mathcal{T}eich(\mathcal{M})=Met(\mathcal{M})/Diff_0(\mathcal{%
M})$ est l'espace de Teichm\"{u}ller. On trouve dans \cite{Nash2}
(p.150) une pr\'{e}cision sur la validit\'{e} de cette
d\'{e}finition qui n'est ad\'{e}quate que pour certaines surfaces,
et qui n\'{e}cessite pour \^{e}tre valable de consid\'{e}rer que
$Met(\mathcal{M})$ ne contient que les m\'{e}triques pour
lesquelles la courbure de Gauss de $\mathcal{M}$ est constante.
Posent probl\`{e}me les surfaces ayant pour rev\^{e}tement
conforme universel la sph\`{e}re de Riemann $\mathcal{S}^2$ ou
$\mathbb{C}$. Les autres cas de rev\^{e}tement $\mathcal{H}$ ne
posent pas de difficult\'{e}. On a donn\'{e} dans \cite{Perrine9}
diff\'{e}rents espaces de Teichm\"{u}ller montrant que piq\^{u}res
et trous ne jouent pas m\^{e}me r\^{o}le sur une surface de
Riemann.

\subsubsection{Groupe de Teichm\"{u}ller des classes d'applications}

Comme $Diff_0(\mathcal{M})$ est un sous-groupe normal de $Diff_{+}(\mathcal{M%
})$, on peut aussi d\'{e}finir le groupe (parfois dit modulaire)
de Teichm\"{u}ller, encore appel\'{e} groupe des classes
d'applications
(mapping class group) $\Gamma _{\mathcal{M}}=Diff_{+}(\mathcal{M})/Diff_0(%
\mathcal{M})$. Le groupe $\Gamma _{\mathcal{M}}$ est un groupe discret,
interpr\'{e}table comme le groupe des composantes connexes du groupe $%
Diff_{+}(\mathcal{M})$. Il est engendr\'{e} par les twists de Dehn (\cite
{Nash2} p.157).\ Les twists de Dehn engendrent le groupe $\Gamma _{\mathcal{M%
}}$ mais ils n'en constituent en g\'{e}n\'{e}ral pas un ensemble minimal de
g\'{e}n\'{e}rateurs. Sont importants ceux qui ne sont pas homotopes \`{a}
l'identit\'{e}, par exemple parce qu'ils font le tour d'une poign\'{e}e de
la surface ou d'une piq\^{u}re. Le groupe $\Gamma _{\mathcal{M}}$ est
isomorphe \`{a} un quotient d'un groupe d'automorphismes du groupe de
Poincar\'{e} (\cite{Ivanov1} (p.\ 17), \cite{Zieschang1} (ch. 2)) ici
not\'{e} $\pi _1(\mathcal{M},*)$ :
\[
\Gamma _{\mathcal{M}}\simeq Aut_{*}(\pi _1(\mathcal{M},*))/Int(\pi _1(%
\mathcal{M},*))=Out_{*}(\pi _1(\mathcal{M},*)),
\]
o\`{u} $Int(\pi _1(\mathcal{M},*))$ est le groupe des automorphismes
int\'{e}rieurs de $\pi _1(\mathcal{M},*)$, et $Aut_{*}(\pi _1(\mathcal{M}%
,*)) $ est le groupe des automorphismes de $\pi _1(\mathcal{M},*)$ induits
par un hom\'{e}omorphisme de $\mathcal{M}$.\ Le groupe $Aut_{*}(\pi _1(%
\mathcal{M},*))$ est contenu dans le groupe de tous les automorphismes $%
Aut(\pi _1(\mathcal{M},*))$.\ Ainsi s'introduit le groupe plus vaste $\Gamma
_{\pi _1(\mathcal{M}_{top},*)}=Out(\pi _1(\mathcal{M},*))$ dont $\Gamma _{%
\mathcal{M}} $ est un sous-groupe. On a regroup\'{e} dans \cite{Perrine9}
tout un ensemble de r\'{e}sultats connus pour les groupes de Poincar\'{e} et
les groupes de classes d'applications, mais dispers\'{e}s dans la
litt\'{e}rature sur ce th\`{e}me \cite{Birman} \cite{Keen2} \cite{Ivanov2}
\cite{Dehn} \cite{Wajnryb} \cite{Gervais}. Dans diff\'{e}rents cas on est
certain de l'\'{e}galit\'{e} $\Gamma _{\mathcal{M}}=\Gamma _{\pi _1(\mathcal{%
M}_{top},*)}$, par exemple\ lorsque le th\'{e}or\`{e}me de Dehn
Nielsen s'applique comme c'est le cas pour les surfaces compactes
\cite{Zieschang} (p. 194).\ Ce th\'{e}or\`{e}me permet
d'expliciter le lien avec la pr\'{e}sentation faite ci-dessus de
la th\'{e}orie de Teichm\"{u}ller par les repr\'{e}sentations. Le
manuscrit de Fenchel et Nielsen \cite{Nielsen1} explicite
d'ailleurs l'hom\'{e}omorphisme qu'induit tout automorphisme
donn\'{e} dans $Aut(\pi _1(\mathcal{M},*))$.

\subsubsection{Lien entre espace de Teichm\"{u}ller et espace des modules}

On a d\'{e}fini plusieurs quotients avec $\mathcal{M}$, l'espace des modules
$\mathcal{M}od(\mathcal{M})=Met(\mathcal{M})/Diff_{+}(\mathcal{M})$,
l'espace de Teichm\"{u}ller $\mathcal{T}eich(\mathcal{M})=Met(\mathcal{M}%
)/Diff_0(\mathcal{M})$, le groupe des classes d'applications $\Gamma _{%
\mathcal{M}}=Diff_{+}(\mathcal{M})/Diff_0(\mathcal{M})$. La
comparaison de leur d\'{e}finition fait appara\^{i}tre l'espace
des modules comme un quotient de l'espace de Teichm\"{u}ller par
le groupe discret des classes
d'applications agissant sur cet espace de fa\c {c}on propre et discontinue (%
\cite{Schneps} p.\ 12)
\[
\mathcal{M}od(\mathcal{M})\simeq \mathcal{T}eich(\mathcal{M})/\Gamma _{%
\mathcal{M}}.
\]
De sorte $\mathcal{T}eich(\mathcal{M})$ peut \^{e}tre consid\'{e}r\'{e}
comme un rev\^{e}tement ramifi\'{e} au-dessus de l'espace des modules $%
\mathcal{M}od(\mathcal{M})$. Cette configuration est comparable
\`{a} celle des groupes fuchsiens agissant sur le rev\^{e}tement
des surfaces de Riemann. Une piste pour d\'{e}velopper son
\'{e}tude \'{e}merge de la comparaison entre espaces de
Teichm\"{u}ller et surfaces de Riemann de rev\^{e}tement conforme
$\mathcal{H}$, car on a :

$\bullet $ L'espace de Teichm\"{u}ller dispose d'une structure topologique (%
\cite{Schneps} p.10).

$\bullet $ Il est muni d'une structure analytique r\'{e}elle
\cite{Abikoff1}.

$\bullet $ C'est une composante d'une vari\'{e}t\'{e} affine
r\'{e}elle d\'{e}finie par des polyn\^{o}mes \`{a} coefficients
rationnels (\cite {Seppala} p.\ 175).

$\bullet $ Il est muni d'une m\'{e}trique naturelle, dite de Weil-Peterson (%
\cite{Nash2} p.\ 157).

$\bullet $ On peut y construire une structure analytique complexe
naturelle \cite{Nag} \cite{Earle}.

$\bullet $ C'est une vari\'{e}t\'{e} k\"{a}hl\'{e}rienne de
courbure n\'{e}gative (\cite{Nash2} p.\ 157, \cite{Ahlfors}).

$\bullet $ Il poss\`{e}de une structure d'espace de Stein (\cite{Imayoshi}
p.\ 171, \cite{Bers1}).

\subsection{Compactification de l'espace de Teichm\"{u}ller}

Etant en g\'{e}n\'{e}ral de dimension sup\'{e}rieure \`{a} 2, les
espaces de Teichm\"{u}ller peuvent \^{e}tre vus comme des
g\'{e}n\'{e}ralisations des surfaces de Riemann. Dans beaucoup de
cas, on a des mod\`{e}les topologiques d'espaces de
Teichm\"{u}ller (\cite{Nash2} (p.153), \cite{Imayoshi} (p.9),
\cite{Nag} (p.111) \cite{Schneps} (p. 18)). Des exemples d'espaces
de Teichm\"{u}ller d\'{e}crits par des \'{e}quations
alg\'{e}briques sont d\`{e}s \`{a} pr\'{e}sent disponibles
(\cite{Keen2} p.\ 1206 relations 4-1 et 4-2). Ils permettent
d'envisager l'existence d'autres \'{e}quations diophantiennes dont
la r\'{e}solution ressemble \`{a} celle de Markoff, et est
intrins\`{e}quement li\'{e}e \`{a} une structure
g\'{e}om\'{e}trique. Un exemple d\'{e}j\`{a} connu de ce type est
donn\'{e} par \cite{Baragar}.\ Mais ce que l'on vient de voir
offre de tr\`{e}s nombreuses autres possibilit\'{e}s. Ce point est
confirm\'{e} par le fait que toute vari\'{e}t\'{e} de Stein est
biholomorphiquement \'{e}quivalente \`{a} une sous-vari\'{e}t\'{e}
analytique complexe de $\mathbb{C}^n$ pour un certain $n$ entier
(\cite{LaurentThiebaut} p.\ 180, \cite{Kaup} p. 269). En
compactifiant une telle vari\'{e}t\'{e}, on fait le lien avec la
g\'{e}om\'{e}trie alg\'{e}brique gr\^{a}ce au th\'{e}or\`{e}me de
Chow (\cite {Serre4} p.29-30). Remarquons que la compactification
d'une surface de Riemann comme $\mathcal{H}$ peut ne plus \^{e}tre
une surface de Riemann. On a observ\'{e} entre les tores
perc\'{e}s conformes hyperboliques et paraboliques quelques-uns
des ph\'{e}nom\`{e}nes intervenant dans la compactification des
espaces de Teichm\"{u}ller.\ L'\'{e}tude de cette compactification
est l'une des perspectives qui ont \'{e}t\'{e} ouvertes par W.\
P.\ Thurston \cite{Fathi}.\ Elle prend ici une signification
particuli\`{e}re dans l'esprit de \cite{Serre4} car elle conduit
inversement \`{a} l'id\'{e}e de consid\'{e}rer toute \'{e}quation
diophantienne que l'on cherche \`{a} r\'{e}soudre comme donn\'{e}e
par un tel processus.\ Les espaces compacts simplement connexes
jouent un r\^{o}le \'{e}quivalent dans les dimensions
sup\'{e}rieures \`{a} celui de la sph\`{e}re de Riemann
$\mathcal{S}^2$.\ Les vari\'{e}t\'{e}s analytiques complexes
compactes et simplement connexes, sont hom\'{e}omorphes \`{a} des
sph\`{e}res (\cite{Massey} p.\ 142).\ Pour m\'{e}moire, la
conjecture de Poincar\'{e} qui transpose ce dernier r\'{e}sultat
aux vari\'{e}t\'{e}s r\'{e}elles de dimension 3 reste toujours
ouverte, sachant qu'elle est r\'{e}solue en dimensions
sup\'{e}rieures \cite {Smale}.

\subsection{Espaces de Stein et domaines de Riemann}

Une \'{e}tude directe des espaces de Stein $\mathfrak{X}$
s'inspirant de celle des surfaces de Riemann constitue une piste
utile pour approfondir la th\'{e}orie de Teichm\"{u}ller. Ces
espaces sont int\'{e}ressants pour avoir suffisamment de fonctions
holomorphes globales pour s\'{e}parer ses points. En notant
$\mathcal{O}(\mathfrak{X})$ la $\mathbb{C}$-alg\`{e}bre unitaire
des fonctions holomorphes de $\mathfrak{X}$ dans $\mathbb{C}$, on
fabrique une
alg\`{e}bre topologique qui est une sous-alg\`{e}bre de Fr\'{e}chet de la $%
\mathbb{C}$-alg\`{e}bre\emph{\ }$\mathcal{C}(\mathfrak{X})$ des
fonctions continues
de $\mathfrak{X}$ dans $\mathbb{C}$. Tout caract\`{e}re continu de l'alg\`{e}bre $%
\chi :\mathcal{O}(\mathfrak{X})\rightarrow $ $\mathbb{C}$ est
d\'{e}fini par un
point de $x\in \mathfrak{X}$ lorsque ce dernier espace est de dimension finie: $%
\mathfrak{X}\subset \mathbb{C}^n$. Et l'application $\chi \in X(\mathcal{O}(\mathfrak{X}%
))\rightarrow x\in \mathfrak{X}$ est un hom\'{e}omorphisme
(\cite{Kaup} p. 268). Cette propri\'{e}t\'{e} est
caract\'{e}ristique des espaces de Stein (\cite {Guichardet0} p.\
72).\ Les surfaces de Riemann ouvertes sont des espaces de Stein
(\cite{Kaup} p. 224), tout comme les espaces complexes qui ne
contiennent qu'un nombre fini de points. Mais les surfaces de
Riemann compactes ne sont pas des espaces de Stein
(\cite{Guichardet0} p. 87).\ Ceci montre clairement que les
espaces de Stein ne sont qu'une g\'{e}n\'{e}ralisation partielle
des surfaces de Riemann m\^{e}me s'ils contiennent les espaces de
Teichm\"{u}ller. Une bonne d\'{e}finition pour englober les
surfaces de Riemann et les espaces de Teichm\"{u}ller dans un
formalisme commun semble plut\^{o}t \^{e}tre celle de domaine de Riemann (%
\cite{Kaup} (p. 38 et p. 96) \cite{Jarnicki} \cite{Grauert}). Elle
correspond aux domaines d'holomorphie simplement connexes
caract\'{e}ris\'{e}s par le fait qu'il existe $f\in
\mathcal{O}(\mathfrak{X})$
non holomorphiquement extensible \`{a} un point se situant hors de $\mathfrak{X}%
\subset \mathbb{C}^n$. Plusieurs autres pistes sont apparues pour
approfondir les r\'{e}flexions pr\'{e}c\'{e}dentes sur la
th\'{e}orie de Markoff :

$\bullet $ L'\'{e}tude des groupes fuchsiens de dimension sup\'{e}rieure,
dans l'esprit de \cite{Apanasov} \cite{Apanasov1} ou \cite{Ratcliffe}.

$\bullet $ La th\'{e}orie de Galois des extensions finies de corps
de fractions $\mathbb{C}(X_1,...X_n)$. Il serait utile de
comprendre si elle a un lien avec les polylogarithmes
(\cite{Waldschmidt4} \cite{Lewin} \cite {Cathelineau}) et comment
l'on peut construire une th\'{e}orie de Galois pour les
\'{e}quations aux d\'{e}riv\'{e}es partielles, ayant
\'{e}ventuellement un lien avec la th\'{e}orie des fonctions
hyperg\'{e}om\'{e}triques g\'{e}n\'{e}ralis\'{e}es \cite{Opdam}.

$\bullet $ La th\'{e}orie de la combinatoire des voies ferr\'{e}es ou
''train tracks'' telle qu'elle est pr\'{e}sent\'{e}e dans \cite{Mosher1}
\cite{Penner} \cite{Mosher}.

\section{Codage des g\'{e}od\'{e}siques}

Une cons\'{e}quence de la th\'{e}orie de Teichm\"{u}ller concerne
le fait que le groupe des classes d'applications poss\`{e}de une
structure que l'on peut d\'{e}crire tout comme celle du groupe de
Poincar\'{e}. Il en d\'{e}coule des cons\'{e}quences pour le
codage des g\'{e}od\'{e}sique d'une surface de Riemann que l'on va
maintenant \'{e}voquer, ainsi que les liens avec les fractions
continues.

\subsection{D\'{e}composition du groupe des classes d'applications}

Le groupe des classes d'applications $\Gamma _{\mathcal{M}}$ se
d\'{e}compose en utilisant les deux op\'{e}rations sur les groupes de somme
amalgam\'{e}e et d'extension HNN (\cite{Serre} \cite{Bass} \cite{Cohen} \cite
{LaHarpe} (III 14) \cite{Hausmann}) :

\begin{proposition}
Pour toute surface de Riemann $\mathcal{M}$ de type conforme
\[
(g,n,m)\notin \{(0,0,0),(0,1,0),(0,2,0),(1,0,0)\},
\]
c'est-\`{a}-dire ayant $\mathcal{H}$ pour rev\^{e}tement conforme,
le groupe des classes d'applications est simplement
d\'{e}composable.
\end{proposition}

A tout groupe simplement d\'{e}composable, on associe un graphe de
d\'{e}composition qui d\'{e}crit tous les composants n\'{e}cessaires et
synth\'{e}tise toutes les indications dont on a besoin pour combiner ces
composants. Le groupe des classes d'applications se d\'{e}compose parce que
la surface de Riemann $\mathcal{M}$ se d\'{e}compose par plombage en
pantalons (\cite{Harvey} (p.\ 312), \cite{Bedford} (article de C.\ Series),
\cite{Ratcliffe} (p. 408), \cite{Seppala} (p. 117)). La d\'{e}monstration
s'effectue en remontant au groupe fuchsien qui d\'{e}finit la surface $%
\mathcal{M}$. Il poss\`{e}de aussi la propri\'{e}t\'{e} d'\^{e}tre
simplement d\'{e}composable (\cite{Harvey} p.\ 312). On est donc
ramen\'{e} \`{a} un probl\`{e}me d'alg\`{e}bre avec un groupe $G$
simplement d\'{e}composable dont on \'{e}tudie le quotient
$Out(G)=Aut(G)/Int(G)$.\ On utilise pour conclure les m\'{e}thodes
de \cite{Pietrowski}. Cette approche vaut pour le groupe de
Poincar\'{e} comme pour le groupe des classes d'applications et
est d\'{e}velopp\'{e}e dans \cite{Vogtmann}. En consid\'{e}rant
des g\'{e}od\'{e}siques de $\mathcal{M}$ dont les longueurs
correspondent aux modules \cite{Imayoshi} et en associant une
valeur dans un groupe $\mathbb{Z}/2\mathbb{Z}$ qui correspond au
sens de parcours de la g\'{e}od\'{e}sique, ainsi qu'une lettre qui
correspond \`{a} un \'{e}l\'{e}ment du groupe $\pi
_1(\mathcal{M},*)$ correspondant \`{a} cette g\'{e}od\'{e}sique,
le graphe de d\'{e}composition permet de reconstruitre
le groupe de Poincar\'{e} $\pi _1(\mathcal{M},*)$ puis $Out(\pi _1(\mathcal{M%
},*))$ et enfin $\Gamma _{\pi _1(\mathcal{M}_{top},*)}$.

\subsection{Codage des g\'{e}od\'{e}siques}

Il exite un ensemble de travaux s'appuyant sur les fractions continues pour
coder les g\'{e}od\'{e}siques ferm\'{e}es des tores perc\'{e}s \cite{Series1}
dont l'extension \`{a} des surfaces de Riemann $\mathcal{M}$ plus
compliqu\'{e}es que les tores perc\'{e}s n'est pas au point \cite{Schmutz2}
malgr\'{e} le grand int\'{e}r\^{e}t de cette question. L'auteur s'est donc
pench\'{e} sur ce sujet en cherchant \`{a} comprendre comment il faudrait
proc\'{e}der pour obtenir une bonne g\'{e}n\'{e}ralisation et des
r\'{e}sultats nouveaux. Le point essentiel qui en r\'{e}sulte est que le
groupe de Poincar\'{e} $\pi _1(\mathcal{M},*)$ et le premier groupe $H_1(%
\mathcal{M},\mathbb{Z})$ de l'homologie singuli\`{e}re de
$\mathcal{M}$ contiennent dans beaucoup de cas l'information
essentielle, ne serait-ce que parce que toute classe de ces
groupes contient alors une g\'{e}od\'{e}sique.

En se limitant dans un premier temps \`{a} un groupe fuchsien de signature $%
(g;r:\upsilon _1,\upsilon _2,...,\upsilon _r;0)$, on peut orienter de fa\c
{c}on coh\'{e}rente les ar\^{e}tes du polygone d\'{e}fini \cite{Katok1} \cite
{Keen} pour d\'{e}crire un domaine fondamental de ce groupe fuchsien.\ Ceci
privil\'{e}gie des sens de parcours sur les lacets g\'{e}od\'{e}siques de la
surface de Riemann $\mathcal{M}$ que l'on consid\`{e}re. A partir de l\`{a},
toute autre g\'{e}od\'{e}sique orient\'{e}e de la surface $\mathcal{H}%
/\Gamma =\mathcal{M}$ peut \^{e}tre cod\'{e}e selon la m\'{e}thode de Morse
et Koebe \cite{Series1}. Chaque fois que progressant dans le sens de la
g\'{e}od\'{e}sique on traverse une g\'{e}od\'{e}sique constituant un
c\^{o}t\'{e} de la dissection canonique, on inscrit la lettre correspondante
avec une puissance $+1$ ou $-1$ qui est le nombre d'intersections
correspondant. Cette convention de signe utilise l'orientation de la surface
$\mathcal{M}$, comme d\'{e}crit dans \cite{Waldschmidt2} (p. 105). Ceci
permet d'associer \`{a} toute g\'{e}od\'{e}sique un mot \'{e}crit comme une
suite doublement infinie de lettres \`{a} la puissance $\pm 1$ prises dans
l'ensemble
\[
\{\overline{A}_1,\overline{B}_1,...,\overline{A}_g,\overline{B}_g,\overline{E%
}_1,...,\overline{E}_r\}.
\]
Toutes les suites ne sont pas possibles.\ Le fait qu'il en existe d'infinies
montre que les suites associ\'{e}es aux g\'{e}od\'{e}siques ne font pas
partie du groupe $\Gamma $ dont les termes s'expriment seulement comme des
mots finis \'{e}crits avec les m\^{e}mes lettres. Le groupe $\Gamma $ est
donc trop petit pour d\'{e}crire toutes les g\'{e}od\'{e}siques de la
surface et on doit imaginer de faire appel pour atteindre cet objectif \`{a}
d'autres op\'{e}rations cat\'{e}goriques que le simple produit libre de
groupes. N\'{e}anmoins les g\'{e}od\'{e}siques ferm\'{e}es correspondent
\`{a} des mots infinis p\'{e}riodiques dont on peut coder la p\'{e}riode
avec les lettres pr\'{e}c\'{e}dentes, d\'{e}signant des \'{e}l\'{e}ments de $%
\Gamma \simeq \pi _1(\mathcal{M},*)$. D'apr\`{e}s ce que l'on
conna\^{i}t sur les tores perc\'{e}s, toute p\'{e}riode finie de
ce type ne permet pas de coder une telle g\'{e}od\'{e}sique
ferm\'{e}e. On sait par contre reconna\^{i}tre les
g\'{e}od\'{e}siques ferm\'{e}es simples, c'est-\`{a}-dire ne se
coupant pas elles-m\^{e}mes dans l'essentiel des cas
\cite{Series1} \cite {Series2}. Ce r\'{e}sultat s'\'{e}tend par
les consid\'{e}rations
pr\'{e}c\'{e}dentes au cas plus g\'{e}n\'{e}ral d'une surface de Riemann $%
\mathcal{M}$ de signature $s$.

Cette approche a un lien avec les groupes d'homotopie et
d'homologie, au moins dans le cas compact o\`{u} le
th\'{e}or\`{e}me de Hilbert indique que toute classe d'homotopie
libre de lacets ferm\'{e}s de $\mathcal{M}$ contient une
g\'{e}od\'{e}sique ferm\'{e}e, et que deux points quelconques de
$\mathcal{M}$ peuvent \^{e}tre reli\'{e}s par une
g\'{e}od\'{e}sique appartenant \`{a} toute classe d'homotopie
donn\'{e}e (\cite{Maurin} p.\ 390). Il reste cependant l\`{a} des
questions \`{a} approfondir. On voit par exemple par ce qui
pr\'{e}c\`{e}de qu'une g\'{e}od\'{e}sique peut \^{e}tre
d\'{e}crite par une suite doublement infinie de telles lettres
(l'analogie est \'{e}vidente avec les fractions continues !), et
qu'un diff\'{e}omorphisme de $\mathcal{M}$ transforme cette
g\'{e}od\'{e}sique en une autre codable de m\^{e}me avec ces
lettres. Ce diff\'{e}omorphisme agit comme un g\'{e}n\'{e}rateur
pseudo-al\'{e}atoire (voir \cite{Cusick3} \cite {Rueppel}
\cite{Sangjing}). Il est connu qu'il puisse ne pas \^{e}tre une
isom\'{e}trie (\cite{Berger1} p.\ 429), ni donc a fortiori un
automorphisme conforme de $\mathcal{M}$. Il existe d'autres
m\'{e}thodes de codage des g\'{e}od\'{e}siques, dont certaines
plus directement tourn\'{e}es vers les fractions continues
\cite{Katok} \cite{Arnoux}. On peut faire
intervenir ces derni\`{e}res en d\'{e}composant toutes les matrices $%
A_1,B_1,...,A_g,B_g,E_1,...E_r$ intervenant en produit de matrices de forme
\[
\left[
\begin{array}{cc}
a & 1 \\
1 & 0
\end{array}
\right] ^{\pm 1},\;a\in \mathbb{N}\backslash \{0\}.
\]
On trouve dans \cite{Lehner} (p. 334) et \cite{Haas} une
\'{e}vocation sommaire des probl\`{e}mes d'approximation
diophantienne li\'{e}s \`{a} ce type de situation. Pour les
approfondir il faut pr\'{e}alablement \'{e}tendre ce qui
pr\'{e}c\`{e}de \`{a} d'autres signatures que celles
privil\'{e}gi\'{e}es ci-dessus, ce qui ne para\^{i}t pas
insurmontable. L'auteur a quelques travaux en cours sur ce
th\`{e}me notamment pour comprendre le lien avec les points de
Weierstrass \cite{Leroy}.

\subsection{Dynamique symbolique}

La dynamique symbolique et sa variante du chaos d\'{e}terministe consiste
\`{a} \'{e}tudier cette situation en poursuivant les travaux fondateurs de
Hadamard (\cite{Dahan} p.\ 396, \cite{Bedford}, \cite{Ruelle}, \cite{Lind},
\cite{Adler}, \cite{Adler1}, \cite{SchmidtK}).\ Les syst\`{e}mes dynamiques
v\'{e}rifiant l'axiome A d'Anosov s'introduisent dans ce contexte, avec le
fait remarquable que les surfaces de Riemann hyperboliques ont toutes un
flot g\'{e}od\'{e}sique ayant cette propri\'{e}t\'{e} (\cite{Anosov} pour le
cas compact, et \cite{Ruelle} (p.\ 171) pour une extension au cas non
compact).\ Ceci permet de classer les hom\'{e}omorphismes entre surfaces de
Riemann avec un important r\'{e}sultat d\^{u} \`{a} W.\ Thurston (cit\'{e}
dans \cite{Perrine2a} ou \cite{Mosher}).

\subsection{Approche ergodique}

Ce th\`{e}me d'\'{e}tude fait le lien avec des sujets aussi importants que
la thermodynamique, la th\'{e}orie ergodique et l'information \cite{Ruelle1}
\cite{Billingsley} \cite{Alseda}, certaines fonctions z\^{e}ta \cite{Parry}
\cite{Terras2}, le d\'{e}compte des nombres premiers et son analogie avec le
comportement de certaines g\'{e}od\'{e}siques \cite{Parry1} \cite{Baladi}
\cite{Bowen} \cite{Watkins} \cite{Kotani} \cite{Hurt}, l'alg\`{e}bre des
corps quadratiques et l'\'{e}valuation de leur nombre de classes \cite
{Sarnak} \cite{Vivaldi}, l'interpr\'{e}tation thermodynamique de la mesure
de Mahler de certains polyn\^{o}mes \cite{SchmidtK} (paragraphe 5.18), la
m\'{e}canique hamiltonienne car le flot g\'{e}od\'{e}sique constitue un
syst\`{e}me hamiltonien sur la vari\'{e}t\'{e} symplectique $\mathcal{D}_2$
des droites \cite{Audin}, le th\'{e}or\`{e}me KAM des tores invariants et
les petits diviseurs \cite{Arnold} \cite{Dodson1} \cite{Yoccoz} \cite{Herman}%
, la dynamique holomorphe et les objets universels \`{a} caract\`{e}re
fractal qu'elle construit \cite{Sullivan} \cite{Yoccoz1}, l'analyse
spectrale de certains op\'{e}rateurs \cite{Moser} \cite{Ruelle2} \cite
{Connes}, les cycles limites et le ph\'{e}nom\`{e}ne de Stokes (16$^{\grave{e%
}me}$ probl\`{e}me de Hilbert), etc...

\section{Ubiquit\'{e} de la fonction \^{e}ta de Dedekind}

Un certain nombre de r\'{e}sultats classiques en th\'{e}orie du
codage de l'information comme la formule de MacWilliams
\cite{Macwilliams} ont un lien avec les surfaces de Riemann. On a
approfondi ce th\`{e}me sachant que les recherches de l'auteur sur
la th\'{e}orie de Markoff ont d\'{e}but\'{e} avec une
pr\'{e}occupation li\'{e}e au codage. Ceci a permis d'identifier
un lien assez remarquable avec la fonction \^{e}ta de Dedekind
dont on a vu qu'elle donne naissance aux \'{e}quations
diophantiennes qui g\'{e}n\'{e}ralisent celle de la th\'{e}orie de
Markoff.\ On a aussi pu pr\'{e}ciser comment l'essentiel des
fonctions transcendantes habituelles s'\'{e}crivent avec la
fonction \^{e}ta de Dedekind qui joue donc un r\^{o}le
fondamental.

\subsection{Les fonctions th\^{e}ta}

Les fonctions th\^{e}ta d\'{e}finies par un r\'{e}seau $\Gamma \subset \mathbb{R%
}^n$ muni de son produit scalaire naturel sont avec $\tau \in \mathcal{H}$
et $q=\exp (2i\pi \tau )$
\[
\theta _\Gamma (\tau )=\sum_{m\in \Gamma }\exp (i\pi \tau <m,m>)=\sum_{m\in
\Gamma }q^{\frac 12<m,m>}=\sum_{r=0}^\infty a_rq^r.
\]
Ce sont les fonctions g\'{e}n\'{e}ratrices des nombres $a_r=Card\{m\in
\Gamma \mid <m,m>=2r\}$. Elles comptent les points du r\'{e}seau $\Gamma $
situ\'{e}s dans une sph\`{e}re de rayon $\sqrt{2r}$ centr\'{e}e \`{a}
l'origine. En d'autres termes, les coefficients du d\'{e}veloppement en
s\'{e}rie de Fourier de ces fonctions donnent le nombre de
repr\'{e}sentations d'un entier par une forme quadratique d\'{e}finie
positive.\ Les formules les plus g\'{e}n\'{e}rales ont \'{e}t\'{e}
donn\'{e}es dans ce domaine par A.\ Malyshev (\cite{Iwaniec} chapitre 11).\
Les fonctions th\^{e}ta d\'{e}finies par un r\'{e}seau unimodulaire pair $%
\Gamma \subset \mathbb{R}^n$ identique \`{a} son dual donnent des
exemples classiques (\cite{Serre1} p.174) de fonctions automorphes
de poids $(n/2)$ pair du groupe $PSL(2,\mathbb{Z})$. Ceci est une
cons\'{e}quence de la formule de Poisson appliqu\'{e}e \`{a} la
fonction r\'{e}elle $\Theta (t)=\theta _\Gamma (it)$,
c'est-\`{a}-dire de la formule de Jacobi (\cite{Moll} p.\ 149).
Cette formule de Poisson a un lien tr\`{e}s profond avec la loi de
r\'{e}ciprocit\'{e} quadratique \cite{Berg}.

\subsection{Lien avec le codage de l'information}

La formule de Poisson peut elle m\^{e}me \^{e}tre
consid\'{e}r\'{e}e comme une formule de trace (\cite{Terras} ch
1.3). Elle donne la formule de MacWilliams sur les polyn\^{o}mes
de poids des codes correcteurs d'erreurs \cite{Macwilliams}.\
L'introduction des fonctions th\^{e}ta permet d'interpr\'{e}ter ce
dernier r\'{e}sultat \cite{Broue} \cite{Broue1} \cite {Milnor3}
\cite{BensonM}. Ceci permet aussi de traduire les relations entre
certains codes importants pour les applications et certains
r\'{e}seaux.\ Par exemple le code de Golay \'{e}tendu correspond
au r\'{e}seau de Leech qui est l'unique r\'{e}seau unimodulaire
pair de $\mathbb{R}^{24}$ sans racine d'apr\`{e}s un r\'{e}sultat
de J.\ H.\ Conway (\cite{Ebeling} p.105). Ceci donne le groupe
simple $M_{24}$, le groupe de Mathieu, dont la simplicit\'{e} peut
\^{e}tre comprise par l'approche galoisienne de la
surface de Riemann correspondante.\ Egalement le polyn\^{o}me de poids $%
W_C(X,Y)$ de tout code $C\subset \mathbb{F}_2^n$ autodual
doublement pair s'\'{e}crit par un th\'{e}or\`{e}me de Gleason
(\cite{Ebeling} 69) comme polynome en $\varphi $ et $\xi $ o\`{u}
\[
W_{\widetilde{H}}(X,Y)=X^8+14X^4Y^4+Y^8=\varphi ,
\]
polyn\^{o}me de poids du code de Hamming \'{e}tendu
$\widetilde{H}\subset \mathbb{F}_2^7$, et
\[
W_{\widetilde{G}}(X,Y)=(X^8+14X^4Y^4+Y^8)^3-42X^4Y^4(X^4-Y^4)^4=W_{%
\widetilde{H}}(X,Y)^3-42\xi ,
\]
polyn\^{o}me de poids du code de Golay \'{e}tendu $\widetilde{G}\subset \mathbb{%
F}_2^{24}$. Pour approfondir l'\'{e}tude du rapport entre les deux codes
cit\'{e}s, et leur lien avec des g\'{e}om\'{e}tries finies comme les
syst\`{e}mes de Steiner, on renvoie \`{a} \cite{Assmus} (\S 7.11 p.\ 284).
Un int\'{e}r\^{e}t de la derni\`{e}re expression pour notre sujet est qu'en
notant $q=\exp (2i\pi \tau )$ et
\[
A=A(\tau )=\sum_{n\in \sqrt{2}\mathbb{Z}}q^{\frac 12n.n}=\sum_{n\in 2\mathbb{Z}%
}q^{\frac 14n.n},\;B=B(\tau )=\sum_{n\in 2\mathbb{Z}+1}q^{\frac
14n.n},
\]
on retrouve la fonction \^{e}ta de Dedekind (\cite{Ebeling} p.67)
:
\[
A^4B^4(A^4-B^4)^4=16q\prod_{n=1}^\infty (1-q^n)^{24}=16\eta (\tau )^{24}.
\]
Ceci a conduit l'auteur \`{a} approfondir l'\'{e}tude des
r\'{e}seaux (\cite {Martinet} \cite{Conway2}) ainsi que les
pavages hyperboliques (\cite {Ebeling} (chapitre 4) \cite{Vinberg}
\cite{Vinberg1} \cite{Magnus1}) pour obtenir des informations sur
certains codes (\cite{Goppa}). Il est reconnu depuis un certain
temps qu'une dualit\'{e} existe entre le codage de l'information
et la quantification, notamment en utilisant des fonctions
th\^{e}ta.\ Les d\'{e}veloppements qui pr\'{e}c\`{e}dent
\'{e}clairent cette observation faite dans \cite{Forney} et que
l'on peut interpr\'{e}ter par recours au groupe d'homologie
$H_2(\mathcal{M},\mathbb{Z})$ et aux formes d'intersection
d\'{e}j\`{a} \'{e}voqu\'{e}es.

Les fonctions th\^{e}ta les plus g\'{e}n\'{e}rales \`{a} plusieurs
variables ont d\'{e}j\`{a} \'{e}t\'{e} introduites en liaison avec
la vari\'{e}t\'{e} jacobienne (voir \cite{Mumford} chapitre 2) et
s'\'{e}crivent avec $u\in \mathbb{C}^g$ et $\mathbf{M\in
}\mathcal{H}_g$
\[
\theta (u,\mathbf{M)=}\sum_{m\in \mathbb{Z}^g}\exp (\pi
i(^tm)\mathbf{M}m+2\pi i(^tm)u).
\]
Elles redonnent $\theta _\Gamma (\tau )=\theta (0,\tau \mathbf{M)}$ avec $%
u=0 $ et $<m,m>=(^tm)\mathbf{M}m$. Elles sont importantes pour d\'{e}crire
diff\'{e}rentes situations physiques telles que la propagation de la chaleur
(\cite{Terras} 1.2 exemple 1, 1.3 exercice 7), la propagation de solitons ou
le comportement de la jonction Josephson (voir l'article de J.\ A.\
Zagrodzinski dans \cite{Planat}).\ Avec $g=1$ et $\mathbf{M}=\tau \mathbf{1}%
_g$ les expressions pr\'{e}c\'{e}dentes donnent la forme plus simple
\'{e}tudi\'{e}e dans le chapitre 1 de \cite{Mumford} avec $u\in \mathbb{C}$ et $%
\tau \in \mathcal{H}$%
\[
\theta (u,\tau \mathbf{)=}\sum_{m\in \mathbb{Z}}\exp (\pi im^2\tau
+2\pi imu).
\]
Pour $u=0$, on obtient $A(\tau )=\theta (0,2\tau \mathbf{)}$ donnant un lien
avec la fonction $\eta $ de Dedekind (\cite{Iwaniec} p.\ 177) qui permet
aussi d'exprimer $B(\tau )$ avec l'expression donn\'{e}e ci-dessus pour $%
16\eta (\tau )^{24}$ :
\[
A(\frac{2r-1}4)=\theta (0,\tau -\frac 12\mathbf{)=}\frac{\eta ^2(z)}{\eta
(2z)}.
\]

\subsection{Lien avec l'\'{e}quation de la chaleur}

La fonction th\^{e}ta $\theta (u,\tau \mathbf{)}$ v\'{e}rifie une
\'{e}quation de la chaleur pour $u$, $\tau =it\in \mathbb{R}$ et
$t$ positif :
\[
\frac \partial {\partial t}\theta (u,it\mathbf{)=}\frac 1{4\pi }\frac{%
\partial ^2}{\partial u^2}\theta (u,it\mathbf{).}
\]
On trouve ainsi une solution fondamentale de l'\'{e}quation de la
chaleur pour $u\in \mathbb{R}/\mathbb{Z}$.\ Cette observation
remonte \`{a} Fourier lui-m\^{e}me (\cite{Fourier} \S 241,
\cite{Jacobi}, \cite{Weil4} p. 28) qui a aussi utilis\'{e}
l'\'{e}quation de la chaleur pour mettre au point ses s\'{e}ries.
Eisenstein a ensuite utilis\'{e} les travaux de Fourier pour
d\'{e}montrer des \'{e}nonc\'{e}s de th\'{e}orie des nombres
relatifs \`{a} la fonction $\zeta $ de Riemann conjectur\'{e}s
ant\'{e}rieurement dans \cite {Euler2}.\ Remarquons que parmi les
produits infinis utilis\'{e}s par Eisentein appara\^{i}t
explicitement une autre solution de l'\'{e}quation de diffusion de
la chaleur, la fonction de Gauss s'\'{e}crivant
\[
T(u,t)=\frac 1{\sqrt{t}}\exp (-\frac{\pi u^2}t).
\]
Cette observation permet comprendre le lien existant entre la th\'{e}orie du
mouvement brownien et les fonctions th\^{e}ta et z\^{e}ta \cite{Yor}.

\subsection{Les quatre fonctions th\^{e}ta habituelles}

Certaines expressions des fonctions th\^{e}ta redonnent, dans l'esprit des
anciens travaux de C.\ G.\ Jacobi, d'autres fonctions automorphes comme par
exemple les fonctions elliptiques. On utilise pour cela les fonctions
th\^{e}ta suivantes qui v\'{e}rifient aussi l'\'{e}quation de la chaleur,
not\'{e}es selon les auteurs
\[
\mathbf{\;}\theta _{jk}(u,\tau \mathbf{)=}\theta \left[
\begin{array}{c}
j \\
k
\end{array}
\right] (u,\tau )=\vartheta _{[2j,-2k]}(\frac u\pi ,\tau )=\sum_{n\in \mathbb{Z}%
}\exp (\pi i(n+j)^2\tau +2\pi i(n+j)(u+k)).
\]
Les fonctions th\^{e}ta permettent de plonger (\cite{Waldschmidt}
p.193) une courbe elliptique dans un espace projectif
$\mathbf{P}^{l-1}(\mathbb{C})$ o\`{u} $l\geq 3$.\ Elles permettent
aussi d'\'{e}crire une telle courbe comme intersection de deux
quadriques gr\^{a}ce \`{a} des relations
classiques dues \`{a} Riemann et Jacobi que l'on retrouve par exemple dans (%
\cite{KumarMurty} ch.\ 7).\ Les fonctions th\^{e}ta sont souvent
pr\'{e}sent\'{e}es comme des g\'{e}n\'{e}ralisations elliptiques de la
fonction exponentielle (par exemple \cite{Weisstein}). Vouloir les utiliser
comme l'exponentielle qui permet de passer d'un groupe de Lie \`{a} une
alg\`{e}bre de Lie \cite{Moore} a ouvert pour l'auteur toute une perspective
de recherches. Usuellement, on restreint \`{a} $4$ le nombre de fonctions
th\^{e}ta utilis\'{e}es gr\^{a}ce aux propri\'{e}t\'{e}s suivantes :
\[
\vartheta _{[2j,-2k]}(z+\frac 12,\tau )=\vartheta _{[2j,-2k-2]}(z,\tau ),
\]
\[
\vartheta _{[2j,-2k]}(z+\frac 12\tau \pi ,\tau )=\exp (-i\pi \tau /4\theta
-iz-ki\pi )\vartheta _{[2j+2,-2k]}(z,\tau ).
\]
Ceci permet de se limiter avec $\mathbf{q}=\exp (i\pi \tau )$ aux quatre
fonctions suivantes qui poss\`{e}dent une d\'{e}composition en produits
infinis (\cite{Chandrasekharan} (ch.V) \cite{Moll} (ch.3) \cite{Perrine9}) :
\[
\vartheta (u,\tau )=2\sum_{n=0}^\infty (-1)^n\mathbf{q}^{(n+\frac 12)^2}\sin
(2n+1)\pi u=\theta (u,\tau ),
\]
\[
\vartheta _1(u,\tau )=2\sum_{n=0}^\infty \mathbf{q}^{(n+\frac 12)^2}\cos
(2n+1)\pi u=\theta (\frac 12-u,\tau ),
\]
\[
\vartheta _2(u,\tau )=1+2\sum_{n=1}^\infty (-1)^n\mathbf{q}^{n^2}\cos (2n\pi
u)=-i\mathbf{q}^{\frac 14}\exp (-i\pi u)\theta (\frac \tau 2-u,\tau ),
\]
\[
\vartheta _3(u,\tau )=1+2\sum_{n=1}^\infty \mathbf{q}^{n^2}\cos (2n\pi u)=%
\mathbf{q}^{\frac 14}\exp (-i\pi u)\theta (\frac{\tau +1}2-u,\tau ).
\]

\subsection{Expressions avec la fonction \^{e}ta de Dedekind}

Si $\vartheta ^{\prime }(u,\tau )$ d\'{e}signe la d\'{e}riv\'{e}e de $%
\vartheta $ par rapport \`{a} $u$, on a une propri\'{e}t\'{e} d'automorphie
avec une racine huiti\`{e}me de l'unit\'{e} $\kappa $ d\'{e}pendant de la
matrice utilis\'{e}e dans $SL(2,\mathbb{Z})$%
\[
\vartheta ^{\prime }(0,\frac{a\tau +b}{c\tau +d})=\kappa (c\tau +d)^{\frac
32}\vartheta ^{\prime }(0,\tau ).
\]
Ceci met en \'{e}vidence un lien avec la fonction $\eta $ de Dedekind telle
que :
\[
\eta (\tau )^{24}=\mathbf{q}^2\prod_{n\geq 1}(1-\mathbf{q}^{2n})^{24},\;\;%
\text{o\`{u} }\mathbf{q}=\exp (i\pi \tau ).
\]
On trouve par exemple les formules suivantes (\cite{Chandrasekharan} p. 80
et p.\ 123, \cite{Knopp} p.\ 46, \cite{Apostol} p. 91, \cite{Grant}, \cite
{WalkerP} p. 161) :
\[
\vartheta ^{\prime }(0,\tau )=2\pi \eta ^3(\tau )=-2\pi \eta (\tau /2)\eta
((\tau +1)/2)\eta (2\tau )=\pi \vartheta _1(0,\tau )\vartheta _2(0,\tau
)\vartheta _3(0,\tau ),
\]
\[
\vartheta (0,\tau )=i\exp (-i\pi \tau /9)\eta (\tau /3),
\]
\[
\;\vartheta _1(0,\tau )=2\frac{\eta ^2(2\tau )}{\eta (\tau )},\;\vartheta
_2(0,\tau )=\frac{\eta ^2(\tau /2)}{\eta (\tau )},\;\vartheta _3(0,\tau )=%
\frac{\eta ^2((\tau +1)/2)}{\eta (\tau )},
\]
\[
\prod_{0\leq u,v<m, \, (u,v)\neq (0,0)} \theta \left[
\begin{array}{c}
1/2+u/m \\
1/2+v/m
\end{array}
\right] (0,\tau )=(-1)^{m-1}m\eta (\tau )^{m^2-1}.
\]
Le lien avec l'invariant modulaire $J$ et la fonction $\mathbf{\lambda }%
_\Lambda $ s'en d\'{e}duit (\cite{Chandrasekharan} p. 85) :
\[
J(\tau )=\frac{(\vartheta _1(0,\tau )^8+\vartheta _2(0,\tau )^8+\vartheta
_3(0,\tau )^8)^3}{54(\vartheta _1(0,\tau )\vartheta _2(0,\tau )\vartheta
_3(0,\tau ))^8}=\frac 4{27}\frac{(1-\mathbf{\lambda }_\Lambda +\mathbf{%
\lambda }_\Lambda ^2)^3}{\mathbf{\lambda }_\Lambda ^2(1-\mathbf{\lambda }%
_\Lambda )^2},\;\;
\]
\[
\mathbf{\lambda }_\Lambda (\tau )=\frac{\vartheta _1(0,\tau )^4}{\vartheta
_3(0,\tau )^4}=16(\eta ^2(2\tau )\eta (\tau /2)\eta ^{-3}(\tau ))^8.
\]
Egalement on peut \'{e}crire avec $\eta $ les fonctions elliptiques de
Jacobi (\cite{Chandrasekharan} p. 100 et p.\ 103 \cite{Moll} ch.3 \cite
{WalkerP} (p. 165)). Avec une constante $c$ on a par (\cite{Husemoller} p.
191)
\[
\wp (u,\mathbb{Z}\oplus \mathbb{Z}\tau )=c-\frac{d^2}{du^2}\log
\vartheta (\frac u{\pi \vartheta _3(0,\tau )^2},\tau ).
\]
Le fait que toutes ces fonctions puissent se d\'{e}duire de $\eta $ montre
l'importance fondamentale de cette fonction aussi utilis\'{e}e pour des
calculs d'approximation \cite{Garvan}. Pour les fonctions $L$ la situation
est plus compliqu\'{e}e mais li\'{e}e \cite{Ericksson}.

\section{Approche hyperg\'{e}om\'{e}trique de la th\'{e}orie de Markoff}

On a approfondi quelques remarques faites par Harvey Cohn dans son \'{e}tude
de la th\'{e}orie de Markoff.

\subsection{Relation avec une fonction elliptique}

Dans son article initial \cite{Cohn2}, Harvey\ Cohn donne la relation
suivante pour interp\'{e}ter g\'{e}om\'{e}triquement la th\'{e}orie de
Markoff avec un r\'{e}seau $\Lambda $ particulier :
\[
1-J(\tau )=\wp ^{\prime 2}(z)=4\wp ^3(z)+1.
\]
Le module $J$ est une fonction automorphe pour le groupe fuchsien
$\Gamma =PSL(2,\mathbb{Z})$ de facteur $\mu =1$ et de poids $0$.\
Harvey\ Cohn dit que les triplets de matrices $(A,B,C)$
associ\'{e}s \`{a} la th\'{e}orie de Markoff classique
d\'{e}terminent un pavage hexagonal du demi-plan de Poincar\'{e}
en $\tau $ et correspondent par cette relation entre $\tau $ et
$z$ \`{a} un pavage quadrilat\'{e}ral par un r\'{e}seau $\Lambda $
du plan complexe en $z$. Il l'illustre g\'{e}om\'{e}triquement sur
une figure o\`{u} apparaissent les matrices not\'{e}es
\[
A_0=\left[
\begin{array}{cc}
1 & 1 \\
1 & 2
\end{array}
\right] ,\;\;B_0=\left[
\begin{array}{cc}
1 & -1 \\
-1 & 2
\end{array}
\right] ,
\]
L'aspect alg\'{e}brique de ces remarques de Harvey Cohn se r\'{e}sume (\cite
{Perrine9} fig. 7.7) en d\'{e}crivant les domaines fondamentaux respectifs
de deux pavages de $\mathcal{H}$, donnant au quotient le tore perc\'{e}.\ Le
premier est un pavage hexagonal $\alpha \beta \gamma \delta \varepsilon
\zeta \eta \theta \iota $. Le second donne un domaine quadrilat\'{e}ral $%
\kappa \lambda \mu \nu \xi $. Le passage entre les deux est
faisable par un jeu de tangram hyperbolique utilisant pour
pi\`{e}ces des morceaux composant le domaine fondamental bien
connu pour le groupe $PSL(2,\mathbb{Z})$. Pour aller vers le
domaine hexagonal, il suffit d'appliquer au domaine modulaire six
matrices de forme
\[
\left[
\begin{array}{cc}
1 & k \\
0 & 1
\end{array}
\right] \;(k=-2,-1,0,1,2,3).
\]
Pour aller vers le domaine quadrilat\'{e}ral, il suffit d'utiliser les deux
demi-domaines modulaires et les six matrices suivantes (\cite{Appel} Tome 2
p.368)
\[
\left[
\begin{array}{cc}
1 & 0 \\
0 & 1
\end{array}
\right] ,\left[
\begin{array}{cc}
0 & -1 \\
1 & 0
\end{array}
\right] ,\left[
\begin{array}{cc}
0 & -1 \\
1 & -1
\end{array}
\right] ,\left[
\begin{array}{cc}
1 & 0 \\
1 & 1
\end{array}
\right] ,\left[
\begin{array}{cc}
1 & -1 \\
1 & 0
\end{array}
\right] ,\left[
\begin{array}{cc}
-1 & 0 \\
1 & -1
\end{array}
\right] .
\]

\subsection{Sph\`{e}re \`{a} trois piq\^{u}res et invariant modulaire}

En r\'{e}alit\'{e}, il existe une autre fa\c {c}on de fabriquer
une surface de Riemann avec le domaine quadrilat\'{e}ral $\kappa
\lambda \mu \nu \xi $ et cette m\'{e}thode a \'{e}t\'{e}
g\'{e}n\'{e}ralis\'{e}e dans \cite {Schmidt}. Il suffit
d'identifier $\kappa \lambda $ et $\xi \nu $ par une
transformation de $a\in PSL(2,\mathbb{Z})$, ainsi que $\mu \lambda
$ et $\mu \nu $ par $b\in PSL(2,\mathbb{Z})$. on fabrique ainsi
une sph\`{e}re \`{a} trois piq\^{u}res correspondant aux points
$0$, $1$, $\infty $. Le calcul explicite peut \^{e}tre fait et
d\'{e}termine pour $a$ et $b$ les matrices
\[
a=\left[
\begin{array}{cc}
1 & 2 \\
0 & 1
\end{array}
\right] ,\;b=\left[
\begin{array}{cc}
1 & 0 \\
2 & 1
\end{array}
\right] .
\]
Ces matrices engendrent le groupe $\Gamma (2)$ qui est libre (\cite{Iversen}
p.154) et d\'{e}terminent une structure g\'{e}om\'{e}trique unique sur $%
\mathcal{H}/\Gamma (2)$. Ce qui pr\'{e}c\`{e}de garantit par le
th\'{e}or\`{e}me de Riemann (\cite{Ford} p.\ 163) l'existence d'une relation
alg\'{e}brique entre $J$ et une fonction automorphe pour le groupe $\Gamma
(2)$.\ Cette relation est usuellement calcul\'{e}e \`{a} partir des
expressions donn\'{e}es pour le cas elliptique
\[
y^2=4x^3-g_2x-g_3=P(x)=4(x-e_1)(x-e_2)(x-e_3).
\]
On pose, \`{a} une permutation pr\`{e}s sur $e_1$, $e_2$, $e_3$%
\[
\nu _{31}=(e_3-e_1)\neq 0,\;\mathbf{x}=\frac{(x-e_1)}{(e_3-e_1)},\;\mathbf{%
\lambda }_\Lambda =\frac{(e_2-e_1)}{(e_3-e_1)},\;\nu ^2=\frac 1{4\nu
_{31}^3},\;\mathbf{y}^2=\nu ^2y^2.
\]
Ceci transforme l'\'{e}quation $y^2=P(x)$ en la forme de Legendre suivante
\[
\mathbf{y}^2=\mathbf{x}(\mathbf{x}-1)(\mathbf{x}-\mathbf{\lambda }_\Lambda )%
\text{ o\`{u} }\mathbf{\lambda }_\Lambda \notin \{0,1\}.
\]
Les permutations possibles sur $e_1$, $e_2$, $e_3$, montrent que deux
courbes elliptiques $E_{\mathbf{\lambda }_\Lambda }$ et $E_{\mathbf{\lambda }%
_\Lambda ^{\prime }}$ obtenues ainsi sont isomorphes si et seulement si on a
\[
\mathbf{\lambda }_\Lambda ^{\prime }\in \{\mathbf{\lambda }_\Lambda ,\frac 1{%
\mathbf{\lambda }_\Lambda },1-\mathbf{\lambda }_\Lambda ,\frac 1{1-\mathbf{%
\lambda }_\Lambda },\frac{\mathbf{\lambda }_\Lambda }{\mathbf{\lambda }%
_\Lambda -1},\frac{\mathbf{\lambda }_\Lambda -1}{\mathbf{\lambda }_\Lambda }%
\}.
\]
Ceci permet de se limiter aux valeurs complexes
\[
\mathbf{\lambda }_\Lambda \in S_4=\{\lambda \mid \lambda \in
\mathbb{C},\;\mid \lambda \mid <1,\;\mid 1-\lambda \mid
<1,\;\Re(\lambda )\geq (1/2)\}.
\]
En inversant les relations pr\'{e}c\'{e}dentes pour d\'{e}duire $e_3$ et $%
e_2 $ on obtient les expressions suivantes montrant que $\mathbf{\lambda }%
_\Lambda $ ne suffit pas \`{a} d\'{e}finir le polyn\^{o}me $P(x)$ mais que
le param\`{e}tre accessoire $\nu _{31}$ est indispensable :
\[
\nu _{31}+\nu _{31}\mathbf{\lambda }_\Lambda =-3e_1,\;4\nu _{31}^2\mathbf{%
\lambda }_\Lambda e_1=(12e_1{}^2-g_2)e_1=8e_1{}^3+g_3,
\]
\[
g_2=\frac{4\nu _{31}^2}3(1-\mathbf{\lambda }_\Lambda +\mathbf{\lambda }%
_\Lambda ^2),\;g_3=\frac{4\nu _{31}^3}{27}(\mathbf{\lambda }_\Lambda +1)(%
\mathbf{\lambda }_\Lambda -2)(2\mathbf{\lambda }_\Lambda -1),
\]
\[
g_2^3-27g_3^2=16\nu _{31}^6\mathbf{\lambda }_\Lambda ^2(1-\mathbf{\lambda }%
_\Lambda )^2.
\]
Ceci donne l'expression de $J$ recherch\'{e}e et tr\`{e}s classique
\[
J=\frac{g_2^3}{g_2^3-g_3^2}=\frac 4{27}\frac{(1-\mathbf{\lambda }_\Lambda +%
\mathbf{\lambda }_\Lambda ^2)^3}{\mathbf{\lambda }_\Lambda ^2(1-\mathbf{%
\lambda }_\Lambda )^2}.
\]
Ainsi $\mathbf{\lambda }_\Lambda $ appara\^{i}t comme $J$ en tant
que fonction d'une variable $\tau \in \mathcal{H}$. En
consid\'{e}rant que $\tau =\omega
_2/\omega _1$ o\`{u} $\omega _1$, $\omega _2$ engendrent le r\'{e}seau $%
\Lambda $, on peut observer l'action sur $\mathbf{\lambda
}_\Lambda (\tau )$ d'une transformation de $PSL(2,\mathbb{Z})$. Il
est facile de voir que si la transformation est dans $\Gamma (2)$,
le groupe de la sph\`{e}re \`{a} trois trous, la valeur de cette
fonction ne change pas (\cite{Ford} p.159). La fonction
$\mathbf{\lambda }_\Lambda (\tau )$ est donc automorphe pour ce
groupe dont un domaine fondamental appara\^{i}t aussi sur la
figure pr\'{e}c\'{e}dente. Comme ce domaine fondamental $\kappa
\lambda \mu \nu \xi $ est constitu\'{e} de copies du domaine
fondamental de $PSL(2,\mathbb{Z})$, on retrouve d'une autre fa\c
{c}on par la m\'{e}thode de Riemann (\cite{Ford}
p.\ 163) l'existence de la relation liant $J$ et $\mathbf{\lambda }_\Lambda $%
. Celle-ci vient d'\^{e}tre calcul\'{e}e. On voit facilement (\cite{Perrine9}
fig.7.8 inspir\'{e}e de \cite{Cohn5}) ce que donne la fonction $\tau
\rightarrow \mathbf{\lambda }_\Lambda (\tau )$. Elle v\'{e}rifie
\[
\mathbf{\lambda }_\Lambda (\tau +1)=\frac{\mathbf{\lambda }_\Lambda (\tau )}{%
\mathbf{\lambda }_\Lambda (\tau )-1},\;\;\mathbf{\lambda }_\Lambda (-\frac
1\tau )=1-\mathbf{\lambda }_\Lambda (\tau ).
\]
Ces conditions mettent en \'{e}vidence deux matrices dont on v\'{e}rifie
ais\'{e}ment qu'elles engendrent le groupe des permutations \`{a} trois
\'{e}l\'{e}ments:
\[
\mathbf{\lambda }_\Lambda \circ S=\left[
\begin{array}{cc}
-1 & 1 \\
0 & 1
\end{array}
\right] \circ \mathbf{\lambda }_\Lambda ,\;\;\mathbf{\lambda }_\Lambda \circ
T=\left[
\begin{array}{cc}
1 & 0 \\
1 & -1
\end{array}
\right] \circ \mathbf{\lambda }_\Lambda .
\]
Ainsi s'introduit un groupe fini de matrices isomorphe au groupe des
permutations de $3$ \'{e}l\'{e}ments avec
\[
\mathbf{S}=\left[
\begin{array}{cc}
-1 & 1 \\
0 & 1
\end{array}
\right] \rightarrow \left(
\begin{array}{ccc}
1 & 2 & 3 \\
2 & 1 & 3
\end{array}
\right) ,\;\;\mathbf{ST}=\left[
\begin{array}{cc}
0 & -1 \\
1 & -1
\end{array}
\right] \rightarrow \left(
\begin{array}{ccc}
1 & 2 & 3 \\
3 & 1 & 2
\end{array}
\right) .
\]
Ce groupe de permutations agit sur les valeurs de $\mathbf{\lambda }_\Lambda
$ avec des orbites \`{a} $6$ \'{e}l\'{e}ments sauf les trois cas suivants :

$\mathbf{\lambda }_\Lambda \in \{1/2,-1,2\}$ soit $\tau $ dans la classe de $%
i$ donnant $J=1$ et la ramification d'ordre $2$ de $J$ (en pratique, deux
droites se coupent sur la figure pr\'{e}c\'{e}dente, ce qui correspond \`{a}
une sym\'{e}trie carr\'{e}e).

$\mathbf{\lambda }_\Lambda \in \{-\rho ,-\rho ^2\}$ soit $\tau $
dans la classe de $\rho =(-1+i\sqrt{3})/2$ donnant $J=0$, la
ramification d'ordre $3$ de $J$ (en pratique, trois droites se
coupent sur la figure pr\'{e}c\'{e}dente, ce qui correspond \`{a}
une sym\'{e}trie hexagonale).

$\mathbf{\lambda }_\Lambda \in \{0,1,\infty \}$ soit $\tau $ dans
la classe de $\infty $ donnant $J=\infty $ hors de $\mathcal{H}$
et de $\mathbb{C}$.

\subsection{L'\'{e}tude hyperg\'{e}om\'{e}trique des relations de H.\ Cohn}

Pour comprendre l'origine de la relation utilis\'{e}e par Harvey\ Cohn dans
\cite{Cohn2} pour interpr\'{e}ter la th\'{e}orie de Markoff, consid\'{e}rons
l'expression
\[
f=\frac 1{27}(\mathbf{\lambda }_\Lambda +1)(\frac 1{\mathbf{\lambda }%
_\Lambda }+1)(1-\mathbf{\lambda }_\Lambda +1)(\frac 1{1-\mathbf{\lambda }%
_\Lambda }+1)(\frac{\mathbf{\lambda }_\Lambda }{\mathbf{\lambda }_\Lambda -1}%
+1)(\frac{\mathbf{\lambda }_\Lambda -1}{\mathbf{\lambda }_\Lambda }+1).
\]
C'est par construction un invariant pour le groupe des permutations de $3$
\'{e}l\'{e}ments appliqu\'{e} dans le plan en $\mathbf{\lambda }_\Lambda $
\cite{Dixmier} exprimable en fonction de $g_2$ et $g_3$.\ En faisant ce
calcul, on trouve facilement (\cite{Ford} p. 160) la premi\`{e}re partie de
l'expression donn\'{e}e par Harvey Cohn
\[
1-f=\frac 4{27}\frac{(1-\mathbf{\lambda }_\Lambda +\mathbf{\lambda }_\Lambda
^2)^3}{\mathbf{\lambda }_\Lambda ^2(1-\mathbf{\lambda }_\Lambda )^2}=J.
\]
On trouve dans \cite{Hunt} (p. 136) une fa\c {c}on de traiter une telle
\'{e}quation. Ce n'est pas la m\'{e}thode utilis\'{e}e ici.\ On veut
plut\^{o}t \'{e}crire $f$ avec une fonction elliptique en $\tau $
particuli\`{e}re.\ Pour cela on identifie les bords du domaine
consid\'{e}r\'{e} dans le plan en $\tau $ avec le groupe $[SL(2,\mathbb{Z}%
),SL(2,\mathbb{Z})]$.\ Ceci donne au quotient un tore perc\'{e}
conforme. Comme le calcul pr\'{e}c\'{e}dent en $\mathbf{\lambda
}_\Lambda $ \'{e}tait li\'{e} \`{a} quelques singularit\'{e}s
pr\`{e}s \`{a} la sph\`{e}re du domaine modulaire et faisait
appara\^{i}tre $J$, de m\^{e}me le tore moins un point est li\'{e}
\`{a} un tore complet dont il s'agit d'utiliser la fonction de
Weierstrass associ\'{e}e. Ceci revient \`{a} travailler \`{a} la
conjonction de deux uniformisations \cite{Mazur1}, une dans
$\mathbb{C}$ puis une dans $\mathcal{H}$.

Dans ses diff\'{e}rents articles (\cite{Cohn1}, \cite{Cohn7}, \cite{Cohn8},
\cite{Cohn1}), Harvey Cohn mentionne en liaison avec la question
\'{e}tudi\'{e}e une autre formule issue de travaux de R.\ Fricke \`{a}
prendre en compte et qui sous-entend une sym\'{e}trie hexagonale
\[
dz=const.\times \frac{dJ}{J^{2/3}(J-1)^{1/2}}.
\]
Il \'{e}voque la difficult\'{e} du passage entre les diff\'{e}rentes
expressions, renvoyant \`{a} \cite{Chudnovski} \cite{Keen5} o\`{u} le
probl\`{e}me est \'{e}tudi\'{e} sous l'aspect d'un param\`{e}tre accessoire
v\'{e}rifiant une \'{e}quation diff\'{e}rentielle de Lam\'{e} (\cite{Yosida2}
p. 110), mais sans conclusions bien nettes. Cette question est li\'{e}e au 22%
$^{i\grave{e}me}$ probl\`{e}me de Hilbert qui est celui de
l'uniformisation num\'{e}rique d'une surface de Riemann, encore
non encore totalement r\'{e}solu aujourd'hui \cite{Seppala1},
m\^{e}me si les \'{e}quations de Lam\'{e} font l'objet d'un regain
d'int\'{e}r\^{e}t aujourd'hui \cite {Arscott} \cite{Waall}. La
derni\`{e}re expression reliant $z$ et $J$ peut \^{e}tre
construite par simple diff\'{e}rentiation. Supposons que l'on ait
\[
1-J(\tau )=\wp ^{^{\prime }2}(z)=4\wp ^3(z)+1,
\]
ceci donne
\[
-dJ=12\wp ^2(z)\wp ^{^{\prime }}(z)dz,\;\;\wp ^{^{\prime
}}(z)=(1-J)^{1/2},\;\;\wp ^2(z)=(J/4)^{2/3}.
\]
D'o\`{u} en rempla\c {c}ant dans l'expression de $-dJ$, et \`{a}
un facteur multiplicatif pr\`{e}s, l'expression donn\'{e}e pour
$dz$. En sens inverse l'int\'{e}gration d'une telle expression
reliant $dz$ et $dJ$ pr\'{e}sente des difficult\'{e}s car elle
d\'{e}pend du chemin consid\'{e}r\'{e}. A un facteur pr\`{e}s, on
trouve une int\'{e}grale hyperg\'{e}om\'{e}trique d\'{e}finie dans
le cas o\`{u} $\Re(c)>\Re(a)>0$,$\;\mid x\mid <1 $, ici $a=(1/3)$,
$c=(5/6)$, $b=0$, ou encore une fonction b\^{e}ta
d\'{e}finie pour $\Re(p)>0$,$\;\Re(q)>0$, ici $p=(1/3)$, $%
q=(1/2) $
\[
F(a,b,c,x)=\frac{\mathbf{\Gamma }(c)}{\mathbf{\Gamma }(a)\mathbf{\Gamma }%
(c-a)}\int_0^1t^{a-1}(1-t)^{c-a-1}(1-tx)^{-b}dt,
\]
\[
B(p,q)=\int_0^1t^{p-1}(1-t)^{q-1}dt=\frac{\mathbf{\Gamma }(p)\mathbf{\Gamma }%
(q)}{\mathbf{\Gamma }(p+q)},\;\;\mathbf{\Gamma }(p)=\int_0^\infty
t^{p-1}\exp (-t)dt
\]
Les difficult\'{e}s d'int\'{e}gration sont illustr\'{e}es dans \cite{Yosida}
(p. 85 - 90) o\`{u} l'on montre comment l'int\'{e}gration sur un double
contour de Pochhammer autour de $[0,1]$ change
\[
\int_0^1J^{p-1}(1-J)^{q-1}dJ
\]
en la multipliant cette valeur par un facteur $(1-\exp (2i\pi p))(1-\exp
(2i\pi q))$. La fonction hyperg\'{e}om\'{e}trique $F(a,b,c,x)$ est solution
de l'\'{e}quation diff\'{e}rentielle \`{a} deux singularit\'{e}s $x=0$ et $%
x=1$, o\`{u} $x\in \mathbb{C}$ :
\[
E(a,b,c):\;\;x(1-x)\frac{d^2F}{dx^2}+(c-(a+b+1)x)\frac{dF}{dx}-abF=0.
\]
Lorsque les param\`{e}tres $a$, $b$, $c$, sont r\'{e}els et tels que $c$, $%
c-a-b$, $a-b$, non entiers, on peut d\'{e}finir sur $\mathcal{D}=\mathbb{C}%
\backslash \{]-\infty ,0]\cup [1,\infty [\}$ l'application de Schwarz :
\[
Sch:J\in \mathcal{D}\longrightarrow
(F(a,b,c,J):J^{1-c}F(a+1-c,b+1-c,2-c,J))\in \mathbf{P}^1(C).
\]
Pour $\mid 1-c\mid =(1/\upsilon _1)$, $\mid c-a-b\mid =(1/\upsilon _2)$, $%
\mid a-b\mid =(1/\upsilon _3)$ strictement plus petits que $1$, l'image de $%
\mathcal{H}$ par cette application est un triangle de la
sph\`{e}re de Riemann avec les angles $(\pi /\upsilon _1)$ en
$Sch(0)$, $(\pi /\upsilon _2) $ en $Sch(1)$, et $(\pi /\upsilon
_3)$ en $Sch(\infty )$. On retrouve ainsi les groupes classiques
de pavages par des isom\'{e}tries des surfaces de Riemann
simplement connexes (\cite{Berger} chapitre 1), avec les trois cas
sph\'{e}rique, euclidien et hyperbolique.\ On peut prolonger
l'application de Schwarz \`{a} $\mathbb{C}\backslash \{0,1\}$ par
le principe de r\'{e}flexion sur le bord de $\mathcal{H}$ et
fabriquer des transformations conformes (\cite{Yosida} p. 78) qui
interpr\`{e}tent le lien entre $J$ et $\mathbf{\lambda }_\Lambda $
montrant le caract\`{e}re d\'{e}terminant de ce qui se passe en
certains points singuliers :
\[
F(1/12,5/12,1,x):x=J(\tau )\in \mathbb{C}\longmapsto \tau \in \mathcal{H}%
/PSL(2,Z)\text{ d'inverse la fonction }J,
\]
\[
F(1/2,1/2,1,x):x=\text{ }\mathbf{\lambda }_\Lambda (\tau )\in \mathbb{C}%
\backslash \{0,1\}\longmapsto \tau \in \mathcal{H}/\Gamma (2)\text{
d'inverse la fonction }\mathbf{\lambda }_\Lambda .
\]
Un tel prolongement permet effectivement de consid\'{e}rer un contour de
Pochhammer et permet de comprendre la nature de la difficult\'{e}
rencontr\'{e}e. Pour les valeurs $a=(1/3)$, $b=0$, $c=(5/6)$ de l'expression
diff\'{e}rentielle de H.\ Cohn entre $dz$ et $dJ$, on a $\mid 1-c\mid =(1/6)$%
, $\mid c-a-b\mid =(1/2)$, $\mid a-b\mid =(1/3)$. Ceci correspond \`{a} un
cas euclidien de cristal plan hexagonal. On trouve dans les travaux de R.\
Dedekind \cite{Dedekind} une approche compl\'{e}mentaire \`{a} ce qui
pr\'{e}c\`{e}de, avec un lien explicite avec la fonction $\eta $.\ Il montre
que la fonction $w(\tau )$ d\'{e}finie \`{a} une constante pr\`{e}s par
\[
w(\tau )=c\frac{J^{\prime }(\tau )^{1/2}}{J(\tau )^{1/3}(1-J(\tau ))^{1/4}},
\]
v\'{e}rifie une \'{e}quation diff\'{e}rentielle hyperg\'{e}om\'{e}trique $%
E((1/12),(1/12),(2/3))$ permettant d'\'{e}crire $w$ en fonction de $J$. La
fonction $\eta $ est elle-m\^{e}me une racine carr\'{e}e de $w$ \`{a} un
coefficient pr\`{e}s (\cite{Chandrasekharan} p. 135 ou \cite{Moll} p. 180)
telle que :
\[
\eta (\tau )^{24}=\frac 1{(48\pi ^2)^3}\frac{J^{\prime }(\tau )^6}{J(\tau
)^4(1-J(\tau ))^3}=-\frac 1{4^4\pi ^6}\frac{\mathbf{\lambda }_\Lambda
^{\prime }(\tau )^6}{\mathbf{\lambda }_\Lambda (\tau )^4(1-\mathbf{\lambda }%
_\Lambda (\tau ))^4}.
\]
Pour aller plus avant, il est n\'{e}cessaire de faire le lien entre ce que
l'on vient de voir et l'\'{e}quation hyperg\'{e}om\'{e}trique perturb\'{e}e
que l'on a mise en \'{e}vidence ci-dessus en liaison avec la
repr\'{e}sentation monodromique d\'{e}finie par les matrices $A_0$ et $B_0$.
Ce point fait l'objet d'un travail en cours de d\'{e}veloppement, pr\'{e}vu
pour \^{e}tre pr\'{e}sent\'{e} \`{a} la cinqui\`{e}me conf\'{e}rence
internationale ''Symmetry in Nonlinear Mathematical Physics'' de Kyiv, en
juin 2003.

\section{Approche par la double uniformisation}

La comparaison de ce que l'on vient de voir avec les r\'{e}sultats du
chapitre pr\'{e}c\'{e}dent sugg\`{e}re que dans certains cas la valeur $%
\mathbf{\lambda }_\Lambda $ puisse \^{e}tre choisie \'{e}gale au module $%
(\mu ^2/\lambda ^2)$ d'un tore perc\'{e} parabolique prolongeable en le tore
$\Lambda $. En effet, l'\'{e}quation $\mathbf{y}^2=\mathbf{x}(\mathbf{x}-1)(%
\mathbf{x}-\mathbf{\lambda }_\Lambda )$ met en \'{e}vidence trois racines $%
\alpha ^{\prime }=1$, $s^{\prime }=0$, $\beta ^{\prime }=$ $\mathbf{\lambda }%
_\Lambda $, et ne d\'{e}finit bien l'\'{e}quation diophantienne de
d\'{e}part qu'au coefficient $\nu _{31}$ pr\`{e}s. De m\^{e}me, le
tore perc\'{e} est d\'{e}fini avec $\alpha =-1$, $s=0$, $\beta
=(\mu ^2/\lambda ^2)$, au coefficient $\lambda $ pr\`{e}s. En
approfondissant ce th\`{e}me, on a construit la propri\'{e}t\'{e}
de double uniformisation des tores perc\'{e}s, et on en a tir\'{e}
les cons\'{e}quences pour la th\'{e}orie de Markoff. Le
r\'{e}sultat essentiel obtenu est la relation profonde qui existe
entre la fonction \^{e}ta de Dedekind et l'op\'{e}rateur de
Laplace-Beltrami d'un tore.\ Ceci explique la d\'{e}composition en
produit infini de la fonction $\eta $. Comme cette fonction est
li\'{e}e \`{a} beaucoup d'autres fonctions transcendantes, ceci
explique l'existence de produits infinis pour toutes ces
fonctions, notamment les fonctions th\^{e}ta.

\subsection{Une construction g\'{e}n\'{e}rale}

En comparant la repr\'{e}sentation des tores de module $(\mu ^2/\lambda ^2)$
du chapitre pr\'{e}c\'{e}dent au plan en $\mathbf{\lambda }_\Lambda $, on
fait en sorte que se correspondent des points de m\^{e}me ordre de
ramification. Ainsi $(\mu ^2/\lambda ^2)=2$ correspond \`{a} une
ramification d'ordre 2 que l'on obtient avec $\mathbf{\lambda }_\Lambda =1/2$%
. De m\^{e}me $(\mu ^2/\lambda ^2)=1$ correspond ainsi \`{a} une
ramification d'ordre 3 que l'on obtient avec $\mathbf{\lambda }_\Lambda
=-\rho ^2$. Ceci assure la coh\'{e}rence avec la ramification de $J$, on
l'observe sur le domaine de cette fonction entre des valeurs correspondantes
qui sont r\'{e}elles et qui valent $J=1$ pour $\mathbf{\lambda }_\Lambda
=1/2 $, ainsi que $J=0$ pour $\mathbf{\lambda }_\Lambda =-\rho ^2=(1+i\sqrt{3%
})/2$.\ On consid\`{e}re inversement des valeurs complexes $\mathbf{\lambda }%
_\Lambda $ se situant sur le bord $[(1/2),-\rho ^2]$ du domaine $S_4$ de
notre figure 7.8 de \cite{Perrine9}.\ Elles correspondent de fa\c {c}on
bijective aux valeurs $J=J(\mathbf{\lambda }_\Lambda )\in [0,1]\subset \mathbb{R%
}$.\ On pose ainsi $(\mu ^2/\lambda ^2)=\mathbf{\beta }(J)\in [1,2]\subset
\mathbb{R}$, avec $\mathbf{\beta }$ bijection croissante de $[0,1]\subset \mathbb{R%
}$ dans $[1,2]\subset \mathbb{R}$. On choisit alors $\Theta _\alpha =\lambda ^2$%
, et on se ram\`{e}ne au cas o\`{u} $\alpha =-1$, $s=0$, $\beta =\mathbf{%
\beta }(J)$, $p=\infty $.\ Ceci normalise le tore perc\'{e} que
l'on consid\`{e}re et donne un tore perc\'{e} parabolique
d\'{e}fini par les matrices $A$ et $B$ suivantes dans
$SL(2,\mathbb{R})$
\[
A=\left[
\begin{array}{cc}
\lambda \sqrt{\mathbf{\beta }(J)} & \lambda \sqrt{\mathbf{\beta }(J)} \\
\dfrac \lambda {\sqrt{\mathbf{\beta }(J)}} & \dfrac{1+\lambda ^2}{\lambda
\sqrt{\mathbf{\beta }(J)}}
\end{array}
\right] ,\;B=\left[
\begin{array}{cc}
\lambda & -\lambda \mathbf{\beta }(J) \\
-\lambda & \dfrac{1+\lambda ^2\mathbf{\beta }(J)}\lambda
\end{array}
\right] .
\]
En r\'{e}alit\'{e}, ces deux matrices sont d\'{e}finies au facteur r\'{e}el $%
\lambda >0$ pr\`{e}s. Il y a tout un ensemble de tores diff\'{e}rents qui
peuvent convenir \`{a} la m\^{e}me valeur $\mathbf{\beta }(J)$, et qui ont
donc des propri\'{e}t\'{e}s communes. Il n'y a aucune raison de se
d\'{e}barrasser ici du coefficient $\lambda $. Une analyse approfondie de
cette situation a \'{e}t\'{e} faite et a conduit au th\'{e}or\`{e}me
''Jugendtraum'' de Kronecker (\cite{Dieudonne2} tome 1, p. 236). Les deux
matrices mises en \'{e}vidence d\'{e}finissent un groupe libre dans $PSL(2,%
\mathbb{R})$ que l'on peut ab\'{e}lianiser pour introduire une
courbe elliptique.\ Les endomorphismes du groupe libre qu'elles
engendrent sont associ\'{e}s \`{a} des polyn\^{o}mes
\cite{Peyriere}. Ceci provient des r\'{e}sultats qui ont
\'{e}t\'{e} d\'{e}montr\'{e}s pour les tores perc\'{e}s conformes
paraboliques. Une m\'{e}thode pour fabriquer une courbe elliptique
associ\'{e}e consiste \`{a} utiliser les calculs de \cite
{Husemoller} (p.179) et \`{a} les prolonger pour la valeur $\mathbf{\lambda }%
_\Lambda \mathbf{=}-\rho ^2$ . En dehors de ce cas singulier qui ne pose
d'ailleurs pas de probl\`{e}me (\cite{Silverman0} ch VI), les valeurs $%
\mathbf{\lambda }_\Lambda $ s\'{e}lectionn\'{e}es sont telles que
\[
\mathbf{\lambda }_\Lambda =\frac 12+i\Im(\mathbf{\lambda }_\Lambda
),\;0\leq \Im(\mathbf{\lambda }_\Lambda )\leq
\frac{\sqrt{3}}2,\;\mid \mathbf{\lambda }_\Lambda \mid <1,\;\mid
1-\mathbf{\lambda }_\Lambda \mid <1.
\]
Ceci permet de bien d\'{e}finir deux p\'{e}riodes engendrant un r\'{e}seau $%
\Lambda $%
\[
\omega _1(\mathbf{\lambda }_\Lambda )=\int_{-\infty }^0\frac{dx}{\sqrt{%
x(x-1)(x-\mathbf{\lambda }_\Lambda )}},\;\;\omega _2(\mathbf{\lambda }%
_\Lambda )=\int_1^\infty \frac{dx}{\sqrt{x(x-1)(x-\mathbf{\lambda }_\Lambda )%
}}.
\]
D'o\`{u} la construction effective d'une fonction elliptique associ\'{e}e
\`{a} ce r\'{e}seau
\[
\wp _\Lambda ^{\prime 2}(z)=4\wp _\Lambda ^3(z)-g_2\wp _\Lambda (z)-g_3,
\]
\[
g_2=\frac{4\nu _{31}^2}3(\mathbf{\lambda }_\Lambda ^2-\mathbf{\lambda }%
_\Lambda +1),\;g_3=\frac{4\nu _{31}^3}{27}(\mathbf{\lambda }_\Lambda +1)(%
\mathbf{\lambda }_\Lambda -2)(2\mathbf{\lambda }_\Lambda -1).
\]
Pour $\mathbf{\lambda }_\Lambda =-\rho ^2$, on obtient $g_2=0$. La
courbe elliptique correspondante, associ\'{e}e au r\'{e}seau
$\mathbb{Z}[\rho ]$, est bien d\'{e}finissable (\cite{Silverman}
p. 102) au moyen de l'\'{e}quation rationnelle $y^2=x^3+1$
utilis\'{e}e ci-dessus pour analyser les informations
diff\'{e}rentielles donn\'{e}es par H.\ Cohn. En effet, cette
derni\`{e}re est issue de l'\'{e}quation $%
y^2=4x^3-3i\sqrt{3}\nu _{31}^3$ obtenue avec les expressions donn\'{e}es
pour $g_2$ et $g_3$.\ Pour $\mathbf{\lambda }_\Lambda =(1/2)$, on obtient $%
g_3=0$. La courbe elliptique correspondante, associ\'{e}e au r\'{e}seau $%
\mathbb{Z}[i]$, est d\'{e}finissable (\cite{Silverman} p. 101) par
l'\'{e}quation $y^2=x^3+x$ d\'{e}ductible cette fois de $%
y^2=4x^3-\nu _{31}^2x$.\ La donn\'{e}e du param\`{e}tre accessoire
$\nu _{31}\neq 0$ permet de retrouver toutes les donn\'{e}es de la
courbe elliptique avec $e_3=e_1+\nu
_{31},\;e_2=e_1+\mathbf{\lambda }_\Lambda \nu _{31}$.\ A une
transformation conforme pr\`{e}s de $\mathbb{C}$ construite par
translation et rotation, on peut normaliser sans changer
$\mathbf{\lambda }_\Lambda $ cette courbe elliptique en se
ramenant \`{a} $e_1=0,\;e_3=\nu _{31}=\parallel \nu _{31}\parallel
\in \mathbb{R}^{+}$. Ayant normalis\'{e} le tore perc\'{e} et le
tore selon deux m\'{e}thodes diff\'{e}rentes, on s'attache alors
\`{a} faire en sorte que le tore perc\'{e} provienne du tore par
simple extraction d'un point, sachant que les probl\`{e}mes de
m\'{e}trique induite restent \`{a} v\'{e}rifier. L'identification
de $e_1$ et $e_3=\nu _{31}$ sur le tore donne un grand cercle que
l'on identifie sur le tore perc\'{e} \`{a} un cercle de m\^{e}me
longueur reliant la piq\^{u}re \`{a} elle-m\^{e}me, et donc
correspondant dans $\mathcal{H}$ \`{a} la g\'{e}od\'{e}sique
$0\alpha $ o\`{u} $\alpha =-1$. Ceci d\'{e}finit la transformation
$\Upsilon $ de $\mathbb{C}$ dans $\mathcal{H}$ avec
\[
\Upsilon (e_1)=-1,\;\Upsilon (e_2)=\infty ,\;\Upsilon (e_3)=0,\;\Upsilon
(e_2+e_3)=\beta .
\]
La translation $t_{e_3}:z\rightarrow z+e_3$ de $\mathbb{C}$
correspond ainsi \`{a} la transformation $A$ de $\mathcal{H}$ avec
$A\circ \Upsilon =\Upsilon \circ t_{e_3}$. L'identification sur
les autres bords avec la m\^{e}me transformation $\Upsilon $ entre
les domaines fondamentaux respectifs et la translation
$t_{e_2}:z\rightarrow z+e_2$ donne $B^{-1}\circ \Upsilon
=\Upsilon \circ t_{e_2}$. Dans les conditions pr\'{e}c\'{e}dentes, lorsque $%
J $ varie sur le segment retenu, $\mathbf{\lambda }_\Lambda $ varie sur
l'arc associ\'{e}, et $\tau $ varie dans son plan complexe. On peut imposer
la contrainte $f(\tau )=1-J(\tau )=\wp _\Lambda ^{\prime 2}(z)$, on obtient
ainsi une relation entre $J$ et $z$ que l'on peut traduire sous forme
diff\'{e}rentielle. Ceci redonne les expressions de Harvey Cohn. Le point
int\'{e}ressant dans cette construction tr\`{e}s g\'{e}n\'{e}rale est que le
domaine fondamental pour le groupe $gp(A,B)$ a une forme tr\`{e}s simple
d\'{e}duite des points $\alpha =-1$, $s=0$, $\beta =$ $\mathbf{\beta }(J)$, $%
p=\infty $. Le nombre $\beta $ \'{e}tant fix\'{e}, ce domaine fondamental
est bien d\'{e}termin\'{e}.\ Il est assez facile de caract\'{e}riser
l'identification de ses bords correspondant \`{a} la donn\'{e}e d'une valeur
$\lambda $, et de comparer ce que donnent des valeurs $\lambda $
diff\'{e}rentes gr\^{a}ce \`{a} une affinit\'{e} ayant pour base le bord de $%
\mathcal{H}$. Pour $J=0$, on trouve seulement $\beta =1$. Mais la
construction que l'on vient de faire pour le bord de $S_4$ est plus
g\'{e}n\'{e}rale et est extensible \`{a} tout $\mathbf{\lambda }_\Lambda \in
S_4$ o\`{u} elle donne des tores perc\'{e}s hyperboliques.

\subsection{Notions attach\'{e}es au tore $\mathcal{T}$}

Pour le tore $\mathcal{T}$ on a
$\mathcal{T}eich(\mathcal{T})=\mathcal{H}$ et
$\mathcal{M}od(\mathcal{T})=\mathcal{H}/PSL(2,\mathbb{Z})$. Chaque
point du domaine fondamental $\mathcal{M}od(\mathcal{T})$
correspond \`{a} une classe d'\'{e}quivalence du tore,
c'est-\`{a}-dire une classe d'isomorphisme de courbe elliptique
\cite{Toubiana} (p.\ 203). On retrouve ainsi la surface modulaire
perc\'{e}e \`{a} l'infini que l'on peut identifier \`{a}
$\mathbb{C}$ en tant que surface de Riemann gr\^{a}ce \`{a}
l'invariant modulaire $J$. On donne dans \cite{Nakahara} (p.\
487-491) une description compl\`{e}te de la m\'{e}trique de
Peterson-Weil dans ce cas. Sur cet exemple existe un op\'{e}rateur
de Laplace-Beltrami $\mathbf{\Delta }$ dont on trouve dans
\cite{Gelfand} (p. 41) la propri\'{e}t\'{e} caract\'{e}ristique
qui est d'\^{e}tre un op\'{e}rateur diff\'{e}rentiel de second
ordre sur $\mathcal{H} $ qui commute avec toutes les
transformations suivantes sur les fonctions $f$ d\'{e}finies sur
le demi plan
\[
T(\psi \left[
\begin{array}{cc}
a & b \\
c & d
\end{array}
\right] )f(z)=f(\frac{az+b}{cz+d}).
\]
Une telle transformation repr\'{e}sente le groupe
$PSL(2,\mathbb{Z})$, voire un groupe plus large comme
$PSL(2,\mathbb{R})$, en tant que groupe d'op\'{e}rateurs sur un
espace fonctionnel dont on peut faire l'analyse harmonique
\cite{Howe}. Ceci est facilit\'{e} par le fait que l'on a, pour
tout $\overline{g}=\psi (\left[
\begin{array}{cc}
a & b \\
c & d
\end{array}
\right] )\in \Gamma _{\mathcal{H}}=PSL(2,\mathbb{Z})$ et pour le laplacien $%
\mathbf{\Delta }$ une relation de commutation $\mathbf{\Delta }T(\overline{g}%
)=T(\overline{g})\mathbf{\Delta }$ qui conduit \`{a} penser \`{a} des
vecteurs propres communs. Pour tout $\tau =\tau _1+i\tau _2\in \mathcal{T}%
eich(\mathcal{T})=\mathcal{H}$, cet op\'{e}rateur est \'{e}crit ici avec un
signe
\[
\mathbf{\Delta }=-\tau _2^2(\frac{\partial ^2}{\partial \tau _1^2}+\frac{%
\partial ^2}{\partial \tau _2^2}).
\]
L'op\'{e}rateur de Casimir qui a des propri\'{e}t\'{e}s comparables au
pr\'{e}cedent est d\'{e}fini dans \cite{Borel} par $\mathcal{C}^{*}=-2%
\mathbf{\Delta }$.\

\subsubsection{Formes automorphes et op\'{e}rateur de Laplace - Beltrami}

Les formes automorphes jouent un r\^{o}le particulier par rapport \`{a} ces
op\'{e}rateurs, notamment parce que les fonctions m\'{e}romorphes sur une
surface de Riemann $\mathcal{H}/\Gamma $ sont donn\'{e}es par les fonctions
m\'{e}romorphes du demi-plan de Poincar\'{e} $\mathcal{H}$ invariantes par $%
\Gamma $, et que l'op\'{e}rateur $\mathbf{\Delta }$ se transporte
lui-m\^{e}me de $\mathcal{H}$ sur les surfaces de Riemann
\cite{RosenbergS}.\ Comme la plupart des \'{e}quations
essentielles de la physique s'expriment en fonction de
l'op\'{e}rateur $\mathbf{\Delta }$ et peuvent concerner des
ph\'{e}nom\`{e}nes relatifs \`{a} des objets mod\'{e}lis\'{e}s par
des surfaces de Riemann (que l'on peut chauffer, \'{e}clairer ou
bien faire vibrer), l'\'{e}tude de cette situation est tr\`{e}s
importante \cite{Hopf} \cite{Safarov}.\ Les formes automorphes se
groupent elles-m\^{e}mes en familles ayant de propri\'{e}t\'{e}s
particuli\`{e}res (formes d'ondes de Maass, formes modulaires
holomorphes, etc...\cite{Bruggeman1}).\ On peut songer \`{a} les
utiliser pour obtenir des valeurs de fonctions particuli\`{e}res,
comme les s\'{e}ries de Fourier ont par exemple \'{e}t\'{e}
utilis\'{e}es par Dirichlet pour d\'{e}montrer un certain nombre
des valeurs de la fonction z\^{e}ta fournies par Euler
\cite{Dirichlet}. Une particularit\'{e} importante est cependant
que dans un certain nombre de cas, le spectre des valeurs propres
de $\mathbf{\Delta }$ poss\`{e}de une partie continue.\ Les choses
sont donc plus compliqu\'{e}es qu'avec un laplacien euclidien
ordinaire.

Plus pr\'{e}cis\'{e}ment \cite{Gelbart}, supposons donn\'{e}e une
fonction automorphe de poids $k$ pour le groupe
$PSL(2,\mathbb{Z})$, avec la condition de d\'{e}finition
\'{e}crite maintenant sous la forme
\[
\forall \gamma \in PSL(2,\mathbb{Z}),\;\forall z\in \mathcal{H}%
,\;f(z)=(cz+d)^{-k}f(\gamma z).
\]
Elle permet la d\'{e}finition d'une fonction sur
$SL(2,\mathbb{R})$ avec l'expression
\[
\Phi _f(\left[
\begin{array}{cc}
a & b \\
c & d
\end{array}
\right] )=(ci+d)^{-k}\;f(\left[
\begin{array}{cc}
a & b \\
c & d
\end{array}
\right] i).
\]
Pour tout $g\in SL(2,\mathbb{R})$ et tout $\gamma \in
SL(2,\mathbb{Z})$ cette
fonction v\'{e}rifie du fait de l'automorphie de $f$%
\[
(C1):\;\;\Phi _f(\gamma g)=\Phi _f(g).
\]
On a aussi pour toute matrice de rotation, et ceci est li\'{e}
\`{a} la structure quotient de $\mathcal{H}\simeq
SL(2,\mathbb{R})/SO(2,\mathbb{R})$
\[
(C2):\;\;\Phi _f(g\left[
\begin{array}{cc}
\cos \theta & -\sin \theta \\
\sin \theta & \cos \theta
\end{array}
\right] )=\exp (ik\theta )\Phi _f(g).
\]
Si l'on suppose que $f$ est holomorphe, on obtient avec le laplacien $%
\mathbf{\Delta }$ de $SL(2,\mathbb{R})$ (dont celui de
$\mathcal{H}$ est l'image) la condition
\[
(C3):\;\;\mathbf{\Delta }\Phi _f=-\frac 14k(k-2)\Phi _f=-\frac{(k-1)^2-1}%
4\Phi _f.
\]
Cette condition se simplifie sous la forme $\mathbf{\Delta }\Phi
_f=-s(s-1)\Phi _f$ si l'on se limite comme dans ce qui pr\'{e}c\`{e}de aux
valeurs $k=2s$ paires. Mais d'autres fonctions propres de $\mathbf{\Delta }$
existent \cite{Howe} \cite{Taylor} \cite{Borel}.\ C'est en affaiblissant
cette condition que Maass a invent\'{e} ses propres formes d'onde \cite
{Maass}. C'est aussi en \'{e}tudiant cette situation que Selberg a
trouv\'{e} sa c\'{e}l\`{e}bre formule g\'{e}n\'{e}ralisant celle de Poisson
cit\'{e}e en 5.2, ainsi que les m\'{e}thodes de Dirichlet pour \'{e}valuer
les sommes de Gauss ou d\'{e}montrer la loi de r\'{e}ciprocit\'{e}
quadratique \cite{Selberg}. Il y a deux conditions suppl\'{e}mentaires
tr\`{e}s importantes
\[
(C4):\;\;\int_{SL(2,\mathbb{R})/SL(2,\mathbb{Z})}\mid \Phi
_f(g)\mid ^2dg<\infty .
\]
Cette premi\`{e}re condition introduit un espace de Hilbert $L^2(SL(2,\mathbb{R}%
)/SL(2,\mathbb{Z}))$ de fonctions de carr\'{e} int\'{e}grable. On
consid\`{e}re aussi
\[
(C5):\;\;\int_{\mathbb{R}/\mathbb{Z}}\mid \Phi _f(\left[
\begin{array}{cc}
1 & x \\
0 & 1
\end{array}
\right] g)\mid ^2dx=0.
\]
Cette seconde condition d\'{e}finit un sous-espace particulier
$L_0^2$ dans le pr\'{e}c\'{e}dent, l'espace des ''formes-pointes''
dans lequel on peut identifier un sous-espace $A_k(\Gamma )$
isomorphe au sous-espace des formes $f\in S_k(\Gamma )$ qui
s'annulent sur les pointes.\ Cet espace est un sous-espace du
$\mathbb{C}$-espace vectoriel $\mathbf{M}_k(\Gamma )$ des
fonctions automorphes.

\subsubsection{Lien avec les repr\'{e}sentations de $SL(2,\mathbb{R})$}

Une cons\'{e}quence importante de ce qui pr\'{e}c\`{e}de est que
l'on peut en d\'{e}duire une repr\'{e}sentation r\'{e}guli\`{e}re
\`{a} droite, unitaire et de dimension infinie de
$SL(2,\mathbb{R})$ dans l'ensemble des op\'{e}rateurs unitaires de
$L_0^2$ :
\[
\text{Pour tous }g,h\in SL(2,\mathbb{R}),\;\;R(g)\Phi _f(h)=\Phi
_f(gh),
\]
o\`{u} pour tout $g\in SL(2,\mathbb{R})$ et $\Phi _f$ bien choisi $\mathbf{%
\Delta }R(g)\Phi _f=R(g)\mathbf{\Delta }\Phi _f$.\ Ceci
d\'{e}compose $R$ avec des sous-espaces invariants pour
$\mathbf{\Delta }$, c'est-\`{a}-dire de fonctions propres de
$\mathbf{\Delta }$, et donc comme somme directe de
r\'{e}pr\'{e}sentations de $SL(2,\mathbb{R})$.\ Celles ci sont au
demeurant toutes connues \cite{Howe}\cite{Taylor}. Ces
repr\'{e}sentations induisent
des repr\'{e}sentations des groupes fuchiens que l'on peut remonter dans $%
SL(2,\mathbb{R})$ par la proposition 1.4.

On trouve aussi (\cite{Terras} chapitre III) des expressions en ''s\'{e}rie
de Fourier'' de $K$-fonctions de Bessel (rempla\c {c}ant les sinusoides)
o\`{u} $x+iy\in \mathcal{H}$
\[
f(x+iy)=\sum_{n\in \mathbb{Z}}a_n\exp (2i\pi
nx)\sqrt{y}K_{it}(2\pi y\mid n\mid ),
\]
\[
K_s(z)=\frac 12\int_0^\infty \exp (-\frac z2(u+\frac 1u)u^{s-1}du,
\]
\[
\frac{t^2+1}4\text{ valeur propre de }\mathbf{\Delta }.
\]
Une conjecture importante due \`{a} Selberg affirme que pour les groupes $%
\Gamma _0(n)$ associ\'{e}s \`{a} la th\'{e}orie de Hecke (\cite{Shimura}
ch3, \cite{Miyake} \S\ 4.5) cette valeur propre qui correspond \`{a} $%
s=(1+it)/2$ est sup\'{e}rieure ou \'{e}gale \`{a} $(1/4)$. On donne dans
\cite{Apostol} un syst\`{e}me fini de g\'{e}n\'{e}rateurs de $\Gamma _0(p)$
pour $p$ premier, ainsi que des formes automorphes associ\'{e}es. Ce que
l'on vient de voir revient \`{a} dire que $t$ est r\'{e}el.

\subsubsection{Lien entre le laplacien d'un tore et la fonction \'{e}ta de
Dedekind}

On consid\`{e}re l'op\'{e}rateur $\mathbf{\Delta }$ sur un tore conforme $%
\mathcal{T}$ d\'{e}fini par un param\`{e}tre complexe $\tau =\tau
_1+i\tau _2\in \mathcal{T}eich(\mathcal{T})=\mathcal{H}$. Pour
repr\'{e}senter ce tore, on utilise \cite{Nakahara} le plan
complexe en $z\in \mathbb{C}$ et des coordonn\'{e}es associ\'{e}es
\`{a} un r\'{e}seau $\Lambda =\mathbb{Z}\xi ^1\oplus \mathbb{Z}\xi
^2$ donn\'{e}es par :
\[
\xi ^1=i\frac{\overline{\tau }z-\tau \overline{z}}{\tau _2},\;\;\xi ^2=i%
\frac{\overline{z}-z}{2\tau _2}.
\]
La m\'{e}trique de $\mathbb{C}$ d\'{e}finit une m\'{e}trique
induite sur le tore \`{a} partir de laquelle on peut calculer la
mesure de Weil-Petersson \`{a} adopter, et avec laquelle le
laplacien du tore peut \^{e}tre \'{e}crit simplement \`{a} partir
du laplacien du demi-plan de Poincar\'{e}. Pour le
laplacien de $\mathcal{H}$, en introduisant pour $i=1,2$ l'op\'{e}rateur $%
\partial _i=\partial /\partial \xi ^i$, on a
\[
\mathbf{\Delta }=-\frac 1{2\tau _2^2}(\mid \tau \mid ^2(\partial _1)^2-2\tau
_1\partial _1\partial _2+(\partial _2)^2).
\]
On trouve alors facilement des fonctions propres de $\mathbf{\Delta }$
v\'{e}rifiant les bonnes conditions au bord du parall\'{e}logramme de la
figure pr\'{e}c\'{e}dente, de fa\c {c}on \`{a} pouvoir en d\'{e}duire des
fonctions sur le tore quotient
\[
\psi _{m,n}(\xi )=\exp (2i\pi (n\xi ^1+m\xi ^2),\;\;m,n\in
\mathbb{Z}.
\]
Les valeurs propres associ\'{e}es sont
\[
\lambda _{m,n}=\frac{2\pi ^2}{\tau _2^2}(m-n\tau )(m-n\overline{\tau })=%
\frac{2\pi ^2}{\tau _2^2}\mid m+n\tau \mid ^2.
\]
Par analogie avec ce que l'on sait pour les op\'{e}rateurs sur les espaces
de dimension finie, le d\'{e}terminant du laplacien $\mathbf{\Delta }$ sur
le tore $\mathcal{T}$ pourrait \^{e}tre envisag\'{e} comme un produit infini
de ces valeurs propres
\[
Det(\mathbf{\Delta })=\prod_{(m,n)\in
\mathbb{Z}^2-(0,0)}\frac{2\pi ^2}{\tau _2^2}\mid m+n\tau \mid ^2.
\]
Mais une telle d\'{e}finition qui fait appara\^{i}tre un produit
infini est insuffisante. On peut cependant la rendre rigoureuse en
introduisant les s\'{e}ries d'Eisenstein $E(\tau ,s)$. On
d\'{e}crit ici la m\'{e}thode pour ce faire telle qu'elle est
donn\'{e}e dans \cite{Nakahara} (p.\ 489). L'\'{e}valuation d'une
telle s\'{e}rie utilise la fonction \^{e}ta de
Dedekind $\eta $. La s\'{e}rie d'Eisenstein est d\'{e}finie pour $\Re%
(s)>1$ par
\[
E(\tau ,s)=\tau _2^s\sum_{(m,n)\in \mathbb{Z}^2\backslash
\{(0,0)}\frac 1{\mid m+\tau n\mid ^{2s}}=\tau _2^sG_{2s}(\tau
),\quad \quad g_2(\tau )=\tau _2^{-1}E(\tau ,1),
\]
elle v\'{e}rifie l'\'{e}quation fonctionnelle
\[
\pi ^{-s}\Gamma (s)E(\tau ,s)=\pi ^{-(1-s)}\Gamma (1-s)E(\tau ,1-s),
\]
et poss\`{e}de une formule limite due \`{a} Kronecker en son p\^{o}le simple
$s=1$ o\`{u} appara\^{i}t $\eta $ et la constante d'Euler $\gamma $%
\[
E(\tau ,s)=\frac \pi {s-1}+2\pi (\gamma -\log (2)-\log (\sqrt{\tau _2}\mid
\eta (\tau )\mid ^2))+O(s-1).
\]
La m\'{e}thode consiste \`{a} utiliser un logarithme et \`{a} n\'{e}gliger
une infinit\'{e} de termes $2\pi ^2$ pour d\'{e}finir seulement le nombre
\[
\frac{\det (\mathbf{\Delta })}{\tau _2}=\exp (-\log \tau _2(1+E(\tau
,0))-E^{\prime }(\tau ,0)).
\]
On utilise alors la formule de Kronecker et des expressions classiques pour
les fonctions $\Gamma $ pour en d\'{e}duire des \'{e}valuations en $s$ des
deux termes \'{e}gaux par l'\'{e}quation fonctionnelle
\[
sE(\tau ,1-s)=-\pi +2\pi s(\gamma -\log 2-\log (\sqrt{\tau _2}\mid \eta
(\tau )\mid ^2)+...),
\]
\[
\pi ^{1-2s}\frac{\Gamma (1+s)}{\Gamma (1-s)}E(\tau ,s)=\pi E(\tau
,0)+(-2(\log \pi +\gamma )E(\tau ,0)+E^{\prime }(\tau ,0)\pi s+...).
\]
La comparaison donne
\[
E(\tau ,0)=-1,\;\;E^{\prime }(\tau ,0)=2(\log 2-\log (\sqrt{\tau _2}\mid
\eta (\tau )\mid ^2),
\]
c'est-\`{a}-dire avec une expression qui pr\'{e}c\`{e}de :
\[
\frac{\det (\mathbf{\Delta })}{\tau _2}=\exp (-E^{\prime }(\tau ,0))=\tau
_2\mid \eta (\tau )\mid ^4.
\]
Cette expression donne une signification particuli\`{e}re \`{a} la fonction
de Dedekind par rapport \`{a} un d\'{e}terminant construit avec
l'op\'{e}rateur de Laplace-Beltrami du tore.\ Elle permet de comprendre
pourquoi cette fonction se d\'{e}compose sous forme d'un produit infini
particulier. En notant ici $q=\exp (2\pi i\tau )=\mathbf{q}^2,$ on retrouve
le produit donn\'{e} dans le commentaire de R.\ Dedekind relatif au fragment
XXVIII de B.\ Riemann \cite{Riemann} (p. 397)
\[
\eta (\tau )^{24}=q\prod_{n\geq 1}(1-q^n)^{24}.
\]
Cette fonction a d\'{e}j\`{a} \'{e}t\'{e} rencontr\'{e}e comme
d\'{e}finissant une forme automorphe de poids 12. Son expression est
li\'{e}e au discriminant $Disc(E_\Lambda )$ de la courbe elliptique $%
E_\Lambda $ attach\'{e}e \`{a} un r\'{e}seau $\Lambda
=\mathbb{Z}\omega _1\oplus \mathbb{Z}\varpi _2$ correspondant
\`{a} un $\mathcal{T}_\Lambda $ pour lequel $\tau =\Im(\omega
_1/\omega _2)$ et
\[
\eta (\tau
)^{24}=g_2^3-27g_3^2=16(e_1-e_2)^2(e_2-e_3)^2(e_3-e_1)^2=Disc(E_\Lambda ),
\]
\[
J=\frac{g_2^3}{g_2^3-27g_3^2}=\frac 1{54}\frac{%
((e_1-e_2)^2+(e_2-e_3)^2+(e_3-e_1)^2)^3}{(e_1-e_2)^2(e_2-e_3)^2(e_3-e_1)^2}.
\]
On a vu avec la branche principale du logarithme et pour une
transformation de $PSL(2,\mathbb{Z})$ d\'{e}finie par une matrice
de $SL(2,\mathbb{Z})$ que l'on avait
\[
\log \eta (\frac{a\tau +b}{c\tau +d})=\log \eta (\tau )+\frac 14\log
(-(c\tau +d)^2)+\pi i\frac{a+d}{12c}-\pi is(d,c).
\]
On peut r\'{e}sumer cette \'{e}galit\'{e} en disant que $\eta $ a une
propri\'{e}t\'{e} d'automorphie de poids $(1/2)$. Mais il faut pour cela
introduire une racine 24$^{i\grave{e}me}$ de l'unit\'{e} permettant
d'\'{e}crire
\[
\eta (\frac{a\tau +b}{c\tau +d})=\chi _\eta (\left[
\begin{array}{cc}
a & b \\
c & d
\end{array}
\right] ).(c\tau +d)^{(1/2)}\eta (\tau ).
\]
On utilise donc d\'{e}sormais la d\'{e}finition de \cite{Kac} (p. 257) plus
satisfaisante que celle que l'on a utilis\'{e}e ant\'{e}rieurement pour les
fonctions automorphes. On dit que $\eta $ est une forme modulaire de poids $%
(1/2)$ et de syst\`{e}me de multiplicateur $P\chi _\eta $, o\`{u}
dans le cas le plus g\'{e}n\'{e}ral $P\chi _\eta :\Gamma
=PSL(2,\mathbb{Z})\rightarrow \mathbb{C}\backslash \{0\}$ est une
fonction telle que pour tout $\gamma \in
\Gamma $, on ait $\mid P\chi _\eta (\gamma )\mid =1$, et si $P:SL(2,\mathbb{Z}%
)\rightarrow PSL(2,\mathbb{Z})$ projection canonique $P\chi _\eta
\circ P=\chi _\eta $. La fonction $g_2$ est quant \`{a} elle une
fonction modulaire de poids $4$ pour un syst\`{e}me de
multiplicateur trivial, d'o\`{u} se
d\'{e}duit avec la modularit\'{e} de poids $12$ du discriminant $%
g_2^3-27g_3^2$ la propri\'{e}t\'{e} de modularit\'{e} de poids $0$ de $J$.
Ces deux derni\`{e}res fonctions peuvent \`{a} leur tour \^{e}tre
consid\'{e}r\'{e}es comme vecteurs propres d'op\'{e}rateurs que l'on peut
expliciter. Le discriminant est ainsi fonction propre des op\'{e}rateurs de
Hecke (\cite{Serre1} p. 168), op\'{e}rateurs qui commutent tous avec le
laplacien ce qui en donne l'analyse spectrale.

On renvoie \`{a} \cite{Rademacher} (ch. 8, 9) pour toutes les
v\'{e}rifications compl\'{e}mentaires des calculs qui pr\'{e}c\`{e}dent. Les
conclusions importantes sont qu'il existe un lien profond entre la fonction
de Dedekind et l'op\'{e}rateur de Laplace-Beltrami du tore $\mathcal{T}$, et
donc aussi celui de $\mathcal{H}$, et que ce dernier est reli\'{e} en
profondeur aux repr\'{e}sentations unitaires dans un espace de Hilbert $%
L_0^2 $ de dimension infinie du plus simple des groupes de Lie non compact $%
SL(2,\mathbb{R})$. On a d'ailleurs vu comment $\mathcal{H}$ admet le quotient $%
PSL(2,\mathbb{R})$ comme groupe d'automorphismes, cette
derni\`{e}re propri\'{e}t\'{e} est donc parfaitement
compr\'{e}hensible. Le passage au tore permet l'apparition d'un
produit infini interpr\'{e}table comme partie
maitrisable du d\'{e}terminant d'un op\'{e}rateur de Laplace-Beltrami $%
\mathbf{\Delta }$. Evidemment une question qui se pose est de
savoir si la technique de r\'{e}surgence de Ecalle \cite{Ecalle}
ne permettrait pas de placer les calculs pr\'{e}c\'{e}dents dans
un cadre plus satisfaisant. Le lien mis en \'{e}vidence dans ce
qui pr\'{e}c\`{e}de entre fonction $\eta $ et un op\'{e}rateur
\cite{Kostant} trouve une application particuli\`{e}re dans la
th\'{e}orie des champs \cite{Bunke}, laissant appara\^{i}tre
l'existence d'une v\'{e}ritable construction fonctorielle pour
cette th\'{e}orie des champs, de port\'{e}e beaucoup plus vaste
que les d\'{e}veloppements classiques qu'ont permis la cyclotomie
et le ''Jugendtraum'' de Kronecker \cite{Landsman}.

\subsubsection{Sommes de Gauss}

La fonction $P\chi _\eta $ peut \^{e}tre \'{e}tudi\'{e}e de fa\c {c}on
directe.\ Elle a un lien profond avec les sommes de Gauss (\cite
{Chandrasekharan} (ch. IX), \cite{Lemmermeyer}) et c'est son comportement
qui permet en r\'{e}alit\'{e} la d\'{e}monstration cyclotomique de la loi de
r\'{e}ciprocit\'{e} quadratique. Au demeurant, c'est dans ce facteur que se
concentrent en exposant d'une puissance les sommes de Dedekind. D'o\`{u}
\'{e}galement le lien entre ces sommes et la r\'{e}ciprocit\'{e}
quadratique. On trouve dans \cite{Knopp} (p. 51) une expression de ce
multiplicateur utilisant le symbole de Jacobi :
\[
\text{si }c\text{ impair }\chi _\eta (\left[
\begin{array}{cc}
a & b \\
c & d
\end{array}
\right] )=\left(
\begin{array}{c}
d \\
\mid c\mid
\end{array}
\right) \exp (\mathcal{F}-3c),
\]
\[
\text{si }c\text{ pair }\chi _\eta (\left[
\begin{array}{cc}
a & b \\
c & d
\end{array}
\right] )=(-1)^{\frac{sgn(c)-1}2\frac{sgn(d)-1}2}\left(
\begin{array}{c}
c \\
\mid d\mid
\end{array}
\right) \exp (\mathcal{F}+3d-3-3cd),
\]
\[
\text{o\`{u} }\mathcal{F}=\frac{\pi i}{12}(a+d)c-bd(c^2-1).
\]
On peut v\'{e}rifier \`{a} partir de l\`{a} que l'on a bien
affaire \`{a} une racine 24$^{i\grave{e}me}$ de l'unit\'{e}. Les
sommes de Gauss sont donn\'{e}es par
\[
G(a,n)=\sum_{k=0}^n\exp (\frac{2i\pi ak^2}n).
\]
Pour $p$ premier impair et $a$ non congru \`{a} $0$ modulo $p$, si $c=1$ ou $%
c=-i$ selon que $p\equiv 1(\mod\,4)$ ou $p\equiv 3(\mod\,4)$,
elles v\'{e}rifient pour la transform\'{e}e de Fourier
discr\`{e}te \cite{Crandall} (p.92) :
\[
\left(
\begin{array}{c}
a \\
p
\end{array}
\right) =\frac{G(a,p)}{\frac 12\sqrt{p}(1+i)(1+i^p)}=\frac c{\sqrt{p}%
}\sum_{k=0}^n\left(
\begin{array}{c}
k \\
p
\end{array}
\right) \exp (\frac{2i\pi ak}p).
\]
On peut en d\'{e}duire l'expression du nombre de classes d'id\'{e}aux d'un
corps quadratique \cite{Hilbert5} (th\'{e}or\`{e}me 114 p.135). Si $p$ et $q$
sont premiers entre eux, la r\'{e}ciprocit\'{e} quadratique se d\'{e}montre
avec $G(p,q)G(q,p)=G(1,pq)$.\ Les sommes de Gauss v\'{e}rifient aussi
l'identit\'{e} de Landsberg-Schaar dont on peut d\'{e}duire la
r\'{e}ciprocit\'{e} quadratique \cite{Moll} p.\ 153 \cite{Binz} :
\[
\frac 1{\sqrt{p}}\sum_{n=0}^{p-1}\exp (\frac{2i\pi n^2q}p)=\frac{\exp (i\pi
/4)}{\sqrt{2q}}\sum_{n=0}^{2q-1}\exp (\frac{-i\pi n^2p}{2q})\;\;(p>0,\;q>0).
\]
Il est remarquable que cette formule soit issue de la trace d'un
op\'{e}rateur d'\'{e}volution longitudinale associ\'{e} \`{a} une
\'{e}quation de Schr\"{o}dinger. On trouve une d\'{e}monstration
dans \cite {Armitage} \`{a} partir d'une \'{e}quation de
Schr\"{o}dinger sur un espace de phase cylindrique, que l'on
modifie pour le rendre toro\"{i}dal, ce qui d'ailleurs
discr\'{e}tise le temps.

\subsubsection{Lien avec la fonction z\^{e}ta de Riemann}

Dans le cadre pr\'{e}sent\'{e} s'introduit \'{e}galement la
fonction z\^{e}ta de Riemann. La s\'{e}rie d'Eisenstein peut
\^{e}tre \'{e}tudi\'{e}e de fa\c {c}on directe en tant que noyau
reproduisant de l'op\'{e}rateur
autoadjoint qui \'{e}tend l'op\'{e}rateur laplacien sur $L^2(\mathcal{M}od(%
\mathcal{T}))$. On a rappel\'{e} dans ce qui pr\'{e}c\`{e}de comment
s'introduisait naturellement cette structure d'espace de Hilbert. Elle
permet la d\'{e}finition d'un autre noyau reproduisant (\cite{Nakahara}
(p.426) ou \cite{Gilkey}), le noyau de chaleur li\'{e} \`{a} l'op\'{e}rateur
elliptique laplacien. Dans le contexte plus g\'{e}n\'{e}ral d'une
vari\'{e}t\'{e} $\mathcal{M}$ plong\'{e}e dans un espace de dimension $D$ ce
noyau est donn\'{e} par l'expression suivante
\[
h(x,y;t)=<x\mid \exp (-t\mathbf{\Delta })\mid y>=\sum_n\exp (-t\lambda
_n)<x\mid n><n\mid y>.
\]
Il v\'{e}rifie l'\'{e}quation de la chaleur qui s'\'{e}crit compte
tenu du choix fait pour le signe du laplacien $\left( \partial
_t+\mathbf{\Delta }\right) h(x,y;t)=0$.\ Il permet de d\'{e}finir
le semi-groupe de la chaleur $\{\exp (-t\mathbf{\Delta });t\geq
0\}$. La transform\'{e}e de Mellin donne une fonction z\^{e}ta
$\zeta (x,y;s)$ qui vaut
\[
\sum_n\frac 1{\mathbf{\Gamma }(s)}\int_0^\infty t^{s-1}\exp (-t\lambda
_n)<x\mid n><n\mid y>dt.
\]
Elle d\'{e}termine la fonction $\zeta _\Delta $ g\'{e}n\'{e}ralis\'{e}e
suivante qui a la forme d'une trace :
\[
\zeta _{\mathbf{\Delta }}(s)=\int_{\mathcal{M}}\zeta (x,x;s)dx=\sum_n\lambda
_n^{-s}.
\]
On \'{e}voque dans \cite{Nakahara} (p.429) et on approfondit dans \cite
{Gilkey} les d\'{e}veloppements de ces calculs vers la d\'{e}finition d'un $%
\eta $-invariant pour certains op\'{e}rateurs elliptiques,
pratiquement une signature d'une forme quadratique d'intersection,
qui d\'{e}bouche sur le th\'{e}or\`{e}me de l'indice
d'Atiyah-Patodi-Singer et des applications importantes en
dimension 4 (\cite{Naber}). Cet invariant est donc une
g\'{e}n\'{e}ralisation de la fonction \^{e}ta de Dedekind pour des
objets plus larges que les surfaces de Riemann \cite{Muller}
\cite{Bismut}. L'application au cas o\`{u} $\mathcal{M}$ est un
tore est facile.\ Elle permet de d\'{e}finir de fa\c {c}on plus
intrins\`{e}que une fonction z\^{e}ta (\cite{Lapidus} p. 229) en
utilisant de fa\c {c}on directe la trace d'une puissance du
laplacien
\[
\zeta _{\mathbf{\Delta }}(s)=tr(\mathbf{\Delta }^{-s})=\sum_{(m,n)\in \mathbb{Z}%
^2\backslash \{(0,0)}\lambda _{m,n}^{-s}.
\]
La conjecture de Riemann \cite{Bombieri} semble correspondre d'une
certaine fa\c {c}on \`{a} ce qui se passe lorsque le tore que l'on
consid\`{e}re est tel que $\tau $ tende vers un nombre entier, ce
qui introduit \`{a} la limite une brisure de sym\'{e}trie
modifiant dramatiquement l'alg\`{e}bre d'op\'{e}rateurs
engendr\'{e}e par $\mathbf{\Delta }$ sur laquelle on travaille. On
peut construire un syst\`{e}me dynamique pour ce cas dont la
fonction de partition soit $\zeta _{\mathbf{\Delta }}$. Il suffit
de suivre la m\'{e}thode de \cite{CohenP} dans son expos\'{e}
tr\`{e}s clair des travaux de \cite{Connes5} permettant de
consid\'{e}rer la fonction $\zeta $ de Riemann elle-m\^{e}me comme
fonction de partition d'un syst\`{e}me dynamique $(A,\sigma _t)$
avec $A$ une $C^{*}$-alg\`{e}bre et $\sigma _t$ un groupe \`{a} un
param\`{e}tre d'automorphismes de $A$. Inversement, le
probl\`{e}me de construire un op\'{e}rateur hermitien qui pourrait
\^{e}tre selon Michael\ Berry \cite{Berry} un hamiltonien
gouvernant un syst\`{e}me m\'{e}canique quantique \`{a}
m\'{e}canique classique sous jacente chaotique et \`{a} temps
irr\'{e}versible correspond \`{a} la conjecture de Hilbert et
Polya \cite{Watkins}. Un tr\`{e}s r\'{e}cent article de Alain
Connes \cite {Connes6} laisse penser que l'hypoth\`{e}se de
Riemann pourrait correspondre comme la formule de Selberg
\cite{Selberg} \`{a} une formule de trace pour un tel hamiltonien
\cite{Fedosov} (theorem 9.5.2 p.\ 307). On peut comparer \`{a} ce
que donnent les th\'{e}ories d'Arakelov de dimensions
sup\'{e}rieures \cite{Lang} (pp. 172-173). Une question importante
para\^{i}t \^{e}tre de bien formaliser dans le contexte
pr\'{e}sent\'{e} la transformation de Mellin, comme une
anti-\'{e}quivalence particuli\`{e}re de cat\'{e}gories, de
vari\'{e}t\'{e}s ab\'{e}liennes vers des alg\`{e}bres
d'op\'{e}rateurs supportant des fonctions $\zeta $. Une autre
piste consiste \`{a} approfondir le lien qui est d\'{e}crit dans
\cite{Chudnovski} entre l'approximation de Ap\'{e}ry de $\zeta
(3)$ et des \'{e}quations de Lam\'{e} que cet article relie
explicitement \`{a} l'\'{e}quation de Markoff.\ Un projet consiste
\`{a} consid\'{e}rer l'op\'{e}rateur $L$ que l'on a introduit
ci-dessus en liaison avec les matrices $A_0$ et $B_0$, \`{a}
consid\'{e}rer des propri\'{e}t\'{e}s d'orthogonalit\'{e}
associ\'{e}es et \`{a} utiliser des m\'{e}thodes analogues \`{a}
celles d\'{e}velopp\'{e}es dans \cite{VanAssche}.

\subsubsection{Gaz de bosons et bruit en $1/f$}

Le formalisme pr\'{e}c\'{e}dent a \'{e}t\'{e} appliqu\'{e} \`{a}
la m\'{e}canique statistique des gaz de bosons.\ Il s'appuie sur
le lien qui a \'{e}t\'{e} \'{e}tudi\'{e} par Ramanujan entre la
fonction \^{e}ta de Dedekind et les partitions d'entiers. Ceci se
mat\'{e}rialise avec la fonction multiplicative $\mathbf{\tau }_R$
de Ramanujan (\cite{Serre1} p. 156, \cite{Chowla} p.\ 57) donnant
\[
\eta (\tau )^{24}=q\prod_{n\geq 1}(1-q^n)^{24}=\sum_{n\geq 1}\mathbf{\tau }%
_R(n)q^n,\;\;\text{o\`{u} }q=\exp (2\pi i\tau )=\mathbf{q}^2=\exp (-h\nu
/kT).
\]
Ceci permet de d\'{e}finir une fonction de partition par mode o\`{u} $p(n)$
nombre de partitions de l'entier $n$ associ\'{e} aussi \`{a} la fonction $%
\eta $ par la formule
\[
Z(q)=q^{\frac 1{24}}\eta (\tau )^{-1}=\prod_{n\geq 1}(\frac
1{1-q^n})=\sum_{n\geq 1}p(n)q^n=\frac{\exp (\pi i\tau /12)}{\eta (\tau )}.
\]
Si $\sigma _k(n)$ d\'{e}signe la somme des puissances $k^{i\grave{e}mes}$
des diviseurs de $n$, on obtient des grandeurs interpr\'{e}tables par
analogie avec la m\'{e}canique statistique
\[
\text{l'\'{e}nergie libre }F=-kT\sum_{n\geq 1}\sigma _{-1}(n)\exp (-nh\nu
/kT),
\]
\[
\text{l'\'{e}nergie interne }E=h\nu \sum_{n\geq 1}\sigma _1(n)\exp (-nh\nu
/kT),
\]
\[
\text{l'entropie }S=k\sum_{n\geq 1}(h\nu /kT\sigma _1(n)+\sigma
_{-1}(n))\exp (-nh\nu /kT).
\]
Sur cette base, les fluctuations d'\'{e}nergie dans un
r\'{e}sonateur \`{a} quartz ont \'{e}t\'{e} \'{e}valu\'{e}es
\cite{Planat4}, faisant appara\^{i}tre un bruit quantique en
$(1/f)$. Au del\`{a} du cas du r\'{e}sonateur \`{a} quartz, il
faudrait creuser le sujet pr\'{e}c\'{e}dent pour montrer comment
donner dans une perspective plus g\'{e}n\'{e}rale une explication
profonde du bruit en $(1/f)$ que l'on rencontre si fr\'{e}quemment
dans la nature. Quelques pistes r\'{e}centes ont commenc\'{e}
\`{a} \^{e}tre explor\'{e}es.\ Elles font le lien avec les sommes
de Ramanujan \cite{Planat9}.

\subsection{Notions attach\'{e}es \`{a} un tore perc\'{e} $\mathcal{T}%
\backslash \{p\}$}

L'espace de Teichm\"{u}ller du tore perc\'{e} est $\mathcal{T}eich(\mathcal{T}%
\backslash \{p\})=\mathcal{H}$.\ On a aussi $\Gamma
_{\mathcal{T}\backslash \{p\}}$ $=GL(2,\mathbb{Z})$. Ce groupe est
not\'{e} $S^{*}L(2,\mathbb{Z})$ pour indiquer qu'il agit dans
$\mathcal{H}$ par transformations conformes et anticonformes. Les
r\'{e}sultats obtenus sur les tores perc\'{e}s paraboliques
permettent de se ramener \`{a} l'action de $SL(2,\mathbb{Z})$ dans
le demi-plan de Poincar\'{e} pour d\'{e}crire au quotient l'espace
des modules $\mathcal{M}od(\mathcal{T}\backslash \{p\})$ gr\^{a}ce
\`{a} la surface modulaire perc\'{e}e. Ces donn\'{e}es
d\'{e}duites de \cite{Nag} (p.\ 153) sont int\'{e}ressantes car
elles ne correspondent pas \`{a} ce qui a \'{e}t\'{e} vu ci-dessus
dans l'\'{e}tude des tores perc\'{e}s conformes paraboliques. On a
donn\'{e}
\[
\mathcal{T}eich(\mathcal{T}\backslash \{p\})\simeq \mathcal{F}(\lambda ,\mu
)=\{(\lambda ,\mu )\mid \lambda >0,\mu >0\},
\]
et l'on a d\'{e}crit la fa\c {c}on dont $\Gamma
_{\mathcal{T}\backslash \{p\}}=GL(2,\mathbb{Z})$ agit dans
$\mathcal{F}(\lambda ,\mu )$.\ Au quotient on identifie bien les
classes d'\'{e}quivalence diff\'{e}omorphe (et donc conforme) sur
le tore perc\'{e}, c'est-\`{a}-dire les modules du tore perc\'{e}.
Ceci correspond au commentaire de la d\'{e}finition 1.6 de \cite
{Schneps} (p.10). Tout se passe comme si $\mathcal{H}$
correspondait \`{a}
un mod\`{e}le topologique de l'espace de Teichm\"{u}ller, et $\mathcal{F}%
(\lambda ,\mu )$ \`{a} un mod\`{e}le g\'{e}om\'{e}trique d\'{e}crit par une
\'{e}quation alg\'{e}brique. Le lien entre ces deux mod\`{e}les a
\'{e}t\'{e} \'{e}tudi\'{e} en d\'{e}tail dans \cite{Keen5}, mais ce travail
devrait \^{e}tre repris \`{a} la lumi\`{e}re des consid\'{e}rations qui
pr\'{e}c\`{e}dent. Il est \'{e}galement tr\`{e}s important de remarquer que
la th\'{e}orie de la r\'{e}duction qui a \'{e}t\'{e} pr\'{e}sent\'{e}e pour
les tores paraboliques, va beaucoup plus loin que ce que donne la seule
action de $\Gamma _{\mathcal{T}\backslash \{p\}}=GL(2,\mathbb{Z})$ sur $%
\mathcal{F}(\lambda ,\mu )$. G\'{e}n\'{e}raliser un tel r\'{e}sultat est
concevable en rentrant dans l'\'{e}tude de la pr\'{e}sentation des groupes
de classes d'applications $\Gamma _{\mathcal{M}}$. Sans aller jusque l\`{a},
on peut indiquer sommairement comment on retrouve les r\'{e}sultats
d\'{e}j\`{a} rencontr\'{e}s au chapitre pr\'{e}c\'{e}dent avec les remarques
formul\'{e}es par \cite{Keen5} (p.\ 203) et issues de \cite{Keen1}. On
traduit ce que dit Linda Keen sous la forme
\[
\pi ^{\prime }(\chi )=\left[
\begin{array}{cc}
0 & -1 \\
1 & 0
\end{array}
\right] \text{ agit sur }\mathcal{F}(\lambda ,\mu )\text{ par }(\lambda ,\mu
)\rightarrow (\frac \mu {\lambda ^2+\mu ^2},\frac \lambda {\lambda ^2+\mu
^2}),
\]
\[
\pi ^{\prime }(\chi ^{\prime })=\left[
\begin{array}{cc}
1 & 1 \\
-1 & 0
\end{array}
\right] \text{ agit sur }\mathcal{F}(\lambda ,\mu )\text{ par }(\lambda ,\mu
)\rightarrow (\frac \mu \lambda ,\frac 1\lambda ).
\]
Ces deux matrices respectivement d'ordre 2 et 3 sont telles que
leurs images par $\psi $ dans $PSL(2,\mathbb{Z})$ engendrent ce
groupe. On peut maintenant consid\'{e}rer que $\pi ^{\prime }$ est
un morphisme d'ab\'{e}lianisation, avec dans le groupe des
automorphismes $Aut(\mathbf{F}_2)$ du groupe libre
\`{a} deux \'{e}l\'{e}ments $\mathbf{F}_2$ engendr\'{e} par $A$ et $B$%
\[
\chi =(B,A^{-1}),\quad \quad \chi ^{\prime }=(AB,B^{-1}).
\]
Il suffit alors de consid\'{e}rer l'action de ces deux automorphismes sur le
triplet
\[
(tr(B^{-1}),tr(A),tr(B^{-1}A^{-1}))=(\frac{1+\lambda ^2+\mu ^2}\mu ,\frac{%
1+\lambda ^2+\mu ^2}\lambda ,\frac{1+\lambda ^2+\mu ^2}{\lambda \mu }),
\]
\[
\chi \text{ donne }(tr(A),tr(B^{-1}),tr(AB^{-1})),\;\;\chi ^{\prime }\text{
donne }(tr(B^{-1}),tr(B^{-1}A^{-1}),tr(A)).
\]
Plus g\'{e}n\'{e}ralement, le groupe $Aut(\mathbf{F}_2)$ agit gr\^{a}ce
\`{a} $\pi ^{\prime }$ sur $\mathcal{F}(\lambda ,\mu )$.\ On a d'ailleurs $%
\pi ^{\prime }(Aut(\mathbf{F}_2))=$ $GL(2,\mathbb{Z})=\Gamma _{\mathcal{T}%
\backslash \{p\}}$. On a d\'{e}velopp\'{e} l'\'{e}tude de cette
situation, expliquant comment le groupe
$\mathbf{T}_3=\mathbf{T}^{*}(\infty ,\infty ,\infty )$
appara\^{i}t ici. Ce groupe a \'{e}t\'{e} mis en \'{e}vidence avec
le triangle curviligne $\mathbf{LMN}$.

\subsection{Interpr\'{e}tation g\'{e}om\'{e}trique de la double
uniformisation}

En comparant les deux cas du tore $\mathcal{T}$ et du tore perc\'{e} $%
\mathcal{T}\backslash \{p\}$, tout se passe comme si on observait
dans l'espace la surface $x^2+y^2+z^2=xyz$ et que l'on
repr\'{e}sente cette configuration dans $\mathcal{H}$. Les formes
quadratiques donnent tout l'espace $\mathbb{R}^3$, puis
projectivement $\mathcal{H},$ et on sait faire agir
$PSL(2,\mathbb{Z})$ sur ces espaces. Dans $\mathbb{R}^3$ on
visualise cette surface, et on la repr\'{e}sente projectivement
par $\mathcal{H}$.\ On trouve ainsi une signification \`{a}
l'action de $GL(2,\mathbb{Z})$ sur cette surface. La r\'{e}duction
porte ainsi une information beaucoup plus profonde que la simple
inclusion d'un objet topologique dans un autre.\ Elle traduit la
fa\c {c}on dont un objet g\'{e}om\'{e}trique est contenu dans un
autre. On trouve ainsi une signification comparable \`{a} ce qui
est expliqu\'{e} dans l'article de B.\ Mazur \cite{Mazur1} sur les
doubles rev\^{e}tements conformes. ''C'est la conjonction de deux
uniformisations (l'une en l'occurrence euclidienne et l'autre
hyperbolique de type arithm\'{e}tique, c'est-\`{a}-dire
p\'{e}riodique par rapport \`{a} un groupe de congruence) qui
cr\'{e}e une structure exceptionnellement riche sur les courbes
elliptiques et entraine des implications profondes pour des
questions arithm\'{e}tiques (en fait \cite{Knapp} (ch.XII) la
conjecture de Shimura Taniyama Weil d\'{e}montr\'{e}e par A.\
Wiles \cite{Wiles}: une courbe elliptique sur les nombres
rationnels poss\`{e}de un fonction z\^{e}ta provenant de formes
modulaires de poids 2).'' Ce que l'on vient de
d\'{e}crire entre le tore $\mathcal{T}$ et le tore perc\'{e} $\mathcal{T}%
\backslash \{p\}$ donne deux uniformisations possibles pour le tore
perc\'{e} conforme.

\section{Approche par le chaos quantique}

Comme on vient d'\'{e}tendre la d\'{e}finition du laplacien \`{a} des tores
perc\'{e}s, une question qui se pose est de savoir s'il existe une
interpr\'{e}tation m\'{e}canique correspondant \`{a} la th\'{e}orie de
Markoff classique, ou aux g\'{e}n\'{e}ralisations qu'on en a donn\'{e}es. Il
faut comprendre si dans ce nouveau contexte le spectre de Markoff pourrait
\^{e}tre le spectre d'un op\'{e}rateur \`{a} construire sur le tore
perc\'{e}. L'id\'{e}e suivie par l'auteur pour \'{e}tudier cette question a
consist\'{e} \`{a} examiner ce que donne la th\'{e}orie du chaos quantique
sur diff\'{e}rents tores perc\'{e}s non conform\'{e}ment \'{e}quivalents
puis \`{a} consid\'{e}rer la m\^{e}me question sur des surfaces de Riemann,
comme le fait \cite{Gutzwiller}, enfin sur des espaces plus complexes.

Pour toute surface de Riemann $\mathcal{M}$ d\'{e}finie par un groupe
fuchsien on a introduit de fa\c {c}on naturelle la g\'{e}om\'{e}trie
symplectique en consid\'{e}rant le premier groupe d'homologie $H_1(\mathcal{M%
},\mathbb{Z})$ et le nombre d'intersections (\cite{Waldschmidt2}
p. 105). Le formalisme de la m\'{e}canique hamiltonienne et de la
quantification s'introduit \`{a} partir de l\`{a} (\cite{MacLane1}
\cite{Gotay} \cite {Fedosov} \cite{DodsonCTJ} \cite{Dodson1}
\cite{Fischer} \cite{Nelson} \cite {Cassa} \cite{Takhadjian}),
avec encore beaucoup de choses \`{a} \'{e}claircir \cite{MacKay}.
Ceci permet de mod\'{e}liser certains probl\`{e}mes de
m\'{e}canique au moyen de telles surfaces de Riemann. On notera
qu'en m\'{e}canique des solides ordinaires le formalisme
hamiltonien se met en place avec un espace de phases de dimension
fini.\ Les choses deviennent un peu plus compliqu\'{e}es d\`{e}s
que l'on aborde des probl\`{e}mes d'hydrodynamique car l'espace
des phases devient de dimension infinie, obligeant \`{a} avoir
recours \`{a} des outils comme les espaces de Hilbert. Mais
m\^{e}me \`{a} ce prix d'autres domaines de la physique ne
rentrent pas facilement dans ce formalisme dont l'un des grands
int\'{e}r\^{e}ts a \'{e}t\'{e} de montrer l'importance de la
topologie pour la physique (voir par exemple \cite{Casetti}
\cite{Mineev}).

\subsection{Quelques exemples\ }

On \'{e}voque ici trois exemples pour illustrer les limites du formalisme
hamiltonien et les voies de son extension.

$\bullet $ La m\'{e}thode du ''scattering inverse'' est utilis\'{e}e pour
int\'{e}grer des \'{e}quations diff\'{e}rentielles non-lin\'{e}aire. Son
interpr\'{e}tation hamiltonienne est due \`{a} L.\ D.\ Fadeev \cite{Fadeev}%
.\ Elle s'applique \`{a} des \'{e}quations tr\`{e}s importantes de
la Physique (Sine-Gordon, Lam\'{e} c'est-\`{a}-dire
Schr\"{o}dinger p\'{e}riodique \`{a} une dimension \cite{Feldman},
Schr\"{o}dinger non lin\'{e}aire, Korteweg-deVries, etc.)
admettant une pr\'{e}sentation hamiltonienne avec des \'{e}tats
dans un espace de Hilbert.\ Certains solitons entrent dans le
domaine couvert par ce d\'{e}veloppement \cite{Remoissenet} qui
d\'{e}passe largement le cadre des seules surfaces de Riemann. On
renvoie pour approfondir le th\`{e}me des solitons \`{a}
\cite{Gesztesy}. Mais les surfaces de Riemann interviennent aussi
dans ce cadre \cite{Dubrovin}.

$\bullet $ Les \'{e}quations de Maxwell classique (dont l'auteur voudrait
formaliser le lien avec la th\'{e}orie de Hodge) r\'{e}gissent la
propagation des ondes et de la lumi\`{e}re.\ Elles n'entrent pas dans le
formalisme hamiltonien sauf \`{a} \'{e}tendre \`{a} une dimension infinie la
dimension de l'espace des phases.\ Elles d\'{e}crivent en effet des
variations de champ \'{e}lectrique et magn\'{e}tique en tout point de
l'espace. La transformation de ces champs transporte de l'\'{e}nergie et
donne en l'absence de charge et de courant une \'{e}quation d'onde qui
d\'{e}crit la propagation de l'onde qui transporte cette \'{e}nergie.
L'\'{e}quation de Schr\"{o}dinger appliqu\'{e}e \`{a} une fonction d'onde
repr\'{e}sentant un photon isol\'{e} donne exactement les \'{e}quations de
Maxwell. Avec un \'{e}lectron, elle donne l'\'{e}quation de Dirac \`{a} la
base comme ces derni\`{e}res de l'\'{e}lectrodynamique quantique \cite
{Penrose}. Le d\'{e}veloppement d'un cadre global commun pour les lois de la
physique que l'on vient d'\'{e}voquer passe donc bien par l'introduction
d'un cadre hilbertien et d'une analyse dans celui-ci de l'\'{e}quation de
Schr\"{o}dinger.

$\bullet $ La th\'{e}orie quantique des champs a \'{e}t\'{e} introduite
\`{a} la suite des travaux d'Einstein sur l'invariance par les
transformations de Lorentz des \'{e}quations de l'\'{e}lectromagn\'{e}tisme
de Maxwell
\[
\nabla (E+iB)=q+ig,\;\frac \partial {\partial t}(E+iB)+i\nabla \times
(E+iB)=j_e+ij_m.
\]
Le souci de rendre ces deux \'{e}quations invariantes par d'autres
transformations $(E+iB)\rightarrow \exp (i\phi )(E+iB)$ a conduit
\`{a} la th\'{e}orie du champ conforme et \`{a} la tentative
d'unifier la gravit\'{e} aux autres forces de la nature par la
th\'{e}orie des cordes. Cette d\'{e}marche a eu un temps fort avec
l'article \cite{Polyakov}.\ En r\'{e}alit\'{e}, cette th\'{e}orie
ne semble avoir qu'un int\'{e}r\^{e}t restreint car il a
\'{e}t\'{e} constat\'{e} que son domaine d'application reste
limit\'{e}. Il est cependant \'{e}tabli que cette th\'{e}orie
admet une pr\'{e}sentation hamiltonienne avec des \'{e}tats dans
un espace de Hilbert, une $C^{*}$-alg\`{e}bre d'op\'{e}rateurs et
un groupe de sym\'{e}tries de jauge, c'est-\`{a}-dire la
g\'{e}om\'{e}trie non commutative d'Alain Connes \cite{Connes}
\cite{Waldschmidt2} (p.548).\ Cette derni\`{e}re devrait permettre
d'\'{e}tendre fonctoriellement le projet sans doute trop restreint
de la th\'{e}orie du champs conforme \cite{Witten} \cite
{Landsman}.\ Une quantification dans cette th\'{e}orie se
d\'{e}duit des remarques qui pr\'{e}c\`{e}dent, dont on trouve les
\'{e}l\'{e}ments essentiels dans \cite{Friedan} \cite{Gawedski}
\cite{Vafa} \cite{Puta} \cite {Nakahara} \cite{Grandati}
\cite{Bott}.

\subsection{L'int\'{e}grale de pas de Feynman\ }

On trouve un expos\'{e} g\'{e}n\'{e}rique de cette question en
coordonn\'{e}es les plus g\'{e}n\'{e}rales dans \cite{Grosche} (p.\ 67-91)
et \cite{Golubeva}.\ Sur une vari\'{e}t\'{e} $\mathcal{M}$ (par exemple une
surface de Riemann compacte) contenue dans un espace de dimension $D$ et
munie d'une m\'{e}trique $ds^2=g_{ab}(\mathbf{q})dq^adq^b$ donn\'{e}e avec
des param\`{e}tres locaux de position $\mathbf{q}=(q^1,...,q^D)$, on peut
consid\'{e}rer l'espace des fonctions de carr\'{e} int\'{e}grable $L^2(%
\mathcal{M})$ pour le produit scalaire
\[
<f_1,f_2>=\int_{\mathcal{M}}\sqrt{\det (g_{ab})}f_1(\mathbf{q})\overline{f_2(%
\mathbf{q})}d\mathbf{q,}
\]
et l'op\'{e}rateur de Laplace Beltrami, appel\'{e} laplacien, o\`{u} $%
(g^{ab})$ inverse de $(g_{ab})$ :
\[
\mathbf{\Delta }=g^{ab}\partial _a\partial _b+(g^{ab}\Gamma
_a+g_a^{ab})\partial _b,\;\;\text{o\`{u} }\Gamma _a=\frac{\partial \log
\sqrt{\det (g_{ab})}}{\partial q^a}.
\]
Les param\`{e}tres d'impulsion, op\'{e}rateurs hermitiens adapt\'{e}s au
produit scalaire introduit, ont une forme particuli\`{e}re :
\[
p_{-a}=-i\hbar (\frac \partial {\partial q^a}+\frac{\Gamma _a}2).
\]
L'op\'{e}rateur associ\'{e} \`{a} l'\'{e}nergie est d\'{e}fini \`{a} partir
de la variable temps :
\[
i\hbar \frac \partial {\partial t}.
\]
L'\'{e}quation de Schr\"{o}dinger (\cite{Ngo} p.\ 45) d\'{e}pendant du temps
pour une particule de masse $m$ se d\'{e}pla\c {c}ant dans un champ
potentiel $V(\mathbf{q})$ ind\'{e}pendant du temps sur la vari\'{e}t\'{e} $%
\mathcal{M}$ s'\'{e}crit alors avec un hamiltonien
\[
i\hbar \frac \partial {\partial t}\psi (\mathbf{q},t)=\left[ -\frac{\hbar ^2%
}{2m}\mathbf{\Delta }+V(\mathbf{q})\right] \psi (\mathbf{q},t)=H\psi (%
\mathbf{q},t).
\]
Dans certains cas elle poss\`{e}de une unique solution g\'{e}n\'{e}rale (%
\cite{Waldschmidt2} p. 549) donn\'{e}e par une int\'{e}grale de Feynman
construite \`{a} partir d'une amplitude de probabilit\'{e} $K(\mathbf{q}",t";%
\mathbf{q}^{\prime },t^{\prime })$ qu'une particule quitte sa position
initiale pour atteindre sa position finale, et gr\^{a}ce \`{a} laquelle on
peut d\'{e}crire l'\'{e}volution dans le temps de la fonction d'onde $\psi $
duale de la particule que l'on consid\`{e}re
\[
\psi (\mathbf{q"},t")=\int_{\mathbb{R}^D}\sqrt{g(\mathbf{q}^{\prime })}K(%
\mathbf{q}",t";\mathbf{q}^{\prime },t^{\prime })\psi (\mathbf{q}^{\prime
},t^{\prime })d\mathbf{q}^{\prime }.
\]
M\^{e}me si le potentiel $V(\mathbf{q})$ est nul ce calcul peut \^{e}tre
fait \cite{Kleinert} en s'appuyant sur les g\'{e}od\'{e}siques de $\mathcal{M%
}$. En supposant le syst\`{e}me global stable et isol\'{e},
c'est-\`{a}-dire dans un \'{e}tat stationnaire, l'\'{e}nergie
totale du syst\`{e}me est une
constante qui est une valeur propre $E$ de $H$ avec laquelle on a $\psi (%
\mathbf{q},t)=\psi (\mathbf{q},0)\exp (-iEt/\hbar )$ et
\[
E\psi (\mathbf{q},0)=\left[ -\frac{\hbar ^2}{2m}\mathbf{\Delta }+V(\mathbf{q}%
)\right] \psi (\mathbf{q},0).
\]

\subsection{Cas de l'oscillateur harmonique quantique}

On trouve une \'{e}quation comparable dans le cas de l'oscillateur
harmonique quantique \`{a} une seule dimension $D=1$, o\`{u} $V(\mathbf{q}%
)=(1/2)m\omega ^2\mathbf{q}^2$ et $\mathbf{\Delta =}\partial
^2/\partial \mathbf{q}^2$, et avec les polyn\^{o}mes de Hermite
(\cite{Perrine9} pp. 295-296)
les seules \'{e}nergies totales possibles $%
E_n=E_0+n\hbar \omega $ et le vecteur ket $\mid n>=\psi
_n(\mathbf{q},0)$ associ\'{e} \`{a} chacune d'elle. Ceci donne
aussi la forme hermitienne \`{a} consid\'{e}rer pour laquelle ces
vecteurs ket forment une base orthonorm\'{e}e de l'espace de
Hilbert des fonctions associ\'{e}es. Sur cet espace s'introduisent
les trois op\'{e}rateurs auto-adjoints qui correspondent aux
observables de position, d'impulsion et d'\'{e}nergie
utilis\'{e}es :
\[
\mathbf{Q}=\sqrt{\frac{m\omega }\hbar }\mathbf{q},\;\mathbf{P}=\frac{\mathbf{%
p}}{\sqrt{m\omega \hbar }},\;\;[\mathbf{P},\mathbf{Q}]=i\neq 0,
\]
\[
H=\hbar \omega (\mathbf{AA}^{*}-\frac 12)\text{ o\`{u} }\mathbf{A=}\frac 1{%
\sqrt{2}}(\mathbf{Q}+i\mathbf{P})\neq \mathbf{A}^{*}.
\]
On a \'{e}galement sur cet espace un op\'{e}rateur unitaire naturel (\cite
{Mackey} p.75) qui s'\'{e}crit
\[
(\frac H{\hbar \omega }+\frac 12+i)(\frac H{\hbar \omega }+\frac 12-i)^{-1},
\]
il est utilisable pour \'{e}tudier l'hypoth\`{e}se de Riemann associ\'{e}e
selon les m\'{e}thodes de \cite{Connes6} et \cite{CohenP}. On peut enfin
d\'{e}velopper (\cite{Perrine9} p.296) une approche statistique de la
distribution des \'{e}tats d'\'{e}nergie $E_n$ lorsque cet oscillateur de
pulsation $\omega =2\pi \nu $ est en contact avec un milieu ext\'{e}rieur
beaucoup plus grand que lui et agissant comme thermostat de temp\'{e}rature
constante $T$. Les \'{e}tats d'\'{e}nergie sont quantifi\'{e}s en $\hbar
\omega =h\nu $, o\`{u} $h$ est la constante de Planck et $\hbar =(h/2\pi )$.

\subsection{Le chaos quantique et les g\'{e}od\'{e}siques}

Ce que l'on vient de r\'{e}sumer pour l'oscillateur harmonique se
g\'{e}n\'{e}ralise en la formulation hamiltonienne que l'on a donn\'{e}e
pour toute vari\'{e}t\'{e}, et donc toute surface de Riemann $\mathcal{M}$.\
Ceci condense de l'information sur sa g\'{e}om\'{e}trie et conduit
naturellement \`{a} une probl\'{e}matique de quantification en
consid\'{e}rant le spectre des valeurs propres associ\'{e} \`{a}
l'op\'{e}rateur apparaissant dans l'\'{e}quation de Schr\"{o}dinger. Une
relation peut \^{e}tre \'{e}tablie avec les orbites g\'{e}od\'{e}siques
p\'{e}riodiques de $\mathcal{M}$ gr\^{a}ce \`{a} la formule de trace issue
des travaux de Selberg \cite{Gutzwiller1} \cite{Watkins}. C'est l'un des
d\'{e}veloppements r\'{e}cents de la th\'{e}orie du chaos quantique. Dans
\cite{Colin} (p.\ 59) on indique que pour d\'{e}crire les
g\'{e}od\'{e}siques de $\mathcal{M}$ on peut consid\'{e}rer un hamiltonien
pseudo-diff\'{e}rentiel $\hbar \sqrt{-\mathbf{\Delta }}$ et se ramener \`{a}
l'\'{e}quation de Schr\"{o}dinger
\[
i\hbar \frac \partial {\partial t}\psi =\hbar \sqrt{-\mathbf{\Delta }}\psi .
\]
Une simplification par $\hbar $ se produit dans cette \'{e}quation et sa
solution est donn\'{e}e par le groupe \`{a} un param\`{e}tre $U(t)=\exp (-t%
\sqrt{-\mathbf{\Delta }})$. Cette remarque conduit \`{a} se poser la
question de la nature g\'{e}om\'{e}trique profonde de la constante de Planck
(\cite{Mendes}, \cite{Fedosov}: ''la constante de Planck pourrait ne prendre
que des valeurs telles que l'indice topologique soit un nombre entier.'').
Dans l'approche statistique associ\'{e}e la fonction de partition quantique
associ\'{e}e est
\[
Z(t)=tr(U(t))=\sum_{n=1}^\infty \exp (-i\mu _nt),
\]
o\`{u} les $\mu _n$ correspondent aux solutions stationnaires de forme $\exp
(-i\mu _nt)\psi _n(\mathbf{q},0)$ avec
\[
\mathbf{\Delta }\psi _n(\mathbf{q},0)=-\lambda _n\psi _n(\mathbf{q}%
,0),\;\;\mu _n=\sqrt{\lambda _n},\;\;\lambda _1=0<\lambda _2\leq ...\leq
\lambda _n\leq ...
\]
Elles se d\'{e}duisent des valeurs propres $\lambda _n$ de l'op\'{e}rateur
de Laplace associ\'{e} \`{a} la vari\'{e}t\'{e} $\mathcal{M}$. Il existe
toute une litt\'{e}rature sur ce sujet, sachant que cet op\'{e}rateur est la
plupart du temps d\'{e}fini comme l'oppos\'{e} de celui que l'on vient
d'utiliser (\cite{RosenbergS} \cite{Safarov} articles de I.\ Chavel pp.
30-75 et M. Shubin pp. 226-283).\

\subsection{Application \`{a} la th\'{e}orie de Markoff}

Lorsque la vari\'{e}t\'{e} $\mathcal{M}$ n'est pas compacte, le
spectre n'a pas de raison d'\^{e}tre discret et peut donc contenir
une partie cantorienne ou une partie continue. On ne voit plus
alors appara\^{i}tre l'\'{e}quivalent de la constante de Planck
comme dans le cas de l'oscillateur harmonique quantique. On a vu
ci-dessus comment l'identit\'{e} de Landsberg-Schaar sur les
sommes de Gauss est issue de la trace d'un op\'{e}rateur
d'\'{e}volution longitudinale associ\'{e} \`{a} une \'{e}quation
de Schr\"{o}dinger \cite{Armitage}. On a indiqu\'{e} comment \`{a}
partir d'un espace de phase cylindrique rendu toro\"{i}dal on
retrouvait la r\'{e}ciprocit\'{e} quadratique intimement li\'{e}e
\`{a} la fonction \^{e}ta de Dedekind, elle m\^{e}me li\'{e}e au
tore. Mais on a vu aussi que cette approche discr\'{e}tise le
temps et fait dispara\^{i}tre l'\'{e}quation de Schr\"{o}dinger
avec un param\`{e}tre temporel continu. Ceci semble indiquer que
pour aller plus loin dans la g\'{e}n\'{e}ralit\'{e} du formalisme
de l'\'{e}quation de Schr\"{o}dinger, il faudrait consid\'{e}rer
les temps comme les autres param\`{e}tres observables.

La question qu'on se pose alors est de savoir si ce formalisme pourrait
interpr\'{e}ter le spectre de Markoff lorsque l'espace des phases $\mathcal{M%
}$ est le tore perc\'{e} parabolique mis en \'{e}vidence par Harvey Cohn
dans \cite{Cohn2}. Il faudrait pour progresser dans cette voie donner une
bonne \'{e}quation de Schr\"{o}dinger \`{a} consid\'{e}rer.\ On devrait
s'assurer que l'on n'est pas alors dans un cas de nombre fini de ses
solutions pour une telle \'{e}quation, le minimum intervenant dans la
th\'{e}orie de Markoff pouvant alors correspondre \`{a} une minimisation de
l'\'{e}nergie. Ce programme de travail de l'auteur n'en est qu'\`{a} ses
d\'{e}buts, de sorte que peu de r\'{e}sultats peuvent encore \^{e}tre
donn\'{e}s quant \`{a} l'approche propos\'{e}e. Une piste pour progresser
dans cette voie pourrait \^{e}tre d'expliciter la formulation hamiltonienne
quantique associ\'{e}e aux oscillateurs \`{a} v\'{e}rouillage de phase de
Michel Planat \cite{Planat}. Il semble bien qu'ils correspondent \`{a} un
espace de phase torique perc\'{e}, constituant donc un mod\`{e}le plus
sophistiqu\'{e} que l'oscillateur quantique \`{a} une dimension. La question
de la d\'{e}g\'{e}n\'{e}rescence discr\`{e}te \'{e}ventuelle de
l'\'{e}quation de Schr\"{o}dinger dans ce cas est un probl\`{e}me
int\'{e}ressant.

\section{Quelques th\`{e}mes de r\'{e}flexion connexes}

\subsection{Liens avec les fibr\'{e}s vectoriels et la $K$-th\'{e}orie}

Pour toute surface de Riemann $\mathcal{M}$ le th\'{e}or\`{e}me
d'uniformisation de Poincar\'{e}, Koebe et Klein a donn\'{e} des domaines $%
\mathcal{U}\subset \mathcal{S}^2$ et des transformations holomorphes
injectives $t$ de $\mathcal{U}$ dans $\mathcal{M}$ telles qu'en tout point $%
x\in \mathcal{U}$, $t$ uniformise localement $\mathcal{M}$ au
point $t(x)$, cartographiant le voisinage de ce point dans
$\mathcal{M}$. Aujourd'hui, ce r\'{e}sultat a finalement
\'{e}t\'{e} pris comme d\'{e}finition des surfaces de Riemann par
H.\ Weyl \cite{Weyl}. Les groupes fuchsiens permettent de traiter
alg\'{e}briquement certaines de ces surfaces, et l'on reconstruit
dans l'alg\`{e}bre des fonctions automorphes associ\'{e}e les
invariants caract\'{e}ristiques. On trouve dans \cite{Milne1} (p.\
53-54) l'id\'{e}e que les facteurs d'automorphie correspondent
\`{a} des cocycles (la cohomologie est l\`{a}!) et cet auteur
montre qu'ils sont en correspondance bijective avec des fibr\'{e}s
vectoriels sur la surface d'une fa\c {c}on qui interpr\`{e}te les
fonctions automorphes de poids $2k$ comme des sections d'un
fibr\'{e} $L_k^{*}$ sur un compactifi\'{e} $\mathcal{M}^c$
d\'{e}termin\'{e} par un facteur d'automorphie canonique (la
$K$-th\'{e}orie appara\^{i}t!). Cette remarque est tr\`{e}s
importante pour comprendre pourquoi la th\'{e}orie de Markoff
d\'{e}termine des fibr\'{e}s exceptionnels et des h\'{e}lices du
plan projectif $P_2(\mathbb{C})$ (voir \cite{Grothendieck1},
\cite{Drezet}, \cite{Rudakov}, \cite{Nogin}, \cite{Nogin1}, \cite
{Gorodentsev}, \cite{Gorodentsev1}, \cite{Drezet1},
\cite{Drezet2}). Il serait d'un grand int\'{e}r\^{e}t d'associer
d'autres fibr\'{e}s et h\'{e}lices aux \'{e}quations
$M^{\varepsilon _{1,}\varepsilon _2}(a,\partial K,u)$ mises en
\'{e}vidence dans le pr\'{e}sent ouvrage, ne serait-ce que pour
mieux comprendre la structure des fibr\'{e}s vectoriels sur
diff\'{e}rents types de vari\'{e}t\'{e}s et les classifier
\cite{Van de Ven} \cite{LePotier} \cite{Sen} \cite{Klyachko}
\cite{Klyachko1} \cite {Ionescu} \cite{Baez1}. On conjecture que
ceci est possible.\ Cette recherche s'inscrit dans la grande
tradition des analogies entre corps de nombres et corps de
fonctions ch\`{e}re \`{a} Andr\'{e} Weil \cite{Weil0}, qui a
conduit aux sch\'{e}mas d'Alexandre Grothendieck \cite{Silverman2}
(A.9), puis \`{a} la cohomologie \'{e}tale pour
g\'{e}n\'{e}raliser la th\'{e}orie de Galois \cite{Milne2}, \`{a}
la g\'{e}om\'{e}trie d'Arakelov \cite{Soule2}, enfin \`{a} la
cohomologie motivique \cite{Levine} et \`{a} la r\'{e}solution de
la conjecture de Langlands sur les corps de fonctions
\cite{Laumon}. Cette approche a permis la r\'{e}solution de
l'hypoth\`{e}se de Riemann pour les courbes de genre quelconque
sur un corps fini par Andr\'{e} Weil \cite{Weil2}, puis pour
toutes les vari\'{e}t\'{e}s sur un corps fini par Pierre Deligne
\cite{Deligne} et \`{a} la r\'{e}solution de la conjecture de
Langlands sur les corps de fonctions \cite{Laumon} \cite
{Soergel}. Un r\'{e}sum\'{e} rapide de la d\'{e}marche historique
se trouve dans \cite{Cartier2} ou \cite{Milne} (p.\ 97-100).\ Pour
d'autres perspectives on renvoie \`{a} \cite{Goss} \cite{Buium}.

Une cons\'{e}quence du projet de recherche que l'on vient d'\'{e}voquer pour
les fibr\'{e}s est de donner une interpr\'{e}tation ''automorphe''
g\'{e}n\'{e}rale des $K$-groupes $K_i(R)$ de la th\'{e}orie de D.\ Quillen.\
L'importance de cette question est clairement mise en lumi\`{e}re dans \cite
{Weibel1} (p.17-18). Quant \`{a} la d\'{e}finition classique des groupes $%
K_i(R)$, on la trouve dans \cite{Rosenberg}, ou plus directement
dans \cite {Arlettaz}. Sur ceux-ci se transposent des
r\'{e}sultats de la th\'{e}orie alg\'{e}brique des nombres comme
le th\'{e}or\`{e}me des unit\'{e}s de Dirichlet \cite{Rosenberg}
(p.\ 288). Dans ces r\'{e}sultats, $R$ d\'{e}signe un anneau
d'entiers d'un corps $F$ extension finie de $\mathbb{Q}$ et il y a
un lien profond entre ces $K$-groupes et la fonction $\zeta _F$ du
corps $F$ \cite{Lichtenbaum} \cite{Weibel1} \cite{Bump}
\cite{Benson}. Il est aussi connu que les fonctions z\^{e}ta sont
li\'{e}es aux sommes de Dedekind et \`{a} la g\'{e}om\'{e}trie
torique qui a \'{e}t\'{e} d\'{e}velopp\'{e}e pour faire un lien
entre la th\'{e}orie des ensembles convexes dans un r\'{e}seau et
la g\'{e}om\'{e}trie alg\'{e}brique \cite {Ziegler} (p. 224)
\cite{Danilov} \cite{Pommersheim2}. Enfin le lien entre la
g\'{e}om\'{e}trie torique et les fonctions automorphes est
clairement explicite dans des travaux tels que \cite{Borisov}
\cite{Cox1} \cite{Cox2}.\ On trouve des d\'{e}veloppements plus
directs sur le lien entre les fonctions z\^{e}ta (ou $L$) et les
sommes de Dedekind dans des travaux tels que \cite{Stevens}
\cite{Sczech}.

\subsection{Lien avec les fonctions z\^{e}ta}

L'apparition des fonctions z\^{e}ta peut se comprendre avec une
remarque faite lors de l'\'{e}vocation des fonctions th\^{e}ta.
Les espaces de fonctions automorphes de poids successifs se
d\'{e}duisant par des exponentiations de groupes, on peut faire
appara\^{i}tre naturellement (\cite
{Dieudonne3} p. 297) les nombres de Bernouilli (ici $\mathbf{b}%
_n=(-1)^{n+1}b_{2n}>0)$ avec une ''demi-formule de Poisson'' qui concerne
des exponentielles successives d'un op\'{e}rateur $d$, et donne la fonction
de partition $Z$ de l'oscillateur harmonique dans la th\'{e}orie de
Boltzmann et Planck en rempla\c {c}ant $d$ par $-(h\nu /kT)$
\[
-\sum_{k\geq 1}\exp (kd)=\frac{\exp (d)}{\exp (d)-1}=\frac 1{1-\exp
(-d)}=d^{-1}+\frac 12+\sum_{n\geq 1}(-1)^{n+1}\mathbf{b}_n\frac{d^{2n-1}}{%
(2n)!}.
\]
Appliqu\'{e}e \`{a} une fonction analytique, une telle formule
donne la formule classique d'Euler et Mac-Laurin
(\cite{Dieudonne3} p.\ 302 \cite {Kac1} ch. 25). Cette formule est
applicable aux structures car elle est de nature fonctorielle
\cite{Gelfand1}. On trouve dans \cite{Tits} une traduction pour
les alg\`{e}bres de Kac-Moody. On sait aussi passer d'une
alg\`{e}bre de Lie \`{a} un groupe de Lie par l'exponentielle qui
transforme des sommes en produits, des traces en d\'{e}terminants
(\cite{Arnold00} p. 116-119).\ On trouve dans \cite{Postnikov}
(p.175) les cons\'{e}quences pour les cat\'{e}gories
correspondantes notamment les \'{e}quivalences de cat\'{e}gories
entre groupes de Lie et alg\`{e}bres de Lie, et dans \cite
{Postnikov} (p. 97) comment l'alg\`{e}bre enveloppante universelle
d'une alg\`{e}bre de Lie poss\`{e}de une structure naturelle
d'alg\`{e}bre de Hopf. Dans \cite{Guichardet} (p.\ 27)
appara\^{i}t la dualit\'{e} entre les groupes alg\'{e}briques
affines et les alg\`{e}bres de Hopf commutatives de type fini, le
cas semi-simple de dimension finie correspondant aux groupes
finis. Le lien avec les cat\'{e}gories tress\'{e}es et les
familles d'arbres est essentiel \cite{Moore} \cite{Larson}. Dans
\cite{Chari} (p.\ 4-5) on indique aussi comment la cat\'{e}gorie
des groupes quantiques devrait \^{e}tre d\'{e}finie comme duale
(c'est-\`{a}-dire anti-\'{e}quivalente) \`{a} celle des
alg\`{e}bres de Hopf.\ Pour d'autres \cite{Majid} les groupes
quantiques ne sont autres que les alg\`{e}bres de Hopf, ce qui ne
satisfait pas l'auteur du pr\'{e}sent texte. Comme il est fait de
fa\c {c}on explicite une relation avec la pr\'{e}sentation
hamiltonienne de la m\'{e}canique et de sa quantification depuis
les travaux de l'\'{e}cole de L.\ D.\ Fadeev \cite{Fadeev}, on est
conduit naturellement \`{a} l'id\'{e}e de comparer les
vari\'{e}t\'{e}s ab\'{e}liennes aux groupes quantiques.
L'introduction de \cite{Chari} rappelle comment se sont
d\'{e}velopp\'{e}s ces travaux de m\'{e}canique \cite{Moyal} pour
d\'{e}boucher sur les travaux de A.\ Connes (\cite{Connes},
\cite{Connes2}) avec lesquels il y a donc une dualit\'{e}
profonde. Dans la derni\`{e}re formule donn\'{e}e l'exponentielle
permet de passer d'un groupe $K_{2k}(\mathcal{M})$ \`{a} un espace
$\mathbf{M}_k(\Gamma )$ dont la dimension est connue
(\cite{Milne1} p.45). La somme de gauche correspond au passage
\`{a} la limite d'une somme de groupes $\mathbf{M}_k(\Gamma )$
pour construire l'alg\`{e}bre gradu\'{e}e $\mathbf{M}(\Gamma )$.
Celle de droite correspond \`{a} une construction particuli\`{e}re
restant \`{a} formaliser de fa\c {c}on pr\'{e}cise (un espace
classifiant). Les groupes $K_{2k}(\mathcal{M})$ sont dans cette
perspective comparables \`{a} des groupes de cohomologie $H^{*}(\mathcal{M},%
\mathbb{Z})$ et donc \`{a} $\mathbf{M}(\Gamma )$. Les conjectures
de Lichtenbaum qui se positionnent dans cette perspective
(\cite{Soule} p.\ 107) s'\'{e}crivent alors avec $k$ pair
\[
\frac{CardK_{2k-2}(\mathcal{M})}{CardK_{2k-1}(\mathcal{M})}=\frac{\mathbf{b}%
_k}k2^r.
\]

\subsection{L'automorphie de la fontion \^{e}ta li\'{e}e au nombre d'or}

L'automorphie de $\eta $ est la propri\'{e}t\'{e}
caract\'{e}ristique de cette fonction \cite{Toyoizumi} qui donne
naissance \`{a} la somme de Dedekind $s$, et qui a comme
cons\'{e}quence l'existence de la th\'{e}orie d\'{e}velopp\'{e}e
dans les chapitres pr\'{e}c\'{e}dents. Cette remarque conduit
\`{a} l'id\'{e}e de regarder dans les travaux qui sont relatifs
\`{a} l'op\'{e}rateur de Lagrange-Beltrami ou dans ceux sur les
repr\'{e}sentations unitaires de dimension infinie des
alg\`{e}bres de Lie comme $SL(2,\mathbb{R})$ o\`{u} l'on pourrait
utiliser les r\'{e}sultats qui ont \'{e}t\'{e} d\'{e}velopp\'{e}s
autour des g\'{e}n\'{e}ralisations de l'\'{e}quation de Markoff.
On trouve dans \cite{Kac} (p.270) mention d'un
r\'{e}sultat qui \'{e}voque nos travaux. Soit $\theta _a(S)=[0,\underline{%
S^{*},a}]$ un alg\'{e}brique de degr\'{e} 2 tel que
$S=(a_0,a_1,...,a_n)$ est une suite telle que $S=S^{*}$. On
consid\`{e}re
\[
f_c(\tau )=q^c\prod_{j=1}^{j=\infty }(1-q^j)^{a_{j-1}}\text{ o\`{u} }q=\exp
(2\pi i\tau ),
\]
Cette expression d\'{e}finit une fonction modulaire au sens de \cite{Kac}
(p. 257) pour un groupe $\Gamma (n)$ si et seulement si on a
\[
c=\frac{(n+2)(a+\sum_{j=0}^na_j)}{24}-\frac
1{4(n+2)}\sum_{j=1}^{n+1}j(n+2-j)a_{j-1}.
\]
Avec le nombre d'or $\theta _1(S)=[0,\underline{1}]$ qui donne $n=0$, la
valeur $c$ que l'on obtient est $c=(1/24)$. On retrouve ainsi la fonction $%
\eta $ de Dedekind. Le lien avec le pentagone que traduit ce
dernier cas appara\^{i}t aussi dans l'identit\'{e} pentagonale
d'Euler (\cite{Euler1} 1748)
cit\'{e}e dans \cite{Moll} p. 143 ou \cite{Kac1} ch. 12, d\'{e}composant $%
\eta $ en s\'{e}rie de Fourier et permettant son interpr\'{e}tation comme
inverse d'une fonction de partition d'un ensemble d'oscillateurs
ind\'{e}pendants de fr\'{e}quences multiples d'une fr\'{e}quence de base :
\[
\sum_{n\in
\mathbb{Z}}(-1)^nq^{\frac{n(3n+1)}2}=\prod_{j=1}^{j=\infty
}(1-q^j).
\]
On peut \'{e}galement pr\'{e}ciser le lien avec le pavage de
Penrose (\cite {Connes} fig. II.3. p.\ 89) qui donne de son
c\^{o}t\'{e} avec la construction de Vaughan Jones une
$C^{*}$-alg\`{e}bre canonique et pour premier indice non entier
d'un facteur de type II$_1$ le nombre d'or (\cite {Connes} p.\
507-508, \cite{Connes4}). La d\'{e}monstration m\^{e}me de ce
dernier r\'{e}sultat montre bien le lien qui existe avec les
fonctions modulaires et les surfaces de Riemann et les noeuds.
Remarquons que la formule donn\'{e}e pour $f_c$ d\'{e}bouche plus
g\'{e}n\'{e}ralement sur la d\'{e}finition de fonctions modulaires
donn\'{e}es par des produits de fonctions $\eta $, ce qui
physiquement correspond \`{a} des ensembles d'oscillateurs
ind\'{e}pendants.\ Pour $n\in \{2,3,4,6,12\}$ on trouve dans
\cite{Shimura} (p.\ 49) de telles expressions pour les surfaces
$X(n)$, tout comme dans \cite{Ligozat} pour les surfaces $X_0(n)$
de genre 1. Il y a l\`{a} un sujet \`{a} creuser pour lequel on
donne quelques r\'{e}f\'{e}rences \cite{Cox} \cite{Kondo}
\cite{MacDonald} \cite {Voskresenskaya} \cite{Saito2}
\cite{Robins} \cite{Okstate} \cite{Martin} \cite{Meyer}
\cite{Robins} \cite{Ligozat} \cite{Hiramatsu} \cite{Mackey} (p.\
366).

\subsection{Lien avec des espaces topologiques plus g\'{e}n\'{e}raux}

Le lien avec les espaces lenticulaires, qui sont eux-m\^{e}mes
li\'{e}s \`{a} la loi de r\'{e}ciprocit\'{e} quadratique
(\cite{Bredon} p. 365 \cite {Sossinsky1} p.\ 108) et plus
g\'{e}n\'{e}ralement \`{a} l'invariant $\eta $ des formes
d'espaces sph\'{e}riques, est approfondi dans \cite{Luck} \cite
{Gilkey} \cite{Gilkey2} \cite{Hilsum}. Ceci donne tout un ensemble
de d\'{e}veloppements d\'{e}bouchant sur des sujets comme la
$K$-th\'{e}orie \'{e}quivariante, les complexes de Koskul,
...\cite{Soergel}. L'invariant \^{e}ta de Dedekind que l'on a
utilis\'{e} pour nos travaux admet en r\'{e}alit\'{e} une
g\'{e}n\'{e}ralisation profonde qui a \'{e}t\'{e} mise en
lumi\`{e}re avec les travaux d'Atiyah, Patodi et Singer vers 1975.
On trouve dans \cite{Muller} une synth\`{e}se sur ce sujet faite
il y a une dizaine d'ann\'{e}es qui met bien en \'{e}vidence le
r\^{o}le des points c\^{o}ne et des bords de surface (la
propagation de la chaleur est perturb\'{e}e par les bords et les
points c\^{o}nes).\ Un lien explicite est
fait avec les travaux de F.\ Hirzebuch (\cite{Hirzebruch}, \cite{Hirzebruch1}%
) qui mettent eux-m\^{e}mes l'accent sur le lien entre
singularit\'{e}s et fractions continues (\cite{Laufer} ch.II,
\cite{Oka} p.95). L'invariant \^{e}ta joue le r\^{o}le d'un
polyn\^{o}me cyclotomique infini, laissant imaginer qu'un nouveau
''Jugendtraum'' plus vaste peut \^{e}tre \'{e}nonc\'{e}, li\'{e}
aux vari\'{e}t\'{e}s ab\'{e}liennes et \`{a} des invariants
combinatoires \`{a} pr\'{e}ciser (\cite{Gabriel} \cite{Hsu}),
\`{a} la g\'{e}om\'{e}trie non commutative \cite{Manin2}, voire
\`{a} une th\'{e}orie du corps de classe non commutative
\cite{Ihara} \cite{Ihara1}. Derri\`{e}re ces sujets se trouvent la
description des singularit\'{e}s isol\'{e}es des surfaces et la
correspondance de McKay \cite{Kapranov} \cite {Ito} \cite{Milnor3}
\cite{Yau1} \cite{Van de Ven1} (p.\ 72-89) \cite{Dimca}
\cite{Lamotke} pour la r\'{e}solution par les courbes
exceptionnelles et les singularit\'{e}s rationnelles A-D-E, la
dualit\'{e} \'{e}trange d'Arnold et la formule de Verlinde, les
diagrammes de Dynkin \cite{Gelfand2} \cite {Draxler}
\cite{Gabriel} \cite{Ponomarev}, les formes quadratiques \cite
{Ebeling1} \cite{Dolgachev1} \cite{Minac}, les noeuds et leur
monodromie \cite{Lines} \cite{Vershinin} \cite{Yosida2}
\cite{Zieschang2}, les modules de Verma et les syst\`{e}mes de
poids \cite{Saito3} \cite{Saito2} \cite {Martin}, la th\'{e}orie
de Galois diff\'{e}rentielle \cite{Gray} \cite{Put} \cite{Kac1}
\cite{Bertrand}, la th\'{e}orie de la repr\'{e}sentation des
alg\`{e}bres de dimension infinie et les cons\'{e}quences qu'elle
a pour l'\'{e}tude de fonctions sp\'{e}ciales utiles \`{a} la
physique \cite{Cahn} \cite{Dyson} \cite{Kac} \cite{Opdam}
\cite{VanAssche} \cite{Varchenko}, les lois de r\'{e}ciprocit\'{e}
plus g\'{e}n\'{e}rales \cite{Fukuhara} \cite {Fukuhara1}
\cite{Fukuhara2} \cite{Brylinski1} \cite{Diaz} \cite{Halbritter}
\cite{Hida} \cite{Iyanaga} \cite{Hiramatsu} \cite{Berg}, une
th\'{e}orie non commutative du corps de classe \'{e}troitement
li\'{e}e \`{a} la cohomologie \cite{Ihara1} \cite{Iyanaga} et
\`{a} la conjecture de Riemann \cite {Beilinson}.

\section{Une perspective globale en guise de conclusion}

On a d\'{e}crit dans ce qui pr\'{e}c\`{e}de plusieurs pistes de
g\'{e}n\'{e}ralisation de la th\'{e}orie de Markoff :

$\bullet $ Par le calcul des fractions continues, on a mis en
\'{e}vidence des \'{e}quations diophantiennes
$M^{s_1s_2}(b,\partial K,u)$ plus g\'{e}n\'{e}rales que
l'\'{e}quation classique de Markoff $M^{++}(2,0,0)$. On a
montr\'{e} comment les r\'{e}soudre, ainsi que le lien avec le
groupe du triangle et $GL(2,\mathbb{Z})$ qui le contient.

$\bullet $ Par l'\'{e}tude g\'{e}om\'{e}trique des tores perc\'{e}s, on a
montr\'{e} que l'\'{e}quation de Markoff $M^{++}(2,0,0)$ permet la
description de tous les tores perc\'{e}s paraboliques. On a \'{e}galement
montr\'{e} que nos \'{e}quations $M^{s_1s_2}(b,\partial K,u)$ apparaissent
dans l'\'{e}tude g\'{e}n\'{e}rale des tores perc\'{e}s et ont un lien avec
des pinceaux de coniques et un groupe libre \`{a} deux g\'{e}n\'{e}rateurs
qui existe dans ce contexte. On a \'{e}galement trouv\'{e} dans ce contexte
d'autres \'{e}quations permettant la description de tous les tores
perc\'{e}s hyperboliques.

$\bullet $ En se limitant aux surfaces de Riemann dont le
rev\^{e}tement conforme est le demi-plan de Poincar\'{e}, on a
montr\'{e} qu'une g\'{e}n\'{e}ralisation naturelle de la
th\'{e}orie de Markoff est la th\'{e}orie de Teichm\"{u}ller.\
Ceci a permis de faire le lien avec des \'{e}quations
diophantiennes plus g\'{e}n\'{e}rales ayant des
caract\'{e}ristiques analogues \`{a} celle de Markoff, et
\'{e}ventuellement plus de variables. On a identifi\'{e} un cadre
plus g\'{e}n\'{e}ral, celui des domaines de Riemann, o\`{u} des
r\'{e}sultats plus g\'{e}n\'{e}raux existent. L'\'{e}quation que
l'on consid\`{e}re appara\^{i}t dans ce contexte comme liant les
caract\`{e}res de la repr\'{e}sentation du groupe de Poincar\'{e}
que l'on consid\`{e}re.

\[
\]

Le pr\'{e}sent chapitre a explor\'{e} ce qui concerne les surfaces de
Riemann, et on y a int\'{e}gr\'{e} dans chaque paragraphe diff\'{e}rentes
perspectives pour des travaux futurs sur lesquelles on ne revient pas ici.
Certains sujets importants ont \'{e}t\'{e} laiss\'{e}s de c\^{o}t\'{e} que
l'on mentionne pour m\'{e}moire :

$\bullet $ L'analyse harmonique non commutative \cite{Gross} et tous ses
d\'{e}veloppements obtenus en consid\'{e}rant les mouvements d\'{e}crits par
des points sur des courbes d'une surface de Riemann. Cette th\'{e}orie
diff\`{e}re de l'analyse harmonique commutative d\'{e}velopp\'{e}e sur la
surface de Riemann dans l'esprit de \cite{Terras} (chapitre 3). Dans
diff\'{e}rents cas, ce mouvement peut \^{e}tre d\'{e}compos\'{e} selon des
mouvements sur des g\'{e}od\'{e}siques correspondant aux g\'{e}n\'{e}rateurs
du groupe de Poincar\'{e} de la surface. Une telle approche peut mener \`{a}
des \'{e}quations diff\'{e}rentielles dont on a laiss\'{e} de c\^{o}t\'{e}
l'\'{e}tude dans ce qui pr\'{e}c\`{e}de.\ Sur les tores perc\'{e}s on
renvoie \`{a} \cite{Cherry} qui s'est inspir\'{e} des travaux originaux de
Poincar\'{e} pour d\'{e}crire les \'{e}quations possibles et \`{a} \cite
{Gray} pour l'approfondissement de ce sujet qui a conduit aux th\'{e}ories
de Picard-Vessiot et Drach ainsi qu'\`{a} une th\'{e}orie de Galois
sp\'{e}cifique.

$\bullet $ Le lien avec la th\'{e}orie des tresses et des noeuds a
\'{e}t\'{e} \`{a} plusieurs reprises \'{e}voqu\'{e}.\ La relation avec les
d\'{e}veloppements qui pr\'{e}c\`{e}dent est assur\'{e}e par une
construction d'Ivanov \cite{Ivanov1}.\ Soit $\mathcal{M}$ une surface de
Riemann poss\'{e}dant un nombre fini de trous. En collant des disques
ferm\'{e}s sur tous les trous de $\mathcal{M}$, on fabrique une surface
compacte $\mathcal{N}$. Les diff\'{e}omorphismes $\mathcal{M}\rightarrow
\mathcal{M}$ donnent des diff\'{e}omorphismes $\mathcal{N}%
\rightarrow \mathcal{N}$, d'o\`{u} un homomorphisme canonique surjectif de $%
\Gamma _{\mathcal{M}}$ dans $\Gamma _{\mathcal{N}}$. Son noyau est le groupe
des tresses $B_n(\mathcal{N})$, o\`{u} $n$ est le nombre de trous de la
surface $\mathcal{M}$. Ceci permet d'expliciter le lien avec l'\'{e}tude des
noeuds rationnels, les ''rational tangles'' de Conway (\cite{Murasugi} ch.9,
\cite{Kauffman2}) li\'{e}s aux fractions continues et qui sont utilis\'{e}s
dans certaines applications \`{a} la recombinaison des enzymes et de l'ADN
\cite{Sumners} \cite{Ernst} \cite{Dessalles} \cite{Kari} \cite{Carbone} \cite
{Salomaa}.

$\bullet $ La th\'{e}orie des dessins d'enfants \cite{Belyi} \cite
{Grothendieck} \cite{Jones} \cite{Luo1} \cite{Waldschmidt2} (p.\
99) a \'{e}t\'{e} tr\`{e}s peu \'{e}voqu\'{e}e. Son
d\'{e}veloppement en dimension sup\'{e}rieure est envisageable.
Son analogie avec diff\'{e}rents travaux d'astronomes sur la forme
cristallis\'{e}e du vide quantique est \'{e}clairante
\cite{Lehoucq} \cite{Thurston1}. Plus g\'{e}n\'{e}ralement
d'ailleurs tous les d\'{e}veloppements qui ont \'{e}t\'{e}
pr\'{e}sent\'{e}s autour des surfaces de Riemann permettent de
comprendre des travaux contemporains de physique qui leur donnent
une nouvelle importance pour les applications \cite{Mineev}
\cite{Davies}.\ On a \'{e}voqu\'{e} le lien avec les solitons
\cite{Moll} (ex. 2, p.\ 91) \cite{Belokolos} \cite{Gesztesy} pour
lesquelles on peut g\'{e}n\'{e}raliser la d\'{e}marche qui
pr\'{e}c\`{e}de. Mais l'invariant \^{e}ta semble poss\'{e}der dans
ce contexte une importance fondamentale, comme s'il \'{e}tait
li\'{e} \`{a} l'\'{e}nergie du vide quantique et \`{a} ses
infinies vibrations \'{e}l\'{e}mentaires, pourquoi pas au bruit en
$1/f$ sous-jacent au bruit de fonds de l'univers cr\'{e}\'{e} par
la singularit\'{e} du Big Bang rendant sa g\'{e}om\'{e}trie
hyperbolique ?
\[
\]

Les probl\`{e}mes que l'on a abord\'{e}s dans le pr\'{e}sent
chapitre concernent essentiellement la th\'{e}orie de
Teichm\"{u}ller sur les surfaces de Riemann et les fonctions
modulaires.\ On a cherch\'{e} \`{a} comprendre comment ils sont
li\'{e}s \`{a} des probl\`{e}mes non r\'{e}solus d'une grande
actualit\'{e}: l'hypoth\`{e}se de Riemann, la conjecture de
Poincar\'{e}, la conjecture de Hodge \cite{Lewis}, la conjecture
de Birch et Swinnerton-Dyer \cite{Wiles2}, l'explication du
d\'{e}faut de masse dans les \'{e}quations de Yang et Mills
(\cite{Nash2} (chapitre VIII) \cite{Nakahara} (chapitre 10)), etc.
C'est pour comprendre le contexte de ces sujets que notre approche
a \'{e}t\'{e} d\'{e}velopp\'{e}e, avec l'id\'{e}e de faire un lien
avec les m\'{e}thodes de l'analyse spectrale. Les relations avec
des espaces de Hilbert et des $C^{*}$-alg\`{e}bres
d'op\'{e}rateurs a \'{e}t\'{e} creus\'{e} m\^{e}me si on reste
loin du compte pour ce qui concerne la pr\'{e}sentation de
l'appareillage math\'{e}matique n\'{e}cessaire \cite {Wells}
\cite{Witten} \cite{Vladimirov}.\ La dimension 2 a \'{e}t\'{e}
privil\'{e}gi\'{e}e parce que l'on a travaill\'{e} essentiellement
sur les surfaces de Riemann.\ Or elle pr\'{e}sente des
diff\'{e}rences qualitatives tr\`{e}s importantes par rapport aux
dimensions sup\'{e}rieures o\`{u} l'on a vu que l'on pouvait aussi
g\'{e}n\'{e}raliser la th\'{e}orie de Markoff.\ Par exemple le
lien donn\'{e} par le th\'{e}or\`{e}me de Dehn-Nielsen entre
hom\'{e}omorphisme et transformation conforme n'est plus si direct
dans les dimensions sup\'{e}rieures \`{a} 2. \
\[
\]

Au terme de ces r\'{e}flexions, ce qui para\^{i}t \`{a} l'auteur
le plus fascinant est le lien avec la nature du calcul
\cite{Feynman} \cite{Benioff} \cite{Shor} \cite{Penrose} et la
th\'{e}orie algorithmique de l'information. L'id\'{e}e qui se
d\'{e}veloppe aujourd'hui est que les calculateurs ont un
mod\`{e}le m\'{e}canique quantique et que ce dernier est le
d\'{e}veloppement naturel du calcul classique, de la m\^{e}me fa\c
{c}on que la m\'{e}canique quantique succ\`{e}de \`{a} la
m\'{e}canique classique. Comme si l'analogie ch\`{e}re \`{a} Weil,
qui a \'{e}t\'{e} cit\'{e}e \`{a} plusieurs reprises \cite{Weil0}
\cite{Deninger}, d\'{e}bouchait sur une interaction beaucoup plus
profonde que l'on pourrait d\'{e}signer par le vocable de
quantification de la logique, d'ailleurs entrevue par John von
Neumann \cite{Birkhoff} et bien d\'{e}crite dans \cite{Weaver}.\
Il reste largement \`{a} formaliser cette analogie que Rolf Berndt
r\'{e}sume dans son panorama des travaux de E.\ K\"{a}hler par les
correspondances suivantes \cite{Berndt}
\[
\text{anneau }\rightarrow \text{ objet,}
\]
\[
\text{homomorphisme }\rightarrow \text{ perception,}
\]
\[
\text{id\'{e}al }\rightarrow \text{ perspective,}
\]
\[
\text{corps de Galois }\rightarrow \text{ oscillateur.}
\]
La derni\`{e}re correspondance avec les oscillateurs peut
surprendre, mais elle a \'{e}t\'{e} entrevue dans ce qui
pr\'{e}c\`{e}de et est clairement apparente dans diff\'{e}rents
travaux tels que \cite{Bismut1} \cite{Sierra} \cite{Prasad}
\cite{Borel} \cite{Rallis} \cite{Przebinda} \cite{Mounier}.\ Elle
permet d'envisager une interpr\'{e}tation quantique de
l'arithm\'{e}tique, le nombre $1$ \'{e}tant repr\'{e}sentable
comme un oscillateur de fr\'{e}quence
$\nu $, le nombre $2$ correspondant \`{a} un oscillateur de fr\'{e}quence $%
2\nu $, et ainsi de suite... On pourrait ainsi comparer la relation
d'incertitude de Heisenberg au r\'{e}sultat bien connu
d'ind\'{e}cidabilit\'{e} de G\"{o}del, et imaginer que les arbres
constituent un moyen privil\'{e}gi\'{e} de concentration de l'information
qui n'est pas ind\'{e}pendant de ces questions. Le dixi\`{e}me probl\`{e}me
de Hilbert pourrait lui-m\^{e}me induire une explication comparable \cite
{Matiiassevitch} (ch.3-4).\

L'analogie de Weil pourrait quant \`{a} elle d\'{e}boucher sur une
compr\'{e}hension plus profonde du codage quantique de l'information \cite
{Benioff} \cite{Benioff1} \cite{Shor} \cite{Delahaye1} \cite{Preskill} \cite
{NielsenM}. Dans le domaine du calcul algorithmique, la quantification est
en effet d\'{e}sormais \`{a} l'oeuvre \cite{Feynman}, comme sont \`{a}
l'oeuvre les solitons dans la transmission \`{a} distance de l'information
et le traitement optique dans certains \'{e}quipements exp\'{e}rimentaux qui
seront utilis\'{e}s dans l'Internet du futur.\ Dans le domaine de la
repr\'{e}sentation, les surfaces de Riemann interviennent dans la
th\'{e}orie de la vision des objets \cite{Sochen} \cite{Schmitter} et de
processus non lin\'{e}aires \cite{Planat} (p. 304).\ La caract\'{e}ristique
d'Euler-Poincar\'{e} et les anneaux de Grothendieck apparaissent dans les
structures alg\'{e}briques les plus g\'{e}n\'{e}rales et les ensembles
d\'{e}finissables \cite{Krajicek}, laissant imaginer la possibilit\'{e}
d'associer fonctoriellement \`{a} chaque objet ainsi structur\'{e} une
surface de Riemann.\ Les limites techniques ressemblent \`{a} celles, plus
fondamentales, qui viennent d'\^{e}tre \'{e}voqu\'{e}es \cite{Lloyd} et qui
ont une r\'{e}sonance dans l'impossibilit\'{e} de pr\'{e}voir le mouvement
de certains syst\`{e}mes m\'{e}caniques \cite{Moore1} \cite{Penrose}
(p.202). Faut-il interpr\'{e}ter l'incertitude de Heisenberg comme une
limite algorithmique impos\'{e}e par les moyens logico-math\'{e}matiques que
nous utilisons pour penser la physique?\ En tout cas le calcul int\'{e}gral
lui-m\^{e}me a des limites qui ont une importance dans ces questions de
calculabilit\'{e} et impactent les r\'{e}sultats de la m\'{e}canique
m\^{e}me \cite{Matiiassevitch} (p. 193), sachant qu'il est concevable sans
d\'{e}pense d'\'{e}nergie et sans accroissement d'entropie physique \cite
{Delahaye1} (p.\ 27). Il y a l\`{a} tout une perspective globale de
r\'{e}flexions concernant la nature informationnelle et vivante de la
math\'{e}matique que l'auteur voudrait approfondir en examinant de plus
pr\`{e}s l'intuition que math\'{e}matique et th\'{e}orie de l'information
sont une seule et m\^{e}me chose.

\[
\]

On conclut sur une pens\'{e}e d'Alexandre Grothendieck qui est
exprim\'{e}e dans son Esquisse d'un Programme.\ Elle r\'{e}sume
\`{a} elle seule la fa\c {c}on dont l'auteur du pr\'{e}sent texte
con\c {c}oit sa propre d\'{e}marche de recherche :

\begin{eqnarray*}
&&\text{''...la d\'{e}marche de la pens\'{e}e qui sonde et qui d\'{e}couvre,
} \\
&&\text{en tat\^{o}nnant dans la p\'{e}nombre bien souvent, avec des
trou\'{e}es de lumi\`{e}re } \\
&&\text{subite quand quelque tenace image fausse, ou simplement
inad\'{e}quate, } \\
&&\text{se trouve enfin d\'{e}busqu\'{e}e et mise \`{a} jour,} \\
&&\text{et que les choses qui paraissaient de guingois se mettent en place,}
\\
&&\text{dans l'harmonie mutuelle qui leur est propre.''}
\end{eqnarray*}
\[
\]

\ \ \ \ \ \ \ \ \ \ \ \ \ \ \ \ \ \ \ \ \ \ \ \ \ \ \ \ \ \ \ \ \ \ \ \ \ \
\ \ \ \ \ \ \ \ \ \ \ \ Metz, f\'{e}vrier 2003.
\[
\]

\tableofcontents
\end{document}